\documentclass[reqno,oneside,10pt]{amsart}

\usepackage{url,amsmath,amscd,amsfonts,amsthm,graphicx,marginnote,amssymb,bm,mathtools,MnSymbol}

\usepackage[utf8]{inputenc}
\usepackage{braket}
\usepackage{color}
\usepackage{thm-restate}

\theoremstyle{plain}
\newtheorem{thm}{Theorem}[section]
\newtheorem{cor}[thm]{Corollary}
\newtheorem{lem}[thm]{Lemma}
\newtheorem{prop}[thm]{Proposition}

\newtheorem{ques}[thm]{Question}
\newtheorem{theoremin}{Theorem}
\newtheorem{propin}[theoremin]{Proposition}
\newtheorem{definin}[theoremin]{Definition}
\theoremstyle{definition}
\newtheorem{defin}[thm]{Definition}
\newtheorem{rem}[thm]{Remark}
\newtheorem{exa}[thm]{Example}

\UseRawInputEncoding

\numberwithin{equation}{section}

\let\frontmatter\relax
\def\mainmatter{\def\baselinestretch{1}\normalfont \setlength{\parskip}{0.5em}}

\usepackage[margin=1.3in]{geometry}

\usepackage{etoolbox}

\makeatletter
\def\l@subsection{\@tocline{2}{0pt}{2.5pc}{5pc}{}}

\renewcommand{\tocsection}[3]{%
  \indentlabel{\@ifnotempty{#2}{\bfseries\ignorespaces#1 #2.~}}\bfseries#3}
\renewcommand{\tocsubsection}[3]{%
  \indentlabel{\@ifnotempty{#2}{\ignorespaces#1 #2.~}}#3}

\makeatother

\makeatletter
\renewcommand{\section}{\@startsection
{section}
{1}
{\z@}
{-\baselineskip}
{0.8\baselineskip}
{\centering\scshape\large}} 

\renewcommand{\subsection}{\@startsection
{subsection}
{2}
{\z@}
{-0.8\baselineskip}
{0.5\baselineskip}
{\normalfont \bf \normalsize}} 

\renewcommand{\subsubsection}{\@startsection
{subsubsection}
{3}
{\z@}
{-0.8\baselineskip}
{0.5\baselineskip}
{\normalfont \it \normalsize}} 

\makeatother

\let\emptyset\varnothing
\DeclareMathOperator*{\Res}{Res}

\usepackage{mathtools}
\newcommand\norder[1]{{\vcentcolon}\!\mathrel{#1}\!{\vcentcolon}}

\newcommand{\ii}{\mathbf{i}}
\newcommand{\bea}{\begin{eqnarray}}
\newcommand{\eea}{\end{eqnarray}}
\newcommand{\beq}{\begin{equation}}
\newcommand{\eeq}{\end{equation}}
\newcommand{\dd}{\mathrm{d}}

\usepackage[plainpages=false]{hyperref}

\definecolor{darkgreen}{rgb}{0.1, 0.8, 0.1}

\begin{document}
\frontmatter

\title[Higher Airy structures, $\mathcal{W}$ algebras and topological recursion]{Higher Airy structures, $\mathcal{W}$ algebras \\ and topological recursion}

\author{Ga\"etan Borot}
\address{Max Planck Institut f\"ur Mathematik, Vivatsgasse 7, 53111 Bonn, Germany.}
\address{Humboldt-Universit\"at zu Berlin, Institut f\"ur Mathematik \& Institut f\"ur Physik, Rudower Chaussee 25, 12489 Berlin, Germany}
\email{gaetan.borot@hu-berlin.de}

\author{Vincent Bouchard}
\address{Department of Mathematical \& Statistical Sciences,
University of Alberta, 632 CAB\\
Edmonton, Alberta, Canada T6G 2G1}
\email{vincent.bouchard@ualberta.ca}

\author{Nitin K. Chidambaram}
\address{Department of Mathematical \& Statistical Sciences,
University of Alberta, 632 CAB\\
Edmonton, Alberta, Canada T6G 2G1}
\address{ Max Planck Institut f\"ur Mathematik, Vivatsgasse 7, 53111 Bonn, Germany.}
\email{kcnitin@mpim-bonn.mpg.de}

\author{Thomas Creutzig}
\address{Department of Mathematical \& Statistical Sciences,
University of Alberta, 632 CAB \\
Edmonton, Alberta, Canada T6G 2G1}
\email{creutzig@ualberta.ca}

\author{Dmitry Noshchenko}
\address{Faculty of Physics, University of Warsaw \\ ul. Pasteura 5, 02-093 Warsaw, Poland}
\email{dsnoshchenko@gmail.com}

\subjclass[2010]{81R10, 14N10, 51P05}

\keywords{$\mathcal{W}$ algebras, topological recursion, Airy structures, enumerative geometry}

\begin{abstract} 
We define higher quantum Airy structures as generalizations of the Kontsevich--Soibelman quantum Airy structures by allowing differential operators of arbitrary order (instead of only quadratic). We construct many classes of examples of higher quantum Airy structures as modules of $\mathcal{W}(\mathfrak{g})$ algebras at self-dual level, with $\mathfrak{g}= \mathfrak{gl}_{N+1}$, $\mathfrak{so}_{2 N }$ or $\mathfrak{e}_N$. 
We discuss their enumerative geometric meaning in the context of (open and closed) intersection theory of the moduli space of curves and its variants. Some of these $\mathcal{W}$ constraints have already appeared in the literature, but we find many new ones. For $\mathfrak{gl}_{N+1}$ our result hinges on the description of previously unnoticed Lie subalgebras of the algebra of modes. As a consequence, we obtain a simple characterization of the spectral curves (with arbitrary ramification) for which the Bouchard--Eynard topological recursion gives symmetric $\omega_{g,n}$s and is thus well defined. For all such cases, we show that the topological recursion  is equivalent to $\mathcal{W}(\mathfrak{gl})$ constraints realized as higher quantum Airy structures, and obtain a Givental-like decomposition for the corresponding partition functions.
\end{abstract}

\maketitle

\newpage

\tableofcontents

\mainmatter

\newpage
\section{Introduction}

\subsection{Motivation}

Virasoro constraints are ubiquitous in enumerative geometry. The general statement goes as follows. Given a particular enumerative geometric context, such as intersection theory on the moduli spaces of curves, or Gromov--Witten theory of a given target space, an interesting object of study is the generating series $F$ for connected descendant invariants and the corresponding generating series $Z = e^F$ for disconnected invariants. The statement of Virasoro constraints is that $Z$ satisfies a collection of differential equations of the form $H_kZ = 0$, where the $H_k$s are differential operators (in the formal coordinates of the generating series $F$) that form a representation of a subalgebra of the Virasoro algebra. In this context, the starting point is a given enumerative theory, and the goal is to show that the generating series $Z$ satisfies Virasoro constraints.

An interesting question is whether the sequence of events can be reversed. Can ``Virasoro-like constraints'' be formulated abstractly such that there always exists a unique solution to the collection of differential equations, in the form of the exponential of a generating series? One may understand the recent of work of Kontsevich and Soibelman \cite{KS} (see also \cite{ABCD}) as providing an answer to this question, in the form of ``quantum Airy structures''. 

Let $V$ be a vector space of dimension $D$ (which may be countably infinite) over $\mathbb{C}$. Using the notation $I = \{1, \ldots, D\}$, let $(x_i)_{i\in I}$ be linear coordinates on $V^*$, and denote by
$$
\mathcal{D}_{T^*V}^{\hbar} \cong \mathbb{C}[\![\hbar, (x_l)_{l \in I},(\hbar \partial_{x_l})_{l \in I}]\!]
$$
the completed algebra of differential operators on $V$. We introduce a grading on $\mathcal{D}_{T^*V}^{\hbar}$ by assigning:
$$
\deg x_l = \deg \hbar \partial_{x_l}  = 1\, ,\qquad \deg \hbar = 2\,.
$$
Then a quantum Airy structure is a collection of differential operators $(H_k)_{k \in I}$ of the form
\begin{equation}\label{eq:form}
H_k = \hbar \partial_{x_k} - P_k\, ,
\end{equation}
where $P_k \in \mathcal{D}_{T^*V}^{\hbar}$ is homogeneous of degree $2$, such that the $H_k$ generate a graded Lie subalgebra of  $\mathcal{D}_{T^*V}^{\hbar}$. That is, there exists scalars $c_{k,l}^m$ such that
\begin{equation}\label{eq:sub}
[H_k, H_l ] = \hbar \sum_{m \in I} c_{k,l}^m H_m\, .
\end{equation}
The crucial theorem proved in \cite{KS} is that for any quantum Airy structure, there exists a unique solution $Z$ to the collection of differential constraints $H_k Z = 0$, $k \in I$, of the following form:
\begin{equation}\label{eq:sol}
Z = \exp\left(\sum_{\substack{g \geq 0,\,\,n \geq 1 \\ 2g - 2 + n > 0}} \frac{\hbar^{g-1}}{n!}   \sum_{\alpha \in I^n} F_{g,n}[\alpha] x_{\alpha_1} \cdots\, x_{\alpha_n}\right)\, .
\end{equation}
It does not say what kind of enumerative invariants the coefficients $F_{g,n}[\alpha]$ are; this depends on the choice of quantum Airy structure. But the existence and uniqueness of a solution to the differential constraints is guaranteed.

There are two key features in the definition of quantum Airy structures that are responsible for existence and uniqueness of a solution. The first one is the particular form of the differential operators $H_i$, which implies that the differential constraints $H_i Z = 0$ translate into a recursive system for the coefficients $F_{g,n}[\alpha]$. The second is the subalgebra property, which, together with the form of the operators, ensures the existence of a solution.

While quantum Airy structures may be understood as an abstract construction of Virasoro-like constraints, they were first introduced in \cite{KS} as generalizations of the topological recursion of  Chekhov, Eynard and Orantin \cite{EO,EORev}. The Chekhov--Eynard--Orantin topological recursion appears rather different from quantum Airy structures or Virasoro constraints \emph{a priori}. It starts from the geometry of a spectral curve $\mathcal{C}$, and constructs an infinite sequence of meromorphic symmetric differentials on $\mathcal{C}^n$ through a period computation. But it turns out that for any admissible spectral curve with simple ramification, the Chekhov--Eynard--Orantin topological recursion can be recast into a quantum Airy structure. In other words, its recursive structure is equivalent to the collection of differential constraints of a quantum Airy structure.

Thus, quantum Airy structures provide a clear conceptual framework behind the Chekhov--Eynard--Orantin topological recursion, and a generalization thereof. In particular, it clarifies the relations between topological recursion, symplectic geometry and deformation quantization. It also incorporates earlier observations of Kazarian about the role of symplectic loop spaces and polarizations in the theory of \cite{EORev} (see also \cite{KZ}) and provides a simpler approach to the relation with the Givental group action and semi-simple cohomological field theories established in \cite{DBOSS}.

Natural generalizations of Virasoro constraints that appear in enumerative geometry are $\mathcal{W}$ constraints. They are known to be satisfied in some contexts, such as intersection theory on the moduli space of curves with $r$-spin structures, and certain Fan--Jarvis--Ruan theories. $\mathcal{W}$ constraints are similar in nature to Virasoro constraints. They consist of a collection of  differential constraints $H_i Z = 0$ for a generating series of disconnected invariants, but where the $H_i$s form a representation of a subalgebra of a $\mathcal{W}$ algebra. Recall that $\mathcal{W}$ algebras are non-linear extensions of the Virasoro algebra, which arise in conformal field theory when the theory contains chiral primary fields of conformal weight $>2$. $\mathcal{W}$ algebras always contain the Virasoro algebra as a subalgebra.

In this paper we provide an answer to the question: Can ``$\mathcal{W}$-like constraints'' be formulated abstractly such that there always exists a unique solution to the collection of differential equations, and that this solution has the form of an exponential of a generating series? 

The answer takes the form of ``higher quantum Airy structures''. We use the same conceptual framework as for quantum Airy structures, but we relax the two conditions on the differential operators. We consider differential operators $(H_k)_{k \in I}$ of the same form as in \eqref{eq:form}, but with $P_k \in \mathcal{D}_{T^*V}^{\hbar}$ a sum of terms of degree $\geq 2$. The subalgebra condition is replaced by the requirement that the left $\mathcal{D}_{T^*V}^{\hbar}$-ideal generated by the $H_k$ is a graded Lie subalgebra of $\mathcal{D}_{T^*V}^{\hbar}$. Concretely, this means that \eqref{eq:sub} is replaced by:
$$
[H_k, H_l ] = \hbar \sum_{m \in I} g_{k,l}^m\,H_m\, , \qquad g_{k,l}^m \in \mathcal{D}_{T^*V}^{\hbar}\, .
$$
Under these conditions, Kontsevich and Soibelman (in \cite{KS}) already proved the existence and uniqueness of a solution to the collection of differential constraints $H_k Z = 0$ of the same form as \eqref{eq:sol}. The goal of this paper is to construct many examples of higher quantum Airy structures from $\mathcal{W}$ algebras and discuss their enumerative meaning.

\subsection{Main results}

Let us now describe the main results of the paper briefly. First, we construct various types of $\mathcal{W}$ constraints. Second, we show that the Bouchard--Eynard topological recursion of \cite{BE2,BE,BHLMR} is equivalent to a previously constructed class of $\mathcal{W}$ constraints. We also proved along the way a property about the modes of the $ \mathcal{W}(\mathfrak{gl}_{N+1}) $ algebra at the self-dual level, which was essential to our construction of $\mathcal{W}$ constraints.

\subsubsection{Higher quantum Airy structures from $\mathcal{W}$ constraints}

Our general recipe to produce higher quantum Airy structures from modules of $\mathcal{W}$ algebras goes as follows. The starting ingredients are a Lie algebra $\mathfrak{g}$ and an element $\sigma$ of the Weyl group of $\mathfrak{g}$. We then consider the principal $\mathcal W$-algebra of $\mathfrak{g}$ at the self-dual level $k=-h^\vee+1$ ($h^\vee$ is the dual Coxeter number of $\mathfrak g$). This vertex operator algebra is denoted by  $\mathcal{W}(\mathfrak{g})$ and we realize it
 as a subalgebra of the Heisenberg vertex operator algebra associated to the Cartan subalgebra $\mathfrak{h}$ of $\mathfrak{g}$. Then:
\begin{enumerate}
\item We construct a $\sigma$-twisted module $\mathcal{T}$ of the Heisenberg vertex operator algebra.
\item Upon restriction to the $\mathcal{W}(\mathfrak{g})$ algebra, we obtain an untwisted module. We realize the modes of the generators of the $\mathcal{W}(\mathfrak{g})$ algebra as differential operators acting on the space of formal series in countably many variables.
\item We pick a subset of modes generating a left ideal which is a graded Lie subalgebra of the algebra of modes. These modes fulfill the second (and hardest to check) condition to be a higher quantum Airy structure.
\item If possible, we conjugate these modes (dilaton shift) to bring them in the form of a higher quantum Airy structure.
\end{enumerate}

Remarkably, following this simple recipe we can construct a large variety of higher quantum Airy structures, including many that have interesting enumerative interpretations. Our general construction reproduces some of the $\mathcal{W}$ constraints that have already appeared in the literature, but most of the higher quantum Airy structures that we obtain are new.

We also note that our construction relies on certain explicit strong generators of the $\mathcal{W}$ algebras that are known in the literature. We discuss this in detail in Section~\ref{sec:examples}.

\noindent {\bf $\mathcal{W}(\mathfrak{gl}_{N+1})$ higher quantum Airy structures: first class.} For clarity let $r:=N+1$. Our first set of examples starts with $\mathfrak{g}=\mathfrak{gl}_{r}$ and $\sigma = (1\,2\,\,\cdots\,\, r)$ --- the Coxeter element of the Weyl group $\mathfrak{S}_{r} $. Theorem \ref{WHAS} is the main result of this construction, which can be summarized as follows:
\begin{theoremin}
\label{THT1} Let $r \geq 2$ and $s \in \{1,\ldots,r + 1\}$ be such that $r = r's - \epsilon$ with $\epsilon \in \{\pm 1\}$ and $r'$ integer. Let 
$$
\mathfrak{d}^i = i - 1 - \Big\lfloor \frac{s(i - 1)}{r} \Big\rfloor,\qquad \tilde{S}_{s} = \big\{(i,k)\quad \big|\quad i \in \{1,\ldots,r\}\quad {\rm and}\quad k \geq \mathfrak{d}^i + \delta_{i,1}\big\}\, .
$$
There exists an (explicit) quantum $r$-Airy structure on $V = \bigoplus_{l > 0} \mathbb{C}\langle x_{l} \rangle$ based on a representation of the subset of modes $(W_{k}^i)_{(i,k) \in \tilde{S}_{s}}$ of the $\mathcal{W}(\mathfrak{gl}_{r})$ algebra generators with central charge $r$ in $\mathcal{D}_{T^*V}^{\hbar}$. We use $Z_{(r,s)}$ to denote its partition function. Its coefficients for $2g - 2 + n = 1$ are
\[
F_{0,3}[l_1,l_2,l_3] = \epsilon r' l_1l_2l_3\,\delta_{l_1 + l_2 + l_3,s},\qquad F_{1,1}[l] = \frac{r^2 - 1}{24}\,\delta_{l,s}.
\]
\end{theoremin}

The case $s=r+1$ (for all $r$) was studied by Bakalov and Milanov \cite{BakalovMilanov1, BakalovMilanov2, Milanov}. $Z_{(r,r + 1)}$ is a generating series for intersection numbers on the moduli space of curves with $r$-spin structure, as explained in Section \ref{Srspin}.
Other choices of $s$ however are new. As explained in the proof of Theorem \ref{WHAS}, the condition that $s$ be coprime with $r$ arises for the dilaton shift to yield differential operators of the right form for a higher quantum Airy structure. The condition that $r = \pm 1 \text{ mod } s$ is necessary and sufficient for the left ideal generated by the subset of modes to be a graded Lie subalgebra (see Theorem~\ref{WTHUT2}). We will come back to this statement, and state the precise result in Section \ref{s:wmodes}.

The enumerative meaning of the cases corresponding to these general values of $ s $ is particularly intriguing. The partition function $Z_{(2,1)}$ corresponds to the Br\'ezin--Gross--Witten tau function of the KdV hierarchy \cite{BGROSS,GWitten}. Further, Norbury constructed in \cite{Norbclass} a cohomology class on the moduli space of curves such that the partition function $Z_{(2,1)}$ generates its descendant invariants. It is then natural to ask whether similar results exist for $r>2$, and for the various allowed values of $s$. It would be  interesting to find an enumerative interpretation for all $Z_{(r,s)}$ since they are the building blocks for the Givental-like decomposition proved in Theorem~\ref{TGIV} below for the Bouchard--Eynard topological recursion. For instance, we can ask: does $Z_{(r,1)}$ coincide with the $r$-Br\'ezin--Gross--Witten tau function? And if it does, what is the analog of the Norbury class such that $Z_{(r,1)}$ becomes the generating series of its descendant invariants? In Section \ref{s:BGW}, we explore these questions in greater detail.

Propositions \ref{t:condition} and \ref{t:conjugated} provide straightforward generalizations of the above construction, by allowing direct sums and conjugations of the quantum Airy structures of Theorem~\ref{THT1}. This easy observation will be necessary to compare the $ \mathcal{W} $ constraints with the Bouchard--Eynard topological recursion.

\noindent {\bf $\mathcal{W}(\mathfrak{gl}_{N+1})$ higher quantum Airy structures: second class.} For our second important class of examples, we keep $\mathfrak{g}=\mathfrak{gl}_{r}$, with $r=N+1$, but  replace the Coxeter element of the Weyl group by an arbitrary automorphism $\sigma$. Although part of our construction is general, we only complete the program in the case $\sigma = (1\,\cdots\,\,r - 1)$. Theorem \ref{t:ropen} realizes these modules as higher quantum Airy structures with half-integer powers of $\hbar$ (which we call ``crosscapped'') and can be summarized as follows.

\begin{theoremin}
	\label{TTH2} Let $r \geq 3$ and $s \in \{1,\ldots,r\}$ dividing $r$. Let
	$$
	\mathfrak{d}^i = i - 1 - \Big\lfloor \frac{s(i - 1)}{r - 1} \Big\rfloor\, ,\qquad \hat{S}_{s} = \big\{(i,k) \quad \big|\quad i \in \{1,\ldots,r\} \quad {\rm and}\quad k \geq \mathfrak{d}^i + \delta_{i,1} + \delta_{i,r}\big\}\,.
	$$
	There exists an (explicit) $1$-parameter family of crosscapped quantum $r$-Airy structures on $V = \bigoplus_{p > 0} \mathbb{C}\langle x_{p}^1 \rangle \oplus \mathbb{C}\langle x_{p}^2 \rangle$ based on a representation of the subset of modes $(W_k^i)_{(i,k) \in \hat{S}_{s}}$ of the $\mathcal{W}(\mathfrak{gl}_{r})$ algebra generators with central charge $r$ into $\mathcal{D}_{T^*V}^{\hbar^{1/2}}$.
\end{theoremin}
In Section~\ref{Sopen}, we speculate that the enumerative geometry interpretation of these quantum Airy structures lies in the open intersection theory developed by Pandharipande, Solomon and Tessler \cite{Tessleropen,SolomonTessler}. Indeed, for $(r,s) = (3,3)$ we can identify them with the $\mathcal{W}(\mathfrak{sl}_3)$ constraints derived by Alexandrov in \cite{Alexandrovopen} for the partition function of the open intersection theory on the moduli space of bordered Riemann surfaces. For higher $r$, do we recover the tau function of the extended $(r-1)$-KdV hierarchy constructed by Bertola and Yang \cite{BertolaYang}? Can it be understood in terms of the open $(r-1)$-spin intersection theory of \cite{Tessleropen2}?

It would be interesting to classify the automorphisms $\sigma$ that can lead to higher quantum Airy structures and the corresponding structures themselves, as we did when $\sigma$ is a $r$ or $(r - 1)$-cycle.

\noindent {\bf $\mathcal{W}(\mathfrak{so}_{2 N})$ higher quantum Airy structures.} Another class of examples is obtained by choosing the Lie algebra $\mathfrak{g} = \mathfrak{so}_{2 N }$ and the Coxeter element $\sigma$ of the Weyl group, which has order $r= 2(N-1)$. The resulting higher quantum Airy structures are presented in Theorem \ref{t:DN}, summarized here:
\begin{theoremin}
	\label{tDintro} Let $N \geq 3$, that is $r = 2(N-1) \geq 4$, and $s = 1 $ or $ r+1$. Let
	$$
	\mathfrak{d}^i = \delta_{s,1} (i - 1 )\, ,\qquad \tilde{S}_{s} = \big\{(i,k)\quad \big|\quad i \in \{2,4,\ldots,2N-2\} \cup \{ N \} \quad {\rm and}\quad k \geq \mathfrak{d}^i \big\}\, .
	$$
	There exists an (explicit) quantum $r$-Airy structure on  $V = \bigoplus_{p > 0} \mathbb{C}\langle x_{2p+1} \rangle \oplus \mathbb{C}\langle \tilde{x}_{2p+1} \rangle$ based on a representation of the subset of modes $(W_{k}^i)_{(i,k) \in \tilde{S}_{s}}$ of the $\mathcal{W}(\mathfrak{so}_{2 N})$ algebra generators with central charge $N$ in $\mathcal{D}_{T^*V}^{\hbar}$. 
\end{theoremin}
 Here, for any $r$ we get two higher quantum Airy structures (where $ s = 1 $ and $ s = r+1 $), corresponding to the well-known subalgebra of modes of Proposition \ref{lem:modesplus} and \ref{lem:modesdelta}. For now, we do not have a construction for more general values of $ s $ for $\mathfrak{so}_{2 N}$, as in Theorem~\ref{THT1} for $\mathfrak{gl}_{N+1}$ (equivalently, the analog of Theorem \ref{WTHUT} for $\mathfrak{so}_{2 N }$). The enumerative meaning of these higher quantum Airy structures is discussed in Section \ref{s:FJRW}, in terms of Fan--Jarvis--Ruan  theory \cite{FJRid}.

\noindent {\bf Exceptional higher quantum Airy structures.} We construct two higher quantum Airy structures starting with the exceptional Lie algebras $\mathfrak{g} = \mathfrak{e}_{N}$ with $N \in \{6,7,8 \}$ and using  the Coxeter element (of order denoted by $ r $) as the automorphism $ \sigma $. Our main result here is Theorem~\ref{thEEE}, which we summarize as follows.
\begin{theoremin}
Let $\mathbb{D} = \{d_1,\ldots,d_N\}$ the set of Dynkin exponents of $\mathfrak{e}_{N}$  (see Section~\ref{SecE}). Let $s \in \{1,r + 1\}$ and denote
$$
\tilde{S}_{s} := \big\{(i,k) \quad \big|\quad i \in \{1,\ldots,N\}\,\,\,{\rm and}\,\,\,k \geq \mathfrak{d}^i\big\}\,\qquad \mathfrak{d}^i = \left\{\begin{array}{lll} 0 & & {\rm if}\,\,s = r + 1 \\ d_i - 1 & & {\rm if}\,\,s = 1 \end{array}\right.\,.
$$
There exists a quantum $r$-Airy structure on $V = \bigoplus_{p \in \mathbb{D} + r\mathbb{N}} \mathbb{C}\langle x_{p} \rangle$ based on a representation of the subset of modes $(W_k^i)_{(i,k) \in \tilde{S}_{s}}$ of the  $\mathcal{W}(\mathfrak{e}_{N})$ algebra generators with central charge $N$ into $\mathcal{D}_{T^*V}^{\hbar}$.
\end{theoremin}
This Airy structure is as (un)explicit as the generators of the $\mathcal{W}(\mathfrak{e}_{N})$ algebra, see Theorem~\ref{thmde}. For $s = r + 1$ it is not new: its partition function coincides with the Fan--Jarvis--Ruan invariants of $E$-type (see Section~\ref{s:FJRW}) and it was already known that it is uniquely determined by $\mathcal{W}$ constraints, see Section~\ref{s:FJRW} for references. We have a new case $s = 1$ whose enumerative geometry interpretation is currently unknown. For simple but non simply-laced Lie algebras, according to a private communication of Di Yang, $\mathcal{W}$ constraints cannot be brought to the form \eqref{eq:form} and therefore cannot yield higher quantum Airy structures.

\subsubsection{Higher quantum Airy structures from topological recursion}
\label{S122} 
The Chekhov--Eynard--Orantin topological recursion \cite{EO, EORev} associates, to the data of a spectral curve \mbox{$\mathcal{S} = (\mathcal{C},x,y,\omega_{0,2})$} satisfying certain conditions, a sequence of meromorphic differentials $(\omega_{g,n})_{2g - 2 + n > 0}$ that generate enumerative invariants. It was shown in \cite{KS,ABCD} that for a given $\mathcal{S}$ with simple ramification, the topological recursion is equivalent to a quantum Airy structure that has countable dimension and whose Lie algebra is isomorphic to a direct sum of subalgebras of the Virasoro algebra. The $F_{g,n}$s for $2g - 2 + n > 0$ encode the coefficients of decomposition of the meromorphic $n$-differentials $\omega_{g,n}$s of \cite{EORev} on a suitable basis of meromorphic $1$-forms. The choice of polarization in the construction of the quantum Airy structure is determined by $\omega_{0,2}$, which is part of the data of the spectral curve. This dictionary was established in detail in \cite{KS,ABCD}, but we should also mention the earlier work of Kostov and Orantin where some of these elements of comparison already appeared \cite{KO10}.

The original formulation of the Chekhov--Eynard--Orantin topological recursion requires the branched cover $x\,:\,\mathcal{C} \rightarrow \mathbb{C}$  to have simple ramification points only, \textit{i.e.} $\dd x$ has simple zeroes. This restriction on the order of the ramification points was lifted in \cite{BE2, BE,BHLMR}. For arbitrary spectral curves, the combinatorial structure of the topological recursion becomes a little more involved; it is now known in the literature as the Bouchard--Eynard topological recursion.

In Section \ref{sec:hasfromtr} we extend the dictionary between topological recursion and Airy structures to arbitrary spectral curves without any restriction on the order of ramifications. Our main results (Theorems~\ref{t:HASseveral} and Theorem~\ref{c:symmetric} in the text) can be summarized as follows.

For each ramification point $p_{\alpha}$, denote $r_{\alpha}$ the order of ramification at $p_{\alpha}$, and introduce a local coordinate $\zeta$ such that $x(z)  - x(p_{\alpha}) = \frac{\zeta^{r_{\alpha}}(z)}{r_{\alpha}}$. Let us consider the series expansion (denoted with $\equiv$) near the ramification points
$$
y(z) \equiv \sum_{l > 0} F_{0,1}\big[\begin{smallmatrix} \alpha  \\ -l \end{smallmatrix}\big]\,\zeta(z)^{l -r_{\alpha}}\,\qquad  z \rightarrow p_{\alpha}\, ,
$$
and introduce
$$
s_{\alpha} := \min\big\{l > 0 \quad |\quad F_{0,1}\big[\begin{smallmatrix} \alpha \\ -l \end{smallmatrix}\big] \neq 0 \,\,\,{\rm and}\,\,\,r_{\alpha} \nmid l\big\}\, .
$$

The statement of Theorem \ref{c:symmetric} can be summarized as follows:
\begin{theoremin}
\label{symB} The Bouchard--Eynard topological recursion is well defined (\textit{i.e.} produces symmetric $\omega_{g,n}$) if and only if $r_{\alpha} = \pm 1\,\,{\rm mod}\,\,s_{\alpha}$ for all $\alpha$ (the $\pm$ could depend on $\alpha$). When this condition is not satisfied, the lack of symmetry is apparent in $\omega_{0,3}$. 
\end{theoremin}

\begin{definin}
\label{def:ADMIN}We say that the spectral curve is admissible when $r_{\alpha} = \pm 1\,\,{\rm mod}\,\,s_{\alpha}$ for all ramification points $p_{\alpha}$.
\end{definin}
Now let us consider the series expansion
$$
\omega_{0,2}(z_1,z_2) \equiv \bigg(\frac{\delta_{\alpha_1,\alpha_2}}{(\zeta(z_1) - \zeta(z_2))^2} + \sum_{l_1,l_2 > 0} \phi_{l_1,l_2}^{\alpha_1,\alpha_2}\,\zeta(z_1)^{l_1- 1} \zeta(z_2)^{l_2 - 1}\bigg)\, \dd \zeta(z_1) \otimes \dd \zeta(z_2)\, , \qquad z_j \rightarrow p_{\alpha_j}\, ,
$$
and introduce for $l > 0$ the meromorphic $1$-forms on $\mathcal{C}$
$$
\dd\xi_{l}^{\alpha}(z) := \Res_{z' = p_{\alpha}} \bigg( \int_{p_{\alpha}}^{z'} \omega_{0,2}(\cdot,z)\bigg) \frac{\dd \zeta(z')}{\zeta(z')^{l + 1}} \, .
$$

Theorem \ref{t:HASseveral} relates the Bouchard--Eynard topological recursion to higher quantum Airy structures, as summarized below:
\begin{theoremin}
\label{TGIV} For any admissible spectral curve, the $\omega_{g,n}$ computed by the Bouchard--Eynard topological recursion can be decomposed as finite sums
$$
\omega_{g,n}(z_1,\ldots,z_n) = \sum_{\substack{\alpha_1,\ldots,\alpha_n  \\ l_1,\ldots,l_n > 0   }} F_{g,n}\big[\begin{smallmatrix} \alpha_1 & \cdots & \alpha_n \\ l_1 & \cdots & l_n \end{smallmatrix}\big]\,\bigotimes_{j = 1}^n \dd\xi_{l_j}^{\alpha_j}(z_j)\, ,
$$
and the generating series
$$
Z =  \exp\bigg(\sum_{\substack{g \geq 0,\,\,n \geq 1 \\ 2g - 2 + n > 0}} \frac{\hbar^{g-1}}{n!} \sum_{\substack{\alpha_1,\ldots,\alpha_n \\ l_1,\ldots,l_n > 0}} F_{g,n}\big[\begin{smallmatrix} \alpha_1 & \cdots & \alpha_n \\ l_1 & \cdots & l_n \end{smallmatrix}\big] \prod_{j = 1}^n x_{l_j}^{\alpha_j}\bigg)
$$
is the partition function of a higher quantum Airy structure based on an (explicit) representation of a subset of modes of the $\bigoplus_{\alpha} \mathcal{W}(\mathfrak{gl}_{r_{\alpha}})$ algebra generators as differential operators.

More precisely, $Z$ satisfies a Givental-like decomposition:
\begin{equation}
\label{Giveqe}
Z =
 \exp\left(\,\,\sum_{\alpha,l} \frac{F_{0,1}\big[\begin{smallmatrix} \alpha \\ -l \end{smallmatrix}\big] + \delta_{l,s_{\alpha}}}{l}\,\partial_{x_{l}^{\alpha}} + \frac{\hbar}{2} \sum_{\substack{\alpha_1,\alpha_2 \\ l_1,l_2 > 0}} \frac{\phi_{l_1,l_2}^{\alpha_1,\alpha_2}}{l_1\,l_2}\,\partial_{x^{\alpha_1}_{l_1}}\partial_{x^{\alpha_2}_{l_2}}\right) \prod_{\alpha} Z_{(r_{\alpha},s_{\alpha})}\big((x_{l}^{\alpha})_{l > 0}\big)\, ,
\end{equation} 
where the $Z_{(r,s)}$s are the partition functions of the quantum Airy structures described in Theorem~\ref{THT1}. 
\end{theoremin} 
The formula \eqref{Giveqe} is a Givental-like decomposition for the Bouchard--Eynard topological recursion. If $r_{\alpha} = 2$, Theorem~\ref{TGIV} was obtained in \cite{Einter} for $s_{\alpha} = 3$ and in \cite{ChekhovNorbury} when $s_{\alpha}$ can take any of the admissible values $1$ or $3$.

Let us comment on our approach as we do not construct the higher quantum Airy structures directly from the topological recursion as in \cite{ABCD}. Rather, we start with the notion of ``higher abstract loop equations'' of Definition \ref{d:hle} which generalizes the one of \cite{BEO13,BSblob} to arbitrary ramifications. We prove in Appendix \ref{a:proof} that if a solution to the higher abstract loop equations exists, then it is uniquely given by the Bouchard--Eynard topological recursion. Thus, it is just as good to take the higher abstract loop equations as starting point. But there is a fundamental reason why we start with the loop equations instead of the topological recursion. It is not too difficult to construct differential operators that produce a recursive structure equivalent to the Bouchard--Eynard topological recursion; but proving the graded Lie subalgebra condition required for existence of a common solution of these differential operators (\textit{i.e.} the symmetry of the $F_{g,n}$) appears quite difficult. While if we start with loop equations, we observe that the resulting differential operators can be identified directly with those coming from modules over $\mathcal{W}(\mathfrak{gl}_r)$ algebras; therefore we can use Theorems~\ref{WTHUT} and \ref{WTHUT2} to prove the graded Lie subalgebra condition. In other words, the loop equations make the algebraic structure of the corresponding higher quantum Airy structure explicit, at the expense of obscuring the recursive structure of the original system of equations.

The identification with higher quantum Airy structures constructed from $\mathcal{W}(\mathfrak{gl}_r)$ algebras has a number of interesting consequences. We are not aware of a direct proof that the Bouchard--Eynard topological recursion produces symmetric differentials for arbitrary spectral curves. An indirect argument exists for spectral curves that appear as limits of family of curves with simple ramification \cite{BE}, but it is not clear which spectral curves precisely satisfy this condition. A consequence of our identification between loop equations and higher quantum Airy structures is that for any admissible spectral curves, a solution to the loop equations exist. It must then be given uniquely by the Bouchard--Eynard topological recursion. It then follows that for all admissible spectral curves the Bouchard--Eynard topological recursion produces symmetric differentials (the announced Theorem~\ref{symB}, which is Theorem~\ref{c:symmetric} in the text).

What is particularly intriguing though is the cases that fail. The admissibility in Definition~\ref{def:ADMIN} is a constraint on the local behavior of $\omega_{0,1} = y\dd x$. While the condition that $s$ is coprime with $r$ is easy to understand from the geometry of spectral curves (it says that $\mathcal{C}$ is locally irreducible at its ramification points), the condition that $r = \pm 1\,\, \text{ mod }\,s$ is rather unexpected and its geometric meaning is mysterious for us. Nonetheless, when it is not satisfied, we show in Proposition~\ref{p:symmetry} that the Bouchard--Eynard topological recursion \emph{does not} in fact produce symmetric differentials. The simplest such case is $(r,s) = (7,5)$. Consequently, we can deduce that the left ideal generated by the appropriate set of modes of the $\mathcal{W}(\mathfrak{gl}_{r})$ algebra cannot be a graded Lie subalgebra of the algebra of modes, and that the collection of differential operators is not a higher quantum Airy structure.

For $r = 3$, Safnuk,  in \cite{Safnukopen}, recast the $\mathcal{W}$ constraints of \cite{Alexandrovopen} for open intersection theory into a period computation, which turns out to be an unusual modification of the topological recursion on the spectral curve $x = y^2/2$. It would be interesting --- but beyond the scope of this article ---  to generalise Safnuk's result and obtain the $F_{g,n}$ of Theorem~\ref{TTH2} by a period computation. The same question could be asked if quantum Airy structures are found for other automorphisms $\sigma \in \mathfrak{S}_r$. It amounts to asking what is the appropriate modification of the topological recursion to treat reducible spectral curves, and whether there will be new conditions of admissibility (like the one we found in Theorem~\ref{symB}). This level of generality may enlighten the geometric meaning of those admissibility constraints.

\subsubsection{Results on $ \mathcal{W} $ algebras}
\label{s:wmodes}
As a side result of our construction, we prove a certain curious property of the algebra of modes of the $\mathcal{W}(\mathfrak{gl}_{r})$ algebra at the self dual level. Let $W^1, \ldots W^r$ be the strong generators of $\mathcal{W}(\mathfrak{gl}_{r})$ with conformal weights $1,\ldots,r$, $\mathcal{A}$ be the suitably completed algebra of modes of $\mathcal{W}(\mathfrak{gl}_{r})$, and $\mathcal{F}_p \mathcal{A}$ be the filtration on $\mathcal{A}$ induced by Li's filtration on $\mathcal{W}(\mathfrak{gl}_{r})$ (see Section \ref{s:subalgebra}). Propositions \ref{lem:modesplus}, \ref{lem:modesdelta} and Theorem \ref{t:sub} can be combined into the following result.
\begin{theoremin}
	\label{WTHUT}Let $r \geq 2$ and $\lambda_1 \geq \cdots \geq \lambda_{p} \geq 1$ such that $\sum_{i = 1}^{p} \lambda_i = r$. For $i \in \{1,\ldots,r\}$ denote
	$$
	\lambda(i) := \min\Big\{m > 0 \quad \Big|\quad \sum_{j = 1}^m \lambda_j \geq i\Big\},\qquad S_{\lambda} = \big\{(i,k)\quad \big| \quad i \in \{1,\ldots,r\} \quad {\rm and}\quad k \geq i - \lambda(i)\big\} \,.
	$$
	The left $\mathcal{A}$-ideal generated by the modes $W_{k}^{i}$ indexed by $(i,k) \in S_{\lambda}$ is a graded Lie subalgebra of $\mathcal{A}$, \textit{i.e.} there exists $g_{(k_1,i_1),(k_2,i_2)}^{(k_3,i_3)} \in \mathcal{A}$ such that
	$$
	\forall (i_1,k_1),(i_2,k_2) \in S_{\lambda}\,,\qquad  [W_{k_1}^{i_1},W_{k_2}^{i_2}] = \sum_{(i_3,k_3) \in S_{\lambda}} g_{(i_1,k_1),(i_2,k_2)}^{(i_3,k_3)}\,W_{k_3}^{i_3}\, \in \mathcal{F}_{i_1 + i_2-2} \mathcal{A}.
	$$
\end{theoremin}
The case $\lambda = (1,\ldots,1)$, which corresponds to $k \geq 0$, gives a well-known Lie subalgebra. It is the one generated by the modes annihilating the vacuum vector. The other cases however seem new.
Some of these Lie subalgebras are used to prove Theorem~\ref{THT1}, thanks to an arithmetic correspondence  established in Proposition \ref{p:partition} in Appendix~\ref{a:rs}, which is summarized here:
\begin{propin}
	\label{WTHUT2} For any $s \in \{1,\ldots,r + 1\}$ such that $r = r's + r''$ with $r'' \in \{1,s - 1\}$, we have the equality $S_{\lambda} = \tilde{S}_{s}$ between the set of modes appearing in Theorem~\ref{THT1} and Theorem~\ref{WTHUT} for the choice
	$$
	\lambda_{1} = \cdots = \lambda_{r''} = r' + 1,\qquad \lambda_{r'' + 1} = \cdots = \lambda_{s} = r'
	$$
	If $r \neq \pm 1\,\,{\rm mod}\,\,s$, there exists a unique sequence $(\lambda_j)_{j}$ such that $S_{\lambda} = \tilde{S}_{s}$ but it is not weakly decreasing and the left $\mathcal{A}$-ideal generated by the modes $(W_{k}^i)_{(i,k) \in \tilde{S}_{s}}$ does not form a subalgebra of $\mathcal{A}$.
\end{propin}

The proof of Theorem~\ref{WTHUT} relies on the construction of a highest weight module whose highest weight vector is annihilated by the modes indexed by $S_{\lambda}$. The existence of such a highest weight module is perhaps unexpected; it relies heavily on our realization of the $\mathcal{W}(\mathfrak{gl}_r)$ algebra as a subalgebra of the Heisenberg vertex operator algebra and on certain embeddings of $
\mathfrak{gl}_{\lambda_1} \oplus \cdots \oplus \mathfrak{gl}_{\lambda_{p}}$ into $\mathfrak{gl}_{r}$. It would be worth investigating this construction further, and see whether it can be generalized to $\mathcal{W}$ algebra of other types. In particular, this would yield generalizations of Theorem~\ref{tDintro}. Note that it is important in the proof for $\mathfrak{gl}_{r}$ that $(\lambda_j)_{j}$ be a weakly decreasing sequence and this is confirmed by the counterexamples mentioned in the last claim in Theorem~\ref{WTHUT2}.

\subsection{Outline}

We start in Section \ref{s:HAS} by defining higher quantum Airy structures. We first propose in Section \ref{S1} a basis-independent definition, starting from the point of view of quantization of classical higher Airy structures, as in \cite{KS}. In Section \ref{S22} we revisit higher quantum Airy structures using bases. We calculate the explicit recursive system satisfied by the coefficients $F_{g,n}$. We also prove a reduction statement to get rid of linear differential operators in higher quantum Airy structures. We introduce crosscapped Airy structures in Section \ref{Scrosscap}, which are related to generating functions in open intersection theory.

In Section \ref{s:WW} we first introduce the background on vertex operators algebras (Section \ref{s:VOA}) and $\mathcal{W}(\mathfrak{g})$ algebras (Section \ref{s:WG}) that will be needed for the construction of our first type of higher quantum Airy structures. We construct in Section \ref{s:subalgebra} a number of left ideals for the algebra of modes that are graded Lie subalgebras (see Propositions \ref{lem:modesplus}, \ref{lem:modesdelta} and Theorem \ref{t:sub}). We then review the concept of twisted modules for vertex operators algebras in Section \ref{s:TM}, in preparation for the next section.

Our construction of higher quantum Airy structures as modules of $\mathcal{W}(\mathfrak{g})$ algebras is proposed in Section \ref{HASW}. The first class of $\mathcal{W}(\mathfrak{gl}_{r})$ higher quantum Airy structures, with the automorphism $\sigma$ given by the Coxeter element of the Weyl group, is explored in Section \ref{SWgg}. The second class of $\mathcal{W}(\mathfrak{gl}_{r})$ higher quantum Airy structures for arbitrary automorphisms $\sigma$ is studied in Section \ref{s:glra}. We introduce the $\mathcal{W}(\mathfrak{so}_{2r })$ higher quantum Airy structures in Section \ref{s:DN}, and the $\mathcal{W}(\mathfrak{e}_r)$ higher quantum Airy structures in Section \ref{s:EN}.

Section \ref{sec:hasfromtr} is devoted to the reconstruction of the higher quantum Airy structures associated to the Bouchard--Eynard topological recursion on arbitrary admissible spectral curves. We study the geometry of local spectral curves in Section \ref{s:local}, and describe the relation with the standard notion of (global) spectral curves. We introduce the Bouchard--Eynard topological recursion and higher abstract loop equations in Section \ref{s:tr}. We then prove that the higher abstract loop equations for local spectral curves with one component are equivalent to $\mathcal{W}(\mathfrak{gl}_r)$ higher quantum Airy structures in Section \ref{s:one}. The general result for local spectral curves with several components is obtained in Section \ref{s:several}.

Finally, Section \ref{Section6} reviews the known and conjectural enumerative geometric interpretations of the higher quantum Airy structures that we construct; we also attempt to summarize the rich history of existing results in this area. We discuss the (closed) $r$-spin intersection theory (Section \ref{Srspin}), higher analogs of the Br\'ezin--Gross--Witten theory (Section \ref{s:BGW}), open $r$-spin intersection theory (Section \ref{Sopen}), and Fan--Jarvis--Ruan theories (Section \ref{s:FJRW}).

We conclude with three appendices. In Appendix \ref{Approot}  we prove, by elementary means, various properties of certain sums over roots of unity that play an important role in our construction of the $\mathcal{W}(\mathfrak{gl}_{r})$ quantum Airy structures. In Appendix \ref{a:rs}, we show that the graded Lie subalgebra property is only satisfied for values of $(r,s)$ such that $r = \pm 1 \text{ mod } s$. When $r = \pm 1 \text{ mod } s$, we show that we get a subalgebra of the intermediate type described in Theorem~\ref{WTHUT2}; and when $r \neq \pm 1 \text{ mod } s$, we prove that there is no symmetric solution to the system of differential equations, and hence the left ideal generated by the set of modes cannot be a graded Lie subalgebra. The proof consists of an explicit computation and check of (lack of) symmetry for $F_{0,3}$ by elementary --- but still lengthy --- arithmetics. We also compute explicitly $F_{1,1}$ in Appendix~\ref{F11comput}. Lastly, in Appendix \ref{a:proof} we prove that if a solution to the higher abstract loop equations that respects the polarization exists, then it is uniquely constructed by the Bouchard--Eynard topological recursion.

\subsubsection*{Recent work}
Since the submission of this manuscript in January 2019, several works based on the present one have appeared in the literature, which we now briefly summarize. There are two direct follow-ups to this article, \cite{BKS} and \cite{BouMas}, which obtain an almost complete classification of admissible twists and dilaton shifts for Airy structures based on $ \mathcal W(\mathfrak{gl}_{N+1}) $ (generalizing Section~\ref{Section422}). The computation of $F_{g,n}$ for these Airy structures is recast in \cite{BKS} as a period computation (generalizing Section~\ref{sec:hasfromtr}), thus giving a definition of spectral curve topological recursion for singular (reducible) spectral curves with certain admissibility conditions. This answers the question posed at the end of Section~\ref{S122}. In a different direction, \cite{BBCC} investigated Whittaker vectors for $ \mathcal W $ algebras of type $ ADE $ in the context of higher Airy structures. In particular, \cite{BBCC} applied this to show that Gaiotto vectors in $\mathcal{N} = 2$ four-dimensional gauge theory (and thus, the Nekrasov instanton partition function) can be reconstructed using topological recursion.

\subsubsection*{Acknowledgments}
We thank Alexander Alexandrov, Todor Milanov, Nicolas Orantin, Ran Tessler and Di Yang for discussions, as well as Nezhla Aghaei, Reinier Kramer, Anton Mellit, Pieter Moree, Aniket Joshi, and Yannik Sch\"uler for comments. We also thank Leonid Chekhov and the other organizers of the workshop ``Combinatorics of the moduli spaces, etc.'' in Moscow, June 2018, where a subset of the authors could meet and work on this project and a preliminary version was presented. The work of G.B. benefited from the support of the Max-Planck-Gesellschaft. The work of D.N. is supported by the TEAM programme of the Foundation
for Polish Science co-financed by the European Union under the
European Regional Development Fund (POIR.04.04.00-00-5C55/17-00). V.B., N.K.C. and T.C. acknowledge the support of the Natural Sciences and Engineering Research Council of Canada.

\newpage

\section{Higher quantum Airy structures} 

\label{s:HAS}

\subsection{A conceptual approach}
\label{S1}

In this section, we provide a conceptual introduction to the concept of higher Airy structures, starting from the point of view of quantization of higher classical Airy structures. We propose a basis-free definition of $\infty$-Airy structures, and its finite counterpart $r$-Airy structures. We offer a basis-dependent and computational approach to higher Airy structures in Section \ref{S22}. Readers who are mostly interested in the computational aspects of Airy structures may prefer to skip directly to Section \ref{S22}.

\subsubsection{Classical picture}

Let $\mathbf{k}$ be a field of characteristic zero and $W$ be a finite-dimensional symplectic $\mathbf{k}$-vector space equipped with the symplectic form $\Omega$. $\mathbf{k}[\![W]\!]$ is the completion of the graded ring of polynomial functions on $W$. It is a Poisson algebra. The projection from $\mathbf{k}[\![W]\!]$ onto its subspace of degree $i$ is denoted $\pi_i$. In fact, we can consider $\pi_1$ as a linear map $\mathbf{k}[\![W]\!] \rightarrow W$ since the subspace of linear functions on $W$ is naturally isomorphic to $W^*$ and can be identified with $W$ itself via the pairing $\Omega$.

\begin{defin}
	\label{Def1} A \emph{classical $\infty$-Airy structure} on $(W,\Omega)$ is the data of a $\mathbf{k}$-vector space $V$ together with a linear map $\lambda\,:\,V \rightarrow \mathbf{k}[\![W]\!]$ such that\begin{itemize}
		\item[$(i)$] $\pi_0 \circ \lambda = 0$.
		\item[$(ii)$] $\mathcal{I} = \pi_1 \circ \lambda\,:\,V \rightarrow W$ is a linear embedding of $V$ as a Lagrangian subspace of $W$.
		\item[$(iii)$] The $\mathbf{k}[\![W]\!]$-ideal generated by ${\rm Im}\,\lambda$ is a Poisson subalgebra of $\mathbf{k}[\![W]\!]$.
	\end{itemize}
	If ${\rm Im}\,\lambda$ is a subspace of the space $\mathbf{k}_{r}[W]$ of polynomial functions of degree at most $r$ for some given integer $r \geq 2$, we will call it a \emph{classical $r$-Airy structure}. 
\end{defin}

For $r = 2$, $(iii)$ is equivalent to requiring that ${\rm Im}\,\lambda$ is a Poisson subalgebra and we recover the Airy structures studied in \cite{KS,ABCD}. Definition~\ref{Def1} formally corresponds to $r = \infty$.

\subsubsection{The quantization problem}
\label{quang}
The Poisson algebra $\mathbb{C}[\![W]\!]$ can be quantized by forming the Weyl algebra $\mathcal{D}_{W}^{\hbar}$. We define it as the completion of the graded associative algebra over $\mathbb{C}[\![\hbar]\!]$ of non-commutative polynomials in elements of $W$ modulo the relations $[w,w'] = \hbar \Omega(w,w')$ for any $w,w' \in W$. The grading is defined by $\deg W = 1$ and $\deg \hbar = 2$.
\begin{defin}\label{def:gli}
	A subspace $\hat{\Lambda} \subset \mathcal{D}_{W}^{\hbar}$ is a \emph{graded Lie subalgebra} if $[L,L'] \in \hbar\cdot \hat{\Lambda}$ for any $L,L' \in \hat{\Lambda}$.
\end{defin}

We have a linear map ${\rm cl}\,:\,\mathcal{D}_{W}^{\hbar} \rightarrow \mathbb{C}[\![W]\!]$ which is a reduction to $\hbar = 0$ and is called the \emph{classical limit}. It is such that
$$
{\rm{cl}}\left(\frac{1}{\hbar}[L,L']\right)  = \{{\rm cl}(L),{\rm cl}(L')\}\,.
$$
Obviously, if $\hat{\Lambda}$ is a graded Lie subalgebra, then ${\rm cl}(\hat{\Lambda})$ is a Poisson subalgebra in $\mathbb{C}[\![W]\!]$. Conversely, given a Poisson subalgebra $\Lambda \subset \mathbb{C}[\![W]\!]$, we may ask whether it can be quantized, \textit{i.e.} whether there exists a graded Lie subalgebra $\hat{\Lambda} \subseteq \mathcal{D}_{W}^{\hbar}$ such that ${\rm cl}(\hat{\Lambda}) = \Lambda$. In this article, we will study the quantization of classical $\infty$-Airy structures in the following sense.

\begin{defin}\label{d:abstract}
	A quantum $\infty$-Airy structure on $ V $ is a linear map $\hat{\lambda}\,:\,V \rightarrow \mathcal{D}_{W}^{\hbar}$ such that ${\rm cl} \circ \hat{\lambda}$ is a classical $\infty$-Airy structure and the left ideal $\hat{\Lambda} = \mathcal{D}_{W}^{\hbar}\cdot {\rm Im}\,\hat{\lambda}$ is a graded Lie subalgebra. As before, if ${\rm Im}\,\hat{\lambda}$ is a subspace of the space of elements of degree at most $ r  $ in $\mathcal{D}_{W}^{\hbar}$, we will call it a \emph{quantum $r$-Airy structure} instead.
\end{defin}

 For $r = 2$, the condition that $\mathcal{D}_{W}^{\hbar}\cdot {\rm Im}\,\hat{\lambda}$ is a graded Lie subalgebra is equivalent to ${\rm Im}\,\hat{\lambda}$  being a graded Lie subalgebra, but this is no longer true for $r > 2$.

\subsubsection{The partition function}

The most important fact about quantum Airy structures is that they determine a partition function via a topological recursion once a polarization is chosen.

\begin{defin}
	A \textit{polarization} of a symplectic vector space $(W,\Omega)$ is a decomposition $W = V \oplus V'$ such that $V$ and $V'$ are Lagrangian. 
	
	If we are given a Lagrangian subspace $V$ of $ W $, a polarization is a choice of a transverse Lagrangian subspace $ V' $. In this case, we say that the \textit{polarization is adapted to $V$}.
\end{defin}

If we choose a polarization, the symplectic pairing gives a canonical identification $V^* \cong V'$ and, therefore, an isomorphism $W \cong T^*V$ of symplectic vector spaces. Here $T^*V = V \oplus V^*$ is equipped with the natural symplectic form defined for $v \in V$ and $\phi \in V^*$ by $\Omega(v,\phi) = \phi(v)$. Therefore, the $\mathbb{C}[\![\hbar]\!]$ algebra $\mathcal{D}_{W}^{\hbar}$ acts faithfully on the space of functions on the formal neighborhood of $0$ in $V$
$$
{\rm Fun}_{V}^{\hbar} = \prod_{d \geq 0} {\rm Sym}^{d}(V^*)[\![\hbar]\!]\,.
$$
Elements $v \in V$ act on $f \in {\rm Fun}_{V}^{\hbar}$ by derivation and $x \in V^*$ by multiplication by linear functions
$$
v \cdot f = \hbar \partial_{v} f,\qquad x \cdot f = xf\,.
$$
and the commutation relations in $\mathcal{D}_{W}^{\hbar}$ are represented by the Leibniz rule. 

The space ${\rm Fun}_{V}^{\hbar}$ is in fact a graded associative algebra, where the grading is specified again by $\deg V^* = 1$ and $\deg \hbar = 2$. Besides, the action of $\mathcal{D}_{W}^{\hbar}$ respects the grading
$$
(\mathcal{D}_{W}^{\hbar})_{d} \cdot ({\rm Fun}_{V}^{\hbar})_{d'} \subseteq ({\rm Fun}_{V}^{\hbar})_{d + d'}\,.
$$
Thanks to this grading, for any elements $L \in \mathcal{D}_{W}^{\hbar}$ and $F \in {\rm Fun}_{V}^{\hbar}$ it is possible to make sense of $e^{-F/\hbar}(L \cdot e^{F/\hbar})$ as an element of ${\rm Fun}_{V}^{\hbar}$. 

Note that any $F \in {\rm Fun}_{V}^{\hbar}$ can be uniquely decomposed as
$$
F = \sum_{g,n \geq 0} \frac{\hbar^{g}}{n!}\,F_{g,n},\qquad F_{g,n} \in {\rm Sym}^{n}(V^*)\,.
$$

\begin{thm}[\cite{KS}, Theorem 2.4.2]
	\label{partif} Let $\hat{\lambda} : V \to \mathcal{D}^{\hbar}_{W}$ be a quantum $\infty$-Airy structure on $ V $ and choose a polarization of $(W,\Omega)$ adapted to the Lagrangian subspace $\mathcal{I}(V)$ of $ W $ (as given by the corresponding classical $\infty$-Airy structure). There exists a unique $F \in {\rm Fun}_{V}^{\hbar}$ such that
	\begin{itemize}
		\item[$(i)$] $e^{-F/\hbar}(\hat{\lambda}(v) \cdot e^{F/\hbar}) = 0$ for any $v \in V$,
		\item[$(ii)$] $F_{g,0} = 0$ for any $g \geq 0$,
		\item[$(iii)$] $F_{0,1} = 0$ and $F_{0,2} = 0$.
	\end{itemize}
\end{thm}

The $F_{g,n}$ are usually called ``amplitudes'' or ``correlation functions'' or ``free energies'', and $Z = e^{F/\hbar}$ is called the ``partition function''. Conditions $(i)-(ii)-(iii)$ imply that the amplitudes are uniquely determined by a recursion on $2g - 2 + n > 0$. The recursive formula is spelled out in Corollary~\ref{cor28}. The main feature to remember about this formula is that its terms are in correspondence with equivalence classes of excisions of embedded $S \mapsto \Sigma_{g,n}$ of smooth surfaces $S$ of genus $h$ with $k$ ordered boundaries into a smooth surface $\Sigma_{g,n}$ of genus $g$ with $n$ ordered boundaries, such that the first boundary component of $S$ coincides with the first boundary component of $\Sigma_{g,n}$. Here, two embeddings $S \mapsto \Sigma_{g,n}$ and $S' \mapsto \Sigma_{g,n}$ are considered equivalent if they are related by a diffeomorphism of $\Sigma_{g,n}$ preserving the ordering of the boundary components of $\Sigma_{g,n}$. Therefore, the number of equivalence classes is finite and they are characterized by the topology of $\Sigma_{g,n} - S$. This justifies the name \emph{topological recursion}.  In the special case of $(r = 2)$-Airy structures, the only terms appearing correspond to excisions of pairs of pants, \textit{i.e.} $(h,k) = (0,3)$. The topological recursion modeled on the excision of surfaces other than pairs of pants first appeared  in \cite{BHLMR}.

\subsubsection{Finite vs. countable dimension}
\label{infinite}
We will encounter vector spaces $V$ which are countable products of finite-dimensional vector spaces.
$$
V = \prod_{p \geq 0} V_{p}\,.
$$
This situation can be handled without difficulty in our discussions, by \emph{defining} tensorial constructions relying on the unambiguous finite-dimensional tensorial constructions. For instance, we agree that the dual is
$$
V^* = \bigoplus_{p \geq 0} V_{p}^*\,,
$$
where $\bigoplus$ is the direct sum as opposed to the direct product $\prod$. Then, the cotangent space $T^*V = V \oplus V^*$ has a well-defined symplectic pairing. We define the tensor product as
$$
V \otimes V' = \prod_{p \geq 0} (V \otimes V')_{p},\qquad (V \otimes V')_{p} = \bigoplus_{q = 0}^{p} V_{q} \otimes V'_{p - q}\,.
$$

\subsubsection{Classical versus quantum Airy structures}

Due to Theorem~\ref{partif}, quantum Airy structures can be considered as initial data for the topological recursion. As many known examples show, the $F_{g,n}$s often have an interpretation in enumerative geometry or topological field theory, \textit{i.e.} count surfaces of genus $g$ with $n$ punctures/boundaries in various instances. Another trend of applications (for $r = 2$) concerns the computation of WKB expansions of sections of holomorphic bundles on curves annihilated by a flat $\hbar$-connection. The beauty of the theory is that all these problems fit in the same universal scheme of the topological recursion. On the other hand, it is not an obvious task to construct quantum Airy structures.

Reversing the usual path from a problem to its solution, we think that it is worth searching for other constructions of quantum Airy structures as they would probably provide solutions to interesting geometric problems. In particular, it is appealing to look for constructions directly from the symplectic or K\"ahler geometry of manifolds and their Lagrangians. We certainly have the possible applications to the moduli space of flat connections on curves in mind.

We now point out that symplectic geometry easily gives rise to \emph{classical} $\infty$-Airy structures. Consider for instance a real symplectic manifold $(X,\Omega_{X})$, which can be assumed to be real-analytic without any loss of generality according to \cite{Kutz}. Let $L_0$ and $L$ be two real-analytic Lagrangian subvarieties, which intersect and are tangent at a point $p \in M$. Take $W = T_{p}X$ with the symplectic form induced by $\Omega := \Omega_{X}|_{p}$, and $V = T_{p}L_0$. A suitable choice of Darboux coordinates gives an analytic isomorphism $f\,:\,U_{X} \rightarrow U_{W}$ from a neighborhood $U_{X} \subseteq X$ of $p$ to a neighborhood $U_{W} \subseteq W$ of $0$, preserving the symplectic structure, such that $f(p) = 0$ and $f(L_0 \cap U_{X}) = V \cap U_{W}$. By the inverse function theorem, upon taking smaller $U$s, there exists a linear map $\lambda_{{\rm an}}$ from $V$ to the space of real-analytic functions on $U_{W}$, realizing $L$ locally as the zero-locus of $\lambda_{{\rm an}}$ 
$$
f(L \cap U_{X}) = \{w \in U_{W} \quad |\quad \forall v \in V,\quad \lambda_{{\rm an}}(v)(w) = 0\}\,.
$$
${\rm Im}\,\lambda_{{\rm an}}$ generates the ideal of the ring of real-analytic functions on $U_{W}$ which vanish on $f(L \cap U_{X})$. In fact, since $L$ is Lagrangian, this ideal is a Poisson subalgebra. The condition $T_{p}\Lambda_0 = T_{p}L$ implies that the linear map $V \rightarrow W$, which associates to $v \in V$ the unique $t_{v} \in W$ such that $\Omega(t_{v},\cdot) = \dd_{0}\lambda_{{\rm an}}(v)$, namely the differential of $\lambda_{{\rm an}}$ at $0 \in W$, is a Lagrangian embedding. Consequently, we obtain a classical $\infty$-Airy structure by taking $\lambda(v)$ to be the formal Taylor series at $0$ of $\lambda_{{\rm an}}(v)$.

In the less frequent situation where there are Darboux coordinates such that $L$ is cut out (locally around $p$) by polynomial equations of degree $\leq r$, we obtain a classical $r$-Airy structure.

It is not easy to exhibit a classical $2$-Airy structure $\lambda$. In fact, this amounts  (see \cite{ABCD}) to finding a collection of functions $(\lambda_i)_{i \in I}$ of the form
\beq
\label{lambdai}\lambda_i = y_i - \sum_{a,b \in I} \big(\tfrac{1}{2} A^i_{a,b}x_ax_b + B^i_{a,b}x_ay_b + \tfrac{1}{2}C^i_{a,b}y_ay_b\big),\qquad \{\lambda_i,\lambda_j\} = \sum_{a \in I} (B^i_{j,a} - B^j_{i,a})\lambda_{a}\,,
\eeq
where $(x_i)_{i \in I}$ is a basis of linear coordinates on $V$, $(y_i)_{i \in I}$ the dual coordinates on $V^*$ such that $\{x_i,y_j\} = \delta_{i,j}$ and $(A^*_{**},B^*_{**},C^*_{**})$ are scalars. The Poisson commutation relations impose an overdetermined system of linear and quadratic constraints on these scalars. However, once a classical $2$-Airy structure has been found, it is fairly easy to describe its possible quantizations. Indeed, such a quantization $\lambda_i$ must be of the form
$$
\hat{\lambda}_i = \hbar\partial_{x_i} - \sum_{a,b \in I} \big(\tfrac{1}{2}A^i_{a,b}x_ax_b + B^i_{a,b}x_a\,\hbar\partial_{x_b} + \tfrac{1}{2}C^i_{a,b}\,\hbar^2\partial_{x_a}\partial_{x_b}\big) - \hbar\,D^i,\qquad [\hat{\lambda}_i,\hat{\lambda}_j] = \sum_{a \in I} \hbar (B^i_{j,a} - B^j_{i,a})\hat{\lambda}_{a}
$$
for some scalars $D^*$. The Lie algebra commutation relations are in fact equivalent to affine constraints for $D^*$. Note that the ``quantum correction'' $D^*$ arises naturally from the ambiguity in the ordering of $x$ and $\hbar\partial_{x}$ to quantize the $B$-terms in \eqref{lambdai}.

The previous example suggests that the difficulty in finding classical $r$-Airy structures decreases with $r$, and disappears for $r = \infty$. On the contrary, the difficulty of quantizing a given classical $r$-Airy structure is absent for $r = 2$, but increases with $r$. Indeed, one has to introduce an increasing number of quantum corrections which, for $r > 2$, must satisfy non-linear constraints in order to lift the Poisson subalgebra condition to a graded Lie subalgebra condition.

\subsection{A computational approach}
\label{S22}

The basis-free definitions of quantum higher Airy structures given in Section~\ref{S1} clarify the geometric context of our work. We are going to restart from scratch and give a roughly equivalent presentation of the setup using bases. It can be read independently of Section~\ref{S1}, some readers may find these more basic definitions easier to grasp, it facilitates the exposition and is closer to the notations of \cite{KS}.

\subsubsection{Basis-dependent definition}

Let $V$ be a $\mathbb{C}$-vector space\footnote{The paper would be equally valid over a field of characteristic $0$.}. We are going to assume that $V$ has finite dimension $D$, but there is no difficulty in adapting it to the case of countably infinite dimension as in Section~\ref{infinite}. Denoting $I = \{1,\ldots,D\}$, let $(y_l)_{l \in I}$ be a basis of $V$ and $(x_l)_{l \in I}$ be the dual basis. We can think of $y$s as linear coordinates on $V^*$ and $x$s as linear coordinates on $V^*$. Then $W = V \oplus V^*$ is equipped with the Poisson bracket
$$
\forall l,m \in I,\qquad \{x_l,y_m\} = \delta_{l,m},\qquad \{x_l,x_m\} = \{y_l,y_m\} = 0\,.
$$

We identify $\mathcal{D}_{W}^{\hbar} \cong \mathbf{k}[\![\hbar, (x_l)_{l \in I},(\hbar \partial_{x_l})_{l \in I}]\!] $ with the completed algebra of differential operators on $V$. We define an algebra grading by assigning
\begin{equation}
\label{eq:hbardeg}
\deg x_l = \deg \hbar \partial_{x_l}  = 1,\qquad \deg \hbar = 2\,.
\end{equation}

\begin{defin} \label{def:HAS}
A \emph{higher quantum Airy structure} on $V$ is a family of differential operators $(H_k)_{k \in I}$ of the form
\begin{equation}
H_k = \hbar \partial_{x_k} - P_k,
\end{equation}
where $P_k \in \mathcal{D}_{W}^{\hbar}$ is a sum of terms of degree $\geq 2$. Moreover, we require that the left $\mathcal{D}_{W}^{\hbar}$-ideal generated by the $H_k$s  forms a graded Lie subalgebra, \textit{i.e.} there exists $g_{k_1,k_2}^{k_3} \in \mathcal{D}_{W}^{\hbar}$ such that
\begin{equation}\label{eq:HAS}
[H_{k_1}, H_{k_2} ] = \hbar \sum_{k_3 \in I} g_{k_1,k_2}^{k_3} H_{k_3}.
\end{equation}
\end{defin}

This definition is a basis-dependent definition that should be compared with the basis-free Definition \ref{d:abstract}. As introduced there, we may define a \emph{quantum $r$-Airy structure} as  a higher quantum Airy structure such that all $P_k$ only have terms of degree $\leq r$.

\begin{rem}
In the particular case where all the $P_k$ are homogeneous of degree equal to $2$, the $g_{k_1,k_2}^{k_3}$ must be scalars, and the $H_k$ generate a graded Lie subalgebra. We then recover the standard definition of quantum Airy structures in \cite{KS}.
\end{rem}

\begin{rem}\label{r:properties}
It is easy to see the two distinctive properties of higher quantum Airy structures from the basis-dependent definition
\begin{enumerate}
\item The operators $H_k$ have a very specific form. There are exactly $D$ operators, and they all start with a linear term of the form $\hbar \partial_{x_k}$. This precise form is what is responsible for the uniqueness of the solution to the constraints $H_k \cdot Z = 0$, as we will see computationally by calculating the resulting topological recursion.
\item The operators satisfy the subalgebra property \eqref{eq:HAS}, which is crucial to ensure that a solution to the constraints $H_k \cdot Z = 0$ exists.
\end{enumerate}
\end{rem}

We can write down an explicit decomposition of the differential operators $H_i$ in monomials. To simplify notation, and anticipating further interpretations, we introduce the operators
\begin{equation}\label{eq:J}
J_l = \hbar\partial_{x_l}, \qquad J_{-l} = lx_l\qquad l \in I .
\end{equation}
We define a new index set $\mathcal{I} = \{-D , \ldots, -1, 1, \ldots, D\}$.

Let $d_k$ (possibly $\infty$) be the maximal degree in $H_k$. We can decompose
\begin{equation}
\label{eq:defC}
H_k = J_k - \sum_{m=2}^{d_k} \sum_{\substack{\ell,j \geq 0 \\ \ell + 2j = m}} \frac{\hbar^{j}}{\ell!} \sum_{\alpha \in \mathcal{I}^{\ell} } C^{(j)}[k| \alpha ]\,\norder{J_{\alpha_1} \cdots\, J_{\alpha_{\ell}}} \,,
\end{equation}
where $\norder{\cdots}$ denotes normal ordering, \emph{i.e.} all the $J_i$ with negative $i$s are on the left. The coefficients $C^{(j)}[k|\alpha]$ are fully symmetric under permutations of $\alpha=(\alpha_1, \ldots, \alpha_{\ell})$. By convention, the product $\norder{J_{\alpha_1}\cdots\, J_{\alpha_{\ell}}}$ is replaced by $1$ when $\ell = 0$.

\begin{rem}
Note that in the quantization framework introduced in Section \ref{S1}, the terms with $j=0$ correspond to the quantization of classical terms with normal ordering, while the terms with $j > 0$ arise as quantum ordering ambiguities.
\end{rem}

\begin{exa}
	To clarify the notation, let us compare with the notation used in \cite{ABCD} for quantum Airy structures, where $d_k=2$ for all $k \in I$. In this case we have
\begin{equation*}
\begin{split}
H_k & =  J_k - \tfrac{1}{2} \sum_{\alpha_1, \alpha_2 \in \mathcal{I}} C^{(0)}[k | \alpha_1, \alpha_2 ]\,\norder{J_{\alpha_1} J_{\alpha_2}} + \hbar\,C^{(1)}[k |\emptyset] \\
&=  \hbar \partial_{x_k} - \bigg(\tfrac{1}{2} \sum_{\alpha_1, \alpha_2 \in I} C^{(0)}[k | - \alpha_1, - \alpha_2 ]\,\alpha_1 \alpha_2 x_{\alpha_1} x_{\alpha_2} +  \sum_{\alpha_1, \alpha_2 \in I} C^{(0)}[k | - \alpha_1,  \alpha_2 ]\, \alpha_1 x_{\alpha_1}\,\hbar\partial_{x_{\alpha_2}}  \\
& \quad + \tfrac{1}{2}  \sum_{\alpha_1, \alpha_2 \in I} C^{(0)}[k | \alpha_1,  \alpha_2 ] \,\hbar^2\partial_{x_{\alpha_1}} \partial_{x_{\alpha_2}} + \hbar C^{(1)}[k |\emptyset]\bigg)\,.
\end{split}
\end{equation*}
	In the notation of \cite{ABCD}, we recognize the tensors 
\begin{equation*}
\begin{split}
C^{(0)}[k| -\alpha_1, - \alpha_2 ] &= \frac{A^k_{\alpha_1, \alpha_2}}{\alpha_1 \alpha_2}  \,, \\
C^{(0)}[k| - \alpha_1, \alpha_2] &= \frac{B^k_{\alpha_1, \alpha_2}}{\alpha_1} \,, \\
C^{(0)}[k| \alpha_1, \alpha_2] &= C^k_{\alpha_1, \alpha_2}\,, \\
C^{(1)}[k|\emptyset] &= D^k\,. 
\end{split}
\end{equation*}
\end{exa}

To a higher quantum Airy structure, we can associate a partition function due to the following key result of Kontsevich and Soibelman (see Theorem \ref{partif}).

\begin{thm} \cite[Theorem 2.4.2]{KS}\label{t:KSthm}
Given a higher quantum Airy structure $(H_k)_{k \in I}$, the system of equations
$$
\forall k \in I,\qquad H_k \cdot Z = 0\,,
$$
has a unique solution of the form
	\begin{equation}\label{eq:ZKS}
	Z = \exp\left(\sum_{\substack{g \geq 0,\,\,n \geq 1 \\ 2g - 2 + n > 0}} \frac{\hbar^{g-1}}{n!}  F_{g,n}\right)\,,\qquad F_{g,n} \in {\rm Sym}^{n}V^*\,.
	\end{equation}
\end{thm}

The existence of partition functions associated with higher quantum Airy structures and the fact that they often have enumerative geometric interpretations (see Section~\ref{Section6}) is essentially the reason why quantum Airy structures are interesting. We can decompose
$$
F_{g,n} =  \sum_{\alpha \in I^n} F_{g,n}[\alpha] x_{\alpha_1} \cdots\, x_{\alpha_n}\,,
$$
where $F_{g,n}[\alpha]$ is fully symmetric under permutations of $\alpha=(\alpha_1, \ldots, \alpha_n)$, and see the $F_{g,n}$ as generating series for the coefficients $F_{g,n}[\alpha]$, which are expected to have an interesting interpretation in enumerative geometry. By applying the differential operators $H_k$ on $Z$, we can obtain the $F_{g,n}[\alpha]$ by induction on $2g - 2 + n > 0$, as we now show explicitly.

\subsubsection{Recursive system}\label{s:recursive}

The set of constraints $H_k \cdot Z = 0$ can be turned into a recursive system for the $F_{g,n}[\alpha]$. Due to the specific form of the differential operators $H_k$ (see Remark \ref{r:properties}), this recursive system is always triangular. And because it is known that a solution to the constraints exists (see Theorem \ref{t:KSthm}), it follows that the recursive system uniquely determines this solution.

Let us  explicitly write down the recursive system satisfied by the $F_{g,n}[\alpha]$. Given a formal series $f$ in $\hbar$, we introduce the notation $[\hbar^g] f$ to denote the coefficient of $f$ of order $g$ in $\hbar$.

\begin{defin}
For $\alpha \in \mathcal{I}^i$ and $\beta \in I^{n - 1}$, we define
\begin{equation}\label{eq:xidef}
\Xi^{(i)}_{g,n}[\alpha| \beta] = [\hbar^{g}]\,\partial_{x_{\beta_1}}\cdots \,\partial_{x_{\beta_{n - 1}}}\big(Z^{-1}\, \norder{J_{\alpha_1} \cdots\, J_{\alpha_i}}\, Z\big) \Big |_{x = 0}\,.
\end{equation}
Notice that $\Xi^{(0)}_{g,n}[\emptyset|\beta] = \delta_{g,0}\delta_{n,1}$.
\end{defin} 
Then we have the following result.

\begin{lem}\label{l:rd}
The system of equations
$$
\forall k \in I,\qquad H_k\cdot Z = 0\,,
$$
implies the following system of equations
\begin{equation}\label{eq:xirec}
\Xi^{(1)}_{g,n}[k|\beta] = \sum_{m=2}^{d_k} \sum_{\substack{\ell,j \geq 0 \\ \ell + 2j = m}} \frac{1}{\ell!} \sum_{\alpha \in \mathcal{I}^{\ell}} C^{(j)}[k| \alpha]\,\Xi^{(\ell)}_{g-j,n}[\alpha | \beta]\,,
\end{equation}
for all $\beta \in I^{n - 1}$, $n \geq 0$, all $g \geq 0$, and all $k \in I$.
\end{lem}

\begin{proof}
Apply the differential operator $\partial_{x_{\beta_2}} \cdots\,\partial_{x_{\beta_n}}$ to $Z^{-1} H_k\cdot Z = 0$, set all $(x_l)_{l \in I}$ to zero and pick the coefficient of order $g $ in $\hbar$.
\end{proof}

We need some more notation. The coefficients $F_{g,n}[\alpha]$ were defined for $2g - 2 + n > 0$ and $\alpha \in I^n$ in \eqref{eq:ZKS}. We extend this definition to $\alpha \in \mathcal{I}^n$, by setting $F_{g,n}[\alpha] = 0$ whenever one of the $\alpha_{l}$ is negative. For $2g - 2 + n = 0$, we introduce
\begin{equation}
\label{F02conv} F_{0,2}[\alpha_1, \alpha_2] =|\alpha_1| \delta_{\alpha_1, - \alpha_2}\,.
\end{equation}

Let $\alpha \in \mathcal{I}^i$ and $\beta \in I^{n - 1}$. The notation $\boldsymbol{\lambda} \vdash \alpha$ means that $\boldsymbol{\lambda}$ is a set partition of $\alpha$, \textit{i.e.} a set of $|\boldsymbol{\lambda}|$ non-empty subsets of $\alpha$ which are pairwise disjoint and whose union is $\alpha$. We denote the elements (sets) of the partition $\boldsymbol{\lambda}$ generically by $\lambda$.  A partition of $\beta$ indexed by $\boldsymbol{\lambda}$ is a map $\boldsymbol{\mu}\,:\,\boldsymbol{\lambda} \rightarrow \mathfrak{P}(\beta)$ such that $(\mu_{\lambda})_{\lambda \in \boldsymbol{\lambda}}$ are possibly empty, pairwise disjoint subsets of $\beta$ whose union is $\beta$. We summarize this notion with the notation $\boldsymbol{\mu} \vdash_{\boldsymbol{\lambda}} \beta$.

Then we have the following result.

\begin{lem}\label{l:deriv}
Let $i,n \geq 1$. For $\alpha \in \mathcal{I}^i$ and $\beta \in I^{n - 1}$, we have
\begin{equation}\label{eq:derFgn}
\Xi^{(i)}_{g,n}[\alpha | \beta]= \sum_{\boldsymbol{\lambda} \vdash \alpha} \sum_{\substack{h\,:\,\boldsymbol{\lambda} \rightarrow \mathbb{N} \\ i + \sum_{\lambda \in \boldsymbol{\lambda}} h_{\lambda} = g + |\boldsymbol{\lambda}|}} \sum_{\boldsymbol{\mu} \vdash_{\boldsymbol{\lambda}} \beta}'' \bigg(\prod_{\lambda \in \boldsymbol{\lambda}} F_{h_{\lambda},|\lambda| + |\mu_{\lambda}|}[\lambda,\mu_{\lambda}]\bigg)\,,
\end{equation}
where the double prime over the summation symbol means that terms with $h_\lambda =0$, $|\mu_\lambda|=0$ and $|\lambda| \leq 2$ are excluded from the sum. In other words, $F_{0,1}$ does not appear in the sum, and $F_{0,2}$ only appears with $|\lambda|=1$ and $|\mu_\lambda|=1$.
\end{lem}

\begin{proof}
For $\alpha \in I^i$, \emph{i.e.} all $\alpha_l > 0$, the identity is straightforward. It involves $F_{0,2}$ only via positive indices, therefore such terms are zero. When some of the $\alpha_l$ are negative, we remember that $J_{\alpha_l} = |\alpha_l|x_{|\alpha_l|}$. Thus one of the $\beta_m$ must be $\beta_m = |\alpha_m|$, otherwise by definition (see \eqref{eq:xidef}) the contribution would be zero. We can include these cases by introducing coefficients $F_{0,2}[\alpha_l, \beta_m]$ that are equal to $|\alpha_l|$ when $\beta_m = - \alpha_l$, and zero otherwise. This is precisely how we defined the $F_{0,2}$ coefficients in \eqref{F02conv}. Thus the formula remains valid with these cases included, as long as the condition enforced by the double primed summation is there.
\end{proof}

\begin{exa}
To clarify the notation, let us write down explicitly what this expression looks like for $i = 1,2,3$.
\begin{equation*}
\begin{split}
\Xi^{(1)}_{g,n}[\alpha_1 | \beta] &= F_{g,n}[\alpha_1, \beta]\,, \\
\Xi^{(2)}_{g,n}[\alpha_1, \alpha_2 | \beta] &= F_{g-1,n+1}[\alpha_1, \alpha_2, \beta] + \sum_{\substack{h_1+h_2=g \\ \mu_1 \sqcup \mu_2 = \beta}}'' F_{h_1,1 + |\mu_1|}[\alpha_1, \mu_1] F_{h_2, 1 + |\mu_2|} [\alpha_2, \mu_2]\,. 
\end{split}
\end{equation*}
Note that the second line is not valid for $(g,n)=(1,1)$, in which case $\Xi^{(2)}_{1,1}[\alpha_1, \alpha_2 | \emptyset] = 0$ because of the double prime condition in the summation (\emph{i.e.} $F_{0,2}[\alpha_1,\alpha_2]$ cannot appear).

Further,
\begin{equation*}
\begin{split}
 & \quad \Xi^{(3)}_{g,n}[\alpha_1, \alpha_2, \alpha_3 | \beta] \\
 & =  F_{g - 2,n + 2}[\alpha_1,\alpha_2,\alpha_3,\beta] + \sum_{\substack{h_1 + h_2 = g - 1 \\ \mu_1 \sqcup \mu_2 = \beta}}^{''} \left(F_{h_1,1 + |\mu_1|}[\alpha_1,\mu_1] F_{h_2,2+|\mu_2|}[\alpha_2,\alpha_3,\mu_2]  \right. \\
 & \quad \left. + F_{h_1,1 + |\mu_1|}[\alpha_2,\mu_1]F_{h_2,2+|\mu_2|}[\alpha_1,\alpha_3,\mu_2] + F_{h_1,1 + |\mu_1|}[\alpha_3,\mu_1]F_{h_2,2 + |\mu_2|}[\alpha_1,\alpha_2,\mu_2] \right) \\
 & \quad + \sum_{\substack{h_1 + h_2 + h_3 = g \\ \mu_1 \sqcup \mu_2 \sqcup \mu_3 = \beta}}^{''} F_{h_1,1 + |\mu_1|}[\alpha_1,\mu_1]F_{h_2,1+|\mu_2|}[\alpha_2,\mu_2] F_{h_3,1+|\mu_3|}[\alpha_3,\mu_3]\,.
 \end{split}
 \end{equation*}
\end{exa}

Substituting \eqref{eq:derFgn} back into \eqref{eq:xirec}, we get the following formula for the coefficients $F_{g,n}[\alpha]$.
\begin{cor} \label{cor28} For all $\beta \in I^{n - 1}$ we have
\begin{equation}
\label{fgnrec} F_{g,n}[k,\beta] = \sum_{\substack{\ell,j \geq 0 \\ 2 \leq \ell + 2j \leq d_k}} \frac{1}{\ell!} \sum_{\alpha \in \mathcal{I}^{\ell}} C^{(j)}[k | \alpha] \sum_{\boldsymbol{\lambda} \vdash \alpha} \sum_{\substack{h\,:\,\boldsymbol{\lambda} \rightarrow \mathbb{N} \\ \ell + j + \sum_{\lambda \in \boldsymbol{\lambda}} h_{\lambda} = g + |\boldsymbol{\lambda}|}} \sum_{\boldsymbol{\mu} \vdash_{\boldsymbol{\lambda}} \beta}'' \bigg(\prod_{\lambda \in \boldsymbol{\lambda}} F_{h_{\lambda},|\lambda| + |\mu_{\lambda}|}[\lambda,\mu_{\lambda}]\bigg)\,.
\end{equation}
\end{cor}

Let us now argue that Corollary \ref{cor28} is a recursive system for the $F_{g,n}[\alpha]$. For each term in the right-hand side, using the constraints under the sums we get
$$
\sum_{\lambda} \big(2h_{\lambda} - 2 + |\lambda| + |\mu_{\lambda}|\big) = 2\big(g + |\boldsymbol{\lambda}| - (\ell + j)\big) - 2|\boldsymbol{\lambda}| + \ell + n - 1 = (2g - 2 + n) + (1 - \ell - 2j)\,.
$$
Since we have $\ell + 2j \geq 2$, we deduce that
\beq
\label{genuscond} \sum_{\lambda} \big(2h_{\lambda} - 2  + |\lambda| + |\mu_{\lambda}|\big) < 2g - 2 + n\,.
\eeq
Since the $F_{0,1}$ terms are absent, all terms in the left-hand side of the inequality are non-negative, hence $2h_{\lambda} - 2 + |\lambda| + |\mu_{\lambda}| < 2g - 2 + n$ for each $\lambda \in \boldsymbol{\lambda}$. In other words,  \eqref{fgnrec} is a recursion on $2g - 2 + n \geq 0$ determining uniquely $F_{g,n}$ starting from the value of $F_{0,2}$ given by \eqref{F02conv}. For instance, the formula gives for $2g - 2 + n = 1$
\begin{equation}
\label{F03AAA}
\begin{split}
 F_{0,3}[k,\beta_1,\beta_2] & =   \beta_1\beta_2\,C^{(0)}[k|-\beta_1,-\beta_2]\,,  \\
F_{1,1}[k] & =  C^{(1)}[k|\emptyset]\,,
\end{split}
\end{equation}
and for $2g - 2 + n = 2$
\begin{equation}
\label{F04AAA} 
\begin{split}
F_{0,4}[k,\beta_1,\beta_2,\beta_3] & = \beta_1\beta_2\beta_3\,C^{(0)}[k|-\beta_1,-\beta_2,-\beta_3] + \sum_{\alpha \in I} \big(\beta_1\,C^{(0)}[k|-\beta_1,\alpha]\,F_{0,3}[\alpha,\beta_2,\beta_3] \\
& \quad + \beta_2\,C^{(0)}[k|-\beta_2,\alpha]\,F_{0,3}[\alpha,\beta_1,\beta_3] + \beta_3\,C^{(0)}[k|-\beta_3,\alpha]\,F_{0,3}[\alpha,\beta_1,\beta_2]\big)\,,
\end{split}
\end{equation}
and
\begin{equation}
\label{F12AAA}
\begin{split}
F_{1,2}[k,\beta] & =  \sum_{\alpha \in I} \beta\,C^{(0)}[k|-\beta,\alpha]\,F_{1,1}[\alpha] + \sum_{\alpha_1,\alpha_2 \in I} \tfrac{1}{2}\,C^{(0)}[k|\alpha_1,\alpha_2]\,F_{0,3}[\beta,\alpha_1,\alpha_2]  \\
& \quad + \beta\,C^{(1)}[k|-\beta]\,.
\end{split}
\end{equation}

\begin{rem}
While \eqref{fgnrec} is recursive, it does not treat $k$ and $\beta_1,\ldots,\beta_{n - 1}$ in a symmetric fashion. In other words, it is not clear from \eqref{fgnrec} that the $F_{g,n}[k,\beta]$ thus constructed are fully symmetric. It could happen that no symmetric solution to \eqref{fgnrec} exists. That is, the recursive system does not justify the existence part of Theorem~\ref{t:KSthm}; it does however imply uniqueness if a solution exists. In fact symmetry cannot hold for general coefficients $C$s. The graded subalgebra property of $(H_k)_{k \in I}$ --- which implies nonlinear relations between the $C$s --- is essential in proving the existence of a solution $Z$ to the constraints, which is equivalent to proving the existence of a symmetric solution to \eqref{fgnrec}.
\end{rem}

\subsubsection{Reduction}\label{sec:reduction}

In general a higher quantum Airy structure $(H_k)_{k \in I}$ may involve linear differential operators. In this section we argue that we can essentially get rid of the linear differential operators. Note that this section is not essential for the rest of the paper.

Let $(H_k)_{k \in I}$ be a higher quantum Airy structure. Assume that $I_{{\rm lin}} \subset I$ is such that $H_k = J_{k}$ for all $k \in I_{{\rm lin}}$. For any $k \in I$, we introduce the reduced differential operator $H_k|_{\rm red}$, which is obtained from $H_k$ by formally setting $J_{m} = 0$ (in the normal-ordered expression for $H_{k}$) whenever $|m| \in I_{{\rm lin}}$. Note that $H_i|_{\rm red} = 0$ for all $i \in I_{{\rm lin}}$. We can think of the $H_k|_{\rm red}$ as differential operators on $V$ or on its subspace
$$
V_{{\rm red}} =\big\{ x \in V\quad | \quad \forall m \in I_{{\rm lin}},\,\,\,x_{m} = 0\big\} .
$$

\begin{lem}
\label{lem:redd} There exists a unique solution to the differential constraints $H_k|_{\rm red} Z = 0$. Moreover, the partition function $Z$, considered as a formal function on $V$, coincides with the unique solution to the differential constraints $H_k Z = 0$.
\end{lem}

In other words, if we are interested in calculating $Z$, we can forget about the linear differential constraints $H_i$ for $i \in I_{{\rm lin}}$, and instead solve the reduced differential constraints $H_k|_{\rm red} Z = 0$ on $V_{{\rm red}}$.

\begin{proof}
Let $\mathcal{J}$ be the left ideal generated by the $H_k$, and let $Z$ be the unique solution to the differential constraints $H_k Z = 0$. It is straightforward to show inductively that for all $k \in I$, $H_k|_{\rm red} \in \mathcal{J}$. Thus, $H_k|_{\rm red} Z = 0$, and hence $Z$ is also a solution to the reduced differential constraints.

To show that it is unique, we look at the form of the differential operators. First, we know that $H_k Z = J_k Z = 0$ for all $k \in I_{\rm lin}$, so $Z$ does not depend on those $x_k$. It follows that $Z$ depends on the same number of variables as the number of non-zero $H_k|_{\rm red}$. Moreover, it is clear that the non-zero $H_k|_{\rm red}$ satisfy the degree 1 condition of quantum higher Airy structures with respect to these variables. Together those imply that the differential constraints $H_k|_{\rm red} Z = 0$ uniquely reconstruct the coefficients $F_{g,n}[\alpha]$ of the partition function by topological recursion. It follows that the solution is unique.
\end{proof}

What we have proven is that there always exists a unique solution to the reduced differential constraints $H_k|_{\rm red} Z = 0$, and that this solution coincides with the unique partition function of the higher quantum Airy structure $(H_k)_{k \in I}$. It is tempting to conclude that the $H_k|_{\rm red}$s thus also form a higher quantum Airy structure. But to claim that we would need to show that the left ideal generated by the reduced $H_k|_{\rm red}$ is a graded Lie subalgebra of the algebra of differential operators on $V_{{\rm red}}$. While we expect this to be true and we prove it in a special case (Lemma~\ref{consred}), we do not  have a complete proof of this fact currently. 

\subsection{Crosscapped Airy structures}
\label{Scrosscap}
A variant of the topological recursion involving $F_{g,n}$ for half-integer $g$ is required in applications to large size expansions in $\beta$-matrix integrals \cite{CE06} and to open intersection theory \cite{Safnukopen,Alexandrovopen}. We can include this variant in the formalism of Airy structures by allowing half-integer powers of $\hbar$, \textit{i.e.} a formal variable $\hbar^{1/2}$ of degree $1$, as follows.

\begin{defin}
A \emph{crosscapped higher quantum Airy structure} on a vector space $V$ equipped with a basis of linear coordinates $(x_k)_{k \in I}$ is a family of differential operators indexed by $k \in I$ of the form $H_k = \hbar\,\partial_{x_{k}} - P_k$ where the terms in $P_k \in \mathcal{D}_{T^*V}^{\hbar^{1/2}}$ have degree $\geq 2$. Moreover, we require that the left $\mathcal{D}_{T^*V}^{\hbar^{1/2}}$-ideal generated by the $H$s forms a graded Lie subalgebra \textit{i.e.} there exists $g_{k_1,k_2}^{k_3} \in \mathcal{D}_{T^*V}^{\hbar^{1/2}}$ such that
$$
\forall k_1,k_2 \in I,\qquad [H_{k_1},H_{k_2}] = \sum_{k_3 \in I} g_{k_1,k_2}^{k_3} H_{k_3}\,.
$$
\end{defin}
The degree condition means that we have a decomposition
$$
H_k = J_{k} -  \sum_{m \geq 2} \sum_{\substack{\ell,\jmath \geq 0 \\ \ell + \jmath = m}} \frac{\hbar^{\jmath/2}}{\ell!} \sum_{\alpha \in \mathcal{I}^{\ell}} C^{(\jmath/2)}[k|\alpha]\,\,\,\norder{J_{\alpha_1}\cdots\,J_{\alpha_{\ell}}}\,.
$$

\begin{prop}
\label{imt}Given  a crosscapped higher quantum Airy structure $(H_k)_{k \in I}$, the system of equations
\beq
\label{gofdugn1111} \forall k \in I,\qquad H_k\cdot Z = 0\,,
\eeq
has a unique solution of the form
\beq
\label{Zcap} Z = \exp\left(\sum_{\substack{g \in \mathbb{N}/2,\,\,n \geq 1 \\ 2g - 2 + n > 0}} \frac{\hbar^{g - 1}}{n!}\,F_{g,n}\right)\,,\qquad F_{g,n} \in {\rm Sym}^n(V^*)\,.
\eeq
given by the recursive system \eqref{fgnrec} where one allows half-integer genera.
\end{prop}

\begin{proof} The proof of existence is a small adaptation of the proof of \cite{KS} and therefore omitted. To prove uniqueness, we repeat the arguments of Section~\ref{s:recursive} to show that \eqref{gofdugn1111} computes the $F_{g,n}$  inductively on $2g - 2 + n > 0$. In fact, this recursive system takes the form \eqref{fgnrec} except that $j$, $g$ and $h_{\lambda}$ can be nonnegative integers or half-integers (but note that $2g - 2 + n$ is always an integer).  The condition \eqref{genuscond} is still valid and implies, as there are no $F_{g_0,n_0}$ with $2g_0 - 2 + n_0 < 0$ in \eqref{Zcap}, that this recursive system determines uniquely all $F_{g,n}$ from the value of $F_{0,2}$ specified by the convention \eqref{F02conv}.
\end{proof}

It is perhaps instructive to write down the value of $F_{g,n}$ given by the recursion. In fact this also gives for $g = 0$ a recursion on $n$, therefore the formula for the $F_{0,n}$ are the same as those of Section~\ref{s:recursive}. Notice that $F_{1/2,1}$ is absent from \eqref{Zcap}. With $2g - 2 + n = 1$, we have a new term $F_{1/2,2}$ while the formulae \eqref{F03AAA} remain unchanged
\begin{equation*}
\begin{split}
F_{0,3}[k,\beta_1,\beta_2] & =  \beta_1\beta_2\,C^{(0)}[k|-\beta_1,-\beta_2]\,,  \\
F_{1/2,2}[k,\beta] & =  \beta\,C^{(1/2)}[k|-\beta]\,,  \\
F_{1,1}[k] & =  C^{(1)}[k|\emptyset] \,.
\end{split}
\end{equation*}
With $2g - 2 + n = 2$, $F_{1,2}$ receives a new contribution compared to \eqref{F12AAA} and we have two new terms with half-integer genus
\begin{equation*}
\begin{split}
F_{0,4}[k,\beta_1,\beta_2,\beta_3] & =  \beta_1\beta_2\beta_3\,C^{(0)}[k|-\beta_1,-\beta_2,-\beta_3] + \sum_{\alpha \in I} \big(\beta_1\,C^{(0)}[k|-\beta_1,\alpha]\,F_{0,3}[\alpha,\beta_2,\beta_3] \\
& \quad + \beta_2\,C^{(0)}[k|-\beta_2,\alpha]\,F_{0,3}[\alpha,\beta_1,\beta_3] + \beta_3\,C^{(0)}[k|-\beta_3,\alpha]\,F_{0,3}[\alpha,\beta_1,\beta_2]\big)\,,
\end{split}
\end{equation*}
\begin{equation*}
\begin{split}
F_{1/2,3}[k,\beta_1,\beta_2] & =  \beta_1\beta_2\,C^{(1/2)}[k|-\beta_1,-\beta_2] + \sum_{\alpha \in I} C^{(1/2)}[k|\alpha]\,F_{0,3}[\alpha,\beta_1,\beta_2] \\
& \quad + \sum_{\alpha \in I}\big(\beta_1\,C^{(0)}[k|-\beta_1,\alpha]\,F_{1/2,2}[\alpha,\beta_2] + \beta_2\,C^{(0)}[k|-\beta_2,\alpha]\,F_{1/2,2}[\alpha,\beta_1]\big)\,,
\end{split}
\end{equation*}
\begin{equation*}
\begin{split}
F_{1,2}[k,\beta] & =  \beta\,C^{(1)}[k|-\beta] + \sum_{\alpha \in I} \beta\,C^{(0)}[k|-\beta,\alpha]\,F_{1,1}[\alpha]  \\
& \quad + \sum_{\alpha_1,\alpha_2 \in I} \tfrac{1}{2}\,C^{(0)}[k|\alpha_1,\alpha_2]\,F_{0,3}[\beta,\alpha_1,\alpha_2]\,, \\
F_{3/2,1}[k] & =  C^{(3/2)}[k|\emptyset] + \sum_{\alpha \in I} \big(C^{(1)}[k|\alpha]\,F_{1/2,1}[\alpha] + C^{(1/2)}[k|\alpha]\,F_{1,1}[\alpha]\big) \\
& \quad  + \sum_{\alpha_1,\alpha_2 \in I}  \tfrac{1}{2}\,C^{(0)}[k|\alpha_1,\alpha_2]\,F_{1/2,2}[\alpha_1,\alpha_2]\,.
\end{split}
\end{equation*}

\newpage

\section{\texorpdfstring{$\mathcal{W}$}{W} algebras and twisted modules}
\label{s:WW}

Our  main construction of higher quantum Airy structures will take the form of $\mathcal{W}$ constraints for some particular modules of $\mathcal{W}$ algebras. $\mathcal{W}$ algebras are vertex operator algebras (VOAs), and hence we introduce some terminology and notation about VOAs and modules over them. 

We are primarily interested in Heisenberg VOAs and $\mathcal{W} $ algebras in this paper. From a conformal field theory point of view, $\mathcal{W}$ algebras arise as the algebra of modes when the CFT includes chiral primary fields of conformal weight $> 2$. Algebraically, they are certain ``non-linear" extensions of the Virasoro algebra; the first examples were constructed in \cite{Z}.

To obtain higher quantum Airy structures we need to construct particular modules for these VOAs. Those will always be obtained by restriction of twisted modules of  Heisenberg VOAs to  $ \mathcal{W} $ algebras. In order to construct a twisted module, we essentially construct fields that have fractional power expansions in formal variables. From the point of view of conformal field theories, these correspond to choosing a branch in the orbifold VOA. 

In this section we introduce VOAs and twisted modules. Along the way we construct a number of interesting left ideals for the algebra of modes of $\mathcal{W}$ algebras that are graded Lie subalgebras. This will prove crucial in the next section to construct higher quantum Airy structures.

\subsection{Vertex operator algebras}
\label{s:VOA}

There are many references on this topic. We mostly follow the presentation of \cite{BakalovMilanov2,Doyon,FLM,FB}. 

\begin{defin}\label{voadef}
	A \emph{vertex operator algebra} (VOA) is a quadruple $ (V,Y, \ket{0}, \ket{w}) $ such that
	\begin{itemize}
		\item $V$ is a $\mathbb{Z}$-graded vector space (the \emph{space of states}) $V = \oplus_{l \in \mathbb{Z}} V_l$ such that $V_l=0$ for $l$ sufficiently negative and $\dim V_l < \infty$ for all $l \in \mathbb{Z}$. If $\ket{v} \in V_l$, we say that the conformal weight of $\ket{v}$ is $l$.
		\item $Y$ is a linear map (the \emph{state-field correspondence})
		$$
		Y(\cdot, z)\,:\, \begin{array}{lcl} V & \longrightarrow & {\rm End}(V)[\![z, z^{-1} ]\!] \\
	\ket{v} & \longmapsto & Y(\ket{v},z) = \sum_{l \in \mathbb{Z}} v_l z^{-l-1} \end{array}\,.
		$$
		$Y(\ket{v}, z)$ is called the \emph{vertex operator} (or \emph{field}) associated to the state $\ket{v}$, and $v_n$ its \emph{modes}.
		\item $\ket{0} \in V$ is the \emph{vacuum state}, which satisfies the vacuum property
		$$
		Y(\ket{0},z) = {\rm id}_V
		$$
		and the creation property 
		$$
		\forall \ket{v} \in V,\qquad Y(\ket{v},z) \ket{0} - \ket{v} \in zV[\![z]\!]\,.
		$$
		\item $\ket{w} \in V$ is the \emph{conformal state}, which satisfies the truncation condition
		$$
		\forall \ket{v} \in V, \qquad v_l \ket{w} = 0 \qquad \text{for $l \in \mathbb{Z}$ sufficiently positive,}
		$$
		and the Virasoro algebra condition, which can be stated as follows. Let $\omega_n$ be the modes of $Y(\ket{w},z)$, and define $L_l = \omega_{l+1}$. Then
		$$
		[L_l, L_m] = (l-m) L_{l+m} + \mathfrak{c}\,\frac{l^3-l}{12} \delta_{l+m,0} {\rm id}_V,
		$$
		where $\mathfrak{c} \in \mathbb{C}$ is the central charge. Further, if $\ket{v}$ is homogeneous of conformal weight $n$, then $L_0 \ket{v} = n \ket{v}$ and we have the derivation property
		$$
		\forall \ket{u} \in V,\qquad Y\left( L_{-1} \ket{u}, z \right) = \frac{{\rm d}}{{\rm d}z} Y(\ket{u}, z).
		$$
		\item 
		Finally, we have the axiom of \textit{locality}.  $\big(Y(\ket{v},z)\big)_{v \in V}$ is a local family of fields; \textit{i.e.}, for $\ket{u},\ket{v} \in V $,
		$$
		(z_1-z_2)^{N_{u,v}} [Y(\ket{u},z_1),Y(\ket{v},z_2)] = 0 \qquad \text{for some } N_{u,v} \in \mathbb{Z}_+,
		$$
		Although innocuous looking, this axiom gives the vertex operator algebra much of its structure. In particular, this is equivalent to the Jacobi identity/Borcherds identity.
			\end{itemize}
\end{defin}

We will often drop the $ Y, \ket{0}$ and  $\ket{w} $ in the definition of a VOA and merely denote it by the underlying space of states $ V $. We note that the mode $ L_0 $ keeps track of the conformal weight of the states.

As the vertex algebra is (usually) non-commutative, we define the notion of normal ordering

\begin{defin}
	We define the \textit{normally ordered product} of two fields $ Y(\ket{u},z) $ and $ Y(\ket{v},z) $ as the following
$$
\norder{ Y(\ket{u},z)Y(\ket{v},w)}  = Y(\ket{u},z)_{-}Y(\ket{v},w) +  Y(\ket{v},w) Y(\ket{u},z)_{+},
$$
where we defined $ Y(\ket{w},z)_+ := \sum_{l >0} w_l z^{-l-1} $ and $ Y(\ket{w},z)_- := \sum_{l \leq 0} w_l z^{-l-1} $.
\end{defin}

\subsection{\texorpdfstring{$\mathcal{W}(\mathfrak{g})$}{W(g)} algebras}
\label{s:WG}

There are various equivalent constructions of $ \mathcal{W}$-algebras. They are defined as the semi-infinite cohomology of affine vertex algebras of level $k\in\mathbb C$ \cite{FF90} associated to a Lie algebra $ \mathfrak{g} $. For generic $k$, they are isomorphic to certain intersections of kernels of screening operators on free field/Heisenberg algebras \cite{FF90, FB, Genra17}, and for the principal $\mathcal{W}$ algebras of simply-laced type there is also a coset realization  \cite{ACL}. Both the coset and screening realizations admit a certain limit where the $\mathcal W$ algebra is described as an orbifold by the compact Lie group $G$ of the Lie algebra $\mathfrak g$. This is the situation we are interested in. In this case, the $\mathcal W$ algebra is a subalgebra of the Heisenberg vertex algebra of rank equal to the rank of  $\mathfrak g$. For $ \mathcal{W} $ algebras of type $ \mathfrak{gl}_{N+1} $, we can also use the quantum Miura transformation, which gives us explicit generators.

We now construct our first example of a VOA, the Heisenberg VOA. Then we explain the construction of $\mathcal{W}$ algebras as subalgebras of the Heisenberg VOA.

\subsubsection{Heisenberg vertex operator algebras }

Let $L$ be a lattice of finite rank equipped with a symmetric non-degenerate bilinear form
$$
\braket{\cdot,\cdot}\,:\,L \times L \to \mathbb{Z}
$$
Define $ \mathfrak{h}:= L \otimes_{\mathbb{Z}} \mathbb{C} $. The bilinear form on $ L $ induces a bilinear form on $ \mathfrak{h} $. We define the \textit{Heisenberg Lie algebra}  $\hat{\mathfrak{h}} $ as the affine Lie algebra
\begin{equation}
\hat{\mathfrak{h}} = \left(\bigoplus_{l \in \mathbb{Z}} \mathfrak{h} \otimes t^l \right) \oplus \mathbb{C}K\,,
\end{equation}
with Lie bracket relations
\begin{equation}
\label{eq:Hlie}
\begin{split}
[\xi_{l} , \eta_{m}] =&  \braket{\xi,\eta} l \delta_{l+m,0} K, \qquad \xi,\eta \in \mathfrak{h}, \quad l,m \in \mathbb{Z}\,, \\
[K,\hat{\mathfrak{h}}] =& 0\,,
\end{split}
\end{equation}
where we introduced the notation $ \xi_{l} := \xi \otimes t^l$, $l \in \mathbb{Z} $ for any $ \xi \in \mathfrak{h} $. 

We define the Weyl algebra $ \mathcal{H}_L $ as the universal enveloping algebra of $ \hat{\mathfrak{h}} $ quotiented by the relation $ K = 1 $. We also define a class of modules over $ \mathcal{H}_L $ called Fock modules as follows. For any $ \lambda \in \mathfrak{h} $, define the Fock module $\mathcal{S}_\lambda $ as the $ \mathcal{H}_L $-module generated by the vector $ \ket{\lambda} $, such that for any $ \xi \in \mathfrak{h} $,
$$
\forall l > 0,\qquad \xi_{l} \ket{\lambda}  = 0\,,  \qquad {\rm and}\qquad  \xi_{0} \ket{\lambda}  = \braket{\xi,\lambda} \ket{\lambda}\,.
$$

If we define $ \mathcal{H}_L^-$ as the subalgebra of $ \mathcal{H}_L $ generated by the negative elements \hbox{$ \{\xi_{l}\,\,|\,\,\xi \in \mathfrak{h},\,\,\,l < 0\}$},  we have the isomorphism $\mathcal{S}_\lambda \cong {\rm Sym}(\mathcal{H}_L^-) \ket{\lambda}$ as vector spaces.

The Fock module $\mathcal{S}_0 \cong {\rm Sym}(\mathcal{H}_L^-) \ket{0}$ admits a vertex operator algebra structure, by which we mean that we can find a quadruple $(\mathcal{S}_0,Y,\ket{0},\ket{w}) $, that satisfies the axioms of Definition~\ref{voadef}. The vacuum state is  $\ket{0}$. The state-field correspondence $Y(\cdot,z)\,:\,\mathcal{S}_0 \to \text{End}(\mathcal{S}_0)[\![z,z^{-1}]\!]$ is defined as
\begin{equation}
 \label{eq:bfields} 
 \begin{split}
Y(\ket{0},z ) &= \text{id}_{\mathcal{S}_0}\,,\\
\forall \xi \in \mathfrak{h},\qquad Y(\xi_{-1} \ket{0}, z) &=  \sum_{l \in \mathbb{Z}} \xi_{l} z^{-l-1}\,.
\end{split}
\end{equation}

States of the form $ \xi^1_{-k_1} \cdots\, \xi^n_{-k_n} \ket{0} $ where $ k_i >0 $ clearly span $ \mathcal{S}_0 $, and the state-field correspondence is defined as
\begin{equation} \label{eq:sfh}
Y(\xi^1_{-k_1} \cdots\, \xi^n_{-k_n} \ket{0},z) = \norder{\frac{1}{(k_1-1)!}\frac{{\rm d}^{k_1-1}}{ {\rm d} z^{k_1-1}} Y(\xi^1_{-1} \ket{0}, z)\,\cdots\, \frac{1}{(k_n-1)!}\frac{{\rm d}^{k_n-1}}{ {\rm d}z^{k_n-1}} Y(\xi^n_{-1} \ket{0}, z)   }\,.
\end{equation}
Finally, if we pick an orthonormal basis $\bar{\xi}^1, \ldots, \bar{\xi}^d$ for $\mathfrak{h}$, we define the conformal vector $\ket{\omega}$ as
\begin{equation}
\ket{\omega} = \frac{1}{2} \sum_{i=1}^d \bar{\xi}^i_{-1} \bar{\xi}^i_{-1} \ket{0}\,.
\end{equation}
Its modes form a Virasoro algebra with central charge $ \mathfrak{c} = \text{dim } \mathfrak{h} = \text{rank } L $. It can be checked that those satisfy the axioms of a VOA.

\begin{defin} We denote the \textit{Heisenberg vertex operator algebra} associated to $ \mathfrak{h}$ by $\mathcal{S}_{0}$.
\end{defin}

\subsubsection{Lattice vertex operator algebras}

From the previous section, one can naturally define the lattice vertex operator algebra associated to $L$, which contains the Heisenberg VOA as a sub-VOA.

The underlying vector space of the lattice VOA is $\mathcal{V}_L := \bigoplus_{\lambda \in L} \mathcal{S}_\lambda $ (recall that $ \mathcal{S}_\lambda $ are the Fock modules defined in the previous section). In particular $\mathcal{S}_{0} \subset \mathcal{V}_L $, and we define the vacuum state $ \ket{0} $ and the conformal state $ \ket{\omega} $  as the ones for the Heisenberg VOA $\mathcal{S}_0$. The state-field correspondence defined earlier ~\eqref{eq:bfields} also holds. It suffices to define the state-field correspondence for the states $ \ket{\lambda}$. (The general prescription is obtained by taking normally ordered products as in ~\eqref{eq:sfh}.) We have\footnote{$V_{\lambda}(z)$ is the standard notation for these operators, here we use bold letters not to confuse them with vector spaces of VOAs also denoted $V$ elsewhere in the text.}
$$
\mathbf{V}_{\lambda}(z) := Y(\ket{\lambda},z) = U_\lambda z^{\lambda_0} \exp\left(-\sum_{l<0} \frac{\lambda_l}{l} z^{-l}\right) \exp\left(-\sum_{l>0} \frac{\lambda_l}{l} z^{-l}\right),
$$
where $ U_\lambda $ is a shift operator
$$
U_\lambda \ket{\nu} = c_{\lambda,\nu} \ket{\nu+\lambda}\qquad {\rm and}\qquad [U_\lambda,\lambda_n] = 0, \ n \neq 0\,,
$$
and $ c_{\lambda,\nu} \in \mathbb{C}^{\times} $ is a (essentially) unique $ 2 $-cocycle. We will also denote the state $\ket{\lambda}$ by ${\rm e}^{\lambda}$.

\begin{defin}
	We denote the \textit{lattice vertex operator algebra} associated to the even lattice $L$ by $\mathcal{V}_{L}$.
\end{defin}
If $L=Q$ is the root lattice of a simple simply-laced Lie algebra $\mathfrak{g}$ then $\mathcal V_Q$ is isomorphic to the simple affine vertex algebra of 
	$\mathfrak{g}$ at level one and is also denoted by $L_1(\mathfrak g)$.

\subsubsection{The \texorpdfstring{$\mathcal{W}(\mathfrak{g})$}{W(g)} algebras }

A standard introduction to $\mathcal W$ algebras is \cite{Arakawa2017}. Let $\mathfrak{g}$ be a simple finite-dimensional Lie algebra. Then to each embedding of $\mathfrak{sl}_2$ in $\mathfrak{g}$ one can associate the $\mathcal{W}$ algebra of $\mathfrak{g}$ at level $k \in \mathbb C$ via quantum Hamiltonian reduction from the affine vertex algebra of $\mathfrak{g}$ at level $k$. The best-known case is the one of the principal embedding of $\mathfrak{sl}_2$ in $\mathfrak{g}$, which we will simply denote by $\mathcal{W}^k(\mathfrak{g})$. Let now $\mathfrak{g}$ be simply-laced. In this case the principal $\mathcal{W}$ algebra can also be realized as a coset \cite[Main Theorem 2]{ACL}, that is, for generic $k$
\[
\mathcal{W}^\ell(\mathfrak g) \cong \left(V_k(\mathfrak g) \otimes L_1(\mathfrak g)   \right)^{\mathfrak{g}[t]}\,, \qquad \ell = -h^\vee + \frac{k+h^\vee}{k+h^\vee+1}\,,
\]
with $h^\vee$ the dual Coxeter number of $\mathfrak g$, and $V_k(\mathfrak g)$ the universal affine vertex algebra of $\mathfrak g$ at level $k$ and $L_1(\mathfrak g)$ its simple quotient at level one. Let $G$ be the compact Lie group whose Lie algebra is $\mathfrak{g}$. In the limit $k\rightarrow \infty$ this coset becomes just the $G$-orbifold of the lattice vertex algebra \cite{CL} and this is the case we are interested in
\[
\mathcal{W}(\mathfrak{g}) := \mathcal{W}^{-h^\vee +1}(\mathfrak g) \cong  L_1(\mathfrak g)^G\,.
\]
$\mathcal{W}^\ell(\mathfrak g)$ and in particular 
$\mathcal{W}(\mathfrak{g}) $ is strongly generated by elements $ W^i $ of conformal weights $ d_i+1 $, where the $ d_i $ are the Dynkin exponents of $ \mathfrak{g}$, see for example \cite[Theorem 15.1.9]{FB}. For generic level it is also freely generated by these fields and the orbifold limit is always a generic point of a deformable family of vertex algebras by \cite{CL}.
\begin{rem}
In summary, the principal $\mathcal{W}$ algebras form a one-parameter family of vertex algebras and we are interested in a very special point, namely the level for which the $\mathcal{W}$ algebra can be realized as a $G$-orbifold inside the lattice vertex algebra (for this $\mathfrak g$ needs to be simply-laced). This level is special for a second reason. $\mathcal{W}$ algebras enjoy Feigin-Frenkel duality \cite{FF91} and our level is the self-dual case, \textit{i.e.} $\mathcal{W}(\mathfrak{g})$ is its own Feigin-Frenkel dual. 
\end{rem}

For completeness we recall the definition of strong generators for a vertex operator algebra:
\begin{defin}
	A vertex operator algebra $V$ is said to be \textit{strongly generated} by elements $(\gamma^i)_{i = 1}^n$ in $V$ if the underlying vector space $ V $ is spanned by
	$$
	\gamma^1_{-k_1}\,\cdots\, \gamma^n_{-k_n} \ket{0}\, , \qquad \text{where } k_i>0\,.
	$$
	In addition, $ V $ is said to be \textit{freely generated} if the above spanning set is a basis for the underlying vector space $ V $.
\end{defin} 

\begin{rem}

If we know the state-field correspondence for the set of strong generators of a vertex operator algebra $ V $,  say \hbox{$\gamma^i = \gamma^i_{-1} \ket{0} $}, we can use the strong reconstruction theorem \cite[Theorem 4.4.1]{FB} to determine the state-field correspondence for the states $ \gamma^1_{-k_1} \cdots\, \gamma^n_{-k_n} \ket{0} $ where  $ k_i>0 $
\begin{equation} 
\label{productfff} Y(\gamma^1_{-k_1} \cdots\, \gamma^n_{-k_n} \ket{0},z) = \norder{\frac{1}{(k_1-1)!}\frac{{\rm d}^{k_1-1}}{ {\rm d} z^{k_1-1}} Y(\gamma^1_{-1} \ket{0}, z)\,\cdots\, \frac{1}{(k_n-1)!}\frac{{\rm d}^{k_n-1}}{ {\rm d}z^{k_n-1}} Y(\gamma^n_{-1} \ket{0}, z)   }\,.
\end{equation}
Hence, we can interpret strong generation as the statement that all fields of the VOA can be obtained as linear combinations of normally ordered products of the fields $ Y(\gamma^i,z) $ where $ i \in \{1,\ldots,n\}$ and their derivatives. 
\end{rem}

\subsubsection{Examples}
\label{sec:examples}
Let us now study some examples of $\mathcal{W}(\mathfrak{g})$ algebras.

\begin{exa}
	The algebra $\mathcal{W}(\mathfrak{sl}_2) $ is isomorphic to the Virasoro vertex algebra with central charge $ \mathfrak{c} =1 $. It is well known that this VOA is strongly generated by a single vector of conformal weight $ 2 $.
\end{exa}

Strictly speaking, we only defined $ \mathcal{W}$ algebras for simple and simply-laced Lie algebras. It is straightforward to construct $ \mathcal{W}$ algebras for direct sums of those. In particular, in the following we will study the algebra $\mathcal{W}( \mathfrak{gl}_{N+1})  := \mathcal{W}(\mathfrak{sl}_{N+1})  \otimes \mathcal S_0$, defined
as the tensor product of $\mathcal{W}(\mathfrak{sl}_{N+1}) $ and a rank one Heisenberg vertex algebra $\mathcal S_0$.

\begin{exa} \label{wgln}
	
	The Lie algebra $  \mathfrak{gl}_{N + 1} $ is the algebra of $(N + 1) \times (N + 1) $ matrices over $ \mathbb{C} $. Its Cartan subalgebra $ \mathfrak{h}$ can be described as the subspace of diagonal matrices. We equip it with the basis $(\chi^i)_{i = 0}^{N}$ where $\chi^i $ is the matrix element that has a  $ 1 $ in the $(i + 1)$th place on the diagonal and $ 0 $ elsewhere. The algebra $\mathcal{W}(\mathfrak{gl}_{N+1}) $ with central charge $\mathfrak{c} = N + 1 $ is strongly freely generated by the following $ N + 1 $ vectors in the Heisenberg VOA $\mathcal{S}_{0}$ associated to $\mathfrak{h}$
\beq
\label{eiWgen}	e_i(\chi^0_{-1},\ldots,\chi^N_{-1}) \ket{0} \qquad i \in \{1,\ldots,N + 1\}\,,
\eeq
	where the $ e_i $ denotes the $ i $-th elementary symmetric polynomial.	The proof of this statement follows immediately from the  Miura transformation, see \cite[Corollary 2.2]{AM} where we take the limit $ \alpha \to 0 $. The result is originally due to \cite{FL}.
\end{exa}

\begin{exa} \label{wdn}
	The Lie algebra $ D_N = \mathfrak{so}_{2N} $ is the Lie algebra of orthogonal $ 2N \times 2N $ matrices over $ \mathbb{C} $. The roots of $  \mathfrak{so}_{2N} $ can be described as $\pm \chi^i\pm \chi^j$ where $(\chi^i)_{i = 1}^{N}$ is an orthonormal basis for the Cartan subalgebra $ \mathbb{C}^N$. The following vectors in $\mathcal{S}_{0}$ strongly generate the algebra $\mathcal{W}(\mathfrak{so}_{2N}) $ with central charge $ \mathfrak{c} = N $.
\begin{equation}
\label{eq:dbasis}  
\begin{split}
	\nu^d &= \bigg(\sum_{i=1}^N {\rm e}^{\chi^i}_{-d}{\rm e}^{-\chi^i}_{-1} + {\rm e}^{-\chi^i}_{-d}{\rm e}^{\chi^i}_{-1}\bigg) \ket{0} \qquad d \in \{2,4,6,\ldots, 2N-2\}\,,\\ 
	\tilde{\nu}^N &= \chi^{1}_{-1} \chi^{2}_{-1}\,\cdots\,\chi^{N}_{-1} \ket{0}\,.
\end{split}
\end{equation}
The conformal weight of these vectors are $ 2,4,\ldots,2N-2$ and $ N $, which are indeed the Dynkin exponents of $ \mathfrak{so}_{2N} $. This statement follows from the results of \cite{ACL,CL}, \textit{i.e.}  from the description of $\mathcal{W}(\mathfrak{so}_{2N}) $ as ${\rm SO}_{2N}$-orbifold of the lattice vertex algebra of $\mathfrak{so}_{2N}$.
	\end{exa}

\begin{rem}\label{rem:weylinv}
	We note the important fact that $\mathcal{W}(\mathfrak{g})$  is invariant under $G$ and hence under the action of the Weyl group of $ \mathfrak{g}$. This remark will be fundamental, in our construction of higher Airy structures as $ \mathcal{W}(\mathfrak{g}) $-modules in Section~\ref{HASW}.
\end{rem}

\subsection{The graded Lie subalgebra property}
\label{s:subalgebra}

In this section we construct a number of left ideals for the algebra of modes of $\mathcal{W}$ algebras that are graded Lie subalgebras. This will be essential for the construction of higher quantum Airy structures from modules of $\mathcal{W}$ algebras in the next section. We refer the reader to Section 3 of \cite{BBCC} for a more thorough treatment of the subalgebra property in the context of VOAs.

\subsubsection{Graded Lie subalgebras and left ideals}

Let $V$ be a vertex operator algebra with finitely many strong and free generators $\gamma^1, \ldots, \gamma^n$ of conformal weights $\Delta_1, \ldots, \Delta_n$, and let $\mathcal A$ be the suitably completed algebra (the current algebra) of modes of $V$.

 Let $\mathcal{F}_p V$ be the subspace of $V$ spanned by elements $\gamma^{i_1}_{-n_1} \cdots \gamma^{i_s}_{-n_s} |0 \rangle$ with $\Delta_{i_1} + \cdots + \Delta_{i_s} \leq p$. Then $\mathcal{F} = \{ \mathcal{F}_p~|~p \in \mathbb{Z} \}$ is a vertex algebra filtration, called ``Li's filtration'' (see Section 3.1.2 in \cite{BBCC}). It induces a filtration on $\mathcal{A}$ denoted by $\mathcal{F}_p \mathcal{A}$.

 The algebra of modes $\mathcal{A}$ is a Lie algebra with respect to the commutator $[\cdot, \cdot]$. Moreover, as shown in Section 3.1.2 of \cite{BBCC} (see Lemma 3.3), for principal $\mathcal{W}$-algebras we have:
 \begin{equation}\label{eq:filt}
 [\mathcal{F}_p \mathcal{A}, \mathcal{F}_q \mathcal{A}] \subseteq \mathcal{F}_{p+q-2} \mathcal{A}.
 \end{equation}


Our goal is to find graded Lie subalgebras of $ \mathcal{A} $, in the following sense. Let $ S $ be a given subset of the modes of the strong generators $\gamma^1, \ldots, \gamma^n$ of $V$, and $\mathcal{A} \cdot S \in \mathcal{A}$ be the left $\mathcal{A}$-ideal generated by  $S$. We say that $\mathcal{A} \cdot S$ is a graded Lie subalgebra of $\mathcal{A}$ if
\begin{equation}\label{eq:subb}
[ \mathcal{A} \cdot S, \mathcal{A} \cdot S ] \subseteq \mathcal{A} \cdot S.
\end{equation}
Because of \eqref{eq:filt}, in our construction of higher quantum Airy structures this condition will become equivalent to the subalgebra condition in Definition \ref{def:HAS} after introducing $\hbar$ as in Section \ref{hbarresc}.\footnote{Perhaps the easiest way to see this is to introduce $\hbar$ from the start in the vertex operator algebra, as in Section 3.1.2 of \cite{BBCC}. In this case, \eqref{eq:filt} becomes $[\mathcal{A}^{\hbar}, \mathcal{A}^{\hbar}] \subseteq \hbar \mathcal{A}^{\hbar}$ for the $\hbar$-rescaled modes, and \eqref{eq:subb} becomes $[\mathcal{A}^{\hbar} \cdot S, \mathcal{A}^{\hbar} \cdot S] \subseteq \hbar \mathcal{A}^{\hbar} \cdot S$.}

We can make this subalgebra property more explicit by introducing an ordering on the set of all modes (\emph{i.e}, the underlying set of $ \mathcal{A} $). We define an ordering such that a mode in $S$ is always greater than a mode not in $S$. We say that elements of the ideal $ \mathcal{A} \cdot S $ are good with respect to $ S $. In particular, $ \gamma  $ is good if the right-most term of every ordered monomial of $ \gamma $ (expressed in terms of the strong generators) is in $ S $.

The following lemma is clear.
\begin{lem}
The left $\mathcal{A}$-ideal generated by the modes in $S$ is a graded Lie subalgebra of $\mathcal{A}$ if and only if for any two modes $\gamma^{i}_{m}, \gamma^{j}_{n} \in S$, one has that $ [\gamma^{i}_{m}, \gamma^{j}_{n}] \in \mathcal{F}_{\Delta_{i}+\Delta_{j}-2} \mathcal{A}$ is good with respect to $S$. 
\end{lem}
The following subsections give examples in an increasing order of complexity. However the idea of construction is always the same. We are looking for a suitable module $\mathcal M_\lambda$ generated by a highest weight vector $\ket{\lambda}$ and such that this highest weight vector is annihilated by a mode if and only if this mode is in the set $S$ of interest (\emph{i.e.}, it is a good mode). It then remains to show that the commutator of two modes in $S$ is still good  and essentially this amounts to showing that a basis of $\mathcal M_\lambda$ is given by all the ordered monomials that are not good.
We start with the case where $\mathcal M_\lambda$ is the vacuum of our vertex algebra. 

\subsubsection{The vacuum subalgebra \texorpdfstring{$\mathcal{A}_{\geq 0}$}{A+}}

Our first subalgebra is  the left ideal generated by all modes of the strong generators of a $\mathcal{W}$ algebra that annihilate the vacuum state $\ket{0}$.

\begin{prop}\label{lem:modesplus}
	Consider a vertex operator algebra $ V $ freely strongly generated by homogeneous states $\gamma^i \in V $ indexed by $i \in \mathcal{I}  $ (where $ \mathcal{I} $ is a finite set), with respective conformal weights $\Delta_i \in \mathbb{Z}$.	Let $ \mathcal{A} $ be the suitably completed algebra of modes of $ V $. Let $S = \{\gamma^i_k \}_{i \in \mathcal{I}, k \geq 0}$, and consider the left ideal $\mathcal{A}_{\geq 0} := \mathcal{A} \cdot S$.  Then, $ \mathcal{A}_{\geq 0} $ is a graded Lie subalgebra of $ \mathcal{A} $. Equivalently, when $ k,k' \geq 0 $,
	\begin{equation}
	[\gamma^i_k,\gamma^{i'}_{k'}] = \sum_{j = 1}^{n}\sum_{p \geq 0} f^{(j,l)}_{(i,k),(i',k')}\gamma^j_p \in \mathcal{F}_{\Delta_i + \Delta_{i'} - 2} \mathcal{A}
	\end{equation} for some $f^{(l,j)}_{(i,k),(i',k')} \in \mathcal{A} $.
	
\end{prop}
\begin{proof}
	We have the following commutation relations which follow from the lo\-ca\-li\-ty axiom/Bor\-cherds identity  \cite[Section 3.3.6]{FB}
\begin{equation}\label{eq:voacomm}
[\gamma^i_k,\gamma^{i'}_{k'}] = \sum_{m \geq 0} \binom{k}{m} (\gamma^i_{m}\gamma^{i'} )_{k+k'-m}\,,
\end{equation}
where $ k,k' \geq 0 $.

	 The assumption on strong generation implies that we can express each $ (\gamma^i_{m}\gamma^{i'} )_{k+k'-m} $ as a finite linear combination of normal ordered monomials in the generators. Let us look at one of these normally ordered terms
	\begin{equation}\label{eq:prodfum}
	\gamma^{b_1}_{p_1} \gamma^{b_2}_{p_2} \cdots \gamma^{b_{L}}_{p_{L}}.
	\end{equation} This monomial could either annihilate the vacuum state $ \ket{0} $ or not. Let us first consider the case where it does. The normal ordering prescription implies that the term furthest to the right, i.e. $ \gamma^{b_{L}}_{p_{L}} $ annihilates the vacuum. In that case, we are done, as $ \gamma^{b_{L}}_{p_{L}}  $ is an element of $ \mathcal{A}_{\geq 0}$.
	
	Now, let us assume that the term~\eqref{eq:prodfum} does not annihilate the vacuum $ \ket{0}$. Then $ p_L < 0 $, and due to the normal ordering prescription, this implies that all the modes appearing in~\eqref{eq:prodfum} are negative modes. We know that $ \gamma^i_k \ket{0} = 0  = \gamma^{i'}_{k'} \ket{0} $ and hence $ [\gamma^i_k,\gamma^{i'}_{k'}] \ket{0}  = 0$. This means that  $ \gamma^{b_1}_{p_1} \gamma^{b_2}_{p_2} \cdots \gamma^{b_{L}}_{p_{L}} \ket{0} $ must cancel with some other terms (which are also normally ordered products of negative modes) in the sum on the right-hand side of \eqref{eq:voacomm} after acting on the vacuum state $\ket{0}$. However, this contradicts the assumption of free generation (which is that  vectors of the form $ \gamma^{b_1}_{p_1} \cdots \gamma^{b_{L}}_{p_{L}} \ket{0} $ where $ p_i < 0 $ form a basis for $ V $), and hence cannot occur.		
\end{proof}

\subsubsection{The subalgebra \texorpdfstring{$\mathcal{A}_{\Delta}$}{Adelta}}

We can now construct another interesting left ideal that is a graded Lie subalgebra of $\mathcal{A}$. In this case, we consider all modes $\gamma_k^i$ of the generators of a $\mathcal{W}$ algebra for $k \geq \Delta_i-1$, where $\Delta_i$ is the conformal weight of $\gamma^i$. The construction is rather straightforward.

\begin{prop}\label{lem:modesdelta}
Consider a vertex operator algebra $ V $  strongly generated by homogeneous states $\gamma^i \in V $ indexed by $i \in \mathcal{I}  $ where $ \mathcal{I} $ is a finite set, with respective conformal weights $\Delta_i \in \mathbb{Z}$. Let $ \mathcal{A} $ denote the suitably completed algebra of modes. Let $S = \{ \gamma^i_k\}_{i \in \mathcal{I}, k \geq \Delta_i - 1}$, and consider the left ideal $ \mathcal{A}_{\Delta} := \mathcal{A} \cdot S$. Then $\mathcal{A}_{\Delta}$  is a graded Lie subalgebra of $ \mathcal{A} $. Equivalently, for $ k \geq \Delta_{i} - 1 $ and $  k' \geq \Delta_{i'} - 1  $, we have
\begin{equation}\label{eq:3.5}
[\gamma^i_k,\gamma^{i'}_{k'}] = \sum_{j = 1}^{n}\sum_{p \geq \Delta_j -1} f^{(j,l)}_{(i,k),(i',k')}\gamma^j_p\, \in \mathcal{F}_{\Delta_i +\Delta_{i'} - 2} \mathcal{A}.
\end{equation}
\end{prop}

\begin{proof}
	Using the strong generation assumption, we can express the commutator~\eqref{eq:3.5} as sums of normally ordered monomials of the form
	\begin{equation}
	\gamma^{b_1}_{p_1} \gamma^{b_2}_{p_2} \cdots \gamma^{b_{L}}_{p_{L}},
	\end{equation} where $ \sum_{i=1}^L (p_i - b_i+1) =  (k - \Delta_i +1 ) +(k'-\Delta_{i'}+1)  $ due to the conformal weight condition. As  $ k \geq \Delta_{i} - 1 $ and $  k' \geq \Delta_{i'} - 1  $, we get	
$$
\sum_{i=1}^L p_i \geq  \sum_{i=1}^L (b_i-1), 
$$
and hence at least one of the $ p_i \geq b_i -1 $. Due to the normal ordering procedure, the last mode on the right $ \gamma^{b_{L}}_{p_{L}} $ will have this property. This gives the statement of the Lemma.	
\end{proof}

\subsubsection{The intermediate subalgebras}
\label{Sinterm}
In fact we can construct many more subalgebras as intermediate cases interpolating between $\mathcal{A}_{\geq 0}$ and $\mathcal{A}_{\Delta}$ for the  $\mathcal{W}(\mathfrak{gl}_{N+1})$ algebras that we described in Example \ref{wgln}. The particular form of the strong generators, namely as elementary symmetric polynomials, is crucial for the construction.

In this subsection, we use a different convention for mode expansion of a field as we find it more convenient. We shift the index of the modes by the conformal weight, \emph{i.e.,} when $ \ket{v} $ has conformal weight $ \Delta_v $
\begin{equation}
Y(\ket{v},z) = \sum_{l \in \mathbb{Z}} \mathsf{v}_l\,z^{-l-\Delta_v}\,.
\end{equation}
The correspondence between the two ways of indexing is $\mathsf{v}_{l} = v_{l + \Delta_v - 1}$.
 
Let us start with the setup. We aim to find one subalgebra in $\mathcal{W}(\mathfrak{gl}_{N+1})$ for each partition of $r := N+1$. So let $\lambda=(\lambda_1, \ldots, \lambda_{p})$ be a fixed partition of $r$, that is the $\lambda_i$ are positive integers such that  $r =\lambda_1 + \lambda_2 + \cdots\, + \lambda_{p}$ and we order them by size, i.e. $\lambda_1 \geq \lambda_2 \geq \ldots \geq \lambda_{p} \geq 1$. Such a partition defines good modes as follows. 
\begin{defin}
We say that $\mathsf{W}^a_{-m}$ is $\lambda$-good if $\lambda(a)-m>0$ where
\[
\lambda(a) :=  \text{min}\{ \, s\,  |  \, \lambda_1 + \cdots\, + \lambda_s \geq a \,\}\,.
\]
Now fix a $\lambda$-order on $\{1, \ldots, r\} \times \mathbb Z$ with the following properties
\begin{enumerate}
\item $(a, -m) > (b, -n)$ if $\mathsf{W}^a_{-m}$ is $\lambda$-good but $\mathsf{W}^b_{-n}$ is not $\lambda$-good. 
\item  $(a, -m) > (b, -n)$ if $\mathsf{W}^a_{-m}$ and $\mathsf{W}^b_{-n}$ are $\lambda$-good  and both $m, n \geq 0$ and $a<b$. 
\item  $(a, -m) > (b, -n)$ if $\mathsf{W}^a_{-m}$ and $\mathsf{W}^b_{-n}$ are $\lambda$-good  and both $m, n \geq 0$ and $a=b$ and $m<n$. 
\end{enumerate}
Let $I = \{ (a_1, -m_1) \geq (a_2, -m_2) \geq  \ldots \geq (a_{\ell}, -m_{\ell}) \}$ be an ordered set. Then we say that 
\[
\mathsf{W}_I := \mathsf{W}^{a_{\ell}}_{-m_{\ell}} \cdots\, \mathsf{W}^{a_2}_{-m_2}\mathsf{W}^{a_1}_{-m_1} 
\]
is an ordered element of the universal enveloping algebra of modes. 
We define the $\lambda$-degree of a mode to be
\[
\text{deg}_\lambda(\mathsf{W}^a_{-m}) = \begin{cases} 2a-1 & \quad \text{if} \ \lambda(a) \neq m \\ 2a & \quad \text{if} \ \lambda(a) = m \end{cases}\,,
\]
and extend this definition to ordered monomials as the sum of the $\lambda$-degrees of the terms. The $\lambda$-degree of a ordered polynomial is then the maximal $\lambda$-degree of its ordered summands. Note that since $\mathsf{W}^a_{-m}$ is a polynomial of degree $a$ in the modes of the Heisenberg vertex algebra it follows immediately that the $\lambda$-degree of any commutator $[\mathsf{W}^a_{-m}, \mathsf{W}^b_{-n}]$ is strictly smaller than $\text{deg}_\lambda(\mathsf{W}^a_{-m})  +\text{deg}_\lambda(\mathsf{W}^b_{-n})$.
\end{defin}
We will call a $\lambda$-order simply an order whenever it is clear which $\lambda$ we are using.
\begin{thm}
\label{t:sub}
Let $\mathcal A$ be the mode algebra of $\mathcal{W}(\mathfrak{gl}_r)$ and $\lambda$ a partition of $r$, then the algebra of $\lambda$-good modes forms a graded Lie subalgebra of the Lie algebra of modes. In addition, there exists a $\mathcal{W}(\mathfrak{gl}_r)$-module $\mathcal M_\lambda$ generated by a highest weight vector $\ket{\lambda}$ such that $\mathsf{W}^a_{-m} \ket{\lambda}=0$ if and only if $\mathsf{W}^a_{-m}$ is a $\lambda$-good mode. 
\end{thm}
\begin{proof}
We first consider the partition $\lambda=(r)$ of $r$. The corresponding $\lambda$-good modes are all non-negative modes $(\mathsf{W}^a_{-m})_{{m \leq 0}}$.
Let $\nu$ be a generic weight of the rank $r$ Heisenberg vertex algebra $\mathcal S_0$ so that via the embedding of $\mathcal{W}(\mathfrak{gl}_r)$ in $\mathcal S_0$ the Fock module $\mathcal S_\nu$ also becomes a $\mathcal{W}(\mathfrak{gl}_r)$-module. For generic weight $\nu$ this is a simple $\mathcal{W}(\mathfrak{gl}_r)$-module and so ordered words in the negative modes acting on the highest weight vector $\ket{v_\nu}$ of $\mathcal S_\nu$ form a basis of $S_\nu$
\[
\mathcal S_\nu = \text{span}_{\mathbb C}\big(\mathsf{W}^{a_\ell}_{-m_{\ell}}\cdots\,  \mathsf{W}^{a_1}_{-m_{1}} \ket{v_\nu} \,\,\, \big|\,\,\, (a_1, -m_1)\geq (a_2, -m_2) \geq \cdots\, \geq (a_\ell, -m_{\ell}), \ m_l>0 \ \text{for} \ l=1, \ldots , \ell\, \big)\,.
\]
We now consider the vector space $\mathcal M$ with above graded PBW-basis but consider the weight as a variable so that $\mathcal M$ can be analytically continued to a module of 
$\mathcal{W}(\mathfrak{gl}_r)$ over the polynomial ring in $r$ variables $\nu_1, \ldots, \nu_r$. Here the $\nu_i$ are the eigenvalues of the zero-modes of $N$ strong generators of the Heisenberg vertex algebra. Then specializing to any weight $\nu$ defines a new module $\mathcal M_\nu$. At generic $\nu$ this module will be simple while at special non-generic points it will be indecomposable but reducible. 
We generically have $\mathcal{M}_\nu \cong \mathcal S_\nu$ but for example $\mathcal M_0 \not\cong \mathcal S_0$.
Denote the highest weight vector of $\mathcal M_0$ by $\ket{0}$. By construction $\mathsf{W}^a_{-m}\ket{0}=0$ if and only if $\mathsf{W}^a_{-m}$ is a $\lambda$-good mode. 
In order to prove that these $\lambda$-good modes form a graded Lie subalgebra of the algebra of modes we have to show that for any two $\lambda$-good modes $\mathsf{W}^a_{-m}$ and $\mathsf{W}^b_{-n}$ the commutator $[\mathsf{W}^a_{-m}, \mathsf{W}^b_{-n}]$ is an ordered polynomial in the modes and the right most term in each summand is $\lambda$-good. Consider an ordered set $I = \{ (a_1, -m_1) \geq (a_2, -m_2) \geq  \cdots\, \geq (a_{\ell}, -m_{\ell}) \}$ so that
\[
\mathsf{W}_I := \mathsf{W}^{a_{\ell}}_{-m_{\ell}} \cdots\, \mathsf{W}^{a_2}_{-m_2}\mathsf{W}^{a_1}_{-m_1} 
\]
is an ordered element of the universal enveloping algebra of modes. We call $\mathsf{W}_I$ a $\lambda$-good monomial if $\mathsf{W}^{a_1}_{-m_1}$ is $\lambda$-good
and say that the index set $ I $ is $\lambda$-good. The PBW-basis on $\mathcal M_0$ is then given by all $\mathsf{W}_I\ket{0}$ such that $I$ is not a $\lambda$-good index set.
 It follows that
\[
[\mathsf{W}^a_{-m}, \mathsf{W}^b_{-n}] = \sum_{I} c_I \mathsf{W}_I = \sum_{I \ \lambda\text{-good}} c_I \mathsf{W}_I  + \sum_{I \ \text{not}\ \lambda\text{-good}} c_I W_I\,.
\]
Acting on $\ket{0}$ and since all $\lambda$-good modes annihilate $\ket{0}$ we have
\[
0 = \sum_{I \ \text{not}\ \lambda\text{-good}} c_I \mathsf{W}_I\ket{0}\,.
\]
Since the $\mathsf{W}_I\ket{0}$ with $I$ not a $\lambda$-good index set form a basis of $\mathcal M_0$ it follows that $c_I=0$ for $I$ not a $\lambda$-good index set. We thus have proven the claim for the partition $\lambda=(r)$. We note that this is precisely the result proved in Proposition~\ref{lem:modesdelta}.

The general case is not much different and can be reduced to this case.  We prove it by induction for $r$. The base case $r=1$ is trivial and just a special case of what we have just proven, since $\mathcal{W}(\mathfrak{gl}_1)$ is the rank one Heisenberg vertex algebra and the only partition of $1$ is $\lambda=(1)$. 

Let $r>1$. The induction hypothesis is that the statement of the Theorem is true for all $r'<r$, i.e. for all partitions $\mu$ of $\mathcal{W}(\mathfrak{gl}_{r'})$ and in addition we require the existence of a $\mathcal{W}(\mathfrak{gl}_{r'})$-module $\mathcal M_\mu$ generated by a highest weight vector $\ket{\mu}$
that is annihilated by all $\mu$-good modes and the $\mathsf{W}_I\ket{\mu}$ with $I$ not $\mu$-good form a basis of $\mathcal M_\mu$. With this notation the module $\mathcal M_0$ is also denoted by $\mathcal M_{(r')}$ and the highest weight vector $\ket{0}$ is denoted by $\ket{(r')}$.

Let $\lambda=(\lambda_1, \ldots, \lambda_{p})$ be a fixed partition of $r$, that is the $\lambda_i$ are positive integers such that  $N =\lambda_1 + \lambda_2 + \cdots\, + \lambda_{p}$ and we order them by size, i.e. $\lambda_1 \geq \lambda_2 \geq \cdots\, \geq \lambda_{p} \geq 1$. Further let $r'=r-\lambda_{p}$ so that $\mu = (\lambda_1, \ldots, \lambda_{\ell-1})$ is a partition of $r'$.
We consider the embedding $\mathcal{W}(\mathfrak{gl}_r)$ in $\mathcal{W}(\mathfrak{gl}_{r'})\otimes \mathcal{W}(\mathfrak{gl}_{\lambda_{p}})$ and the module $\mathcal{M}_{\mu} \otimes \mathcal{M}_{(\lambda_{p})}$.
We want to prove that via this embedding as $\mathcal{W}(\mathfrak{gl}_r)$-modules  $\mathcal{M}_\lambda \cong \mathcal{M}_{\mu} \otimes \mathcal{M}_{(\lambda_{p})}$.

We denote the strong generators of $\mathcal{W}(\mathfrak{gl}_{N})$ by $(\mathsf{W}^a)_{a = 1}^r$, and the ones of $ \mathcal{W}(\mathfrak{gl}_{r'}) \otimes \mathcal{W}(\mathfrak{gl}_{\lambda_{p}})$ by $(\mathsf{Z}^b)_{b = 1}^{r'}$ and $(\mathsf{Y}^c)_{c = 1}^{\lambda_{p}}$. Then due to the realization of the strong generators of the $\mathcal W$ algebras in terms of normally ordered elementary symmetric polynomials of Heisenberg vertex algebra fields we immediately have that 
\[
\mathsf{W}^a(z) = \mathsf{Z}^a(z) + \sum_{d=1}^{a-1} \norder{\mathsf{Z}^{a-d}(z) \mathsf{Y}^d(z)}+ \mathsf{Y}^a(z)\,,
\]
where we note that many of these terms on the right may not appear. For instance $\mathsf{Z}^a(z)=0$ for $a>r'$ and $\mathsf{Y}^a(z)=0$ for $a>\lambda_{p}$.
Hence
\[
\mathsf{W}^a_{-m} = \mathsf{Z}^a_{-m} \otimes 1+ \sum_{d=1}^{a-1}\sum_{n\in \mathbb Z} \mathsf{Z}^{a-d}_{n-m} \otimes \mathsf{Y}^d_{-n} + 1 \otimes \mathsf{Y}^a_{-m}\,,
\]
and of course $\mathsf{Z}^a_{-m}=0$ for $a>r'$ and $\mathsf{Y}^a=0_{-m}$ for $a>\lambda_{p}$.
The $\mu$-degree of $ \mathcal{W}(\mathfrak{gl}_{r'})$ lifts to a degree map on $ \mathcal{W}(\mathfrak{gl}_{r'}) \otimes \mathcal{W}(\mathfrak{gl}_{\lambda_{p}})$ by saying that the $\mathsf{Y}^a_{-m}$ all have $\mu$-degree zero. 
Let $\ket{\lambda}:= \ket{\mu} \otimes \ket{(\lambda_p)}$. Then a straightforward verification tells us that 
\[
\mathsf{W}^a_{-m} \ket{\lambda} =0 \qquad \text{if and only if} \ \mathsf{W}^a_{-m} \ \text{is} \ \lambda\text{-good}\,.
\]
In particular, if $\mathsf{W}^a_{-m}$  is not $\lambda$-good then its leading degree summand is $\mathsf{Z}^a_{-m} \otimes 1$ if $a\leq r'$ and 
$\mathsf{Z}^{r'}_{-\lambda(a)} \otimes \mathsf{Y}^{a-r'}_{-m+\lambda(a)}$ if $a>r'$. In either case the leading degree summand does not annihilate $\ket{\lambda}$ which is equivalent to saying that $\mathsf{Z}^a_{-m} \otimes 1$ if $a\leq r'$ is not $\mu$-good and $\mathsf{Z}^{r'}_{-\lambda(a)}$ and $\mathsf{Y}^{a-r'}_{-m+\lambda(a)}$ are neither $\mu$-good respectively $(\lambda_p)$-good if $a>r'$.
Let $I = \{ (a_1, -m_1) \geq (a_2, -m_2) \geq  \cdots\, \geq (a_{\ell}, -m_{\ell}) \}$ be an ordered set with ordered monomial  $\mathsf{W}_I := \mathsf{W}^{a_{\ell}}_{-m_{\ell}} \cdots\, \mathsf{W}^{a_2}_{-m_2}\mathsf{W}^{a_1}_{-m_1}$. Let $s$ satisfy  $s=\ell$ if $a_{\ell} \leq {r'}$, $s=0$ if $a_1>r'$ and otherwise defined such that
$a_s\leq r'$ but $a_{s+1}>r'$.
It follows that the projection of $\mathsf{W}_I$ on leading $\mu$-degree, which we denote by $\mathsf{X}_I$, is
\[
\mathsf{X}_I =           \mathsf{Z}^{r'}_{-\lambda(a_{\ell})}        \cdots\,  \mathsf{Z}^{r'}_{-\lambda(a_{s+1})}     \mathsf{Z}^{a_s}_{-{m_s}} \cdots\, \mathsf{Z}^{a_2}_{-{m_2}}\mathsf{Z}^{a_1}_{-{m_1}}\otimes  \mathsf{Y}^{a_{\ell}-r'}_{-m_{\ell}+\lambda(a_{\ell})}  \cdots\, \mathsf{Y}^{a_{s+1}-r'}_{-m_{s+1}+\lambda(a_{s+1})}\,.
\]
Looking back at our requirements on the order of modes we see that the first factor is $\mu$-ordered and the second one is $(\lambda_p)$-ordered. 
Consider a polynomial of type 
\[
\sum_{I \ \text{not}\ \lambda\text{-good}} c_I \mathsf{W}_I\,.
\]
 Assume that it annihilates $\ket{\lambda}$. In particular, the leading $\mu$-degree summands annihilate $\ket{\lambda}$ and hence
\begin{equation*}
\begin{split}
 0 &=\sum_{I \ \text{not}\ \lambda\text{-good}} c_I \mathsf{X}_I \ket{\lambda}  \\
 &= \sum_{I \ \text{not}\ \lambda\text{-good}} c_I\  \mathsf{Z}^{r'}_{-\lambda(a_{\ell})}        \cdots\,  \mathsf{Z}^{r'}_{-\lambda(a_{s+1})}     \mathsf{Z}^{a_s}_{-{m_s}} \cdots\, \mathsf{Z}^{a_2}_{-{m_2}}\mathsf{Z}^{a_1}_{-{m_1}}\ket{\mu} \otimes  \mathsf{Y}^{a_{\ell}-r'}_{-m_{\ell}+\lambda(a_{\ell})}  \cdots\,\mathsf{Y}^{a_{s+1}-r'}_{-m_{s+1}+\lambda(a_{s+1})}\ket{(\lambda_p)}\,. 
 \end{split}
 \end{equation*}
 By the induction hypothesis, non-good monomials acting on the highest weight vector form a basis of $\mathcal M_\mu$ respectively $\mathcal M_{(\lambda_p)}$ and hence all $c_I=0$. We thus have constructed the claimed module $\mathcal M_\lambda$. 
 
It is now easy to show that for any two $\lambda$-good modes $\mathsf{W}^a_{-m}$ and $\mathsf{W}^b_{-n}$ the commutator $[\mathsf{W}^a_{-m}, \mathsf{W}^b_{-n}]$ is an ordered polynomial in the modes and the right-most term in each summand is $\lambda$-good. 
We have
\[
[\mathsf{W}^a_{-m}, \mathsf{W}^b_{-n}] = \sum_{I \ \lambda\text{-good}} c_I\mathsf{W}_I  + \sum_{I \ \text{not}\ \lambda\text{-good}} c_I \mathsf{W}_I\,.
\]
Acting on $\ket{\lambda}$ and since all $\lambda$-good modes annihilate $\ket{\lambda}$ we have
\[
0 = \sum_{I \ \text{not}\ \lambda\text{-good}} c_I \mathsf{W}_I\ket{\lambda}\,.
\]
Since we just proved that all the $\mathsf{W}_I\ket{\lambda}$ where $I$ is not a $\lambda$-good index set form a basis of $\mathcal M_\lambda$, it follows that $c_I=0$ for $I$ not a $\lambda$-good index set. This finishes the proof of the Theorem.  
\end{proof}

\subsection{Twisted modules}
\label{s:TM}

In preparation for the construction of higher quantum Airy structures in the next section, we now introduce twisted modules for the Heisenberg VOAs. Those will restrict to  interesting modules for the $\mathcal{W}$ algebras realized as subalgebras of the Heisenberg VOAs.

\subsubsection{Definitions}

Let us  define automorphisms of vertex operator algebras.
\begin{defin} An automorphism $ \sigma $, of finite order $ r $, of a vertex operator algebra $V$ is an automorphism $\sigma\,:\,V \to V$ on the (vector) space of states, with $\sigma^r={\rm id}_V$, which preserves the vacuum state $\ket{0}$ and the conformal state $\ket{w}$, and such that for any $\ket{v} \in V$ it acts as
	$$
	\sigma Y(\ket{v},z) \sigma^{-1} = Y(\sigma \ket{v}, z)\,.
	$$
\end{defin}

Given such an automorphism, we define the notion of a twisted module.

\begin{defin} A \emph{$\mathbb{Z}$-graded $ \sigma $-twisted $V$-module $W$} is a $\mathbb{Z}$-graded vector space $W = \bigoplus_{l \in \mathbb{Z}} W_l$ such that  $W_l=0$ for $l$ sufficiently negative and $\dim W_l < \infty$ for all $l \in \mathbb{Z}$, with a linear map
	
	$$
	Y_{\sigma}(\cdot, z)\,:\, \begin{array}{lcl} V & \longrightarrow & {\rm End}(W)[\![z^{1/r}, z^{-1/r} ]\!] \\
	 \ket{v}& \longmapsto & Y_{\sigma}(\ket{v},z) = \sum_{l \in \frac{1}{r} \mathbb{Z}} v_l z^{-l-1}\end{array}\,.
	$$
	
	We require that the vacuum property, creation property and the Virasoro algebra condition hold for $W$ and $Y_{\sigma}(\cdot, z)$. In addition, we require the following conditions.
	\begin{itemize}
		\item The \textit{monodromy} around $ z = 0 $ is given by the action of $ \sigma $, namely if $ \sigma \ket{v} = e^{2\ii\pi q/r}\ket{v} $ we have
$$
		v_{a} = 0  \text{ unless }  a \in q/r +\mathbb{Z} \,.
$$

		\item $\big(Y_{W}(\ket{v},z)\big)_{v \in V}$ is a \textit{local family of fields}; \textit{i.e.} for $\ket{u},\ket{v} \in V $,
		\begin{equation}
		(z_1-z_2)^{N_{u,v}} [Y_{\sigma}(\ket{u},z_1),Y_{\sigma}(\ket{v},z_2)] = 0 \qquad \text{for some } N_{u,v} \in \mathbb{Z}_+\,.
		\end{equation} 
		
		\item We have a \emph{product formula}
		\beq
		\label{prodid} \frac{1}{k!} \frac{{\rm d}^{k}}{{\rm d}z^k_1} \Big\{(z_1-z_2)^N Y_{\sigma}(\ket{u},z_1) Y_{\sigma}(\ket{v},z_2) \ket{w}\Big\}\Big|_{z_1=z_2 = z} = Y\big(u_{N-1-k}\ket{v},z\big)\ket{w}
		\eeq for all $ \ket{u},\ket{v} \in V $, $ \ket{w} \in W $ and where $ N = N_{u,v}  $ is chosen from the locality axiom.

	\end{itemize}
\end{defin}

In the above definition if we set $ \sigma = {\rm id} $, we get the usual notion of (untwisted) modules. The idea of twisted modules is to introduce  fields that have expansions in fractional powers of $z$. In physics this formalizes the notion of orbifold CFTs. Intuitively, we are working on the branched covering $z = \zeta^r$, by rewriting the fields as expansions in $\zeta$ (or fractional powers of $z$). However, we have to be careful about the normal ordering in this context. In physics terms, the operator product expansion (OPE) of the fields changes, and the product formula~\eqref{prodid} captures this precisely. This product formula (and easy corollaries) will be very useful in our $\mathcal{W}$ algebra computations.

\begin{rem}
	Note that since $\sigma(\ket{\omega}) = \ket{\omega}$, the conformal field has a mode expansion
	\begin{equation}
	Y_W(\ket{\omega},z) = \sum_{l \in \mathbb{Z}} L_l\,z^{-l-1}.
	\end{equation}
	with only integer powers of $z$.
\end{rem}

\subsubsection{Twisted modules of the Heisenberg VOA}

\label{secsigmadef}

Given an automorphism $ \sigma $ of $ \mathfrak{h}$, we  define a $ \sigma $-twisted $\mathcal{S}_0 $-module as follows. We define the $ \sigma $-twisted Heisenberg Lie algebra $ \hat{\mathfrak{h}}_\sigma  $ and define  a $ \hat{\mathfrak{h}}_\sigma  $-module called the twisted Fock module, denoted $\mathcal{T}$. The latter carries the structure of a $ \sigma $-twisted module over the Heisenberg vertex operator algebra $\mathcal{S}_{0}$.

Here is the detailed construction. Let $\sigma$ be an automorphism of the Cartan subalgebra $ \mathfrak{h} \subset \mathfrak{g} $ of finite order $r$
$$
\braket{\sigma(\xi), \sigma(\eta)} = \braket{\xi,\eta}, \qquad \sigma^r = {\rm id}_{\mathfrak{h}}\,.
$$
Any such automorphism lifts to an automorphism of $\mathcal{S}_{0}$ which we also denote by $\sigma$. We note that $ \mathfrak{h} $  admits an orthonormal basis of eigenstates for the action of $ \sigma $.

We extend the automorphism $ \sigma $  to $ \mathfrak{h}[\![t^{1/r},t^{-1/r}]\!] \oplus \mathbb{C} K $ as follows. Given $ \xi \in \mathfrak{h} $, we use the notation $ \xi_n := \xi \otimes t^n $ where $ n \in \frac{1}{r}\mathbb{Z} $ as before. The action of $ \sigma $ is then 
$$
\sigma(\xi_l) = \sigma(\xi) \otimes e^{2\ii \pi l} t^{l}, \qquad \sigma(K) = K, \qquad l \in \tfrac{1}{r}\mathbb{Z}\,.
$$

The $ \sigma $-twisted Heisenberg algebra is the subspace of $\sigma$-invariant elements
$$
\hat{\mathfrak{h}}_{\sigma} := \big(\mathfrak{h}[\![t^{1/r},t^{-1/r}]\!] \oplus \mathbb{C} K\big)^{\sigma}\,.
$$
The algebra $ \hat{\mathfrak{h}}_{\sigma}$ is generated by the elements $\xi_l$ such that $ \xi $ is diagonal under the action of $ \sigma $, and the central element $ K $, with the following Lie bracket relations 
\beq\label{eq:tbracket}
[\xi_l, \eta_m] = l\delta_{l+m,0} \braket{\xi,\eta} K\,,\qquad [K,\hat{\mathfrak{h}}_\sigma] = 0\,.
\eeq
We also introduce its negative part
$$
\hat{\mathfrak{h}}^-_{\sigma} = \bigoplus_{l \in  \frac{1}{r}\mathbb{Z}_{<0}} \mathfrak{h}_{\sigma} \otimes t^l\,.
$$
\begin{defin}
\label{S0gi}Let $\mathcal{T} = {\rm Sym}(\mathfrak{h}^{-}_{\sigma})\ket{0}$ be the $ \hat{\mathfrak{h}}_{\sigma}$-module such that $K\ket{0} = \ket{0}$ and $\xi_{l}\ket{0} = 0$ for  $\xi \in \mathfrak{h}$ and $l > 0$.
\end{defin}
We would like to give $\mathcal{T}$  the structure of a $ \sigma $-twisted module of the Heisenberg VOA $\mathcal{S}_{0}$ as follows. Let $ \xi \in \mathfrak{h} $ be a diagonal element, \textit{i.e.} $ \sigma(\xi) = e^{-2 \ii \pi p} \xi $ for some $ p \in \{0,\frac{1}{r},\ldots,\frac{r - 1}{r}\} $. Then the state-field correspondence for the module is defined as follows
\begin{equation}
 \label{eq:tfields}
\begin{split}
Y_{\sigma}(\ket{0},z) &= \text{id}_{\mathcal{T}}\,,\\
Y_{\sigma}(\xi_{-1}\ket{0},z) &= \sum_{n\in p + \mathbb{Z}} \xi_{n}\,z^{-n-1} \,.
\end{split}
\end{equation}
It is easy to check that this gives $\mathcal{T}$ the structure of a $ \sigma $-twisted module over $\mathcal{S}_{0}$.

\begin{rem}
	The state-field correspondence for general elements in $ \mathcal{T}$ can be computed using the state-field correspondence for the states $ \xi_{-1}\ket{0} $~\eqref{eq:tfields} and the product formula for twisted modules ~\eqref{prodid}.  
\end{rem}

\subsection{Introducing \texorpdfstring{$\hbar$}{h}}
\label{hbarresc}
From now on, it is convenient to rescale the Killing form by some formal parameter $\hbar^{1/2}$, and base change to the field $ \mathbb{C}_{\hbar^{1/2}} := \mathbb{C}(\!(\hbar^{1/2})\!)$. In other words, we have a new Heisenberg VOA (still denoted  $\mathcal{S}_0$) in which the commutation relations read
\begin{equation}\label{eq:hbracket}
[\xi_l,\eta_m] = \hbar\,l\,\langle \xi,\eta\rangle\,\delta_{l + m,0}\,.
\end{equation}
The reason to write $\hbar^{1/2}$ instead of $\hbar$ is to match with the convention \eqref{eq:sol} adopted in \cite{KS,ABCD} for the partition functions of quantum Airy structures. The construction of  Section~\ref{secsigmadef} can still be applied to define a $\sigma$-twisted module again denoted $\mathcal{T}$. The only notable modification compared to the previous sections is that in Propositions~\ref{lem:modesplus} and \ref{lem:modesdelta} a factor of $\hbar$ appears in the right-hand side of the commutation relations, and that in the reconstruction of the state-field correspondence, one should include a factor of $\hbar^{1/2}$ per each $\partial_{z}$. In particular, the Lie subalgebras constructed in Section~\ref{s:subalgebra} become graded Lie subalgebras.

In all our examples except Section~\ref{s:glra}, only integer powers of $\hbar$ will remain the end of the day and we could effectively work with $\mathbb{C}_{\hbar} \subset \mathbb{C}_{\hbar^{1/2}}$.

\newpage

\section{Higher quantum Airy structures from \texorpdfstring{$ \mathcal{W} $}{W} algebras}
\label{HASW}

This section gives a general prescription to produce higher quantum Airy structures starting with a Lie algebra $ \mathfrak{g} $ and an element $ \sigma $ of the Weyl group  of $ \mathfrak{g} $. 
\begin{enumerate}
\item We construct a $\sigma$-twisted module $\mathcal{T}$  of the Heisenberg VOA associated to the Cartan subalgebra $ \mathfrak{h} $ of $\mathfrak{g} $. 
\item Upon restriction to the $ \mathcal{W} $ algebra  $\mathcal{W}(\mathfrak{g})$ (which is a sub-VOA of the Heisenberg VOA), the module becomes untwisted. The underlying vector space of $\mathcal{T}$ is the space of formal series in countably many variables, and elements of $\mathcal{W}(\mathfrak{g})$ act as differential operators (of order at most $\mathrm{rank}(\mathfrak{g})$) in those variables. 
\item In Section \ref{s:subalgebra}, we constructed a number of ideals that are graded Lie subalgebras of the Lie algebra of modes. We pick one of these subalgebras from the algebra of modes of the $ \mathcal{W} $ algebra module $ \mathcal{T} $. These modes fulfill the second (and hardest to check) condition to be a higher quantum Airy structure.
\item A further conjugation of these modes (a.k.a dilaton shift) allows us to realize the first condition about degree $1$ terms, thereby producing quantum $\mathrm{rank}(\mathfrak{g})$-Airy structures.
\end{enumerate}

 We apply this program in detail for $\mathfrak{gl}_{N + 1}$ (type $A_N$) and $\mathfrak{so}_{2N}$ (type $D_N$) for different choices of the Weyl group element $ \sigma $.

\subsection{The \texorpdfstring{$\mathcal{W}(\mathfrak{gl}_{N+1})$}{W(gl(N + 1))} Airy structures}
\label{SWgg}
\subsubsection{The twisted module \texorpdfstring{$\mathcal{T}$}{T} for the Heisenberg VOA}
\label{Sfgggfsg}
Recall Example~\ref{wgln}. The Cartan subalgebra $ \mathfrak{h} \subset \mathfrak{gl}_{N+1} $ has a basis given by $ \chi^{i} $ where $i \in \{0,\ldots, N\}$, with the following bilinear form
$$
\braket{\chi^i,\chi^j} =  \delta_{i,j}\,.
$$
We shall focus on the automorphism $ \sigma $ of the Cartan subalgebra $ \mathfrak{h} $ induced by the Coxeter element of the Weyl group $\mathfrak{S}_{N+1} $, namely
$$
\chi^0 \longrightarrow \chi^1 \longrightarrow \cdots\, \longrightarrow \chi^{N} \longrightarrow \chi^0.
$$
This automorphism has order
$$
r = N+1\,.
$$
We define a primitive $ r $-th root of unity $ \theta := e^{2\ii \pi /r} $, which will appear throughout the section.  Applying a discrete Fourier transform, we can define a basis $ (v^a)_{a = 0}^{r - 1}$ of $ \mathfrak{h} $ that is diagonal under the action of $ \sigma $
\beq \label{eq:vln}
v^a := \sum_{j=0}^{N} \theta^{-aj}\chi^{j} \qquad a \in \{0,\ldots,r - 1\}\,.
\eeq
Then we indeed have $\sigma(v^a) = \theta^{a}v^a$. Note that
\beq
\label{Scagln} \langle v^a,v^{b} \rangle = r\delta_{r|a + b}\,.
\eeq
where the notation $\delta_{r|k}$ means $1$ if $k$ is divisible by $r$, and $0$ otherwise. We observe that $v^a \otimes t^{-a/r - k - 1}$ is invariant under $\sigma$ for $k \in \mathbb{Z}$. Hence we can represent the $\mathcal{S}_{0}(\mathfrak{gl}_{N + 1})$-twisted module
$$
\mathcal{T}(\mathfrak{gl}_{N + 1}) \cong \mathbb{C}_{\hbar^{1/2}}[x_1,x_2,x_3,\ldots]\,,
$$
with the fields
$$
v^a(z) :=Y_\sigma\big(v^a_{-1} \ket{0},z\big) =  \sum_{k \in a/r + \mathbb{Z}} J_{rk}z^{-k-1}\,.
$$
We also recall the differential operators defined in \eqref{eq:J}
$$
\forall l > 0,\qquad J_l = \hbar \partial_{x_l},\qquad J_{-l} = lx_{l}\,.
$$
$ J_0 $ has not been defined before: we set it equal to a scalar $J_0 = Q$. The differential operators $ J_l$ satisfy the expected bracket relations
$$
[J_{rk},J_{rk'}] = \hbar\,r\delta_{r|a+b}\,k\delta_{k + k',0} = \hbar\,\langle v^a,v^b\rangle k\delta_{k + k',0} \qquad {\rm for}\,\,k \in a/r + \mathbb{Z}\,\,{\rm and}\,\,k' \in a'/r + \mathbb{Z}\,,
$$
therefore we do have an equivalent description of the twisted module introduced in Section~\ref{secsigmadef}. We also stress that the normal ordering of the modes carries over to this realization as the standard normal ordering on differential operators, with derivatives on the right and multiplication by variables on the left.

Via restriction, we can now consider $\mathcal{T}$ as a module for the subvertex algebra $\mathcal{W}(\mathfrak{gl}_{N+1}) \subset \mathcal{S}_{0}(\mathfrak{gl}_{N + 1})$. 

\begin{rem} \label{rem:notation}
Even though $\mathcal{T}$ is not a twisted module for the subvertex algebra $\mathcal{W}(\mathfrak{gl}_{N+1})$, we will slightly abuse notation and still refer to the fields associated to the generators of $\mathcal{W}(\mathfrak{gl}_{N+1})$ as ``twist fields''. We will use the notation
$$
	\xi(z) := Y_{\sigma}\big(\xi_{-1}\ket{0},z\big)
$$
for the twist fields, where $\sigma$ is the automorphism of the Heisenberg VOA used to construct the twisted module.
	\end{rem}

\subsubsection{Computing the twist fields of the generators of \texorpdfstring{$\mathcal{W}(\mathfrak{gl}_{N+1})$}{W(gl(N + 1))}}

From Example~\ref{wgln}, we know that the elementary symmetric polynomials
$$
e_i(\chi^0_{-1},\ldots,\chi^{N}_{-1}) \ket{0} \in \mathcal{S}_{0}\qquad i \in \{1,\ldots, N + 1\}
$$
are a set of strong generators for $ \mathcal{W}(\mathfrak{gl}_{N+1})$, and we are going to compute the modes of their twist fields.

Let us introduce some notation and prove some essential lemmas now. We first want to express the twist field corresponding to the state $ v^{a_1}_{-1} v^{a_2}_{-1} \cdots\, v^{a_i}_{-1} \ket{0} $ in terms of the twist fields $ v^l(z) $.
If $A = (a_j)_{j = 1}^i$ is a finite sequence, we use $\mathcal{P}(A)$ to denote the set of unordered, pairwise disjoint subsequences of length $2$ of $A$. If $B \in \mathcal{P}(A)$, we use  $|B|$ to denote the number of pairs appearing in $B$, and $A \setminus B$ the subsequence of $A$ where one has removed the elements that appear in the pairs appearing in $B$. Of course $|B| \leq \lfloor i/2 \rfloor$. For instance, the elements $B$ of $\mathcal{P}(a_1,a_2,a_3,a_4)$ such that $|B| = 2$ are
$$
\{(a_1,a_2),(a_3,a_4)\}\,,\qquad \{(a_1,a_3),(a_2,a_4)\}\,,\qquad \{(a_1,a_4),(a_2,a_3)\} \,.
$$

\begin{lem}\label{lem:p} Let $ A = (a_k)_{k = 0}^i$ where $a_k \in \{0,\ldots,N\}$. The twist field $ Y_\sigma(v^{a_1}_{-1} v^{a_2}_{-1} \cdots\, v^{a_i}_{-1} \ket{0},z) $ can be expressed as the following normally ordered product
$$
	Y_\sigma\big(v^{a_1}_{-1} v^{a_2}_{-1} \cdots\, v^{a_i}_{-1} \ket{0},z\big) = \sum_{B \in \mathcal{P}(A)} (\hbar z^{-2})^{|B|}  \prod_{\{b_1,b_2\} \in B} \frac{b_1b_2\,\delta_{b_1 +b_2,r}}{2r}\,\, \norder{\prod_{l\in A \setminus B} v^l(z)}\,.
$$
\end{lem}  
\begin{proof}
	This is an application of the product formula~\eqref{prodid}. If $ i =2 $ in the above expression, we choose $ N=2 $ in the product formula to get
\begin{equation}
\label{eq:l49}
		Y_\sigma(v^{a_1}_{-1} v^{a_2}_{-1} \ket{0},z) = \frac{1}{2}\frac{\dd^2}{\dd z_1^2} \Big\{(z_{1}-z_2)^2 v^{a_1}(z_1) v^{a_2} (z_2)\Big\}\Big|_{z_1=z_2 = z}\,.
\end{equation}
We compute the OPE of the twist fields, \textit{i.e.} we express their product in terms of the normally ordered products
\begin{equation*}
\begin{split}
v^{a_1}(z_1) v^{a_2}(z_2) & =  \norder{	v^{a_1}(z_1) v^{a_2}(z_2)}  \,\,\,+ \sum_{\substack{k_1 \in a_1/r + \mathbb{Z} \\ k_2 \in a_2/r + \mathbb{Z}}} [v^{a_1}_{k_1},v^{a_2}_{k_2}]\,z_1^{-k_1 - 1}z_2^{-k_2 - 1}\,\delta_{k_1 > 0} \delta_{k_2 < 0} \\
& =  \norder{v^{a_1}(z_1)v^{a_2}(z_2)} \,\,\, + \sum_{\substack{k_1 \in a_1/r + \mathbb{Z} \\ k_1 > 0}} \hbar k_1 \langle v^{a_1},v^{a_2} \rangle\,z_1^{-k_1 - 1}z_2^{k_1 - 1}\,,
\end{split}
\end{equation*}
and the scalar product is given by \eqref{Scagln}. Notice that we can extend the sum to $k_1 = 0$. Let us write $k_1 = a_1/r + k_1'$ for $k_1' \in \mathbb{Z}$.  The condition $k_1 \geq 0$ is then equivalent to $k_1' \geq 0$. We therefore obtain
\begin{equation*}
\begin{split}
v^{a_1}(z_1) v^{a_2}(z_2) & =  \norder{	v^{a_1}(z_1) v^{a_2}(z_2)} + \sum_{k_1' \geq 0} \hbar (\delta_{a_1,0}\delta_{a_2,0} + \delta_{a_1 + a_2,r}) r(a_1/r + k'_1)\,z_1^{- a_1/r - 1 - k_1'}z_2^{a_1/r + k_1' - 1}  \\
& =  \norder{	v^{a_1}(z_1) v^{a_2}(z_2)} \,\,\,+\hbar (\delta_{a_1,0}\delta_{a_2,0} + \delta_{a_1 + a_2,r})r\,\partial_{z_2}\bigg(\frac{(z_2/z_1)^{a_1/r}}{z_1 - z_2}\bigg)\,.
\end{split}
\end{equation*}

Inserting this result into~\eqref{eq:l49}, we get
\begin{equation*}
\begin{split}
	 Y_\sigma(v^{a_1}_{-1} v^{a_2}_{-1} \ket{0},z) &=\norder{v^{a_1}(z) v^{a_2}(z)}  \,\,\,+ \hbar(\delta_{a_1,0}\delta_{a_2,0} + \delta_{a_1 + a_2,r})\,\frac{a_1(r - a_1)}{2r z^2}\\
	 & = \norder{v^{a_1}(z) v^{a_2}(z)} \,\,\, + \hbar\,\delta_{a_1 + a_2,r}\,\frac{a_1a_2}{2rz^2}\,. 
\end{split}
\end{equation*}
The general formula follows from an easy induction argument.
\end{proof}

\begin{defin}
\label{psirootsum} We introduce certain sums over $r$-th roots of unity which we encounter throughout our computations
\begin{equation}
\label{eq:psidef}
\Psi^{(j)} (a_{2j+1},\ldots, a_i)  := \frac{1}{i!} \sum_{\substack{m_1, \ldots , m_{i}=0 \\ m_{l} \neq m_{l'}}}^{r-1} \left( \prod_{l'=1}^{j}\frac{\theta^{m_{2l' - 1}+m_{2l'}}}{(\theta^{m_{2l'}} - \theta^{m_{2l' - 1}})^2}   \prod_{l=2j+1}^{i}\theta^{-m_{l} a_{l}}\right)\,.
\end{equation}
In the special case where $ j = 0 $, we drop the $ (0) $ \textit{i.e.} $ \Psi(a_1,\ldots,a_i) := \Psi^{(0)}(a_1,\ldots,a_i) $. 
\end{defin}

Note that we prove several properties of these sums over roots of unity in Appendix \ref{Approot}.

\begin{defin} \label{DeM0} We introduce the twist fields
$$
W^i(z) := r^{i-1}Y_\sigma \big(e_i(\chi^0_{-1},\ldots,\chi^N_{-1}) \ket{0},z\big)\,,\qquad i \in \{1,\ldots,r\}\,.
$$
\end{defin}
The scalar prefactor $ r^{i-1} $ is just a convenient normalization. Let us express the twist fields in terms of the Heisenberg twist fields $ v^l(z) $.

\begin{prop} \label{prop:twistfields} We have for any $i \in \{1,\ldots,r\}$
$$
	W^i(z)  = \frac{1}{r}  \sum_{\substack{a_{2j+1}, \ldots, a_i = 0 \\ j \leq \lfloor i/2 \rfloor }}^{r - 1} \frac{i!}{2^{j}\,j! (i-2j)!}  \Psi^{(j)} (a_{2j + 1},\ldots, a_i) 
	(\hbar z^{-2})^{j}\,\,\, \norder{\prod_{l = 2j + 1}^{i} v^{a_{l}}(z)}\,.
$$
\end{prop}
\begin{proof}
	
	We express the elementary symmetric polynomials  $ e_i(\chi^0,\ldots,\chi^N) $  in terms of the basis $(v^i)_{i = 0}^{r - 1}$ of $ \mathfrak{h} $. Inverting~\eqref{eq:vln}, we get
$$
\chi^i = \frac{1}{r} \sum_{a=0}^{r - 1} \theta^{-ia}v^{a},
$$
and plugging it into the expression for the elementary symmetric polynomials gives
$$
	e_i(\chi^{0},\ldots, \chi^{N}) = \frac{1}{r^i}   \sum_{a_1, \ldots, a_i = 0}^{r - 1} \Psi(-a_1,\ldots,-a_i) \,v^{a_1} v^{a_2}\cdots\,v^{a_i}\,.
$$ 
We observe that $\Psi(-a_1,\ldots,-a_i) = \Psi(a_1,\ldots,a_i)$. Now, we use Lemma ~\ref{lem:p} to compute the twist fields associated to $ e_i(\chi^0_{-1},\ldots,\chi^N_{-1}) \ket{0} $. 
	\begin{equation}
	\label{eq:l415}
	\begin{split}
	W^i(z) &=  \frac{1}{r}    \sum_{a_1, \ldots, a_i = 0}^{r - 1} \Psi(a_1,\ldots, a_i) Y_\sigma\big(v^{a_1}_{-1} v^{a_2}_{-1} \cdots\,v^{a_i}_{-1} \ket{0},z\big)   \\ 
	&=\frac{1}{r}   \sum_{a_1, \ldots, a_i = 0}^{r - 1} \Psi(a_1,\ldots, a_i) \sum_{B \in \mathcal{P}(a_1,\ldots,a_i)} (\hbar z^{-2})^{|B|} \prod_{\{b_1,b_2\} \in B} \frac{b_1b_2\,\delta_{b_1 + b_2,r}}{2r} \,\,\, \norder{\prod_{l\in A\setminus B} v^l(z)} \,.
	\end{split}
	\end{equation}

We would like to separate the sums over the $2j$ indices appearing in the pairs and the others, for $j \in \{0,\ldots,\lfloor i/2\rfloor\}$. As $A$ is an ordered set, we first need to identify the subset $J \subseteq \{1,\ldots,i\}$ of cardinality $|J| = 2j$ which correspond the indices of elements of $A$ that appear in $B$. For fixed $j$, there are $\frac{i!}{(2j)!(i - 2j)!}$ such $J$s and the corresponding terms in the sum \eqref{eq:l415} are all equal. For instance, they are equal to the case $\{1,\ldots,2j\}$. $B$ now corresponds to a choice of a pairing between elements of $J$. The sum over the values $a_k \in \{0,\ldots,r - 1\}$ for $k \in J$ will not depend on the choice of pairing $B$. There are $(2j - 1)!!$ such pairings. It is enough to consider the single pairing $B = \{(1,2),(3,4),\ldots,(2j - 1,2j)\}$ provided we multiply our sums by
$$
\frac{i!}{(2j)!(i - 2j)!}\cdot (2j - 1)!! = \frac{i!}{2^{j}j!(i - 2j)!}\,.
$$
Consequently,
$$
W^i(z) = \frac{1}{r} \sum_{a_1,\ldots,a_{i} = 0}^{r - 1} \sum_{j = 0}^{\lfloor i/2 \rfloor} \frac{i!\,(\hbar z^{-2})^{j}}{2^{j}\,j!(i - 2j)!}\,\Psi(a_1,\ldots,a_{i})\,\prod_{l' = 1}^{j} \frac{a_{2l' - 1}a_{2l'}\,\delta_{a_{2l' - 1} + a_{2l'},r}}{2r}\,\,\,\norder{\prod_{l = 2j + 1}^{i} v^{a_{l}}}\,.
$$
The claim follows by performing the sum over $a_{1},\ldots,a_{2j}$ using Lemma~\ref{Ajsum} proved in Appendix \ref{Approot}.
\end{proof}

\begin{defin}
\label{DeM1} We define the modes $ W^i_k $ of the twist field $ W^i(z) $ as 
$$
W^i(z) = \sum_{k \in \mathbb{Z} } W^i_k z^{-k-1}\,,
$$
We observe that the expression for the modes $W_k^i$ only involve integer powers of $\hbar$
\end{defin}

We extract the expression for the modes from Proposition ~\ref{prop:twistfields}.

\begin{cor} \label{cor:twist} We have
\begin{equation} \label{eq:Wmodes}
	W^i_k =\frac{1}{r}  \sum_{j = 0}^{\lfloor i/2 \rfloor}  \frac{i!\,\hbar^{j}}{2^{j}\,j!(i-2j)!} \sum_{\substack{p_{2j + 1},\ldots,p_i \in \mathbb{Z} \\ \sum_{l} p_{l} = r(k-i+1)}}  \Psi^{(j)} (p_{2j+1}, p_{2j+2}, \ldots , p_i)  \,\,\,\norder{\prod_{l = 2j + 1}^{i} J_{p_{l}} }\,,
\end{equation}
		where for cases such that $j = i/2$ the condition $\sum_l p_l = r (k-i+1)$ is understood as the Kronecker delta condition $\delta_{k,i-1}$.
\end{cor}

\begin{proof}
	We start with Proposition ~\ref{prop:twistfields} and compute the residue
\begin{equation*}
\begin{split}
	W^i_k &= \frac{1}{r}  \sum_{j = 0}^{\lfloor i/2 \rfloor} \sum_{a_{2j+1}, \ldots, a_i=0}^{r - 1}  \frac{i!\,\hbar^{j}}{2^{j}\,j! (i-2j)!}\, \Psi^{(j)} (a_{2j+1},\ldots, a_i)
	\displaystyle \Res_{z = 0}\bigg(\dd z\,z^{k-2j}\,\,\norder{\prod_{l = 2j + 1}^{i} v^{a_{l}}(z)}\bigg) \\
	&= \frac{1}{r}  \sum_{j = 0}^{\lfloor i/2 \rfloor} \sum_{a_{2j + 1},\ldots,a_i = 0}^{r - 1} \frac{i!\,\hbar^{j}}{2^j\,j! (i-2j)!} \sum_{\substack{k_{l} \in a_{l}/r +\mathbb{Z} \\ \sum_{l} k_{l} = k-i+1} }   \Psi^{(j)}(rk_{2j + 1},\ldots, rk_i)\,\,\norder{\prod_{l=2j+1   }^{i} J_{rb_l}}\,.
	\end{split}
	\end{equation*}
To get to the second line, we used that $\Psi^{(j)}$ is a $r$-periodic function of each of its arguments, because they appear as powers of $r$-th roots of unity. Summing over $a_{2j + 1},\ldots,a_i$ amounts to summing over $p_{l} = r k_{l} \in \mathbb{Z}$ with the only constraint $\sum_{l} p_{\beta} = r(k-i+1)$. Note that in the case where $j=i/2$, the condition $\sum_l p_l = r (k-i+1)$ becomes the delta condition that $k=i-1$.
\end{proof}

It is easy to compute the $\Psi^{(j)}(a_{2j + 1},\ldots,a_i)$ for low values of $i$ (see Lemma~\ref{LEmg}). For instance, we have the linear and quadratic operators for $r \geq 2$
\begin{equation}
\label{W1k}
\begin{split}
W^1_{k} & =  J_{kr}\,, \\
W^2_{k} & =  \frac{1}{2} \sum_{\substack{p_1,p_2 \in \mathbb{Z} \\ p_1 + p_2 = r(k-1)}} \big(r\delta_{r|p_1}\delta_{r|p_2} -1\big)\,\,\,\norder{J_{p_1} J_{p_2}} - \frac{(r^2 - 1)\hbar}{24} \delta_{k,1}\,.
\end{split}
\end{equation}
For $r \geq 3$ we have the cubic operator
\begin{equation}
\label{W3k}
\begin{split}
W^3_{k} & =  \frac{1}{6} \sum_{\substack{p_1,p_2,p_3 \in \mathbb{Z} \\ p_1 + p_2 + p_3 = r(k-2)}} \big(r^2\delta_{r|p_1}\delta_{r|p_2}\delta_{r|p_3} - r\delta_{r|p_1} - r\delta_{r|p_2} - r\delta_{r|p_3} + 2\big)\,\,\,\norder{J_{p_1}J_{p_2}J_{p_3}}  \\
 & \quad - \frac{(r - 2)(r^2 - 1)\hbar}{24}\,J_{r(k-2)}\,,
\end{split}
\end{equation}
and so on.

\subsubsection{The higher quantum Airy structures}

\label{mafm}

We are ready to prove one of our main results. As noted in Example \ref{wgln}, we know that the $\mathcal{W}(\mathfrak{gl}_{N+1})$ vertex algebra with central charge $\mathfrak{c} = N+1$ is strongly freely generated by the states $e_i(\chi^0_{-1},\ldots,\chi^{N}_{-1}) \ket{0}$. Thus we can use the construction of Section \ref{s:subalgebra} to obtain a number of left ideals for the algebra of modes of the twist fields $W^i(z)$ that are graded Lie subalgebras. This gives us the second condition that is required to obtain a higher quantum Airy structure.
For the first condition, we need to modify the modes $W_k^i$ so as to create a term of degree $1$ of the form $J_{p}$  for some $p > 0$ --- which acts as a derivation on $\mathcal{T}(\mathfrak{gl}_{N+1})$. This can be achieved via the following operation.
\begin{defin}
	We define the \textit{dilaton shift} as a conjugation of the differential operators $ W^i_k $
$$
	H^i_k := \hat{T}_{s} W^i_k \hat{T}_{s}^{-1},\qquad \hat{T}_{s} := \exp\Big(-\frac{J_{s}}{s\hbar}\Big)\,.
$$
\end{defin}
We note here that by the Baker--Campbell--Hausdorff formula, conjugating by $\hat{T}_s$ is equivalent to shifting $J_{-s} \to J_{-s} - 1$ in the modes $W^i_k$.

We then construct the following class of higher quantum Airy structures

\begin{thm}\label{WHAS} Let $ r\geq 2$, and $s \in \{1, \ldots, r+1 \}$ be such that $ r  = \pm 1 \mod s$. Let
$$
\mathfrak{d}^i := i-1-  \Big\lfloor \frac{s(i-1)}{r} \Big\rfloor \, .
$$
 Assume $J_0 = Q = 0$. The family of differential operators
\beq
\label{QHAS2W} H_k^i = \hat{T}_{s}W^i_{k}\hat{T}_{s}^{-1}\qquad i \in \{1,\ldots,r\},\qquad k \geq \mathfrak{d}^i + \delta_{i,1}\, , 
\eeq
forms a quantum $r$-Airy structure on the vector space $V = \bigoplus_{p > 0} \mathbb{C}\langle x_p \rangle $ equipped with the basis of linear coordinates $(x_p)_{p > 0}$. 
\end{thm}

\begin{proof}
We note that the $ W^i_k $ defined in \eqref{eq:Wmodes} is a differential operator on $\mathbb{C}_{\hbar^{1/2}}[\![x_1,x_2,x_3,\ldots]\!]$ which is a linear combination of terms of degree $i + 2j$ for $j \in \{0,\ldots,\lfloor i/2 \rfloor\}$ using the notion of degree introduced in \eqref{eq:hbardeg}. We need to check the two conditions of Definition~\ref{def:HAS} for the differential operators $H_k^i$. 

First, we note that the algebra of modes of a VOA-module has the same Lie algebraic structure as the modes of the VOA itself. Further, conjugating by $\hat{T}_s$ does not change the algebra of the modes. Then, the graded Lie subalgebra condition for higher quantum Airy structures follows directly from Section \ref{s:subalgebra}. In the case $s=r+1$, the indicated $W^i_k$ form the graded Lie subalgebra $\mathcal{A}_{\geq 0}$; in the case $s=1$, they form the graded Lie subalgebra $\mathcal{A}_{\Delta}$;  and for the remaining values of $1 < s < r$ such that $ r \pm 1 = 0 \mod s $, we prove in Appendix~\ref{a:rs} that we get a partition of $ r $ (as in Section \ref{s:subalgebra} --- see Theorem \ref{t:sub}), and they form a graded Lie subalgebra $\mathcal{A}_{(r,s)}$. The only subtlety here is that in these subalgebras, the mode $W^1_0=Q$ is always present; since it is a scalar, to be part of a higher quantum Airy structure we must set $Q=0$.

To check the second condition about the form of the operators $H^i_k$, we need to identify the terms of degree at most $1$ in $H^i_{k}$. To start, let us assume that $s \in \mathbb{Z}$ arbitrary.  We first examine the terms of degree $1$. Clearly, since $J_0 = Q = 0$, a term $\norder{\prod_{l = 2j + 1}^i J_{p_{l}}}$ will contribute if and only if $j = 0$ and there is some $l_0$ such that for any $l \neq l_0$ we have $p_{l} = -s$. The constraint on the sum of $p$s imposes $p_{l_0} = rk + (s-r)(i - 1)$. We therefore obtain using the $r$-periodicity of $\Psi$ in each argument
\beq
\label{HKI2}H_k^i = \frac{i}{r}\,(-1)^{i - 1}\,\Psi\big(\underbrace{-s,-s,\ldots,-s}_{i - 1\,\,{\rm times}},(i - 1)s\big)\,\,J_{rk + (s-r)(i - 1)} + O(2)\,,
\eeq
where $O(2)$ indicates terms of degree $\geq 2$. Thus, we see that we need to choose $ (k,i) $ such that 
\begin{equation}\label{eq:cond}
rk + (s-r)(i - 1) > 0.
\end{equation}
In addition, we need to check that the prefactor in equation~\eqref{HKI2} involving $ \Psi $ is always  non-zero. Before we do that, let us consider the terms of degree 0. A term $\norder{\prod_{l = 2j+ 1}^i J_{p_{l}}}$ will contribute in degree $ 0 $ if and only if $j = 0$  and $p_{l} = -s$ for all $l$. The constraint on the sum of $p$ imposes $rk + i(s-r) + r =0$. We see from condition~\eqref{eq:cond} that 
\[
	rk + i(s-r) + r > s.
\] Thus, we choose $ s > 0 $ to ensure that no terms of degree $ 0 $ appear. When $ s > 0 $, the prefactor involving $\Psi$ is evaluated in Lemmas~\ref{Psicom} and \ref{Psicom2} and shown to be never zero. In particular, for $s$ coprime to $r$, we get
$$
H_k^i = J_{rk + (s - r)(i - 1)} + O(2)\,.
$$

Let us introduce the set $\mathcal{I}_{r,s} =\{(i,k)\,\,|\,\, 1 \leq i \leq r\,\,{\rm and}\,\, k \geq \mathfrak{d}^i + \delta_{i,1}\}$ and the map 
	\begin{equation}
	\label{gofugnh} \Pi_{s}\,:\,\begin{array}{ccl} \mathcal{I}  &\longrightarrow & \mathbb{Z} \\ 
	 (i,k) & \longmapsto & rk + (s-r)(i - 1) \end{array}\,.
	\end{equation}
We obtain a higher quantum Airy structure if $\Pi_s$ is a bijection onto $\mathbb{Z}_{+}$, \textit{i.e.} if each $J_{p} = \hbar \partial_{x_p}$ with $p > 0$ appears exactly in one operator $H_k^i$ for $(i,k) \in \mathcal{I}_{r,s}$. It is easy to see that the non-empty fibers of $\Pi_s$ have cardinality $d = {\rm gcd}(r,s)$. In other words, when $r$ and $s$ are not coprime, the same $\hbar \partial_{x_p}$ will appear as degree one term in two different operators $H^k_i$, which cannot happen in higher quantum Airy structure. Let us now assume that $r$ and $s$ are coprime, so that $\Pi_s$ is injective. We can rewrite the condition~\eqref{eq:cond} as the condition $k > i-1 - \frac{s}{r} (i-1)$. For $i=1$, this is $k \geq 1$. For $i \geq 2$, since $s$ is coprime with $r$ and $2 \leq i \leq r$, it follows that $k > i-1 - \frac{s}{r} (i-1)$ if and only if $k \geq i-1-  \lfloor \frac{s(i-1)}{r} \rfloor$.  Therefore $\Pi_{s}(\mathcal{I}_{r,s}) = \mathbb{Z}_{+}$.
\end{proof}
From the last paragraph of the proof, we see that the first condition to be a higher quantum Airy structure restricts the allowed values of $ s $ to be positive integers that are coprime to $ r $. The  second condition to be a higher Airy structure, or equivalently the subalgebras of modes that we identified in Section \ref{s:subalgebra}, imposes the stronger constraint that $ r = \pm 1 \mod s $.

\begin{rem}
For completeness, we compute $F_{0,3}$ and $F_{1,1}$ for all these higher quantum Airy structures in Appendix \ref{a:rs}. In fact, we do a little bit more; we calculate $F_{0,3}$ for all choices of $s$ that are coprime with $r$. The result is that $F_{0,3}$ is indeed well defined and symmetric for $r = \pm 1 \mod s$, as expected; however, it cannot be symmetric when $r \neq \pm 1 \mod s$ (see Proposition \ref{p:symmetry}). In other words, when $r \neq \pm 1 \mod s$, the $H_k^i$ cannot form a higher quantum Airy structure, since a solution $Z$ to the differential constraints $H_k^i Z = 0$ does not exist. Given that for any $s$ coprime with $r$ the $H_i$ have the right form to be a higher quantum Airy structure, it follows that the left ideal generated by the $H_i$ is a graded Lie subalgebra if and only if $r = \pm 1 \mod s$.
\end{rem}

Let $F_{g,n}$ be the coefficients of the partition function of the Airy structure of Proposition~\ref{WHAS}. We can derive from it the following basic properties.
\begin{lem}
For $2g - 2 + n > 0$ and $p_1,\ldots,p_n > 0$, we have
\begin{itemize}
\item Homogeneity:
\[
\sum_{i = 1}^{n} p_i \neq s(2g - 2 + n) \quad \Longrightarrow \quad F_{g,n}[p_1,\ldots,p_n] = 0.
\]
\item If $r|p_m$ for some $m$, then $F_{g,n}[p_1,\ldots,p_n] = 0$.
\item Dilaton equation
\[
F_{g,n + 1}[s,p_1,\ldots,p_n] = s(2g - 2 + n) F_{g,n}[p_1,\ldots,p_n].
\]
\item More generally, for $d \geq -\delta_{s,r + 1}$, we have the Virasoro constraints
\begin{equation}
\begin{split}
& \quad�F_{g,n + 1}[s + dr,p_1,\ldots,p_n] \\
&  = \sum_{m = 1}^n p_m F_{g,n}[k_i + dr,p_1,\ldots,\widehat{p_i},\ldots,p_n] \\
{} & \quad�+ \frac{1}{2} \sum_{l = 1}^{dr - 1} \bigg(F_{g - 1,n + 2}[l,dr - l,p_1,\ldots,p_n] + \sum_{\substack{J \sqcup J' = \{p_1,\ldots,p_n\} \\ h + h' = g}} F_{h,1 + |J|}[l,J]F_{h',1 + |J'|}[dr - l,J']\bigg),
\end{split}
\end{equation}
with the convention that the insertion of a negative index is zero. The dilaton equation is $d = 0$, and for $s = r + 1$ we also have the string equation.
\end{itemize}
\end{lem}

\subsubsection{Reduction to \texorpdfstring{$\mathfrak{sl}_{N+1}$}{sl(r)}}
\label{sec:explic}

The quantum $r$-Airy structures of Theorem~\ref{WHAS} always contain $H^1_{k} = J_{kr}$ for $k > 0$. Hence their partition function $Z$ is independent of the variables $x_{kr}$ for $k > 0$. Let us define the reduced operators by the formula
\begin{equation}
\begin{split}
W_k^i|_{\rm red} & = W_k^i|_{J_{kr} = 0\,\,k \in \mathbb{Z}}  \\
& = \frac{1}{r} \sum_{j = 0}^{\lfloor i/2 \rfloor} \frac{i!\,\hbar^{j}}{2^{j}\,j!\,(i - 2j)!} \sum_{\substack{p_{2j + 1},\ldots,p_{i} \in \mathbb{Z}\setminus r\mathbb{Z} \\ \sum_{l} p_{l} = r(k - i + 1)}} \Psi^{(j)}(p_{2j + 1},\ldots,p_{i})\,\norder{\prod_{l = 2j + 1}^{i} J_{p_{l}}}\,.
\end{split}
\end{equation}
As the dilaton shift in Theorem~\ref{WHAS} does not affect the modes indexed by $k$ divisible by $r$, we also have
$$
H_k^i|_{\rm red} = \hat{T}_{s} W_{k}^{i}|_{\rm red} \hat{T}_{s}^{-1} = H_k^i|_{J_{pr} = 0\,\,p \in \mathbb{Z}}\,.
$$
Although we do not know a general reason for $H_k^i|_{{\rm red}}$ to be a quantum Airy structure itself, for this particular case we can check that it is indeed the case. We also reprove Lemma~\ref{lem:redd} in this particular case.

\begin{lem}
\label{consred} Let us consider a quantum $r$-Airy structure from Theorem~\ref{WHAS}. Its partition function is equivalently characterized by the constraints $J_{kr}\cdot Z = 0$ for any $k > 0$ and 
\beq
\label{Regsfsun} J_{kr}\cdot Z = 0\,\,\,k > 0,\qquad {\rm and}\qquad H_{k}^i|_{{\rm red}}\cdot Z = 0,\qquad i \in \{2,\ldots,r\},\qquad k \geq \mathfrak{d}^i + \delta_{i,1}\,.
\eeq 
Moreover, the family of operators $H_{k}^{i}|_{\rm red}$ indexed by $i \in \{2,\ldots,r\}$ and $k \geq \mathfrak{d}^i + \delta_{i,1}$ forms a quantum $r$-Airy structure on the vector space with basis of linear coordinates $(x_{p})_{p \in \mathbb{N} \setminus r\mathbb{N}}$.
\end{lem}

\begin{proof} As a preliminary, we are going to show that $H_k^i$ can be expressed solely in terms of the reduced operators. Since the dilaton shift does not affect the modes $(J_{kr})_{k \in \mathbb{Z}}$ it is enough to prove this property for $W_k^i$ instead of $H_k^i$, and the result will follow by conjugation. We can always decompose
$$
W_k^i = \sum_{\substack{\ell,m \geq 0 \\ \ell + m \leq i}} \sum_{\substack{a_1,\ldots,a_\ell > 0 \\ b_1,\ldots,b_m > 0}} J_{-ra_1}\cdots J_{-ra_{\ell}}\,\Upsilon^i_{k,\mathbf{a},\mathbf{b}}\,J_{rb_1}\cdots J_{rb_{m}}\,.
$$ 
where the $\Upsilon^i_{k,\mathbf{a},\mathbf{b}}$ do not involve the modes $J_{rp}$ for $l \in \mathbb{Z}$. Using the expressions~\eqref{eq:Wmodes} for the operators $W^i_k$, and   using the $r$-periodicity of $\Psi^{(j)}$ with respect to any of its entries, we get:
\begin{equation*}
\begin{split}
\Upsilon^{i}_{k,\mathbf{a},\mathbf{b}} & = \frac{1}{r} \sum_{j = 0}^{\lfloor (i - \ell - m)/2 \rfloor} \sum_{\substack{p_{2j + 1},\ldots,p_{i - \ell - m} \in \mathbb{Z}\setminus r\mathbb{Z} \\ \sum_{l} p_{l} = r(k - i + 1 + \sum_{l} a_{l} - \sum_{l'} b_{l'})}}  \frac{i!\,\hbar^{j}}{2^{j}\,j!(i - \ell - m - 2j)!}  \\
& \qquad \qquad \qquad \times \Psi^{(j)}(p_{2j + 1},\ldots,p_{i - \ell - m},\underbrace{0,\ldots,0}_{\ell + m\,\,{\rm times}})\,\, \norder{\prod_{l = 2j + 1}^{i - \ell - m} J_{p_{l}}}\,.
\end{split}
\end{equation*}
We also used that $J_{-m_{l}r}$ are always  on the left (resp. $J_{n_{l'}r}$ are on  the right) of a normal ordered expression, so we can remove them outside the normal ordering. The   $\Psi^{(j)}$ with the $ 0 $s in them is evaluated using the Lemma~\ref{lem:Psizero} proved in the Appendix to get
\begin{equation*}
\begin{split}
\Upsilon^i_{k,\mathbf{a},\mathbf{b}} & = \frac{1}{r} \frac{(r - i + \ell + m)!}{(r - i)!}\,\frac{(i - \ell - m)!}{i!}  \\
&\times \sum_{j = 0}^{\lfloor (i - \ell - m)/2 \rfloor} \sum_{\substack{p_{2j + 1},\ldots,p_{i - \ell - m} \in \mathbb{Z}\setminus r\mathbb{Z} \\ \sum_{l} p_{l} = r(k - i + 1 + \sum_{l} m_{l} - \sum_{l'} n_{l'})}} \frac{i!\,\Psi^{(j)}(p_{2j + 1},\ldots,p_{i -\ell - m})}{2^{j}\,j!(i - \ell - m - 2j)!}\,\,\norder{\prod_{l = 2j + 1}^{i - \ell - m} J_{p_{l}}} \,,
\end{split}
\end{equation*}
and therefore
\beq
\label{linearc} H^i_{k} = \sum_{\substack{\ell,m \geq 0 \\ \ell + m \leq i}} \sum_{\substack{a_1,\ldots,a_{\ell} > 0 \\ b_1,\ldots,b_{m} > 0}} \frac{(r - i + \ell + m)!}{(r - i)!}\,J_{-ra_1}\cdots J_{-ra_{\ell}}\,H^{i - \ell - m}_{k + \sum_{l} (a_{l} - 1) - \sum_{l'} (b_{l'} + 1)}|_{{\rm red}} \,J_{rb_1}\cdots J_{rb_{m}}\,.
\eeq

Now consider the constraints $H_k^i \cdot Z = 0$ for $i \in \{1,\ldots,r\}$ and $k \geq \mathfrak{d}^i + \delta_{i,1}$. They contain $H^1_{k} \cdot Z = J_{kr}\cdot Z = 0$ for all $k > 0$ so the partition function is independent of $x_{rm}$ for $m > 0$. As a result, for $i \geq 2$ and $k \geq \mathfrak{d}^i$, the coefficient of $x_{rb_1}\cdots\, x_{rb_m}$ in $H_k^i \cdot Z$ is proportional to $\Upsilon^i_{k,\mathbf{a},\emptyset}\cdot Z$ therefore to $H^{i - \ell}_{k + \sum_{l} (a_{l} - 1)}|_{\rm red}$. Since $m_{l} > 0$, we get the family of constraints
\beq  
\label{hfgfsg}H_k^i|_{{\rm red}}\cdot Z = 0,\qquad i \in \{2,\ldots,r\}\,,\qquad k \geq \mathfrak{d}^i\,.
\eeq 
Conversely, the constraints \eqref{hfgfsg} together with $J_{rk}\cdot Z = 0$ for $k > 0$ imply, by reconstructing the linear combinations \eqref{linearc}, that $H_k^i\cdot Z = 0$ for $i \in \{1,\ldots,r\}$ and $k \geq \mathfrak{d}^i + \delta_{i,1}$.

For the last statement, let
$$
V = \bigoplus_{p > 0} \mathbb{C}\langle x_{p} \rangle\,\qquad V_{{\rm red}} = \bigoplus_{p \in \mathbb{N}\setminus r\mathbb{N}} \mathbb{C}\langle x_{p} \rangle\,,
$$ 
and consider the Weyl algebras $\mathcal{D}_{T^*V_{{\rm red}}}^{\hbar} \subset \mathcal{D}_{T^*V}^{\hbar}$ of differential operators on $V_{\rm red}$ and $V$. Let $\mathcal{J}^+$ be the graded subalgebra of $\mathcal{D}_{T^*V}^{\hbar}$ generated by the $J_{pr}$ for $\pm p > 0$. We have a canonical decomposition
$$
\mathcal{D}_{T^*V}^{\hbar} = \mathcal{J}^{-} \mathcal{D}_{T^*V_{\rm red}}^{\hbar} \mathcal{J}^{+}\,,
$$
and a natural projection $\rho\,:\,\mathcal{D}_{T^*V}^{\hbar} \rightarrow \mathcal{D}_{T^*V_{\rm red}}^{\hbar}$. By definition
$$
H_k^i|_{\rm red} = \rho(H_k^i)\,.
$$
We denote $\mathcal{H}_{\rm red}$ --- respectively $\mathcal{H}$ --- the subspace spanned by $H_k^i|_{\rm red}$  for $i \in \{2,\ldots,r\}$, respectively $i \in \{1,\ldots,r\})$ --- and $k \geq \mathfrak{d}^i + \delta_{i,1}$ over the field $\mathbb{C}(\!(\hbar)\!)$. The graded Lie subalgebra condition for these $H_k^i$ translates into
$$
[\mathcal{H},\mathcal{H}] = \hbar\,\mathcal{D}_{T^*V}^{\hbar}\cdot \mathcal{H}\,.
$$
Let us apply the projection $\rho$ to this equation. We get on the right-hand side $\hbar\,\mathcal{D}_{T^*V}^{\hbar}\cdot \mathcal{H}_{\rm red}$. On the left-hand side, we have to take into account that if $j^{\pm}_1,j^{\pm}_2 \in \mathcal{J}^{\pm}$ and $h_1,h_2 \in \mathcal{D}_{T^*V_{\rm red}}^{\hbar}$ 
$$
\big[j^+_1h_1j^{-}_1,j^{+}_2h_2j^{-}_{2}\big] - j^+_1\,[j^-_1,j_2^+]\,h_1h_2 j_2^-  - j_2^+[j_1^+,j_2^-]h_2h_1 j_1^{-}
= j_1^+j_2^+[h_1,h_2]j_1^-j_2^-\,.
$$
After applying $\rho$ we find a result of the form
$$
\rho\big[j^+_1h_1j^{-}_1,j^{+}_2h_2j^{-}_{2}\big] - \hbar\,c\,h_1h_2 + \hbar\,c'\,h_1h_2 =   [h_1,h_2]\,,
$$
for some $c,c' \in \mathbb{C}[\![\hbar]\!]$. Therefore
$$
[\mathcal{H}_{\rm red},\mathcal{H}_{\rm red}] = \hbar\,\mathcal{D}_{T^*V_{\rm red}} \mathcal{H}_{\rm red}\,,
$$
which proves that the ideal generated by $\mathcal{H}_{\rm red}$ is a graded Lie subalgebra. As it is already clear that for any $p \in \mathbb{N}\setminus r\mathbb{N}$ there exists a unique $(k,i)$ such that $H_k^i|_{\rm red} = \hbar\,\partial_{x_{p}} + O(2)$, this proves the claim.
\end{proof}

\subsubsection{Arbitrary dilaton shifts and changes of polarization}

In this subsection, we will construct deformations of the quantum $r$-Airy structures of Theorem~\ref{WHAS}, by exploring more general conjugations. Although this may seem superfluous, these examples will appear naturally in the next section when we study higher quantum Airy structures coming from general spectral curves.

We first introduce more general dilaton shifts. We would like to conjugate the modes $ W^i_k $ in \eqref{eq:Wmodes} by an operator of the form
$$
\hat{T} := \exp\left(\frac{1}{\hbar} \sum_{l > 0} \frac{\tau_{l}}{l} J_l \right)\,,
$$
where $\tau_l$ are scalars. This simultaneously shifts $J_{-l} \rightarrow J_{-l} + \tau_l$ for all $l > 0$. 

\begin{prop} \label{t:condition} Let $r \geq 2$. Denote
$$
s := \min\{l > 0\,\,|\,\,\tau_{l}\ \neq 0\,\,\,{\rm and}\,\,\,r \nmid l \},\qquad \mathfrak{d}^i := i-1- \Big \lfloor \frac{s(i-1)}{r} \Big\rfloor
$$
and assume that $1 \leq s \leq r + 1$ and $r = \pm 1 \mod s$. The family of differential operators
\beq
\label{fisgubf}H^i_k = (-\tau_{s})^{1 - i}\,\hat{T}W^i_{k}\hat{T}^{-1}\qquad i \in \{1,\ldots,r\},\qquad k \geq \mathfrak{d}^i + \delta_{i,1}  ,
\eeq
forms a quantum $r$-Airy structure up to a change of basis of linear coordinates.
\end{prop}
\begin{proof}
We need to show that the two conditions in Definition~\ref{def:HAS} are satisfied. Since $\tau_s \neq 0$, we define $  \tilde{\tau}_{q} $ as
$$
\tau_{q} = \tau_{s}(\delta_{s,q} + \tilde{\tau}_{q})\,,
$$
so that $\tilde{\tau}_{q} = 0$ for $q \leq s$. We compute as in the proof of Theorem~\ref{WHAS} that (up to rescaling by constants)
$$
H_k^i = \frac{i}{r}(-1)^{i - 1} \sum_{\substack{p \in \mathbb{Z},\,\,q_2,\ldots,q_{i} \geq s \\ p = r(k - i + 1) + \sum_{l = 2}^{i} q_{l}}} \Psi(-q_2,\ldots, -q_{i},p) \bigg[ \prod_{l=2}^{i} (\delta_{s,q_{l}} + \tilde{\tau}_{q_l})\bigg]\,J_{p}   + O(2)\,.
$$
Therefore we can write
$$
H_k^i = \sum_{p \geq \Pi_s(i,k)} L_{\Pi_s(i,k),p}\,J_{p} + O(2)\,,
$$
where $\Pi_{s}$ was defined in \eqref{gofugnh} and is a bijection between the set of indices $(i,k)$ considered in \eqref{fisgubf} and the set of positive integers. $(L_{a,b})_{a,b > 0}$ is an upper triangular matrix with diagonal entries $1$ (this value comes from Lemma~\ref{Psicom2} for $r$ and $s$ coprime). Let us perform the change of basis on linear coordinates
$$
y_{b} = \sum_{m \geq 0} (-1)^{m} \sum_{b = a_0 > a_1 > \ldots > a_{m - 1} > a_{m} > 0} \bigg[ \prod_{l = 0}^{m - 1} L_{a_{l},a_{l + 1}} \bigg]\,x_{a_m}\,.
$$
For any $b > 0$ the right-hand side is well defined as it is a finite sum (using the upper-triangularity of $L$). By construction we have
$$
H_{k}^i = \sum_{p \geq \Pi_s(i,k)} L_{\Pi_s(i,k),p}\,\hbar \partial_{x_p} + O(2) = \frac{\partial}{\partial y_{\Pi_s(i,k)}} + O(2)\,.
$$
Notice that these expressions make sense using the prescriptions for vector spaces of countable dimension described in Section~\ref{infinite}, \textit{i.e.} $\partial_{y_p}$ are elements of $V$ and linear coordinates are elements of the dual. We therefore have checked the first condition of Definition~\ref{def:HAS}.

The graded Lie subalgebra condition which holds for the operators of Theorem~\ref{WHAS} is preserved after conjugation. Hence we obtain a higher quantum Airy structure.
\end{proof}

Another conjugation that will appear in the next section is the change of polarization. We would like to conjugate our modes with an operator of the form
$$
\hat{\Phi} := \exp  \bigg( \frac{1}{2\hbar} \sum_{l,m > 0} \frac{\phi_{l,m}}{l\,m}\,J_{l}J_{m}\bigg)\,,
$$
where $\phi_{l,m} = \phi_{m,l}$ are scalars. Using the Baker-Campbell-Hausdorff formula, we see that it shifts the modes as
\beq
\label{shifteq}\forall a > 0, \qquad J_{-a} \longrightarrow J_{-a}  + \sum_{l > 0} \frac{\phi_{a,l}}{l} J_l\,.
\eeq
and leaves  $J_a$ invariant if $a > 0$.

\begin{prop}\label{t:conjugated}
Under the same conditions as in Proposition \ref{t:condition}, the family of differential operators
$$
H_k^i = (-\tau_s)^{1 - i}\,\hat{\Phi}\,\hat{T}W^i_k\hat{T}^{-1}\,\hat{\Phi}^{-1},\qquad i \in \{1,\ldots,r\},\qquad k \geq \mathfrak{d}^i+ \delta_{i,1}\,,
$$ 
forms a quantum $r$-Airy structure up to a change of basis of linear coordinates.
\end{prop}

\begin{proof}
The graded Lie subalgebra condition is stable under conjugation. We are going to argue that
\beq
\label{phigj}\hat{\Phi}\,(\hat{T}W_k^i\hat{T}^{-1})\,\hat{\Phi}^{-1} = (\hat{T}W_k^i\hat{T}^{-1}) + O(2)\,.
\eeq
This will automatically imply that the $(-\tau_s)^{1 - i}\,\hat{\Phi}\hat{T}W_k^i\hat{T}^{-1}\hat{\Phi}^{-1}$ satisfy the first condition in Definition~\ref{def:HAS}, hence form a quantum $r$-Airy structure.

We observe that the operation \eqref{shifteq} respects the degree. It  replaces $J_{-l}$s by $J_m$s. If the result is not normal ordered anymore, normal ordering creates a new term where  two $J$s are replaced by a $\hbar$ (which is still of the same degree). As there is no term of degree $1$ of the form $J_{-l}$ with $l > 0$ in $H^i_k$ we get the claimed \eqref{phigj}.
\end{proof}

\subsection{\texorpdfstring{$\mathcal{W}(\mathfrak{gl}_{N+1})$}{W(gl(N + 1))} Airy structures for other automorphisms}

\label{s:glra}

\subsubsection{The twisted module}

We come back to the $\mathcal{W}(\mathfrak{gl}_{N+1})$ algebra, but now we construct twisted modules for an arbitrary automorphism $\sigma$, consisting of $d \geq 2$ disjoint cycles of order $r_1,\ldots,r_{d}$ which sum to $r:=N+1$. We relabel  the basis elements of $ \mathfrak{h} $

$$
\chi^{\mu,i} := \chi^{i - 1 + \sum_{\nu < \mu} r_{\nu}}\qquad \mu \in \{1,\ldots,d\},\qquad i \in \{1,\ldots,r_{\mu} \}\,,
$$  such that
$$
\sigma(\chi^{\mu,i}) = \chi^{\mu,i + 1\,\,{\rm mod}\,\,r_{\mu}}\,.
$$

We then introduce the basis of eigenvectors indexed by $\mu \in \{1,\ldots,d\}$ and $a \in \{0,\ldots,r_{\mu} - 1\}$
$$
v^{\mu,a} = \sum_{j = 0}^{r_{\mu} - 1} \theta_{r_{\mu}}^{-aj}\,\chi^{\mu,j},\qquad \theta_{r_{\mu}} = e^{2\ii \pi/r_{\mu}}\,.
$$
which are diagonal under the $ \sigma $ action
$$
\sigma(v^{\mu,a}) = \theta_{r_{\mu}}^{a}v^{\mu,a},\qquad \langle v^{\mu,a},v^{\nu,b} \rangle = \delta_{\mu,\nu}\,r_{\mu}\,\delta_{r_{\mu}|a + b}\,.
$$
Hence we can represent the $\mathcal{S}_{0}(\mathfrak{gl}_{N+1})$-twisted module
$$
\mathcal{T}(\mathfrak{gl}_{N+1}) \cong \mathbb{C}_{\hbar^{1/2}}[x_{1}^1,x_{1}^2,\ldots,x_{1}^{d},x_{2}^1,x_{2}^{2},\ldots,x_{2}^{d},x_{3}^{1},\ldots]\,,
$$
with the fields
$$
v^{\mu,a}(z) = \sum_{k \in a/r_{\mu} + \mathbb{Z}} J^{\mu}_{kr_{\mu}}\,z^{- k - 1}\,,
$$
and
$$
J^{\mu}_{l} = \left\{\begin{array}{lll} \hbar\,\partial_{x_{l}^{\mu}} & & l > 0 \\ Q^{\mu} & & l = 0 \\ -l\,x_{-l}^{\mu} & & l < 0 \end{array}\right. \,.
$$
where $Q^{\mu}$ now are arbitrary scalars. Upon restriction, it becomes an untwisted $\mathcal{W}(\mathfrak{gl}_{N+1})$-module.

\begin{rem}
In contrast to Section~\ref{Sfgggfsg}, we  have the freedom to take $J^{\mu}_{0} = Q^{\mu} \neq 0$ in our construction of Airy structures, provided $Q^{\mu}$ is equal to $0$ modulo terms of positive degree. Scalars with this property in $\mathbb{C}_{\hbar^{1/2}}$ must be $O(2)$. Note that $J_{l}$ for $l \neq 0$ are elements of $\mathcal{D}_{T^*V}^{\hbar}$ of degree $1$. It is therefore natural to replace the base field with $\mathbb{C}_{\hbar^{1/2}}$ to allow scalars of degree $1$. So we construct crosscapped Airy structures as defined in Section~\ref{Scrosscap} rather than usual Airy structures.
\end{rem}

We are going to compute the modes $W_{k}^i$  of the twist fields associated to the strong generators of $\mathcal{W}(\mathfrak{gl}_{N+1})$. The result is expressed in terms of the modes $W_k^{\mu,i}$ of the $\mathcal{W}(\mathfrak{gl}_{r_{\mu}})$-module constructed in Section~\ref{SWgg} via twisting by the Coxeter element of $\mathfrak{gl}_{r_{\mu}}$, which are  according to Corollary~\ref{eq:Wmodes}
$$
W_{k}^{\mu,i} = \frac{1}{r_{\mu}} \sum_{j = 0}^{\lfloor i/2 \rfloor} \frac{i!\,\hbar^{j}}{2^{j}j!(i - 2j)!} \sum_{\substack{p_{2j + 1},\ldots,p_i \in \mathbb{Z} \\ \sum_{l} p_{l} = r_{\mu}(k - i + 1)}} \Psi^{(j)}_{r_{\mu}}(p_{2j + 1},p_{2j + 2},\ldots,p_i)\,\,\,\norder{\prod_{l = 2j + 1}^{i} J^{\mu}_{p_{l}}}\,,
$$
where we have use the  notation $\Psi^{(j)}_{r_{\mu}}$ for $\Psi^{(j)}$ to insist  that we choose the $r_{\mu}$-th roots of unity for its definition.

\begin{lem}
\label{ojngdh}We have
\beq
\label{gfofig111}W_k^i = r^{i - 1} \sum_{M \subseteq \{1,\ldots,d\}} \sum_{\substack{1 \leq i_{\mu} \leq r_{\mu}\,\,\mu \in M \\ \sum_{\mu} i_{\mu} = i}} \sum_{\substack{\mathbf{k} \in \mathbb{Z}^{M} \\ \sum_{\mu} k_{\mu} = k + 1 - |M|}} \prod_{\mu \in M} \frac{1}{r_{\mu}^{i_{\mu} - 1}}\,W^{\mu,i_{\mu}}_{k_{\mu}}\,.
\eeq
\end{lem}
\begin{proof}
We express the strong generators of the $\mathcal{W}(\mathfrak{gl}_{N+1})$ by grouping the basis elements that belong to the same cycle  $\sigma$ together
$$
e_i(\chi^0,\ldots,\chi^{r - 1}) = \sum_{\substack{i_1,\ldots,i_{d} \geq 0 \\ \sum_{\mu} i_{\mu} = i}} \prod_{\mu = 1}^{d} e_{i_{\mu}}(\chi^{\mu,1},\ldots,\chi^{\mu,r_{\mu}})\,.
$$
Now we compute the fields associated to these generators in our twisted module. We note that the modes corresponding to different $\mu$ commute, and for each $\mu$ we recognize (up to the factor  $r^{i_{\mu} - 1}$) the fields associated with the $\mathfrak{gl}_{r_{\mu}}$ generators in Definitions~\ref{DeM0}-\ref{DeM1}. Therefore
$$
W^i(z) = r^{i - 1} \sum_{\substack{i_1,\ldots,i_d \geq 0 \\ \sum_{\mu} i_{\mu} = i}} \prod_{\mu = 1}^{d} \frac{1}{r_{\mu}^{i_{\mu} - 1}}\,W^{\mu,i_{\mu}}(z)\,,
$$ 
where by convention $W^{\mu,0}(z) = 1$. Collecting the coefficient of $z^{-k - 1}$ entails the claim.
\end{proof}

\subsubsection{Higher quantum Airy structures}
\label{Section422} 
It seems rather tedious to find all the dilaton shifts of \eqref{gfofig111} that could lead to  higher quantum Airy structures.   Instead, we focus on the case $\sigma$ is a cycle of length $r - 1$, that is
$$
r_1 = r - 1,\qquad r_2 = 1\,.
$$ 
According to Lemma~\ref{ojngdh}, we have

\begin{equation}
\label{WWWWW}
\begin{split}
W_k^{1} &=  J_{r_1k}^{1} + J_{k}^{2}\,, \\
(r_1/r)^{i - 1}\,W_{k}^{i} & =   W_{k}^{1,i} + \sum_{\substack{k_1,k_2 \in \mathbb{Z} \\ k_1 + k_2 = k - 1}} r_1\,W_{k_1}^{1,i - 1}\,J^{2}_{k_2} \qquad i \in \{2,\ldots,r_1\}\,,  \\
 (r_1/r)^{r - 1}\,W_{k}^{r} & =  \sum_{\substack{k_1, k_2 \in \mathbb{Z} \\ k_1 + k_2 = k - 1}} r_1 \,W_{k_1}^{1,r_1}J^2_{k_2} \,.
\end{split}
\end{equation}
Let $s \in \{1, \ldots, r_1+1 \}$ with $s$ coprime with $r_1$. We perform a dilaton shift $J^1_{-s} \rightarrow J^1_{-s} - 1$, \textit{i.e.} we define
\beq
\label{eq:dshiftr1}H_k^i = (r_1/r)^{i - 1}\exp\bigg(- \frac{J_{s}}{\hbar s}\bigg)W_{k}^i\exp\bigg(\frac{J_{s}}{\hbar s}\bigg)\,.
\eeq
 
\begin{thm}
\label{t:ropen} Let $s \in \{1,\ldots, r\}$ such that $s|r$, and let
$$
\mathfrak{d}^i := i - 1 - \Big\lfloor \frac{s(i - 1)}{r - 1}\Big\rfloor\,.
$$ 
Let $q \in \mathbb{C}$ and assume that $Q^1 = \hbar^{1/2}q = -Q^2$. The family of operators 
$$
\begin{array}{cllll}
H_k^{1} + H_{r-1-s+k}^{r} & = & J_{(r-1)k}^{1} + O(2) & \qquad & k \geq 1 \\
H_{k}^i & = & J_{(r-1)(k - i+1) + s(i - 1)}^1 + O(2) &\qquad & k \geq \mathfrak{d}^i,\qquad i \in \{2,\ldots,r-1 \} \\
-H_{k}^{r} & = & J^2_{k + s - r+1} + O(2) & \qquad & k \geq (r - s)
\end{array}\,.
$$
forms a crosscapped higher quantum Airy structure on $\bigoplus_{p > 0} \big(\mathbb{C}\langle x_p^1 \rangle \oplus \mathbb{C}\langle x_p^2  \rangle\big)$.
\end{thm}

\begin{proof}
We first check the subalgebra property. In the case $s=r_1+1 = r$, the indicated $W^i_k$ form the graded Lie subalgebra $\mathcal{A}_{\geq 0}$; in the case $s=1$, they form the graded Lie subalgebra $\mathcal{A}_{\geq 0}$.  For the remaining values of $1 < s < r_1$, we need to find for what values of $(s,r_1)$ do we get a partition of $r$ (as in Theorem \ref{t:sub}). Using Proposition \ref{p:partition}, replacing $r \rightarrow r_1$, we know that the modes with $i \in \{1, \ldots, r_1\}$ generate an ideal that is a graded Lie subalgebra if $r_1 = \pm 1 \mod s$. What we need to check is that the left ideal generated by adding the modes $W^{r_1+1}_k$ with $k \geq r_1 + 1 - s$ is still a graded Lie subalgebra. For this to be the case, we need to show that the enlarged set of modes still correspond to a partition.

For $W^{r_1}_k$ and $s<r_1$, the condition is $k \geq r_1 -1 - s + \lceil \frac{s}{r_1} \rceil = (r_1-1) - (s-1)$. Moreover, for $W^{r_1+1}_k$ the condition that we want is $k \geq r_1 - (s-1)$. In the notation of Theorem \ref{t:sub}, this means that we are adding one to the last part of the partition corresponding to the subalgebra generated by the modes with $1 \leq i \leq r_1$. This will remain an ordered partition only if all other parts of the original partition are at least one larger than the last part. Looking again at Proposition \ref{p:partition}, we see that this will be the case precisely when $r_1 = r' s + r''$ with $r'' = s -1$. In other words, $r_1 = -1 \mod s$, or equivalently $s|r$. Therefore we conclude that the left ideal generated by the modes $W^i_k$, with $i \in \{1,\ldots, r_1+1 \}$ and satisfying the condition above, is a subalgebra if and only if $s|r$.

For all these cases, as the $W^i_k$ satisfy the graded Lie subalgebra condition, so do the $H^k_i$ with the same indices. Now as usual we need to be careful with zero modes. For $i = 1$ it is clear that
\beq
\label{H0110}H_k^1 = W_k^1 = J_{r_1k}^{1} + J_{k}^2\,.
\eeq
The graded Lie subalgebras contain the mode $H_{0}^1$ which is equal to the scalar $Q^1 + Q^2$. We must assume $Q^1 + Q^2 = 0$ if we desire to have a higher quantum Airy structure. Due to the condition on the degrees, we fix $Q^1 = \hbar^{1/2}q = -Q^2$ for some $q \in \mathbb{C}$. We can drop the zero differential operator $H_0^1$ and deduce that $H_k^i$ for $i \in \{1,\ldots,r\}$ and $k \geq \mathfrak{d}^i + \delta_{i,1} + \delta_{i,r}$ still satisfy the graded Lie subalgebra condition.

It remains to check that the degree $1$ condition holds. We start by computing the result of the shift $J_{-s} \rightarrow J_{-s} - 1$ in $W_k^{1,i}$. Compared to Section~\ref{mafm}, the fact that $Q^1= \hbar^{1/2}q$ gives an extra term
$$
\exp(-\partial_{x_s^1}) r^{-(i - 1)}W_{k}^{1,i}\exp(\partial_{x_s^1}) = -\frac{\delta_{i,r_1}\,\delta_{r_1s + r_1(k - i + 1),0}}{r_1} + \frac{\hbar^{1/2}q}{r_1}\delta_{k,0}\delta_{i,1} + \,J_{r_1(k - i + 1) + s(i - 1)}^1 + O(2)\,.
$$
In particular for $i < r_1$ the degree $0$ term (the first term) vanishes. We now consider $i \in \{2,\ldots,r_1\}$ and compute $H_k^i$ modulo $O(2)$. We get from \eqref{WWWWW}
\beq
\label{Hgfgmi}H_k^i = -\frac{\delta_{i,r_1}\,\delta_{s + k - i + 1,0}}{r_1} + \,J_{r_1(k -i + 1) + s(i - 1)}^1 + O(2)\,.
\eeq
For $s \in \{1, \ldots r_1 + 1 \}$ coprime with $r_1$  and $k \geq \mathfrak{d}^i$ we see that the degree $0$ term in \eqref{Hgfgmi} is absent.  Under these conditions, we have
$$
H_k^i =\,J_{r_1k  + (s - r_1)(i - 1)}^1 + O(2)\,,
$$
which involves a $J_p^1$ with $p > 0$. Finally, we compute $H_k^{r}$ modulo $O(2)$ from \eqref{WWWWW} and find
$$
H_k^{r} = -J_{k + s - r_1}^{2} + O(2)\,.
$$
For $s \in \{1, \ldots r_1 + 1 \}$ coprime with $r_1$  and $k \geq r_1+1-s$ we see that $k+s-r_1 \geq 1$, and hence the $J$ appearing there is a derivation. We then see that
$$
\begin{array}{cllll}
H_k^{1} + H_{r_1-s+k}^{r} & = & J_{r_1k}^{1} + O(2) & \qquad & k \geq 1 \\
H_{k}^i & = & J_{r_1(k - i+1) + s(i - 1)}^1 + O(2) &\qquad & k \geq \mathfrak{d}^i,\qquad i \in \{2,\ldots,r_1 \} \\
-H_{k}^{r} & = & J^2_{k + s - r_1} + O(2) & \qquad & k \geq r_1+1 - s
\end{array}\,.
$$
forms a quantum $r$-Airy structure, which is the claim.
\end{proof}

We will discuss the enumerative geometry interpretation (through open intersection theory) of the associated partition function in Section~\ref{Sopen}. Note that we can easily formulate and prove an analog of Proposition~\ref{t:conjugated} to describe more general higher quantum Airy structures obtained from the ones of Proposition~\ref{t:ropen} by further dilaton shifts and changes of polarization.

\subsection{The \texorpdfstring{$\mathcal{W}(\mathfrak{so}_{2N})$}{W(so(2N))} Airy structures}

\label{s:DN}

\subsubsection{The twisted module \texorpdfstring{$\mathcal{T}$}{T}}

Recall Example~\ref{wdn}. The roots of the Lie algebra of $ D_N $ type can be described as $( \pm \chi_i\pm \chi_j) $ where $ \chi_i $ is an orthonormal basis for $ \mathbb{C}^N $. Let $ \sigma $ be the Coxeter element of the Weyl group, defined by the following action
$$
\chi_1 \to \chi_2 \to \cdots \to \chi_{N-1} \to -\chi_1 \to -\chi_2 \to \cdots\, -\chi_{N-1} \to \chi_1 \qquad {\rm and}\qquad  \chi_N \to - \chi_N\,.
$$
This element has order $ r = 2(N-1) $. We  define a basis $(v^1,v^3,v^5, \ldots, v^{r-1}, \tilde{v}) $ of $ \mathfrak{h} $ that is diagonal under the action of $ \sigma $ as follows
\begin{equation}
\label{eq:vlnd}
\begin{split}
v^a &= \sqrt{2} \sum_{j=0}^{\frac{r}{2} -1 } \theta^{-aj} \chi_{j+1} \qquad a \in \{1,3,5,\ldots,r-1\}\,, \\
\tilde{v} &= \sqrt{2}\,\chi_N\,,
\end{split}
\end{equation}
It has the property $\sigma(v^a) = \theta^{a}v^a$ and $\sigma(\tilde{v}) = -\tilde{v}$, and the inner product for any $a,b \in \{1,3,\ldots,r - 1\}$
\beq
\label{evaluop}\langle v^a,v^b \rangle = r\delta_{r|a + b}\,,\qquad \langle \tilde{v},v^a \rangle = 0\,,\qquad \langle \tilde{v},\tilde{v} \rangle = 2\,.
\eeq

We can therefore define
$$
\mathcal{T}(\mathfrak{so}_{2N}) = \mathbb{C}_{\hbar^{1/2}}[\![x_1,x_3,x_5,\ldots,\tilde{x}_1,\tilde{x}_3,\ldots]\!]
$$
with
\begin{equation}
	\label{eq:vtoJd}
	v^a(z) := \sum_{k \in a/r + \mathbb{Z}} J_{rk}z^{-k-1},\qquad \tilde{v}(z) = \sum_{k \in 1/2 + \mathbb{Z}} \widetilde{J}_{2k}\,z^{-k - 1}\,,
	\end{equation}
	where we take for any odd $l > 0$
	$$
	J_{l} = \hbar\partial_{x_l},\qquad J_{-l} = lx_{l},\qquad \widetilde{J}_{l} = \hbar\partial_{\tilde{x}_l},\qquad \widetilde{J}_{-l} = l\tilde{x}_l\,.
	$$
The evaluation of the pairing \eqref{evaluop} shows that these assignments reproduce the desired commutation relations of the Heisenberg algebra. Thus $\mathcal{T}(\mathfrak{so}_{2N})$ is a twisted $\mathcal{S}_{0}$-module, which we restrict to obtain a $\mathcal{W}(\mathfrak{so}_{2N})$-module.

\subsubsection{Twist fields for the generators}

$\mathcal{W}(\mathfrak{so}_{2N})$ is freely and strongly generated by $\nu^{d}$ for $d \in \{2,4,6,\ldots,2N - 2\}$ and $\tilde{\nu}^N$. The corresponding fields can be computed using the following lemma.

 \begin{lem}\cite[Lemma 3.7]{BakalovMilanov2} \label{lem:dgen}
For all $ d \geq 1 $, $i \in \{1,\ldots,N\}$ and $\varepsilon \in \{-1,1\}$, we have
$$
Y_\sigma({\rm e}^{\varepsilon \chi^i}_{-d}{\rm e}^{-\varepsilon\chi^i}_{-1},z ) = \sum_{\ell=0}^{d} \frac{\hbar^{\ell/2}}{z^{\ell}}\, c_{i}^{(\ell)}\,S_{d-\ell} (\varepsilon\chi^i,z)\,,
$$
where for any $v \in \mathbb{C}^{N}$
$$
S_n(v,z) := \norder{\frac{1}{n!} \left(\hbar^{1/2}\partial_z  + v(z) \right) ^n 1}
$$
are the Fa\`a di Bruno polynomials, and $c$s are scalars such that $c_{i}^{(0)} = 1$.
 \end{lem}
The main ingredient of the proof in \cite{BakalovMilanov2} is the product formula~\eqref{prodid}. Our version only differs by specialization to the orthonormal basis vectors $\chi^i$, and inserting suitable powers of $\hbar$. These merely keep track of the conformal weights. For instance, we would like $ \partial_z $ to have degree $ 1 $ and hence we add a $ \hbar^{1/2} $. The $ \hbar^{\ell/2}$ comes from the product formula.

\begin{rem}
\label{rereDD}	$ Y(\nu^d,z) $ is a field of conformal weight $ d $. The $ \hbar $ grading keeps track of this weight. In particular, the half-integer powers of $ \hbar $ will vanish in $ Y_\sigma(\nu^d,z) $. This is easy to see using that $\nu^{d}$ is invariant under $ \chi^i \to -\chi^i $.
\end{rem}

We define the modes of the twist fields $ Y(\nu^d,z) $ and $ Y(\tilde{\nu}^N,z) $ by
$$
 Y_\sigma(\nu^d,z) = \sum_{k \in \mathbb{Z} } W^d_k z^{-k-1}\,,\qquad  Y_\sigma(\tilde{\nu}^N,z) = \sum_{k \in \mathbb{Z} } \widetilde{W}^N_k z^{-k-1}\,.
$$

\subsubsection{Higher quantum Airy structures}

We can then apply Propositions ~\ref{lem:modesplus} and \ref{lem:modesdelta} to get  graded Lie subalgebras by considering the ideals generated by certain subsets of the modes. As in the $\mathfrak{gl}_{N + 1}$ case, we need to perform a dilaton shift in order to get a higher quantum Airy structure. In this section we will only consider the subalgebras of Propositions ~\ref{lem:modesplus} and \ref{lem:modesdelta}, since we have not constructed the more general intermediate subalgebras analogous to Theorem \ref{t:sub} for $\mathcal{W}(\mathfrak{so}_{2N})$.

\begin{defin}
The dilaton shifted modes of the module $\mathcal{T}(\mathfrak{so}_{2N})$ are defined as follows
\begin{equation*}
\begin{split}
H_k^d & = \gamma_d\,\hat{T}_{s}W_{k}^{d}\hat{T}_{s}^{-1}\qquad\quad d \in \{2,4,\ldots,2(N - 1)\}\,, \\
\widetilde{H}_{k}^{N} & = \tilde{\gamma}_{N,s}\,\hat{T}_{s}\widetilde{W}_{k}^{N}\hat{T}_{s}^{-1}\,,
\end{split}
\end{equation*}
with constants
$$
\gamma_{d} = d^{-1}2^{d/2}(N - 1)^{d - 1},\qquad \tilde{\gamma}_{N,s} = (-1)^{s(N - 2)/2}2^{(N - 1)/2}(N - 1)^{N - 1}\,,
$$
and we recall that $\hat{T}_{s} = \exp\big(-\frac{J_{-s}}{\hbar s}\big)$.
\end{defin}

For $s = 1$ or $r + 1$, we obtain in this way quantum $r$-Airy structures.

 \begin{thm}
 \label{t:DN}
 	Let $N \geq 3$, that is $r = 2(N - 1) \geq 4$.  The family of differential operators
\begin{equation*}
\begin{split}
	H_k^d = \gamma_d\,\hat{T}_{r + 1} W_k^d \hat{T}_{r + 1}^{-1} &\qquad d \in \{2,4,\ldots,2(N - 1)\},\,\,\,\,\,k \geq 0\,, \\
	\widetilde{H}_k^{N} = \tilde{\gamma}_{N,r + 1}\,\hat{T}_{r + 1}\widetilde{W}_{k}^{N} \hat{T}_{r + 1}^{-1} &\qquad  k \geq 0
	\end{split}
	\end{equation*}
	forms a quantum $r$-Airy structure on the vector space $V = \bigoplus_{p > 0} \mathbb{C}\langle x_{2p+1} \rangle \oplus \mathbb{C}\langle \tilde{x}_{2p+1} \rangle$. The same is true for the family of differential operators
\begin{equation*}
\begin{split}
	H_k^d = \gamma_d\,\hat{T}_{1} W_k^d\hat{T}_{1}^{-1} & \qquad  d \in \{2,4,\ldots,2(N - 1)\},\qquad k \geq d - 1 \\
	 \widetilde{H}_{k}^{N} = \tilde{\gamma}_{N,1}\,\hat{T}_{1} \widetilde{W}_{k}^{N}\hat{T}_{1}^{-1} & \qquad  k \geq N - 1\,.
\end{split}
\end{equation*}
 \end{thm}

\begin{proof}
We first look at the modes of $ Y_\sigma(\nu^d,z) $ for $d \in \{2,4,\ldots,2N - 2\}$ and study the terms that can contribute in degree $ \leq 1 $ after a dilaton shift $J_{-s} \rightarrow J_{-s} + 1$ for some $s \in 2\mathbb{Z} + 1$. From Lemma~\ref{lem:dgen}, we see that
$$
Y_{\sigma}(\nu^d,z) = \sum_{i = 1}^{N - 1} (\chi^i(z))^{d} + O(\hbar)\,.
$$
Now, we implement the change of basis \eqref{eq:dbasis}. The inverse change of basis is
$$
\chi^i = \frac{1}{\sqrt{2}(N - 1)} \sum_{\substack{1 \leq a \leq r - 1 \\ a\,\,{\rm odd}}} \theta^{a(i - 1)}\,v^{a},\qquad \chi^N = \frac{\tilde{v}}{\sqrt{2}}\,.
$$
We note that the modes $W_k^d$ of $\nu^d$ are homogeneous differential operators of degree $d$. The mode $J_{-s} = x_{s}/s$ that appears in the dilaton shift is only present in a single $v^{s}(z)$ as coefficient of $z^{s/r - 1}$. So
$$
\hat{T}_{s}^{-1}\chi^i(z)\hat{T}_{s} = \frac{\theta^{-(i - 1)s}\,z^{s/r - 1}}{\sqrt{2}(N - 1)} + \chi^i(z)\,.
$$
Consequently, the terms of degree $0$ and $1$ will be
\begin{equation*}
\begin{split}
&�\quad \frac{d\,H_k^d}{2^{d/2}(N - 1)^{d - 1}}  + O(2) \\
& =  [z^{-(k + 1)}] \bigg(\frac{z^{(s/r - 1)d}\,\delta_{r|s}}{2^{d/2}(N - 1)^{d - 1}} + \frac{d}{2^{(d - 1)/2}(N - 1)^{d - 1}} \sum_{i = 1}^{N - 1} \theta^{-(i - 1)s(d - 1)} z^{(s/r - 1)(d - 1)}\,\chi^i(z) \bigg)  \\
& =  [z^{-(k + 1)}]\,\bigg(\frac{d}{2^{d/2}(N - 1)^{d}} \sum_{i = 1}^{N - 1} \sum_{\substack{1 \leq a \leq r - 1 \\ a\,\,{\rm odd}}} \sum_{m \in \mathbb{Z}} \theta^{(i - 1)(a - s(d -1))}\,J_{a + rm}\,z^{-a/r - m - 1}\,z^{(s/r - 1)(d - 1)}\bigg) \\ 
& =  \frac{d\,J_{r(k - d + 1) + (d - 1)s}}{2^{d/2}(N - 1)^{d - 1}}\,,  
\end{split}
\end{equation*}
In the third line, we dropped the term $\delta_{r|s}$ because $s$ is odd and $r$  is even. So there is no degree $0$ term and we obtain $H_k^d = J_{r(k - d + 1) + (d - 1)s} + O(2)$. This explains the choice of the constant prefactor in the definition of $H_k^d$.

Now, we consider the modes of the twist field $ Y_\sigma(\tilde{\nu}^N,z) $. We use a similar argument as above, using the product formula~\eqref{prodid} instead of Lemma~\ref{lem:dgen}
\begin{equation*}
\begin{split}
\frac{(-1)^{-s(N - 2)/2}\,\widetilde{H}_{k}^i}{2^{(N - 1)/2}(N - 1)^{N - 1}} + O(2) & =  [z^{-(k + 1)}]\,\bigg(\prod_{i = 1}^{N - 1} \frac{\theta^{-s(i - 1)}\,z^{s/r - 1}}{\sqrt{2} (N - 1)}\bigg)  \sum_{m \in 1/2 + \mathbb{Z}} \widetilde{J}_{2m}\,z^{-m - 1} + O(2) \\ 
& =  \frac{(-1)^{-s(N - 2)/2}}{2^{(N - 1)/2}(N - 1)^{N - 1}}\,\widetilde{J}_{2k - s + r}\,,
\end{split}
\end{equation*}
hence $\widetilde{H}^N_{k} =  \widetilde{J}_{2k - s + r} + O(2)$.

First, we consider  the graded Lie subalgebra $\mathcal{A}_{\geq 0}$ corresponding to the modes $k \geq 0$ (see Proposition~\ref{lem:modesplus}). If we want these modes to contain each $J_{p}$ and $\widetilde{J}_{p}$ (for $p > 0$ odd) exactly once, we must choose $s = r + 1$, and we do obtain a quantum $r$-Airy structure.

Second, if we  focus on the graded Lie subalgebra $\mathcal{A}_{\Delta}$ corresponding to the modes $H_k^d$ with $k \geq d - 1$ and $\widetilde{H}_k^{N}$ with $k \geq N - 1$ (see Proposition~\ref{lem:modesdelta}), the same condition forces us to choose $s = 1$ and we obtain a quantum $r$-Airy structure in this case as well.

\end{proof}

As usual, we note that we can easily formulate and prove an analog of Proposition~\ref{t:conjugated} to describe more general higher quantum Airy structures obtained from the ones of Theorem \ref{t:DN} by further dilaton shifts and changes of polarization.

\subsection{The exceptional types}
\label{s:EN}
\label{SecE}
\subsubsection{The twisted module}

We consider the simple complex Lie algebra $\mathfrak{e}_{N}$ with $N \in \{6,7,8\}$, together with a Coxeter element of the Weyl group $\sigma$. Its order is denoted by $r$. $\sigma$ acts on the Cartan subalgebra with simple eigenvalues $(\theta_{r}^{d_a - 1})_{a = 1}^{N}$. We can always order $2 \leq d_1 < \cdots < d_N \leq r$, and we have $d_{N + 1 - a} + d_{a} = r + 2$ for any $a \in \{1,\ldots,N\}$. Let $\mathbb{D} = \{d_1 - 1,\ldots,d_N - 1\}$ 
$$
\begin{array}{|c|c|c|}
\hline
N & r & d_1,\ldots,d_N \\
\hline
6 & 12 & 2,5,6,8,9,12 \\
\hline
7 & 18 & 2,6,8,10,12,14,18 \\
\hline
8 & 30 & 2,8,12,14,18,20,24,30 \\
\hline
\end{array}
$$
We can find a basis of eigenvectors $(v^a)_{a = 1}^N$ such that
$$
\sigma(v^a) = \theta_{r}^{d_{a} - 1}\,v^a,\qquad \langle v^a,v^b \rangle = r\delta_{a + b,r}\,.
$$
We obtain a $\sigma$-twisted module for the Heisenberg algebra
$$
\mathcal{T}(\mathfrak{e}_{N}) = \mathbb{C}_{\hbar^{1/2}}[\![(x_l)_{l \in \mathbb{D} + r\mathbb{N}}]\!]
$$
by assigning
$$
v^a(z) = \sum_{p \in (d_a - 1)/r + \mathbb{Z}} J_{rp}\,z^{-p - 1}\,,
$$
where $J_{l} = \hbar \partial_{x_l}$ for $l > 0$, $J_{0} = 0$ and $J_{l} = -l\,x_{-l}$ for $l < 0$. Via restriction, we get an untwisted $\mathcal{W}(\mathfrak{e}_N)$-module over $\mathbb{C}_{\hbar^{1/2}}$.
  
\subsubsection{The generators and their twist fields}
 
The VOA $\mathcal{W}(\mathfrak{e}_N)$ is strongly and freely generated by elements $w_i$ of conformal weight $d_i$ for $i \in \{1,\ldots,N\}$. Although there is no canonical choice making the generators particularly simple, we will rely on the following structural result.

\begin{thm} \cite{FF1} and \cite[Lemma 3.4]{Yangunique}
\label{thmde}One can choose generators of the form\footnote{The discrepancy with the notations in \cite{Yangunique} comes from the fact that we use a dual basis.}
$$
w^i = (v^{N}_{-1})^{d_i - 1}v^{N + 1 - i}_{-1} + \sum_{m = 2}^{d_i} (v^N_{-1})^{d_i - m}\,P_{i,m}(v^{1}_{-1},\ldots,v^{N - 1}_{-1}) + \tilde{w}^{i} \,,
$$
where $P_{i,d}$ is a homogeneous polynomial of degree $d_i - d$ and $\tilde{w}_{i}$ belongs to the left ideal generated by the $\chi_{-n}$ for $\chi \in \mathfrak{h}$ and $n \geq 2$.
\end{thm}

The fields associated to $w_i$ are denoted
$$
Y_{\sigma}(w^i,z) = \sum_{k \in  \mathbb{Z}} W^i_{k}\,z^{- k - 1}\,.
$$

\begin{rem}\label{remEE} As $\mathfrak{e}_{N}$ is a subalgebra of $\mathfrak{so}_{2N}$, and the generators of the $\mathcal{W}(\mathfrak{so}_{2N})$ algebra after $\hbar$-rescaling indicated in Section~\ref{hbarresc} only involve integer powers of $\hbar$ (see Remark~\ref{rereDD}), the same must be true for the generators of $\mathcal{W}(\mathfrak{e}_{N})$.
\end{rem}

\begin{defin}
The dilaton shifted modes of $\mathcal{T}(\mathfrak{e}_{N})$ are defined as
$$
H_k^i := \hat{T}_{s}^{-1}W_k^i\hat{T}_{s},\qquad \hat{T}_{s} = \exp\bigg(-\frac{J_{s}}{ \hbar s}\bigg)\qquad i \in \{1,\ldots,N\},\qquad k \in \mathbb{Z}\,.
$$
\end{defin}

For $s = 1$ or $r + 1$ we obtain in this way quantum $r$-Airy structures. The one with $s = r + 1$ was anticipated in \cite{Yangunique}, while the one for $s = 1$ is new.

\begin{thm}
\label{thEEE} Let $\epsilon \in \{0,1\}$ and denote $s:= 1 + \epsilon r$. The family of differential constraints
$$
H_k^i\qquad i \in \{1,\ldots,N\},\qquad k \geq (1 - \epsilon)(d_i - 1)
$$
forms a quantum Airy structure on the vector space $V = \bigoplus_{i = 1}^{N} \bigoplus_{l > 0} \mathbb{C}\langle x_{rl + d_{i - 1}} \rangle$.
\end{thm}

\begin{proof}
 The formulas \eqref{productfff}-\eqref{prodid} show that we have for $k_1,\ldots,k_n > 0$ and $\chi^{i_1},\ldots,\chi^{i_n} \in \mathfrak{h}$
$$
Y_{\sigma}\big(\chi_{-k_1}^{i_1}\cdots \chi_{-k_n}^{i_n}\ket{0},z\big) = \big(\delta_{\sum_{j} (k_j - 1),0} + \hbar^{1/2}\delta_{\sum_{j} (k_j - 1),1}\big)\,\,  \norder{ \prod_{j = 1}^{n} \frac{\dd^{k_j - 1}}{\dd z^{k_j - 1}} \chi^{i_1}(z)} + O(2)\,,
$$
where the $O(2)$ arises from terms involving $\hbar^{m}$ with $m \geq 1$. Following Theorem~\ref{thmde} the mode $W_k^i$ is a sum of three parts, which we study them up to $O(2)$. The first part is
$$
\sum_{p_1,\ldots,p_{d_i} \in \mathbb{Z}} \delta\bigg(-(k + 2 - d_i) + \sum_{j = 1}^{d_i} (p_j + 1)\bigg)\,\norder{J_{d_i - 1 + rp_{1}}\,J_{r - 1 + rp_2}\cdots\,J_{r - 1 + rp_{d_i}}\,J_{d_i - 1 + rp_{d_i}}} + O(2)\,,
$$
where $\delta(n) : =\delta_{n,0}$ is the Kronecker delta. The second part is a sum, over $m \in \{2,\ldots,d_i\}$ and $a_1,\ldots,a_m \in \{1,\ldots,N - 1\}$ of an expression which up to $O(2)$ is proportional to
$$
\sum_{p_1,\ldots,p_{d_i} \in \mathbb{Z}} \delta\bigg( - (k + 1 + m - d_i) -\frac{d_i}{r} + \sum_{j = 1}^m \frac{d_{a_j}}{r} + \sum_{j = 1}^{d_i} (p_j + 1)\bigg)\,\norder{J_{d_{a_1} - 1 + rp_{1}}\cdots\,J_{d_{a_m} - 1 + rp_{m}}\,J_{r - 1 + rp_{m + 1}}\cdots\,J_{r - 1 + rp_{d_i}}}\,.
$$
The third part is a sum, over $\ell \in \{1,\ldots,d_i - 1\}$, $b_1,\ldots,b_{\ell - 1} > 0$ and $a_1,\ldots,a_q \in \{1,\ldots,N\}$ such that if we set $b_{q} := 2$ we have $\sum_{j = 1}^{q} b_j = d_i$, of an expression which up to $O(2)$ is proportional to
$$
\hbar^{(d_i - \ell)/2} \sum_{p_1,\ldots,p_{\ell} \in \mathbb{Z}} \delta\bigg(-(k + 1) + \sum_{j = 1}^{\ell} \frac{d_{a_j} - 1}{r} + p_j + b_j\bigg)\,\prod_{j = 1}^{\ell} \frac{\Gamma\big(\frac{d_{a_j} - 1}{r} + p_j + b_j\big)}{\Gamma\big(\frac{d_{a_j} - 1}{r} + p_j + 1\big)} \,\,\norder{ \prod_{j = 1}^{\ell} J_{d_{a_j} - 1 + rp_j}} + O(2)\,.
$$
Due to the power of $\hbar$ in prefactor, up to $O(2)$, we only have to take into account the terms where $\ell = d_i - 1$ and $b_{1} = \cdots = b _{d_i - 2} = 1$.

For a fixed $\epsilon \in \{0,1\}$ we set $s = 1 + \epsilon r$ and  apply the shift $J_{-s} \rightarrow J_{-s} + 1$ to obtain $H_k^i$. We want to identify the terms of degree $0$ and $1$. In degree $0$, there is no contribution from the first and second parts, because they do not come from a monomial of the form $(v^{N}(z))^{q}$ for some $q > 0$, and since the third part has at least a power of $\hbar^{1/2}$ it does not contribute either. We now turn to degree $1$ terms. The first part contributes for $p_2 = \cdots = p_{d_i} = - (1 + \epsilon)$ and yields
$$
J_{s(d_i - 1) + r(k + 1 - d_i)} + O(2)\,.
$$
The second part remains $O(2)$ since it is at least quadratic in the non-shifted $J_{l}$. The third part could contribute when $\ell = d_i - 1$, $b_1 = \cdots = b_{d_i - 2} = 1$ and $(a_j,p_j) = (N,-1 - \epsilon)$ for all $j \in \{1,\ldots,d_i - 1\}$. But the Kronecker delta would impose that $r|(d_i - 1)$ which is never possible. Hence
$$
H_k^i = J_{\Pi_s(i,k)} + O(2),\qquad \Pi_{s}(i,k) := s(d_i - 1) + r(k + 1 - d_i)\,.
$$

Let us denote
$$
\tilde{S}_{s} = \big\{(i,k) \quad \big|\quad i \in \{1,\ldots,N\}\,\,\,{\rm and}\,\,\,k \geq (1 - \epsilon)(d_i - 1)\big\}\qquad {\rm with}\,\,s = 1 + r\epsilon\,.
$$
If $s = r + 1$, we know from Lemma~\ref{lem:modesplus} that the subset of modes $W_k^i$ indexed by $(i,k) \in \tilde{S}_{r + 1}$ generate a graded Lie subalgebra $\mathcal{A}_{\geq 0}$. Therefore, so do the corresponding $H_k^i$. As we remark that $\Pi_{r + 1}$ is a bijection between $\tilde{S}_{r + 1}$ and $\mathbb{D} + r\mathbb{N}$, for each independent linear coordinate $(x_{p})_{p \in \mathbb{D} + r\mathbb{N}}$ on $V$ we have a unique operator $H_k^i$ containing the derivation $\hbar \partial_{x_{p}}$ as its degree $1$ term. Therefore we have obtained a quantum Airy structure. For $s = 1$ we reach a similar conclusion if we use Lemma~\ref{lem:modesdelta} and the graded Lie subalgebra $\mathcal{A}_{\Delta}$ instead.

In fact, due to Remark~\ref{remEE} we could have ignored the third part of the generators right from the start, as it could only be a $O(2)$.

\end{proof}

\newpage

\section{Higher quantum Airy structures from higher abstract loop equations}\label{sec:hasfromtr}

In this section, we show that the Bouchard--Eynard topological recursion of \cite{BE2, BE, BHLMR} for admissible spectral curves with arbitrary ramification yields higher quantum Airy structures. More precisely, we study higher abstract loop equations (whose unique, polarized solution is constructed by the Bouchard--Eynard topological recursion) and construct the associated higher quantum Airy structures. We then show that they coincide with the general dilaton-shifted, polarization changed modules for direct sums of $ \mathcal{W}(\mathfrak{gl}_{r}) $ algebras. The precise dilaton shift and change of polarization involved here is dictated by the local expansion of the spectral curve data around the ramification points.

\subsection{Geometry of local spectral curves}

\label{s:local}

\subsubsection{Local spectral curve with one component}

Let us first introduce the notion of local spectral curves. We start with the complex vector space
\begin{equation}\label{eq:V}
V_z = \big\{ \omega \in \mathbb{C}(\!(z)\!)\,\dd z \quad \big|\quad \Res_{z = 0}\,\,\omega(z) = 0 \big\}\,,
\end{equation}
equipped with the symplectic pairing
\begin{equation}\label{eq:symp}
\Omega_z(\dd f_1, \dd f_2 ) = \Res_{z = 0}\,\, f_1(z)\, \dd f_2(z)\,.
\end{equation}
Let $V_z^+$ be the Lagrangian subspace $V_z^+ = \mathbb{C} [\![ z ]\!] \dd z \subset V_{z}$. Let us define a basis $(\dd \xi_{l})_{l > 0}$ for $V_z^+$ with $\dd\xi_{l}(z) = z^{l-1}\dd z$.

\begin{defin}\label{d:lsc1}
A local spectral curve with one component consists of the data of the symplectic space $V_z$ over $\mathbb{C}$, with a Lagrangian subspace $V_z^+$ and
\begin{itemize}
\item An integer $r \geq 2$. We use it to consider the group action $G \times V_z \to V_z$ with $G = \mathbb{Z} / r \mathbb{Z}$ and $r \geq 2$, such that the generator $\rho$ of $G$ acts as
$$
\rho \cdot \dd f(z)  \longmapsto \dd f(\theta z)\,,
$$
where $\theta = e^{2\ii \pi/r}$ is a primitive $r$-th root of unity. 
\item A one-form $\omega_{0,1} \in V_z^+$. We write its expansion as 
$$
\omega_{0,1}(z) = \sum_{l > 0}^\infty \tau_l\, \dd\xi_{l}(z)\,.
$$
\item A choice of polarization, \textit{i.e.} a Lagrangian subspace $V_z^- \subset V_z$ complementary to $V_z^+$, with basis $(\dd\xi_{l})_{l < 0}$ such that
$$
\forall l,m \in \mathbb{Z}_{\neq 0} \setminus\{0\},\qquad \Omega_z( \dd\xi_l, \dd\xi_m) = \frac{1}{l}\,\delta_{l + m,0}\,.
$$
\end{itemize}
\end{defin}

\begin{defin}\label{d:admissible1}
We say that a local spectral curve is \emph{admissible}\footnote{This admissibility requirement will become clear when we construct the associated higher quantum Airy structures. Remarkably, it turns out that the Bouchard--Eynard topological recursion constructs symmetric differentials only for admissible spectral curves.} if
$$
s := \min\big\{l > 0 \quad | \quad \tau_{l} \neq 0\,\,\,{\rm and}\,\,\,r \nmid l\big\} 
$$
satisfies $1 \leq s \leq r + 1$ and $r = \pm 1\,\,{\rm mod}\,\,s$.
Notice that this congruence implies that $r$ and $s$ are coprime. If $s=r+1$, we say the spectral curve is \emph{regular}, while it is \emph{irregular} if $s < r$.
\end{defin}

\begin{rem}
In the standard topological recursion formalism of Chekhov--Eynard--Orantin \cite{EO}, one would need to choose $r=2$. This requirement was dropped in the Bouchard--Eynard topological recursion \cite{BE, BHLMR}.
\end{rem}

Note that the basis $\dd\xi_l(z)$ for $V_z^+$ is an eigenvector (with eigenvalue $\theta^l$) for the action of the generator of $G$, but it may not be the case for the polarization basis $\dd\xi_{-l}(z)$.

The choice of polarization can be encoded in terms of a formal bidifferential. For $l > 0$, we can write
$$
\dd\xi_{-l}(z) = \frac{\dd z}{z^{l+1}} + \sum_{m > 0} \frac{\phi_{l,m}}{l}\,\dd\xi_{m}(z)\,,
$$
for some coefficients $\phi_{l,m}$. The requirement that $V_{z}^{-}$ is Lagrangian, namely
$$
\forall l,m > 0,\qquad \Omega_z (\dd\xi_{-l}(z), \dd\xi_{-m}(z) ) = 0
$$
imposes that
$$
\phi_{l,m} = \phi_{m,l}\,.
$$
We then introduce the formal bidifferential
\begin{equation}\label{eq:w021}
\omega_{0,2}(z_1, z_2) = \frac{\dd z_1 \otimes \dd z_2}{(z_1-z_2)^2} + \sum_{l,m > 0} \phi_{l,m}\dd\xi_{l}(z_1) \otimes \dd\xi_{m}(z_2)\,.
\end{equation}
which is symmetric under exchange of $z_1$ and $z_2$. This is not an element in $V_{z_1} \otimes V_{z_2}$ but
$$
\omega_{0,2}(z_1, z_2) - \frac{\dd z_1 \otimes \dd z_2}{(z_1-z_2)^2}  \in V_{z_1}^+ \otimes V_{z_2}^+\,.
$$
In any case, for any $l > 0$ we can write
\begin{equation*}
\begin{split}
\dd\xi_{-l}(z') &= \Omega_z \left( \omega_{0,2}(z, z'), \frac{\dd z}{z^{l+1}} \right )  \\ 
&= \Res_{z = 0}  \left(\int^z_0 \omega_{0,2}(\cdot, z') \right) \frac{\dd z}{z^{l+1}}\,,
\end{split}
\end{equation*}
where in the first line the symplectic pairing acts on the variable $z$. In other words, for $|z_1|<|z_2|$, we can write the expansion
$$
\omega_{0,2}(z_1,z_2) \underset{|z_1|<|z_2|}{\approx} \sum_{l > 0} l\,\dd\xi_l(z_1) \otimes \dd\xi_{-l}(z_2)\,.
$$

\begin{defin}\label{d:standardone}
We call \emph{standard polarization} the choice of $V_z^- = z^{-1} \mathbb{C} [z^{-1}]\dd z$, with basis $\dd\xi_{-l}(z) = z^{-l-1} \dd z$. In this case, the bidifferential is simply
$$
\omega_{0,2}(z_1, z_2) = \frac{\dd z_1 \otimes \dd z_2}{(z_1-z_2)^2} \,.
$$
\end{defin}

\subsubsection{Projection map}

We will also need to define a few important maps associated to the group action. Pick an element $\dd f \in V_z$, and denote by $G \cdot \dd f$ the orbit of $\dd f$, that is
\begin{equation*}
\begin{split}
G \cdot \dd f(z) &= \big\{ g \cdot \dd f(z) \quad \big|\quad g \in G \} \\
&= \big\{ \dd f(z), \dd f(\theta z), \ldots, \dd f(\theta^{r-1} z) \big\} \,.
\end{split}
\end{equation*}
There is a natural averaging map
\begin{equation*}
\begin{split}
\mu_z\,:\quad V_z \,\,\,& \longrightarrow \,\,\, V_{z} \\ \qquad \quad \quad \dd f(z) & \longmapsto \sum_{\dd g \in G \cdot \dd f} \dd g(z) = \sum_{k=0}^{r-1} \dd f(\theta^k z)\, .
\end{split}
\end{equation*}
We want to extend this map to tensor products of $V_z$. Let us consider first the case of $V_{z_1} \otimes V_{z_2}$. There is a natural group action of $G \times G$ on $V_{z_1} \otimes V_{z_2}$, with each factor of $G$ acting individually on $V_{z_1}$ and $V_{z_2}$ respectively, and a diagonal group action of $G$. Let us denote by $G^{(2)} := (G \times G) \setminus G$ the set $G \times G$ minus the diagonal embedding of $G$.  For $\omega(z_1, z_2) \in V_{z_1} \otimes V_{z_2}$, let
$$
G^{(2)} \cdot \omega = \big\{ g \cdot \omega \quad \big|\quad  g \in G^2 \}\,,
$$
and define the averaging map
\begin{equation*}
\begin{split}
\mu_{z_1,z_2} \,: \quad  V_{z_1} \otimes V_{z_2}\,\,\, & \longrightarrow \,\,\,V_{z_1} \otimes V_{z_2}  \\
\omega(z_1,z_2) & \longmapsto \sum_{\lambda \in G^{(2)} \cdot \omega} \lambda(z_1,z_2) = \frac{1}{2} \sum_{\substack{m_1,m_2 = 0 \\ m_1 \neq m_2}}^{r - 1} \omega(\theta^{m_1} z_1, \theta^{m_2}z_2)\,.
\end{split}
\end{equation*} 
We also define a specialization map
\begin{equation*}
\begin{split}
\sigma_{z_1,z_2|t}\,:\quad V_{z_1} \otimes V_{z_2}\,\,\,& \longrightarrow  \mathbb{C}[t^{-1},t]\!] \\
 \omega(z_1, z_2) & \longmapsto \frac{\omega(t, t)}{\left(\dd\xi_r (t) \right)^2}\,. 
\end{split}
\end{equation*}
It is then easy to see that the composition, which we call ``projection map'',
$$
P_{z_1,z_2 |t} := \sigma_{z_1, z_2|z} \circ \mu_{z_1,z_2}\,:\,V_{z_1} \otimes V_{z_2} \longrightarrow \mathbb{C}[t^{-r},t^{r}]\!] 
$$
maps elements of $V_{z_1} \otimes V_{z_2}$ to Laurent series that are invariant under the group action $G$.
 
 The generalization to more than two variables straightforward. 
\begin{defin}\label{d:projection1}
Let $\mathbf{z} = (z_l)_{l = 1}^i$. For $i \in \{1,\ldots,r\}$ we define the averaging map
\begin{equation*}
\begin{split}
\mu_{\mathbf{z}}\,:\quad  \bigotimes_{l=1}^i V_{z_l}  \,\,\, & \longrightarrow\,\,\, \bigotimes_{l=1}^i V_{z_l} \\
\omega(\mathbf{z}) \,\,\,\,& \longmapsto \frac{1}{i!} \sum_{\substack{m_1,\ldots, m_i = 0 \\ m_l \neq m_{l'}}}^{r - 1}\omega(\theta^{m_1} z_1, \ldots, \theta^{m_i}z_i)\,,
\end{split}
\end{equation*}
and the specialization map
\begin{equation*}
\begin{split}
\sigma_{\mathbf{z} |t}\,:\,\quad \bigotimes_{l=1}^i V_{z_l}\,\,\, & \longrightarrow\,\,\,  \mathbb{C}[t^{-1},t]\!] \\ \omega(\mathbf{z})\,\,\,\, & \longmapsto \frac{\omega(t, \ldots, t)}{\left(\dd \xi_r (t) \right)^{i}}\,.
\end{split}
\end{equation*}
We define the projection map
$$
P_{\mathbf{z}|t} := \sigma_{\mathbf{z} |t} \circ \mu_{\mathbf{z} }\,:\,\bigotimes_{l=1}^i V_{z_l} \longrightarrow  \mathbb{C}[t^{-r},t^r]\!]\,,
$$
which maps elements of $\bigotimes_{l=1}^i V_{z_l}$ to Laurent series that are invariant under the group action.
\end{defin}

We would like to extend this map to objects that involve the formal bidifferential $\omega_{0,2}$ defined in \eqref{eq:w021}. $\omega_{0,2}(z_1,z_2)$ does not live in $V_{z_1} \otimes V_{z_2}$, and in fact the specialization map $\sigma_{z_1,z_2|t}$ is not well defined on $\omega_{0,2}(z_1,z_2)$ due to the pole on the diagonal. However, it is easy to see that the projection map $P_{z_1,z_2|t}$ is well defined on $\omega_{0,2}(z_1,z_2)$. Similarly, the projection map $P_{\mathbf{z}|t}$ is well defined on objects that may involve factors of $\omega_{0,2}(z_l,z_{l'})$ for $l \neq l'$.

\begin{rem}
We notice that the projection map $P_{\mathbf{z}|t}$ is invariant under permutations of $z_1, \ldots, z_i$.
\end{rem}

\subsubsection{Local spectral curves with \texorpdfstring{$c$}{c} components}

We can generalize the notion of local spectral curves by, roughly speaking, taking $c$ copies of the symplectic space $V_z$. Let $V_z$ be as in \eqref{eq:V}. For $c \geq 1$, we define the larger symplectic space
\begin{equation}\label{eq:Vell}
\mathcal{V}_{z} = \mathbb{C}^{c} \otimes V_z\,.
\end{equation}
We equip $\mathbb{C}^{c}$ with a scalar product $\cdot$ such that the standard basis $(e_i)_{i = 1}^{c}$ is orthonormal, namely $e_i \cdot e_j = \delta_{i,j}$. We define the symplectic pairing on $\mathcal{V}_z$ as being given by
\begin{equation}\label{eq:sympell}
\Omega_z( u_1 \otimes \dd f_1,  u_2 \otimes \dd f_2 ) = u_1\cdot u_2\,\Res_{z = 0} f_1(z) \dd f_2(z)\,,
\end{equation}
where $u_1,u_2 \in \mathbb{C}^\ell$.  For $\alpha \in \{1,\ldots,c\}$, we write $\mathcal{V}^{(\alpha)}_z = e_{\alpha} \otimes V_z \subseteq \mathcal{V}_z$.

Let us define the Lagrangian subspace $\mathcal{V}_z^+ = \mathbb{C}^c \otimes \mathbb{C}[\![ z ]\!] \dd z \subset \mathcal{V}_z$, with basis $\dd\xi_{\alpha,l}(z)$ with $l > 0$ and $\alpha \in \{1,\ldots,c\}$ and given by
\begin{equation}
\dd\xi_{\alpha,l}(z) := e_{\alpha} \otimes \dd\xi_{l}(z) = e_{\alpha} \otimes z^{l-1} \dd z\,.
\end{equation}

\begin{defin}\label{d:lsc}
A local spectral curve with $c$ components consists of the data of the symplectic space $\mathcal{V}_z$ together with its Lagrangian subspace $\mathcal{V}_z^+$, and
\begin{itemize}
\item A family of integers $r_{\alpha} \geq 2$ for $\alpha \in \{1,\ldots,c\}$. We use them to consider the group action $G \times \mathcal{V}_z \to \mathcal{V}_z$, with $G = G_1 \times \ldots \times G_\ell$ and $G_{\alpha} = \mathbb{Z} / r_{\alpha} \mathbb{Z}$. It is such that the generator $\rho_{\alpha}$ of $G_{\alpha}$ acts only on $\mathcal{V}_z^{(\alpha)}$ as
\begin{equation}
\rho_{\alpha} \cdot \dd\xi_{\alpha,l}(z) = \dd\xi_{\alpha,l}( \theta_{\alpha} z)\,,
\end{equation}
where $\theta_{\alpha}$ is a primitive $r_{\alpha}$-th root of unity.
\item A one-form $\omega_{0,1} \in \mathcal{V}_z^+$. We write its expansion as
$$
\omega_{0,1}(z) = \sum_{\alpha=1}^c \sum_{l > 0} \tau_{l}^{\alpha}\,\dd\xi_{\alpha,l}(z)\,.
$$
\item A choice of polarization, that is, a choice of Lagrangian subspace $\mathcal{V}_z^- \subset \mathcal{V}_z$ complementary to $\mathcal{V}_z^+$, with basis $\dd\xi_{\alpha,-l}(z)$, with $l > 0$ and $\alpha \in \{1,\ldots,c\}$ such that
$$
\forall \alpha,\beta \in \{1,\ldots,c\},\quad \forall l,m \in \mathbb{Z}\qquad \Omega_z (\dd\xi_{\alpha,l}, \dd\xi_{\beta,m}) = \frac{1}{l} \delta_{\alpha,\beta} \delta_{l +m,0}\,.
$$
\end{itemize}
\end{defin}

\begin{defin}\label{d:admissible}
We say that a spectral curve is \emph{admissible} if for each $\alpha \in \{1,\ldots,c\}$,
$$
s_{\alpha} := \min\big\{l > 0 \quad |\quad \tau_{l}^{\alpha} \neq 0\,\,\,{\rm and}\,\,\,r_{\alpha} \nmid l\big\}
$$
satisfies $1 \leq s_{\alpha} \leq r_{\alpha} + 1$ and $r_{\alpha} = \pm 1\,\,{\rm mod}\,\,s_{\alpha}$ (the sign could depend on $\alpha$). We say that the spectral curve is \emph{regular at $\alpha$} if $s_\alpha=r_\alpha+1$, while we say that it is \emph{irregular at $\alpha$} if $s_\alpha < r_\alpha$.

\end{defin}

As before, the choice of polarization is nicely encoded in terms of a formal bidifferential. For $l >  0$ and $\alpha \in \{1,\ldots, c\}$, we can write
$$
\dd\xi_{\alpha,-l}(z) = \frac{e_{\alpha} \otimes \dd z}{z^{l+1}} + \sum_{\beta=1}^c \sum_{m >  0} \frac{\phi_{l,m}^{\alpha,\beta}}{l} \,\dd\xi_{\beta,m}(z)\,,
$$
for some coefficients $\phi_{l,m}^{\alpha,\beta}$. The requirement that $\mathcal{V}_{z}^{-}$ is Lagrangian imposes the symmetry $\phi_{l,m}^{\alpha,\beta} = \phi_{m,l}^{\beta,\alpha}$. We define the formal bidifferential
\begin{equation}\label{eq:bd}
\omega_{0,2}(z_1,z_2) =  \sum_{\alpha=1}^c  \frac{(e_{\alpha} \otimes \dd z_1) \otimes  (e_{\alpha} \otimes \dd z_2)}{(z_1-z_2)^2} +\sum_{\alpha,\beta =1}^c \sum_{l,m > 0} \phi_{l,m}^{\alpha,\beta}\,\dd\xi_{\alpha,l}(z_1) \otimes \dd \xi_{\beta,m}(z_2)\,.
\end{equation}
As before, $\omega_{0,2}(z_1,z_2)$ is not in $\mathcal{V}_{z_1} \otimes \mathcal{V}_{z_2}$, but
$$
\bigg(\omega_{0,2}(z_1,z_2) -  \sum_{\alpha=1}^c  \frac{(e_{\alpha} \otimes \dd z_1)\otimes(e_{\alpha} \otimes \dd z_2)}{(z_1-z_2)^2}\bigg) \in \mathcal{V}^+_{z_1} \otimes \mathcal{V}^+_{z_2}\,.
$$
Then, for $l > 0$ and $\alpha \in \{1,\ldots,c\}$, we can write
\begin{equation}\label{eq:bb}
\dd\xi_{\alpha,-l}(z') = \Omega_z \left( \omega_{0,2}(z,z'), e_{\alpha}\otimes \frac{\dd z}{z^{l+1}} \right )\,.\end{equation}
In other words, for $|z_1|<|z_2|$,
$$
\omega_{0,2}(z_1,z_2) \underset{|z_1|<|z_2|}{\approx} \sum_{\alpha=1}^c \sum_{l > 0} l\,\dd\xi_{\alpha,l}(z_1)\otimes \dd\xi_{\alpha,-l}(z_2)\,.
$$

\begin{defin}
We call \emph{standard polarization} the choice of $\mathcal{V}_z^- = \mathbb{C}^c \otimes z^{-1} \mathbb{C} [z^{-1} ] \dd z$, with basis $\dd\xi_{\alpha,-l}(z) = e_{\alpha} \otimes z^{-l-1} \dd z$ for $l > 0$. In this case, the bidifferential is simply
$$
\omega_{0,2}(z_1,z_2) =  \sum_{\alpha=1}^c  \frac{(e_{\alpha}\otimes \dd z_1)\otimes (e_{\alpha} \otimes \dd z_2)}{(z_1-z_2)^2}\,.
$$
\end{defin}

\subsubsection{Projection map}

We can also generalize the construction of the averaging, specialization and projection maps for each $G_{\alpha}$.

\begin{defin}
Let $\mathbf{z} = (z_l)_{l = 1}^i$. For $\alpha \in \{1,\ldots,c\}$ and $i \in \{1,\ldots,r_{\alpha}\}$, we define the averaging map
\begin{equation*}
\begin{split}
\mu^{(\alpha)}_{\mathbf{z}}\,:\,\qquad\qquad \bigotimes_{l=1}^i \mathcal{V}_{z_l}\,\,\, & \longrightarrow \,\,\,\bigotimes_{l=1}^i \mathcal{V}_{z_l}^{(\alpha)}  \\  
\left( \bigotimes_{l=1}^i e_{\alpha_l} \right) \otimes \omega(\mathbf{z}) & \longmapsto  \bigg(\prod_{l=1}^i \delta_{\alpha_l, \alpha} \bigg)\,e_{\alpha}^{\otimes i} \otimes \bigg(\frac{1}{i!} \sum_{\substack{m_1,\ldots, m_i=0 \\ m_{l} \neq m_{l'}}}^{r-1} \omega(\theta^{m_1}_{\alpha} z_1, \ldots, \theta^{m_i}_{\alpha}z_i)\bigg)\,,
\end{split}
\end{equation*}
and the specialization map
\begin{equation*}
\begin{split}
\sigma^{(\alpha)}_{\mathbf{z} |t}\,:\,\quad \bigotimes_{l=1}^i \mathcal{V}_{z_l}^{(\alpha)}\,\,\, & \longrightarrow \mathbb{C}[t^{-1},t]\!] \\
 e_{\alpha}^{\otimes i} \otimes \omega(\mathbf{z}) & \longmapsto  \frac{\omega(t, \ldots, t)}{\left( \dd \xi_{\alpha,r_{\alpha}} (t) \right)^{i}}\,.
\end{split}
\end{equation*}
We define the projection map
$$
P^{(\alpha)}_{\mathbf{z} | t} := \sigma^{(\alpha)}_{\mathbf{z} | t} \circ \mu^{(\alpha)}_{\mathbf{z}}\,:\,\, \bigotimes_{l=1}^i \mathcal{V}_{z_l}  \longrightarrow \mathbb{C}[t^{-r_\alpha},t^{r_\alpha}]\!]\,,
$$
which maps elements of $\bigotimes_{l=1}^i \mathcal{V}_{z_l} $ to Laurent series that are invariant under the group action $G_\alpha$.
\end{defin}
As before, we observe that this projection map is well defined on $\omega_{0,2}(z_1,z_2)$. It is also invariant under permutations of $z_1, \ldots, z_i$.

\subsubsection{Relation with global spectral curves}

The topological recursion of Chekhov--Eynard--Orantin \cite{EO}, and the Bouchard--Eynard topological recursion \cite{BE2,BE, BHLMR}, were not presented in terms of local spectral curves. Let us now briefly show that the notion of spectral curves used in these papers ---which we here call ``global'' --- is a special case of the local spectral curves defined above.

\begin{defin}
\label{D59} A \emph{global spectral curve} is a quadruple $(\mathcal{C}, x, y,B)$, where
\begin{itemize}
\item $\mathcal{C}$ is a Riemann surface;
\item $x$ is a meromorphic function on $\mathcal{C}$. We denote by $R \subset \mathcal{C}$ the set of ramification points of $x$ that are zeros of $\dd x$. For any $p_{\alpha} \in R$, we let $r_{\alpha}$ be its order. We assume $R$ is finite;
\item $y$ is a meromorphic function on $\mathcal{C}$.;
 \item $B$ is a meromorphic symmetric bidifferential on $\mathcal{C} \times \mathcal{C}$, whose only singularity is a double pole on the diagonal with biresidue $1$.
\end{itemize}
\end{defin}

\begin{defin}
We say that a global spectral curve is \emph{admissible} if for each $p_{\alpha} \in R$, either $y$ has a pole of order $r_\alpha-s_{\alpha}$ with $s_{\alpha} \in \{1,\ldots, r_\alpha -1\}$ and
 $r_\alpha = \pm 1 \mod s_\alpha$ (in which case we say that the curve is \emph{irregular} at $p_\alpha$)\footnote{Note that it implies that $s_{\alpha}$ is coprime with $r_{\alpha}$, which is the condition for the plane curve
$$
\big\{(\tilde{x},\tilde{y}) \in \mathbb{C}^2\,\,\,|\,\,\,\exists q \in \mathcal{C}\,\,\,\,\,\, (x(q),y(q)) = (\tilde{x},\tilde{y})\big\}
$$
to be irreducible locally at $q = p_{\alpha}$.}, or $y(p_{\alpha})$ is finite and $\dd y(p_{\alpha}) \neq 0$ (in which case we say that the curve is \emph{regular} at $p_\alpha$).
\end{defin}

To recover the structure of a local spectral curve, we first need to construct the symplectic space $\mathcal{V}$. Here, we consider the space of meromorphic residueless one-forms on $\mathcal{C}$ with poles only on $R$, which takes the structure of $\mathcal{V}$ in \eqref{eq:Vell} after expanding in local coordinates near the critical points.

More precisely, we replace $\mathcal{C}$ by the union of small disks $U_{\alpha}$ around the $p_{\alpha} \in R$. On each $U_{\alpha}$, we define a local coordinate $\zeta$ such that 
$$
x \big |_{U_{\alpha}} (\zeta) = \frac{ \zeta^{r_{\alpha}} }{r_{\alpha}}+ x(p_{\alpha})\,.
$$
Then we can think of one-forms on $\mathcal{C}$ with poles on $R$ as the sum of their formal Laurent expansions on the $U_{\alpha}$ in terms of the local coordinates $\zeta$, and $\mathcal{V}$ then exhibits the structure in \eqref{eq:Vell} with $z \to \zeta$.

The choice of one-form $\omega_{0,1} \in \mathcal{V}$ is naturally given by $\omega_{0,1} = y \dd x$, after expanding locally on the $U_{\alpha}$ in terms of $\zeta$. The admissibility condition matches with the analogous condition in the definition of local spectral curves.

The group actions $G_{\alpha} \times \mathcal{V} \to \mathcal{V}$ indexed by $\alpha \in \{1,\ldots,c\}$ with $G_{\alpha} = \mathbb{Z} / r_{\alpha} \mathbb{Z}$, are naturally given by the deck transformations $\zeta \mapsto \theta_{\alpha}\zeta$ in the local coordinates on each $U_{\alpha}$.

Finally, the choice of polarization is given by the choice of bidifferential $B$. Indeed, if we define, for $\alpha \in \{1,\ldots,c\}$ and $l > 0$, the one-forms $\dd\xi_{\alpha,-l}(z)$ on $\mathcal{C}$ by
$$
\dd\xi_{\alpha,-l}(z') = \Res_{z = p_\alpha} \bigg(\int_{p_{\alpha}}^z B(\cdot, z') \bigg)\frac{\dd\zeta(z)}{\zeta(z)^{l+1}}\,,
$$
we see that this has the same form as \eqref{eq:bb} after expanding $B(\cdot, z')$ in local coordinates such that the expansion near $z' \rightarrow p_{\beta}$ gives the coefficient of $e_{\beta}$.
\begin{exa}
\label{rAiryC}Perhaps the simplest regular spectral curve is the so-called \emph{$r$-Airy curve}, which corresponds to the choice
$$
\mathcal{C} = \mathbb{C}\,, \qquad x = \frac{z^r}{r}\,, \qquad y=-z\,, \qquad B(z_1,z_2) = \frac{\dd z_1 \otimes \dd z_2}{(z_1-z_2)^2}\,.
$$
We observe that the corresponding local spectral curve has standard polarization.
\end{exa}

\begin{exa}
\label{rBesselC}
There is a natural family of irregular spectral curves, which we will call the $(r,s)$ spectral curves. They are indexed by an integer $s \in \{1,\ldots,r-1\}$ such that $r = \pm 1 \mod s$,
$$
\mathcal{C} = \mathbb{C}\,, \qquad x = \frac{z^r}{r}\,,  \qquad y=-\frac{1}{z^{r-s}}\,, \qquad B(z_1,z_2) = \frac{\dd z_1 \otimes \dd z_2}{(z_1-z_2)^2}\,.
$$
We call the extreme case $s = 1$ the $r$-Bessel curve. The corresponding local spectral curves again have standard polarization.
\end{exa}

\begin{exa}
In fact, the more general spectral curve given by
$$
\mathcal{C} = \mathbb{C}\,, \qquad x = \frac{z^r}{r}\,, \qquad y=\sum_{l > 0}^\infty \tau_l z^{l-r}\,, \qquad B(z_1,z_2) = \frac{\dd z_1 \otimes \dd z_2}{(z_1-z_2)^2}\,,
$$
with the condition that the first non-zero term (apart maybe from $\tau_{r}$) in the expansion of $y$ is $\tau_s$ with $s \in \{1,\ldots, r+1\}$ such that $r = \pm 1 \mod s$, corresponds to the general local spectral curve with one component in standard polarization. The most general local spectral curve with one component in arbitrary polarization is associated with the bidifferential of the form
$$
B(z_1,z_2) =  \frac{\dd z_1 \otimes \dd z_2}{(z_1-z_2)^2 } + \sum_{l,m > 0} \phi_{l,m}\, z_1^{l-1} z_2^{m-1} \dd z_1  \otimes \dd z_2\,.
$$
\end{exa}

Global spectral curves are particular examples of local spectral curves. However, local spectral curves are slightly more general. Since modules for $\mathcal{W}$ algebras naturally give rise to the more general structure of local spectral curves, in the following we will reformulate the Bouchard--Eynard topological recursion and higher abstract loop equations in the language of local spectral curves.

\subsection{Bouchard--Eynard topological recursion and higher abstract loop equations}

\label{s:tr}

Let us now review the construction of the Bouchard--Eynard topological recursion\footnote{In these references it was called ``generalized topological recursion''.} of \cite{BE2,BE,BHLMR} and its relation with the higher analog of the abstract loop equations of \cite{BSblob}. We will reformulate everything now in the slightly more general language of local spectral curves. It should be clear to the reader that the global formulation of \cite{BE2,BE,BHLMR} is a particular case of the local formulation given below.

\subsubsection{Notation and definitions}

Consider an admissible local spectral curve with $c$ components, as in Definition \ref{d:lsc}, with symplectic space $\mathcal{V}_z = \mathbb{C}^c \otimes V_z$. The aim of topological recursion is to construct a sequence of ``multilinear differentials''
$$
\omega_{g,n} \in \bigotimes_{j=1}^n \mathcal{V}^-_{z_j}\qquad g \geq 0\,\,{\rm and}\,\,n \geq 1\,\,\,{\rm such}\,\,{\rm that}\,\,2g-2+n > 0\,,
$$
which are invariant under the natural action of the permutation group $\mathfrak{S}_{n}$. In terms of the choice of polarization basis $\dd\xi_{\alpha,-l}(z)$ indexed by $\alpha \in \{1,\ldots, c\}$ and $l > 0$ for $\mathcal{V}^-(z)$, the $\omega_{g,n}$ have an expansion of the form
\begin{equation}\label{e:expansion}
\omega_{g,n}(\mathbf{z}) = \sum_{\alpha_1,\ldots, \alpha_n=1}^c \sum_{l_1, \ldots, l_n > 0} F_{g,n} \left[\begin{smallmatrix}\alpha_1 & \alpha_2 & \ldots & \alpha_n \\ l_1 & l_2 &\ldots & l_n \end{smallmatrix} \right] \,\bigotimes_{m=1}^n \dd \xi_{\alpha_m, - l_m}(z_m)\,,
\end{equation}
with $\mathbf{z} = (z_1,\ldots,z_n)$. The $F_{g,n} \left[\begin{smallmatrix}\alpha_1 & \alpha_2 & \ldots & \alpha_n \\ l_1 & l_2 & \ldots & l_n \end{smallmatrix} \right]$ are scalar coefficients, which are symmetric under the action of the permutation group $\mathfrak{S}_{n}$. In general they encode interesting enumerative invariants, see Section~\ref{Section6}.

To define topological recursion we also need to define $\omega_{0,1}$ and $\omega_{0,2}$. We let $\omega_{0,1}$ be the one-form $\omega_{0,1} \in \mathcal{V}_z^+$ given in the data of a local spectral curve, that is,
\begin{equation}\label{e:exp01}
\omega_{0,1}(z) = \sum_{\alpha=1}^c \sum_{l > 0} \tau_l^{\alpha} \dd\xi_{\alpha,l}(z)\,.
\end{equation}
We will also need to define formal Laurent series $y_{\alpha}(z)$ as follows
$$
y_{\alpha}(z) = \sum_{l > 0} \tau_l^{\alpha} z^{l-r_{\alpha}}\,,
$$
so that we have the expansion when $z \rightarrow p_{\alpha}$
$$
\omega_{0,1}(z) =  \sum_{\alpha=1}^c y_{\alpha}(z) \dd\xi_{\alpha,r_{\alpha}}(z)\,.
$$
As for $\omega_{0,2}$, we take it to be the bidifferential that encapsulates the choice of polarization
\beq
\label{e:exp02} \omega_{0,2}(z_1,z_2) = \sum_{\alpha = 1}^c \frac{(e_{\alpha} \otimes \dd z_1) \otimes (e_{\alpha} \otimes \dd z_2)}{(z_1-z_2)^2} + \sum_{\alpha,\beta = 1}^c \sum_{l,m > 0} \phi_{l,m}^{\alpha,\beta} \dd\xi_{\alpha,l}(z_1) \dd\xi_{\beta,m}(z_2)\,.
\eeq

To define the Bouchard--Eynard topological recursion, we use the notation in Section \ref{s:recursive}, which we recall here. Let $A$ be a set of cardinality $i$, and $B$ a set of cardinality $n - 1$. The notation $\mathbf{L} \vdash A$ means that $\mathbf{L}$ is a set partition of $A$, \textit{i.e.} a set of $|\mathbf{L}|$ non-empty subsets of $A$ which are pairwise disjoint and whose union is $A$. We denote generically by $L$ the elements (sets) of the partition $\mathbf{L}$.  A partition of $B$ indexed by $\mathbf{L}$ is a map $M\,:\,\mathbf{L} \rightarrow \mathfrak{P}(B)$ such that $(M_L)_{L \in \mathbf{L}}$ are possibly empty, pairwise disjoint subsets of $B$ whose union is $B$. We summarize this notion with the notation $\mathbf{M} \vdash_{\mathbf{L}} B$.

Just as in Lemma \ref{l:deriv}, we define the objects

\begin{defin}\label{d:combin}
Let $A$ and $B$ be finite sets of coordinates with cardinality $i$ and $n - 1$ respectively. Then we define
$$
\mathcal{E}^{(i)}_{g,n}(A |B)= \sum_{\mathbf{L} \vdash A} \sum_{\substack{h\,:\,\mathbf{L} \rightarrow \mathbb{N} \\ i + \sum_{L \in \mathbf{L}} h_{L} = g + |\mathbf{L}|}} \sum_{\boldsymbol{\mu} \vdash_{\mathbf{L}} B} \bigg(\bigotimes_{L \in \mathbf{L}} \omega_{h_{L},|L| + |\mu_{L}|}(L,\mu_{L})\bigg)\,,
$$
In the tensor product here and below it is assumed that the corresponding tensor factors are put in the place respecting the natural order in $(\mathbb{C}^{c})^{i} \otimes (\mathbb{C}^{c})^{\otimes (n - 1)}$ associated with $A$ and $B$ coordinates. We also define
$$
\mathcal{R}^{(i)}_{g,n}(A | B)= \sum_{\mathbf{L} \vdash A} \sum_{\substack{h\,:\,\mathbf{L} \rightarrow \mathbb{N} \\ i  + \sum_{L \in \mathbf{L}} h_{L} = g + |\mathbf{L}|}} \sum_{\boldsymbol{\mu} \vdash_{\mathbf{L}} B}' \bigg(\bigotimes_{L \in \mathbf{L}} \omega_{h_{L},|L| + |\mu_{L}|}(L,\mu_{L})\bigg)\,,
$$
where the prime over the summation symbol means that terms that include $\omega_{0,1}$ are excluded from the sum. Finally, for future use, we define
$$
\widetilde{\mathcal{R}}^{(i)}_{g,n}(A | B)= \sum_{\mathbf{L} \vdash A} \sum_{\substack{h\,:\,\mathbf{L} \rightarrow \mathbb{N} \\ i + \sum_{L \in \mathbf{L}} h_{L} = g + |\mathbf{L}|}} \sum_{\boldsymbol{\mu} \vdash_{\mathbf{L}} B}'' \bigg(\bigotimes_{L \in \mathbf{L}} \omega_{h_{L},|L| + |\mu_{L}|}(L,\mu_{L})\bigg)\,,
$$
where the double prime over the summation symbol means that terms with $h_L=0$, $|\mu_L| = 0$ and $|L| \leq 2$ are excluded from the sum. In other words, $\omega_{0,1}$ does not appear in the sum, and $\omega_{0,2}$ only appears when one of the entry comes from $A$ and the other one from $B$.

\end{defin}

We first give two useful combinatorial lemmas relating these three objects. First, we want to relate $\mathcal{E}^{(i)}_{g,n}(A | B)$ and $\mathcal{R}^{(i)}_{g,n}(A | B)$ by extracting the $\omega_{0,1}$ contributions. 
\begin{lem} \cite[Lemma 3.18]{BE2}
\label{l:er}
For all $g,i \geq 0$ and $n \geq 1$
$$
\mathcal{E}^{(i)}_{g,n}(A | B) = \sum_{j=0}^i \sum_{\substack{\gamma \subseteq A \\ |\gamma| = j}}  \left( \bigotimes_{l=1}^i \omega_{0,1}(\gamma_l) \right)\mathcal{R}^{(i-j)}_{g,n} (A \setminus \gamma | B)\,.
$$
\end{lem}

Let us now relate $\mathcal{R}^{(i)}_{g,n}(A | B)$ and $\widetilde{\mathcal{R}}^{(i)}_{g,n}(A | B)$ by extracting contributions from $\omega_{0,2}$ with both entries coming from $A$. We get by straightforward combinatorics
\begin{lem}\label{l:rtilde}
$$
\mathcal{R}^{(i)}_{g,n}(A | B) = \sum_{\substack{\ell,j \geq 0 \\ \ell+2j = i}}\sum_{\substack{\gamma \subseteq A \\ |\gamma| = 2j}} \left( \sum_{\substack{\mathbf{L} \vdash \gamma \\  \forall L \in \mathbf{L}\,\,|L| = 2}} \bigotimes_{L \in \mathbf{L}} \omega_{0,2}(L) \right) \widetilde{\mathcal{R}}^{(\ell)}_{g-j,n} (A \setminus \gamma | B)\,,
$$
\end{lem}

\subsubsection{Bouchard--Eynard topological recursion}

We can now state the Bouchard--Eynard topological recursion formula, which recursively constructs the correlators $\omega_{g,n}$ \cite{BE2,BE,BHLMR}.
\begin{defin}\label{d:BE}
Let $\mathbf{z} = (z_2,\ldots,z_n)$. The correlators $\omega_{g,n}$ are recur\-si\-ve\-ly def\-ined by the \emph{Bou\-chard-Eyn\-ard to\-po\-lo\-gi\-cal re\-cur\-sion}
\begin{multline}
\omega_{g,n}(z_1, \mathbf{z}) = \sum_{\alpha = 1}^c \Res_{t = 0}\left(  \int^t_{0} (e_{\alpha}\cdot \omega_{0,2})(\cdot, z_1) \right)\sum_{i=1}^{r_{\alpha}-1} \frac{(-1)^{i+1}}{i!}  \\
\times \sum_{\substack{m_1, \ldots, m_{i}=1 \\ m_l \neq m_{l'}}}^{r_{\alpha} - 1}  \left( \prod_{l=1}^{i} \frac{1}{(y_{\alpha}(t) - y_{\alpha}(\theta^{m_{l}}_{\alpha} t))} \right) \frac{e_{\alpha}^{\otimes (i + 1)}\cdot \mathcal{R}^{(i+1)}_{g,n}(t, \theta^{m_{1}}_{\alpha} t, \ldots, \theta^{m_{i}}_{\alpha} t | \mathbf{z})}{(\dd\xi_{\alpha,r_{\alpha}}(t))^{i}}\,,\label{eq:BE}
\end{multline}
where the scalar product with $e_{\alpha}^{\otimes (i + 1)}$ only acts on the first $(i + 1)$-tensor factors in $\mathcal{R}^{(i + 1)}_{g,n}$.
\end{defin}

\begin{rem}
Note that $z_1$ plays a special role in the higher topological recursion formula. It is \emph{a priori} not obvious that the correlators $\omega_{g,n}$ constructed by \eqref{eq:BE} are fully symmetric. Symmetry was argued in \cite{BE2} indirectly, only for spectral curves that arise as limits of families of curves with simple ramification points. It is however not clear to us which spectral curves precisely satisfy this condition.  A proof of symmetry directly from the Bouchard--Eynard recursion formula is at the moment not known. As we will see, our identification of this recursive formula with higher quantum Airy structures in fact implies symmetry of the correlators for all admissible spectral curves (Definitions \ref{d:lsc} and \ref{d:admissible}) as a corollary.
\end{rem}

\subsubsection{Higher abstract loop equations}

Instead of extracting the higher quantum Airy structures corresponding to the Bouchard--Eynard topological recursion directly from the recursion formula, in this section we will rather take as starting point the higher abstract loop equations. As we will see, the loop equations give rise directly to the $\mathcal{W}(\mathfrak{gl}_{r}) $ quantum Airy structures constructed in Section~\ref{HASW}.

Let us consider as usual a local spectral curve with $c$ components.
\begin{defin}\label{d:hle}
Let $\mathbf{z} = (z_1, \ldots, z_n)$ and $\mathbf{w} = (w_1, \ldots, w_i)$.
We call \emph{higher abstract loop equations} the statement that, for all $g \geq 0$, $n \geq 1$, $2g-2+n > 0$, $\alpha \in \{1,\ldots, c\}$ and $i \in \{1,\ldots, r_i\}$,
\begin{equation}\label{e:ale}
P^{(\alpha)}_{\mathbf{w} | t} \left( \mathcal{E}^{(i)}_{g,n}(\mathbf{w} | \mathbf{z}) \right)  \in t^{-r_\alpha \mathfrak{d}_{\alpha}^i} \, \mathbb{C} [\![ t^{r_{\alpha}} ]\!] \otimes \mathcal{V}^-_{z_2}\otimes \ldots \otimes \mathcal{V}^-_{z_n}\,,
\end{equation} 
where
$$
\mathfrak{d}_{\alpha}^i := i - 1 - \Big\lfloor \frac{s_{\alpha}(i - 1)}{r_{\alpha}}\Big\rfloor\,.
$$
The key information here is that it is a formal series in $t^{r_{\alpha}}$ with either no negative terms if the spectral curve is regular at $\alpha$ ($\mathfrak{d}^i_{\alpha}= 0$), or starting at $t^{- r_{\alpha} \mathfrak{d}^i_\alpha}$ if the spectral curve is irregular at $\alpha$. It is also necessarily $G_{\alpha}$-invariant by construction, which is the reason why it is a series in $t^{r_{\alpha}}$.
\end{defin}

While the higher abstract loop equations do not appear to be recursive \emph{a priori}, one can show that if a solution that respects the polarization (that is, such that $\omega_{g,n} \in \mathcal{V}^-_{z_2} \otimes \ldots \otimes \mathcal{V}^-_{z_n}$ for all $2g-2+n > 0$) exists, then it is uniquely constructed by the Bouchard--Eynard topological recursion of the previous subsection. The proof of this statement follows arguments similar to those presented in \cite{BSblob} for $r = 2$, and in \cite{BSShad,BE} for general $r$. For completeness, we provide a proof in Appendix \ref{a:proof} (Proposition~\ref{theeeq}). Existence of a solution is however not obvious, but it will follow for admissible spectral curves as a corollary of the results of this section.

Our goal for the rest of this section is to show that solving higher abstract loop equations is equivalent to calculating the partition function of a higher quantum Airy structure, more precisely of the form of the $\mathcal{W}(\mathfrak{gl}_r)$ Airy structures constructed in the previous section. To do so, we will recast the loop equations in the form of the recursive structure in Section \ref{s:recursive}, and construct the corresponding differential operators. We then show that those are the same as the ones obtained from the $\mathcal{W}(\mathfrak{gl}_r)$ modules of the previous section.

\begin{rem}
We could have started with the Bouchard--Eynard topological recursion of Definition \ref{d:BE} instead of the higher abstract loop equations, and recast them as being obtained by the action of a sequence of differential operators acting on a partition function $Z$. This would have been more in line with what was done in \cite{ABCD,KS}. However, to show that the differential system thus obtained is a higher quantum Airy structure, one would then need to show that the left ideal generated by the differential operators is a graded Lie subalgebra. This appears to be very difficult to prove in general. By starting with the higher abstract loop equations, we circumvent this obstacle, since we can identify the differential operators that we obtain with the $\mathcal{W}(\mathfrak{gl}_r)$ modules constructed in the previous section, and use the ideals constructed in Section \ref{s:subalgebra} to prove the subalgebra property.
\end{rem}

\subsection{Local spectral curves with one component}

\label{s:one}

Let us now focus on local spectral curves with one component for clarity. We will start with the higher abstract loop equations and reconstruct the constraints it gives on the coefficients of $\omega_{g,n}$ in the form of a higher quantum Airy structure. We will then identify them with the $\mathcal{W}(\mathfrak{gl}_{r}) $ higher quantum Airy structure of Proposition~\ref{t:conjugated}. 

\subsubsection{Reconstructing the higher quantum Airy structure}

\begin{prop}\label{t:ZZ}
For any local spectral curve with one component, the higher abstract loop equations for $\omega_{g,n} \in {\rm Sym}^n(V_{z}^-)$ with $2g - 2 + n > 0$ are equivalent to a system of differential equations
$$
\forall i \in \{1,\ldots,r\},\,\,\, \forall k \geq \mathfrak{d}^i + \delta_{i,1},\qquad \mathcal{H}^i_k\cdot Z = 0\,,
$$
with $\mathfrak{d}^i = i-1-  \big\lfloor \frac{s(i-1)}{r} \big\rfloor$,
for the partition function
$$
	Z = \exp\left(\sum_{\substack{g \geq 0,\,\,n \geq 1 \\ 2g - 2 + n > 0}} \frac{\hbar^{g-1}}{n!} \sum_{\mathbf{b} \in (\mathbb{Z}_{> 0})^n} F_{g,n}[ \mathbf{b} ]\,x_{b_1} \cdots\, x_{b_n}\right)\,,
$$
constructed from the coefficients of the expansion in \eqref{e:expansion}. The differential operators read 
$$
\mathcal{H}^i_k = \sum_{m=1}^{i} \sum_{\substack{\ell,j \geq 0 \\ \ell + 2j = m}} \frac{\hbar^{j}}{\ell!} \sum_{\mathbf{a} \in (\mathbb{Z}_{\neq 0})^{\ell} } D^{(j)}_i[k | \mathbf{a} ]\,\norder{J_{a_1} \cdots\, J_{a_{\ell}}}\,,
$$
where
\begin{equation}
\label{eq:Dcoeff}
\begin{split}
D^{(j)}_i[k |\mathbf{a}] & =  \frac{1}{(i-\ell-2j)!} \sum_{a_{\ell+1}, \ldots, a_{i-2j} \in \mathbb{Z}_{\neq 0}} \left( \prod_{l=\ell+1}^{i-2j} F_{0,1}[a_l] \right)C^{(j)}[k |\mathbf{a}, a_{\ell+1}, \ldots, a_{i-2j}]\,, \\
C^{(j)}[k |\mathbf{a}] & =  \frac{(\ell+2j)!}{j!\,2^{j}} \Res_{t = 0} \left(\dd \xi_{r(k+1)}(t)  P_{\mathbf{w} | t} \Big(\bigotimes_{l=1}^{\ell} \dd\xi_{-a_l}(w_l) \bigotimes_{l'=1}^{j} \omega_{0,2}(w_{\ell+2l'-1}, w_{\ell+2l'}) \Big) \right)\,.
\end{split}
\end{equation}
\end{prop}

\begin{proof}
For local spectral curves with one component, the higher abstract loop equation is the statement that
$$
P_{\mathbf{w} | t} \left( \mathcal{E}^{(i)}_{g,n}(\mathbf{w} | \mathbf{z}) \right)  \in t^{-r \mathfrak{d}^i}  \mathbb{C} [\![ t^{r} ]\!] \otimes V_{z_2}^- \otimes \ldots \otimes V_{z_n}^-\,,
$$
for $i \in \{1,\ldots,r\}$ and $\mathfrak{d}^i = i-1-  \lfloor \frac{s(i-1)}{r} \rfloor$. This is equivalent to requiring that 
$$
\left[ \Res_{t = 0} \,\dd\xi_{r (k+1)}(t)\,P_{\mathbf{w} | t} \left( \mathcal{E}^{(i)}_{g,n}(\mathbf{w} | \mathbf{z}) \right) \right]_{\mathbf{b}}= 0\,,
$$
for all $k \geq \mathfrak{d}^i$, $\mathbf{b} = (b_2, \ldots, b_n) \in ( \mathbb{Z}_{> 0})^n$ and $2g-2+n > 0$. Here we introduced the notation $[ \cdots ]_{\mathbf{b}}$ which extracts the coefficient of the basis vector $\otimes_{l = 2}^{n} \dd\xi_{-b_l}(z_l)$ in $\otimes_{l = 2}^{n} V_{z_l}^-$.

Let us first evaluate
$$
\left[ \Res_{t = 0}\, \dd\xi_{r(k+1)}(t) \left( P_{\mathbf{w} | t} \left( \widetilde{\mathcal{R}}^{(i)}_{g,n}(\mathbf{w} | \mathbf{z}) \right)  \right) \right]_{\mathbf{b}}\,,
$$
with $ \widetilde{\mathcal{R}}^{(i)}_{g,n}(\mathbf{w} | \mathbf{z})$ defined in Definition \ref{d:combin}.

\begin{lem}\label{l:onep}
$$
\left[ \Res_{t = 0} \dd\xi_{r(k+1)}(t)\, P_{\mathbf{w} | t} \left( \widetilde{\mathcal{R}}^{(i)}_{g,n}(\mathbf{w} | \mathbf{z}) \right) \right]_{\mathbf{b}}
= \frac{1}{i!} \sum_{\mathbf{a} \in (\mathbb{Z}_{\neq 0})^i} C^{(0)}[k|\mathbf{a}] \, \Xi^{(i)}_{g,n}[\mathbf{a} | \mathbf{b}]\,,
$$
where $\Xi^{(i)}_{g,n}[\mathbf{a} | \mathbf{b}]$ was defined in \eqref{eq:derFgn}. Here, we introduced the coefficients
$$
C^{(0)}[k|\mathbf{a}] = i! \Res_{t = 0} \left( \dd\xi_{r(k+1)}(t)  P_{\mathbf{w} | t} \Big( \bigotimes_{l=1}^i \dd\xi_{-a_l}(w_l)  \Big) \right)\,,
$$
with $\mathbf{a} = (a_1, \ldots, a_i)$.
\end{lem}

\begin{proof}
Recall from Definition \ref{d:combin} that
$$
\widetilde{\mathcal{R}}^{(i)}_{g,n}(A | B)= \sum_{\mathbf{L} \vdash A} \sum_{\substack{h\,:\,\mathbf{L} \rightarrow \mathbb{N} \\ i + \sum_{L \in \mathbf{L}} h_{L} = g + |\mathbf{L}|}} \sum_{\boldsymbol{\mu} \vdash_{\mathbf{L}} B}'' \bigg(\bigotimes_{L \in \mathbf{L}} \omega_{h_{L},|L| + |\mu_{L}|}(L,\mu_{L})\bigg)\,,
$$
where the double prime over the summation symbol means that terms with $h_L=0$, $|\mu_L| = 0$ and $|L| \leq 2$ are excluded from the sum.  For $2g-2+n > 0$, we have an expansion
\begin{equation*}
\begin{split}
\omega_{g,n}(\mathbf{z}) =& \sum_{\mathbf{a} \in (\mathbb{Z}_{> 0})^n} F_{g,n}[\mathbf{a}]  \bigotimes_{l=1}^n \dd\xi_{-a_l}(z_l)  \\
=:&\sum_{\mathbf{a} \in (\mathbb{Z}_{\neq 0})^n} F_{g,n}[\mathbf{a}]  \bigotimes_{l=1}^n \dd\xi_{-a_l}(z_l)\,,
\end{split}
\end{equation*}
where in the second line we extended the summation to all non-zero integers by setting the coefficients to be zero whenever one of the $a_i$s is negative.

For $\omega_{0,2}$, since we are not including contribution where both entries come from $A$, after projection with $P_{\mathbf{w} | t}$ we know one of the entry of $\omega_{0,2}$ must be found among the $w_{l}$s, and thus project to $z$, while the second must be found in the $z_l$s. Thus we can use the following expansion for $\omega_{0,2}$,
$$
\omega_{0,2}(w_1,z_2) \underset{|w_1|<|z_2|}{\approx} \sum_{a > 0} a\,\dd\xi_{a}(w_1) \otimes \dd\xi_{-a}(z_2)\,,
$$
and a similar one when the role of $1$ and $2$ is exchanged. In both situations we can do the replacement
$$
\omega_{0,2}(w_1,z_2) \longleftarrow \sum_{\mathbf{a} \in (\mathbb{Z}_{\neq 0})^2} F_{0,2}[\mathbf{a}]\, \dd\xi_{-a_1}(z_1) \otimes \dd\xi_{-a_2}(z_2)\,,
$$
where we introduced the coefficients $F_{0,2}[a_1,a_2] = |a_1| \delta_{a_1 + a_2,0}$. With this notation, and recalling \eqref{eq:derFgn}, we can write
$$
\widetilde{\mathcal{R}}^{(i)}_{g,n}(\mathbf{w} | \mathbf{z}) 
= \sum_{\substack{\mathbf{a} \in (\mathbb{Z}_{\neq 0})^i \\ \mathbf{b} \in (\mathbb{Z}_{\neq 0})^{n - 1}}} \Xi^{(i)}_{g,n}[\mathbf{a}| \mathbf{b}] \, \bigotimes_{l=1}^i \dd\xi_{-a_l}(w_l)   \bigotimes_{m=2}^n \dd\xi_{-b_m}(z_m)\,.
$$
It thus follows that
$$
\left[ \Res_{t = 0} \,\dd\xi_{r(k+1)}(t) \, P_{\mathbf{w} | t} \left( \widetilde{\mathcal{R}}^{(i)}_{g,n}(\mathbf{w} | \mathbf{z}) \right)\right]_{\mathbf{b}} =  \sum_{\mathbf{a} \in (\mathbb{Z}_{\neq 0})^i}  \Xi^{(i)}_{g,n}[\mathbf{a} | \mathbf{b}] \Res_{t = 0} \left( \dd\xi_{r(k+1)}(t)\,P_{\mathbf{w} | t} \Big(\bigotimes_{l=1}^i \dd\xi_{-a_l}(w_l)  \Big) \right )\,,
$$
and the lemma is proven.
\end{proof}

Let us now re-introduce contributions from $\omega_{0,2}$ with the two entries coming from $\mathbf{w}$. 

\begin{lem}\label{l:twop}
$$
\left[ \Res_{t = 0} \dd\xi_{r(k+1)}(t) \,P_{\mathbf{w} | t} \left( \mathcal{R}^{(i)}_{g,n}(\mathbf{w} | \mathbf{z}) \right) \right]_{\mathbf{j}}
=  \sum_{\substack{\ell,j\geq 0 \\ \ell + 2j = i}}  \frac{1}{\ell!} \sum_{\mathbf{a} \in (\mathbb{Z}_{\neq 0})^{\ell} }\,C^{(j)}[k |\mathbf{a}]  \,\Xi^{(\ell)}_{g-j,n}[\mathbf{a}| \mathbf{b}]\,,
$$
with the coefficients defined as
\begin{equation}
\label{Cdeffff} C^{(j)}[k |\mathbf{a}] =\frac{(\ell+2j)!}{j! \,2^{j}} \Res_{t = 0} \left(\dd\xi_{r(k+1)}(t)\, P_{\mathbf{w} | t} \Big(\bigotimes_{l=1}^{\ell} \dd\xi_{-a_l}(w_l) \bigotimes_{m=1}^{j} \omega_{0,2}(w_{\ell+2m-1}, w_{\ell+2m}) \Big) \right)\,.
\end{equation}
\end{lem}

\begin{proof}
Recall from Lemma \ref{l:rtilde} that
$$
\mathcal{R}^{(i)}_{g,n}(A | B) = \sum_{\substack{\ell,j \geq 0 \\ \ell+2j = i}}\sum_{\substack{\gamma \subseteq A \\ |\gamma| = 2j}} \bigg( \sum_{\substack{\mathbf{L} \vdash \gamma \\ \forall L \in \mathbf{L}\,\,|L| = 2}} \bigotimes_{L \in \mathbf{L}} \omega_{0,2}(L) \bigg) \otimes \widetilde{\mathcal{R}}^{(\ell)}_{g-j,n} (A \setminus \gamma | B)\,.
$$
Thus
$$
P_{\mathbf{w} | t} \left( \mathcal{R}^{(i)}_{g,n}(\mathbf{w} | \mathbf{z}) \right)  = \sum_{\substack{\ell,j \geq 0 \\ \ell +2j = i}}  P_{\mathbf{w} | t} \left( \sum_{\substack{\gamma \subseteq \mathbf{w} \\ |\gamma| = 2j}} \bigg( \sum_{\substack{\mathbf{L} \vdash \gamma \\ \forall L \in \mathbf{L}\,\,|L| = 2}} \bigotimes_{L \in \mathbf{L}} \omega_{0,2}(L) \bigg) \otimes \widetilde{\mathcal{R}}^{(\ell)}_{g-j,n} (\mathbf{w} \setminus \gamma | \mathbf{z}) \right)\,.
$$
Since $P_{\mathbf{w} | t}$ is invariant under permutations of the $w_l$s, the order of the $w_l$s in this expression does not matter. So all terms that only differ by permutations of the $w_l$s will give the same result after acting with the projection operator. So we can order the $w_l$s once and for all. We simply need to count the number of terms for a given $j$. We first need to pick a subsequence $\gamma$ of $\mathbf{w}$ of length $2j$: there are $\frac{i!}{(2j)! (i-2j)!}$ ways to do so. Then, we need to pick a set partition $\mathbf{L}$ of $\gamma$ with parts that all have cardinality two. The number of ways of doing so is $(2j-1)\cdot (2j-3)\, \cdots\, 1 = \frac{(2j)!}{2^{j}\,j!}$ . Thus we end up with
$$
P_{\mathbf{w} | t} \left( \mathcal{R}^{(i)}_{g,n}(\mathbf{w} | \mathbf{z}) \right)  =\sum_{\substack{\ell,j \geq 0 \\ \ell+2j = i}} \frac{i!}{\ell ! j! 2^{j}} P_{\mathbf{w} | t} \left( \widetilde{\mathcal{R}}^{(\ell)}_{g-j,n} (w_{1}, \ldots, w_{\ell} | \mathbf{z}) \otimes  \bigotimes_{l'=1}^{j} \omega_{0,2}(w_{\ell+2l'-1}, w_{\ell + 2l'})\right)\,.
$$
As before, we can write
$$
\widetilde{\mathcal{R}}^{(\ell)}_{g-j,n}(w_{1}, \ldots, w_{\ell} | \mathbf{z}) 
= \sum_{\substack{\mathbf{a} \in (\mathbb{Z}_{\neq 0})^{\ell} \\ \mathbf{b} \in (\mathbb{Z}_{\neq 0})^{n - 1} }} \Xi^{(\ell)}_{g-j,n}[\mathbf{a}| \mathbf{b}] \,\bigotimes_{l=1}^{\ell} \dd\xi_{-a_l}(w_l)  \bigotimes_{m=2}^n \dd\xi_{-b_m}(z_m)\,.
$$
Therefore,
\begin{equation*}
\begin{split}
\left[ \Res_{t = 0} \dd\xi_{r(k+1)}(t) \left( P_{\mathbf{w} | t} \left( \mathcal{R}^{(i)}_{g,n}(\mathbf{w} | \mathbf{z}) \right)  \right) \right]_{\mathbf{b}} = \sum_{\substack{\ell,j \geq 0 \\ \ell + 2j = i}} \frac{i!}{\ell ! j! 2^{j}}  \sum_{\mathbf{a} \in (\mathbb{Z}_{\neq 0})^{i} }  \Xi^{(i)}_{g-j,n}[\mathbf{a}| \mathbf{b}] \\ \times \Res_{t = 0} \left( \dd\xi_{r(k+1)}(t) P_{\mathbf{w} | t} \left(\bigotimes_{l=1}^{\ell} \dd\xi_{-a_l}(w_l)   \right)  \otimes \bigotimes_{l'=1}^{j} \omega_{0,2}(w_{\ell+2l'-1}, w_{\ell+2l'}) \right)\,,
\end{split}
\end{equation*}
and the lemma is proven.
\end{proof}

Finally we need to re-introduce the contributions from $\omega_{0,1}$. Recall that we can write
\beq
\label{o01f01} \omega_{0,1}(z) = \sum_{a > 0} \tau_a\,\dd\xi_a(z) =: \sum_{a \in \mathbb{Z}_{\neq 0}} F_{0,1}[a]\,\dd\xi_{-a}(z)\,,
\eeq
where we defined the coefficients to be $F_{0,1}[a] = 0$ and $F_{0,1}[-a] = \tau_a$ for all $a > 0$. 
Then
\begin{lem}\label{l:threep}
$$
\left[ \Res_{t = 0} \dd\xi_{r(k+1)}(t) \left( P_{\mathbf{w} | t} \left( \mathcal{E}^{(i)}_{g,n}(\mathbf{w} | \mathbf{z}) \right)  \right) \right]_{\mathbf{b}} = \sum_{m=1}^{i}   \sum_{\substack{\ell,j \geq 0 \\ \ell +2j = m}} \frac{1}{ \ell!}
\sum_{\mathbf{a} \in (\mathbb{Z}_{\neq 0})^{\ell} }D^{(j)}_i[k |\mathbf{a}]  \, \Xi^{(\ell)}_{g-j,n}[\mathbf{a}| \mathbf{b}]\,,
$$
where we defined the coefficients
\begin{equation}
\label{Dellfff} D^{(j)}_i[k |\mathbf{a}] := \frac{1}{(i-\ell-2j)!} \sum_{a_{\ell+1}, \ldots, a_{i-2j} \in \mathbb{Z}_{\neq 0}} \left( \prod_{l =\ell+1}^{i-2j} F_{0,1}[a_l] \right)C^{(j)}[k |\mathbf{a}, a_{\ell+1}, \ldots,a_{i-2j}]\,.
\end{equation}
\end{lem}

\begin{proof}
Recall from Lemma \ref{l:er} that
$$
\mathcal{E}^{(i)}_{g,n}(A | B) = \sum_{m=0}^i \sum_{\substack{\gamma \subseteq A \\ |\gamma| = m}}  \left( \prod_{l=1}^i \omega_{0,1}(\gamma_l) \right) \otimes \mathcal{R}^{(i-m)}_{g,n} (A \setminus \gamma | B)\,.
$$
Note that for $2g-2+n > 0$, the term with $m=i$ does not contribute, so we can terminate the sum over $m$ at $i-1$.
Thus
$$
P_{\mathbf{w} | t} \left( \mathcal{E}^{(i)}_{g,n}(\mathbf{w} | \mathbf{z}) \right)  = \sum_{m=0}^{i-1} P_{\mathbf{w} | t}  \left( \sum_{\substack{\gamma \subseteq \mathbf{w} \\ |\gamma| = m}}  \bigg(\bigotimes_{l=1}^m \omega_{0,1}(\gamma_l) \bigg) \otimes \mathcal{R}^{(i-m)}_{g,n} (\mathbf{w} \setminus \gamma | \mathbf{z}) \right)\,.
$$
As before, we use the argument that the projection operator is invariant under permutations of the $w_l$s to re-order the entries in the argument. Thus all terms contribute the same. The number of terms is the number of ways to choose $m$ elements in $\mathbf{w}$, which is given by $\frac{i!}{m! (i - m)!}$. So we get
$$
P_{\mathbf{w} | t} \left( \mathcal{E}^{(i)}_{g,n}(\mathbf{w} | \mathbf{z}) \right)  = \sum_{m=0}^{i-1} \frac{i!}{m! (i-m)!}\, P_{\mathbf{w} | t} \left( \mathcal{R}^{(i-m)}_{g,n} (w_1, \ldots, w_{i-m} | \mathbf{z})\otimes  \bigotimes_{l=i-m+1}^{i} \omega_{0,1}(w_l)\right)\,.
$$
Using Lemma~\ref{l:twop} we know that
\begin{equation*}
\begin{split}
P_{\mathbf{w} | t} \left( \mathcal{R}^{(i-m)}_{g,n}(w_1, \ldots, w_{i-m} | \mathbf{z})\bigotimes_{l=i-m+1}^i \omega_{0,1}(w_l) \right)  \\
=  \sum_{\substack{\ell,j \geq 0 \\ \ell+2j = i-m}} \frac{(i - m)!}{\ell!\,j!\, 2^{j}} 
P_{\mathbf{w} | t} \left(\widetilde{\mathcal{R}}^{(\ell)}_{g-j,n} (w_{1}, \ldots, w_{\ell} | \mathbf{z})  \bigotimes_{l'=1}^{j} \omega_{0,2}(w_{\ell+2l'-1}, w_{\ell+2l'}) \bigotimes_{l=i-m+1}^i \omega_{0,1}(w_l) \right)\,.
\end{split}
\end{equation*}
We now have as usual
$$
\widetilde{\mathcal{R}}^{(\ell)}_{g-j,n}(w_{1}, \ldots, w_{\ell} | \mathbf{z}) 
= \sum_{\substack{\mathbf{a} \in (\mathbb{Z}_{\neq 0})^{\ell} \\ \mathbf{b} \in (\mathbb{Z}_{\neq 0})^{n - 1}}} \Xi^{(\ell)}_{g-j,n}[\mathbf{a}| \mathbf{b}] \,\bigotimes_{l=1}^{\ell} \dd\xi_{-a_l}(w_l) 
 \bigotimes_{m=2}^n \dd\xi_{-b_m}(z_m)\,.
$$
Therefore,
\begin{equation*}
\begin{split}
\left[ \Res_{t = 0} \dd\xi_{r(k+1)}(t) \, P_{\mathbf{w} | t} \left( \mathcal{E}^{(i)}_{g,n}(\mathbf{w} | \mathbf{z}) \right) \right]_{\mathbf{b}} = \sum_{m=0}^{i-1}   \sum_{\substack{\ell,j \geq 0 \\ \ell+2j = i-m}} \frac{i!}{m!\, \ell! \,j! \,2^{j}}  \sum_{\mathbf{a} \in (\mathbb{Z}_{\neq 0})^{\ell} }  \Xi^{(\ell)}_{g-j,n}[\mathbf{a}| \mathbf{b}] \\
\times \Res_{t = 0} \left( \dd \xi_{r(k+1)}(t) P_{\mathbf{w} | t} \left(\bigotimes_{l=1}^{\ell} \dd\xi_{-a_l}(w_l)   \bigotimes_{l''=\ell+1}^{\ell+m} \omega_{0,1}(w_{l''})    \bigotimes_{l'=1}^{j} \omega_{0,2}(w_{\ell+m+2l'-1}, w_{\ell+m+2l'}) \right)   \right)\,.
\end{split}
\end{equation*}
Expanding $\omega_{0,1}$ as in \eqref{o01f01} we can write
\begin{equation*}
\begin{split}
 \Res_{t = 0} &\left( \dd\xi_{r(k+1)}(t) P_{\mathbf{w} | t} \Big(\bigotimes_{l=1}^{\ell} \dd\xi_{-a_l}(w_l)   \bigotimes_{l''=\ell+1}^{\ell+m} \omega_{0,1}(w_{l''})    \bigotimes_{l'=1}^{j} \omega_{0,2}(w_{\ell+m+2l'-1}, w_{\ell+ m +2l'}) \Big)  \right) \\
=&\!\!\! \sum_{\substack{a_{\ell+1}, \ldots, a_{\ell+m}  \\ \in \mathbb{Z}_{\neq 0}}} \left( \prod_{l''=\ell+1}^{\ell+m} F_{0,1}[a_{l''}] \right)  \Res_{t = 0} \left( \dd \xi_{r(k+1)}(t) \,P_{\mathbf{w} | t} \Big(\bigotimes_{l=1}^{\ell+m} \dd\xi_{-a_l}(w_l)       \bigotimes_{l'=1}^{j} \omega_{0,2}(w_{\ell + m +2l'-1}, w_{\ell + m +2l'}) \Big)  \right)  \\
=&\!\!\! \sum_{\substack{a_{\ell+1}, \ldots, a_{\ell+m} \\  \in \mathbb{Z}_{\neq 0}}} \left( \prod_{l'' = \ell+1}^{\ell+m} F_{0,1}[a_{l''}] \right) \frac{j!\, 2^{j}}{(\ell + m + 2j)!} \,C^{(j)} [k|\mathbf{a}, a_{\ell+1}, \ldots, a_{\ell+m}] \,, 
\end{split}
\end{equation*}
with $\mathbf{a} \in (\mathbb{Z}_{\neq 0})^{\ell}$ by definition of $C^{(j)}$ in \eqref{Cdeffff}. After introducing a new ind\-ex $m'=i-m$ to re\-write the sum over $m \in \{0,\ldots, i-1\}$ as a sum over $m' \in \{1,\ldots,i\}$, we obtain
\begin{equation*}
\begin{split}
\left[ \Res_{t = 0} \dd\xi_{r(k+1)}( t) \,P_{\mathbf{w} | t} \left( \mathcal{E}^{(i)}_{g,n}(\mathbf{w} | \mathbf{z}) \right)\right]_{\mathbf{b}} = \sum_{m'=1}^{i}   \sum_{\substack{\ell,j \geq 0 \\ \ell +2j = m'}} \frac{1}{(i-m')!\,\ell!}  \\
\times \sum_{\mathbf{a} \in (\mathbb{Z}_{\neq 0})^{\ell}}  \Xi^{(\ell)}_{g-j,n}[\mathbf{a}| \mathbf{b}]\sum_{a_{\ell+1}, \ldots, a_{i-2j} \in \mathbb{Z}_{\neq 0}} \left( \prod_{l''=\ell+1}^{i-2j} F_{0,1}[a_{l''}] \right) C^{(j)} [k|\mathbf{a}, a_{\ell+1}, \ldots, a_{i-2j}]\,. 
\end{split}
\end{equation*}
We recognize the coefficients $D^{(j)}_i$ introduced in \eqref{Dellfff} and the lemma is proven.
\end{proof}
 
We can now finish the proof of Theorem~\ref{t:ZZ}. From Lemmas \ref{l:onep}, \ref{l:twop} and \ref{l:threep}, we find that the higher abstract loop equations for local spectral curves with one component hold if and only if
$$
\sum_{m=1}^i  \sum_{\substack{\ell,j \geq 0 \\ \ell + 2j = m}} \frac{1}{ \ell!}  
\sum_{\mathbf{a} \in (\mathbb{Z}_{\neq 0})^{\ell} } D^{(j)}_i[k |\mathbf{a}] \,\Xi^{(\ell)}_{g-j,n}[\mathbf{a}| \mathbf{b}]  = 0\,,
$$
for $i \in \{1,\ldots, r\}$, $k \geq \mathfrak{d}^i$ and $\mathbf{b}\in ( \mathbb{Z}_{> 0})^n$. Just as in Lemma~\ref{l:rd} this is equivalent to the claimed system of differential equations.
\end{proof}

\subsubsection{Identification with the \texorpdfstring{$\mathcal{W}(\mathfrak{gl}_{r})$}{W(gl(r))} quantum Airy structure}

We now relate the differential operators appearing in Proposition~\ref{t:ZZ} with the $\mathcal{W}(\mathfrak{gl}_{r})$ quantum Airy structure of Proposition~\ref{t:conjugated}. 

\begin{thm}\label{1cccc}
Under the conditions of Propositions~\ref{t:ZZ}, we have for any $i \in \{1,\ldots,r\}$ and $k \geq \mathfrak{d}^i + \delta_{i,1}$ the identification
$$
\mathcal{H}_{k}^i = - r\,\hat{\Phi} \hat{T}  W^i_k \hat{T}^{-1}\,\hat{\Phi}^{-1}\, .
$$
Here, $W^i_{k}$ are defined in \eqref{eq:Wmodes} and we use the dilaton shift and change of polarization defined in terms of the coefficients of expansion \eqref{e:exp01}-\eqref{e:exp02} of $\omega_{0,1}$ and $\omega_{0,2}$
$$ 
\hat{T} =  \exp \left(\frac{1}{\hbar} \sum_{a > 0} \frac{F_{0,1}[-a]}{a}\,J_a \right)\,,\qquad \hat{\Phi} = \exp \left(\frac{1}{2 \hbar} \sum_{l,m > 0}\frac{\phi_{l,m}}{l\,m} J_l J_m \right)\,.
$$ 
In particular, for admissible spectral curves, where $1 \leq s \leq r+1$ and $r = \pm 1\,\,{\rm mod}\,\,s$, the $\mathcal{H}_k^i$ form a $\mathcal{W}(\mathfrak{gl}_r)$ higher quantum Airy structure as in Proposition \ref{t:conjugated}. The coefficients $F_{g,n}$ of the partition function of this $\mathcal{W}(\mathfrak{gl}_r)$ higher quantum Airy structure in the basis $(x_{l})_{l > 0}$ coincide with the coefficients of the expansion \eqref{e:expansion} of the unique $\omega_{g,n} \in (V_{z}^{-})^{\otimes n}$ solution to the higher abstract loop equations \eqref{e:ale}.
\end{thm}

\begin{proof} We first concentrate on the case of standard polarization, that is
$$
\omega_{0,2}(z_1,z_2) = \frac{\dd z_1 \otimes \dd z_2}{(z_1-z_2)^2}\,.
$$
We are going to evaluate the coefficients $C^{(j)}[k |\mathbf{a}]$ and $D^{(j)}_i[k |\mathbf{a}]$ and recognize the coefficients of the higher quantum Airy structure of Proposition~\ref{t:condition}. Recall in particular from Definition~\ref{psirootsum} the sums $\Psi^{(j)}(a_1,\ldots,a_i)$ over $r$-th roots of unity.

\begin{lem}
For a local spectral curve with one component in standard polarization,
$$
C^{(j)}[k|\mathbf{a}]=  \frac{(\ell+2j)!}{j!\,2^{j}}\,\Psi^{(j)} (\mathbf{a})\,\delta_{r(\ell+2j-k-1) + \sum_{l=1}^{\ell} a_{l}, 0}\,,
$$
with $\mathbf{a} \in (\mathbb{Z}_{\neq 0})^{\ell}$.
\end{lem}

\begin{proof}
Recall that 
$$
C^{(j)}[k |\mathbf{a}] =\frac{(\ell+2j)!}{j!\,2^{j}} \Res_{t = 0} \left(\dd \xi_{r(k+1)}(t)  P_{\mathbf{w} | t} \left(\bigotimes_{l=1}^{\ell} \dd\xi_{-a_l}(w_l) \bigotimes_{l'=1}^{j} \omega_{0,2}(w_{\ell+2l'-1}, w_{\ell+2l'}) \right) \right)\,.
$$
For a curve in standard polarization, this simplifies to
$$
C^{(j)}[k |\mathbf{m}] =\frac{(\ell+2j)!}{j!\,2^{j}} \Res_{t = 0} \left( t^{r (k+1)-1} \dd t\, P_{\mathbf{w} | t} \left(\prod_{l=1}^{\ell} w_l^{-a_l-1}  \prod_{l'=1}^{j} \frac{1}{(w_{\ell+2l'-1}-w_{\ell+2l'})^2}  \bigotimes_{l''=1}^i \dd w_{l''}  \right) \right)\,.
$$
By definition of the projection operator and using Definition~\ref{psirootsum} for $\Psi^{(j)}$, we can write
\begin{equation*}
\begin{split}
C^{(j)}[k |\mathbf{a}] &= \frac{(\ell+2j)!}{j!\,2^{j}} \,\Psi^{(j)} (\mathbf{a})\, \Res_{t = 0} \left( t^{rk + r-1-\sum_{l=1}^{\ell} a_l  - r(\ell+2j) }\,\dd t \right) \\\
&=  \frac{(\ell+2j)!}{j!\,2^{j}} \,\Psi^{(j)}(\mathbf{a})\, \delta_{r(\ell+2j - k -1) + \sum_{l=1}^{\ell} a_l, 0}\,.
\end{split}
\end{equation*}
\end{proof}

We can then calculate the coefficients $D^{(j)}_i[k |\mathbf{a}]$.
\begin{lem}
For a local spectral curve with one component in standard polarization,
\begin{equation*}
\begin{split}
D^{(j)}_i[k |\mathbf{a}] & =  \frac{i!}{(i-\ell-2j)! j!\,2^{j} }  \\
&\quad \qquad\qquad \times\sum_{a_{\ell+1}, \ldots, a_{i-2j} \in \mathbb{Z}_{\neq 0}} \left( \prod_{l=\ell + 1}^{i-2j} F_{0,1}[a_l] \right) \Psi^{(j)} (\mathbf{a}, a_{\ell+1}, \ldots, a_{i-2j})\,\delta_{r (i-k-1) + \sum_{l=1}^{i-2j} a_l, 0}\,,
\end{split}
\end{equation*}
with $\mathbf{a} = (a_1, \ldots, a_{\ell}) \in (\mathbb{Z}_{\neq 0})^{\ell}$.
\end{lem}

\begin{proof}
Recall that
$$
D^{(j)}_i[k |\mathbf{a}] := \frac{1}{(i-\ell-2j)!} \sum_{a_{\ell+1}, \ldots, a_{i-2j} \in \mathbb{Z}_{\neq 0}} \left( \prod_{l=\ell+1}^{i-2j} F_{0,1}[a_l] \right)C^{(j)}[k |\mathbf{a}, a_{\ell+1}, \ldots, a_{i-2j}]\,.
$$
Thus, using the previous Lemma,
\begin{equation*}
\begin{split}
D^{(j)}_i[k |\mathbf{a}] & =  \frac{i!}{(i-\ell-2j)! j!\,2^{j} }  \\
& \quad \qquad \qquad \times\sum_{a_{\ell+1}, \ldots, a_{i-2j} \in \mathbb{Z}_{\neq 0}} \left( \prod_{l=\ell+1}^{i-2j} F_{0,1}[a_l] \right) \Psi^{(j)} (\mathbf{a}, a_{\ell+1}, \ldots, a_{i-2j}) \delta_{r (i-k-1) + \sum_{l=1}^{i-2j} a_l, 0}\,. 
\end{split}
\end{equation*}
\end{proof}

Combining Proposition \ref{t:ZZ} with the two previous lemmas, we find that
\begin{equation*}
\begin{split}
\mathcal{H}^i_k & =  \sum_{m=1}^{i} \sum_{\substack{\ell,j \geq 0 \\ \ell+ 2j = m}} \frac{\hbar^{j}}{\ell!} \frac{i!}{(i-\ell-2j)! j!\, 2^{j}}  \\
& \quad \qquad \qquad \times \sum_{\mathbf{a} \in (\mathbb{Z}_{\neq 0})^{i-2j} } \left( \prod_{l=\ell+1}^{i-2j} F_{0,1}[a_l] \right) \Psi^{(j)} (\mathbf{a})\,\delta_{r (i-k-1) + \sum_{l'=1}^{i-2j} m_{l'}, 0}\,\norder{J_{a_1} \cdots\, J_{a_{\ell}}}\,.
\end{split}
\end{equation*}
This can be simplified by writing the sum in a more symmetric way. Instead of extracting $(a_{\ell+1}, \ldots, a_{i-2j})$, we can sum over all ways of extracting $i-2j-\ell$ $a$s out of the $i-2j$ ones. We then need to multiply by the factor $\frac{(i-2j-\ell)! \ell!}{(i-2j)!}$ to avoid over-counting. We get
$$
\mathcal{H}^i_k = \sum_{m=1}^{i} \sum_{\substack{\ell,j \geq 0 \\ \ell + 2j = m}} \hbar^{j} \frac{i!}{(i-2j)! j!\,2^{j}}  \\
\sum_{\substack{\mathbf{a} \in (\mathbb{Z}_{\neq 0})^{i-2j} \\ \sum_{l=1}^{i-2j} m_l  =r(k-i+1)} }  \Psi^{(j)} (\mathbf{a}) \left(  \sum_{\substack{\mathbf{c} \subseteq \mathbf{a} \\ |\mathbf{c}| = i-m}}\prod_{l'=1}^{i-m} F_{0,1}[c_{l'}] \right) \,\norder{J_{a_1} \cdots\, J_{a_{\ell}}}\,,
$$
where in the cases that $i=2j$ the condition $\sum_l a_l = r(k-i+1)$ is understood as the delta condition $\delta_{k,i-1}$.

Let us now compare to the dilaton-shifted $W_k^i$s. Recall that conjugation by $\hat{T}$ is equivalent to the shift $J_{-a} \longrightarrow J_{-a} + F_{0,1}[-a]$. It results in that the coefficient of $\norder{J_{a_1} \cdots\, J_{a_{\ell}}}$ in $H^i_k$ should be the sum of all possible ways of starting with a term of the form $\norder{J_{a_1} \cdots\, J_{a_{\ell'}}}$ with $\ell'>\ell$ in $W^i_k$, and replacing the extra $J_a$s by $F_{0,1}[a]$. This is exactly what the formula for $H^i_k$ does, up to a global prefactor $-r$.

We now turn to the general polarization, \textit{i.e.} we have a basis for $V_{z}^-$
$$
\dd\xi_{-l}(z) = \frac{\dd z}{z^{l+1}}  + \sum_{m >0} \frac{\phi_{l, m}}{l}\,\dd\xi_{m}(z)\,,
$$
for some symmetric coefficients $\phi_{l,m}$, and a formal bidifferential
$$
\omega_{0,2}(z_1, z_2) = \frac{\dd z_1 \otimes \dd z_2}{(z_1-z_2)^2} + \sum_{l,m > 0} \phi_{l,m} \dd\xi_{l}(z_1) \otimes \dd\xi_{m}(z_2)\,.
$$

Looking at the definition of $D^{(j)}_i [k | \mathbf{a}]$, \eqref{eq:Dcoeff} and following the same argument as for the case of standard polarization, it is clear that the relation between the $D^{(j)}_i[k | \mathbf{a}]$ and the $C^{(j)} [k | \mathbf{a}]$ is given by the dilaton shift $J_{-a} \rightarrow J_{-a} + F_{0,1}[-a]$. Thus all we need to check here is the effect of the  change of polarization. Recall the conjugation by $\hat{\Phi}$ amounts to the shift
\begin{equation}\label{eq:shiftJ}
\forall a > 0,\qquad J_{-a} \longrightarrow J_{-a}  + \sum_{l > 0} \frac{\phi_{a,l}}{l} J_l\,.
\end{equation}

Suppose that we start with the $W^i_k$ and do a polarization conjugation. Let us denote the coefficients of $W^i_k$ by $C_{{\rm st}}^{(j)}[k | \mathbf{a}]$. Every factor of $J_{-a}$ with $a > 0$ gets shifted as above. If we turn this around, this means that if we are calculating the coefficient of a term in the conjugated operator $\hat{\Phi}W^i_k \hat{\Phi}^{-1}$ that has a factor of $J_m$ with $m > 0$, say $C^{(j)}[k | \ldots , m,  \ldots]$, then this coefficient will get a contribution of the form
$$
C^{(j)}_{{\rm st}}[k | \ldots, m, \ldots]+ \sum_{l > 0} \frac{\phi_{l,m}}{m}\,C^{(j)}_{\rm st}[k | \ldots, - l, \ldots]\,.
$$
Now, since $C^{(j)}[k | \ldots , m,  \ldots]$ comes with a factor of $\dd \xi_{-m}(w)$ in the residue definition, we see that this change of coefficients is implemented by doing a change of basis
$$
\dd \xi_{-m}(w) \mapsto \dd \xi_{-m}(w) + \sum_{l > 0} \frac{\phi_{m,l}}{m}\,\dd \xi_l(w)\,,
$$
which is precisely what a change of polarization does. More precisely, if we start with the standard polarization, for which $\dd\xi_{-m}(w) = w^{-m-1}\dd w$, then this shift implements the basis definition for a general polarization
$$
\dd \xi_{-m}(w) = w^{-m-1} \dd w+ \sum_{l > 0} \frac{\phi_{m,l}}{m}\,\dd\xi_l(w)
$$
in the definition of the coefficients \eqref{eq:Dcoeff}.  However, one needs to be careful. After shifting as in \eqref{eq:shiftJ}, the $J$s may not be normal ordered anymore. Normal ordering thus will produce extra contributions. This will happen whenever we are shifting the first factor in expressions of the form $J_{-a} J_{-b}$. After the shift, this becomes
$$
J_{-a} J_{-b} \longrightarrow J_{-a} J_{-b} + \sum_{l > 0} \frac{\phi_{a,l}}{l}\,J_l J_{-b}\,,
$$
which is not normal ordered anymore. Normal ordering produces an extra contribution
$$
J_{-a} J_{-b} \longrightarrow J_{-a} J_{-b} + \sum_{l > 0}  \frac{\phi_{a,l}}{l}\,\,\norder{J_lJ_{-b}} + \hbar \phi_{b,a}\,.
$$
This means that the coefficient of, say, $C^{(j)}[k|\ldots]$ with $j \geq 1$, will get an extra contribution of the form $\phi_{b,a} C^{(j-1)}[k | \ldots, -b, -a]$. But since $C^{(j-1)}[k | \ldots, -b, -a]$ comes with a factor of $\dd \xi_{b}(w_1)\otimes\dd\xi_a(w_2)$ in the residue definition, these extra contributions are precisely accounted for by replacing the bidifferential $\frac{\dd z_1 \otimes \dd z_2}{(z_1-z_2)^2}$ in standard polarization by the new bidifferential 
$$
\omega_{0,2}(z_1, z_2) = \frac{\dd z_1 \otimes \dd z_2}{(z_1-z_2)^2} + \sum_{l,m > 0} \phi_{l,m} \dd\xi_{l}(z_1) \otimes \dd\xi_{m}(z_2)
$$
in the definition of the coefficients \eqref{eq:Dcoeff}. We conclude that the definition of the coefficients \eqref{eq:Dcoeff} precisely implements a change of polarization from standard polarization to arbitrary polarization, hence the resulting operators are the result of conjugation by $\hat{\Phi}$. Combining with conjugation by $\hat{T}$ (note that $\hat{T}$ and $\hat{\Phi}$ commute), we obtain the statement of the theorem in full generality.
\end{proof}

\subsection{Local spectral curves with several components}

\label{s:several}

We now give the general result for local spectral curves with $c$ components. Let $R = \{1,\ldots,c\}$, and define $W_{\alpha,k}^i$ to be copies indexed by $\alpha \in R$ of the differential operators \eqref{eq:Wmodes} representing the modes of the generators of the $\mathcal{W}(\mathfrak{gl}_{r_{\alpha}})$ algebra, involving variables $(x_{\alpha,a})_{a > 0}$.
 
\begin{thm}\label{t:HASseveral} 
For any spectral curve with $c$ components as defined in Definition \ref{d:lsc}, the higher abstract loop equations for $\omega_{g,n} \in {\rm Sym}^n(\mathcal{V}_{z}^{-})$ with $2g - 2 + n > 0$ is equivalent to a system of differential equations
$$
\forall \alpha \in R\,,\,\,\,\,\, i \in \{1,\ldots,r_{\alpha}\}\,,\,\,\,\,\, \forall k \geq \mathfrak{d}_{\alpha}^i+ \delta_{i,1}\,,\qquad  \mathcal{H}^i_{\alpha,k}\cdot Z = 0\,,
$$
where $\mathfrak{d}_\alpha^i = i-1-  \big\lfloor \frac{s_\alpha(i-1)}{r_\alpha} \big\rfloor$, for the partition function
$$
Z = \exp\left(\sum_{\substack{g \geq 0,\,\,n \geq 1 \\ 2g - 2 + n > 0}} \frac{\hbar^{g-1}}{n!} \sum_{\boldsymbol{\alpha} \in R^n} \sum_{\mathbf{b} \in (\mathbb{Z}_{> 0})^n} F_{g,n}\left[ \begin{smallmatrix} \boldsymbol{\alpha} \\ \mathbf{b} \end{smallmatrix} \right] x_{\alpha_1, b_1} \cdots\, x_{\alpha_n, b_n}\right)\,,
$$
constructed from the coefficients of the expansion in \eqref{e:expansion}. The differential operators read
$$
\mathcal{H}^i_{\alpha,k} = -r_{\alpha}\,\hat{\Phi}\hat{T}W_{\alpha,k}^i\hat{T}^{-1}\hat{\Phi}^{-1}\,,
$$
where the dilaton shift and change of polarization are given by
\begin{equation*}
\begin{split}
\hat{T} &= \exp \left(\frac{1}{\hbar}\sum_{\alpha=1}^c \sum_{a > 0} \frac{F_{0,1} \left[\begin{smallmatrix} -\alpha \\ a\end{smallmatrix} \right]}{a} J_{\alpha,a}  \right)\,,\\
\hat{\Phi} &= \exp \left(\frac{1}{2 \hbar} \sum_{\alpha,\beta=1}^c \sum_{l,m > 0} \frac{\phi_{l,m}^{\alpha, \beta}}{l\,m}\,J_{\alpha,l} J_{\beta,m} \right)\,.
\end{split}
\end{equation*}
In particular, for admissible spectral curves, where $1 \leq s_{\alpha} \leq r_{\alpha} + 1$ and $r_{\alpha} = \pm 1\,\,{\rm mod}\,\,s_\alpha$, these $\mathcal{H}^i_{\alpha,k}$ form a higher quantum Airy structure, isomorphic to those for the $\bigoplus_{\alpha} \mathcal{W}(\mathfrak{gl}_{r_\alpha})$ algebra.
\end{thm}

\begin{proof} First, we consider the case of standard polarization, that is
\begin{equation}
\omega_{0,2}(z_1,z_2) = \sum_{\alpha=1}^c \frac{(e_{\alpha} \otimes \dd z_1) \otimes (e_{\alpha} \otimes \dd z_2)}{(z_1-z_2)^2}\,.
\end{equation}
Since the $\dd\xi_{\alpha,l}(z) \in \mathcal{V}_z^{(\alpha)}$ for all $l \in \mathbb{Z}_{\neq 1}$, we can really think of a local spectral curve with $c$ components in standard polarization as $c$ spectral curves with one component. The result then follows directly from Theorem~\ref{1cccc}.

For general polarization, the proof follows the exact same lines as for spectral curves with one component, except that we have to keep track of multi-indices. We therefore omit it.
\end{proof}

\begin{rem}
It is straightforward to reformulate Theorem \ref{t:HASseveral} as a Givental-like decomposition formula for the partition function $Z$. This is presented in the statement of Theorem \ref{TGIV} in the introduction.
\end{rem}

One direct consequence of the identification between high\-er abs\-tract loop equa\-tions and higher quan\-tum Airy struc\-tu\-res is the sym\-metry of the me\-ro\-mor\-phic diffe\-ren\-tials cons\-tructed by the Bouchard--Eynard topological recursion.

\begin{thm}
\label{c:symmetric}
For arbitrary admissible local spectral curves, as defined in Definitions \ref{d:lsc} and \ref{d:admissible}, the Bouchard--Eynard topological recursion from Definition \ref{d:BE} produces symmetric differentials $\omega_{g,n}$.
\end{thm}

\begin{proof}
This follows directly from Appendix \ref{a:proof}. There, we show that if a polarized solution to the higher abstract loop equations exists, then it is uniquely constructed by the Bouchard--Eynard topological recursion. Thus a polarized solution exists if and only if the Bouchard--Eynard topological recursion produces symmetric differentials. But Theorem \ref{t:HASseveral} implies that a polarized solution to the higher abstract loop equations does indeed exist for arbitrary admissible local spectral curves, and hence the Bouchard--Eynard must produce symmetric $\omega_{g,n}$.
\end{proof}

\begin{rem}
It should be emphasized here that the admissibility condition on $s_\alpha$ and $r_{\alpha}$  (see Definition \ref{d:admissible}) is crucial. In fact, unexpectedly, when this condition is not satisfied, the Bouchard--Eynard topological recursion does not produce symmetric differentials. This is proven in Proposition \ref{p:symmetry}. Indeed, for choices of $s_\alpha$ and $r_\alpha$ that are coprime but such that $r_\alpha \neq \pm 1 \mod s_\alpha$, our identification between the structure of the higher abstract loop equations and the differential equations produced by the $\mathcal{H}^i_{\alpha,k}$ is still valid. The question is whether there exists a solution to the differential constraints $ \mathcal{H}^i_{\alpha,k}\cdot Z = 0$, or, equivalently, a polarized solution to the higher abstract loop equations. It turns out that the answer is no. It is argued in Proposition \ref{p:symmetry} that there cannot be a symmetric solution to the differential constraints. Correspondingly, this means that the Bouchard--Eynard topological recursion cannot produce symmetric differentials in these cases\footnote{This can also be checked symboli\-ca\-lly on Mathe\-matica for va\-rious exam\-ples of spec\-tral cur\-ves with ${\rm gcd}(r,s) = 1$ that do not meet the ad\-mi\-ssi\-bi\-li\-ty requirement, the simplest ones being
$$(r,s) = (7,5), (8,5), (9,7), (10,7), (11,7), (11,8),\,{\rm etc.}
$$}, otherwise it would construct a polarized solution to the higher abstract loop equations. In the context of higher quantum Airy structures, this implies that for those choices of $s_\alpha$ and $r_\alpha$ the left ideal generated by the $\mathcal{H}^i_{\alpha,k}$ is not a graded Lie subalgebra.
\end{rem}

\newpage

\section{\texorpdfstring{$\mathcal{W}$}{W} constraints and enumerative geometry}

\label{Section6}

In this section, we review the currently known relations between $\mathcal{W}$ constraints and enumerative geometry. We view their possible extension to new cases as a motivation to study higher quantum Airy structures, and formulate new questions raised by our work. The leitmotiv is that for each instance of generating series appearing in one of the following situations
\begin{itemize}
\item[$(i)$] intersection numbers of interesting classes on $\overline{\mathcal{M}}_{g,n}$,
\item[$(ii)$] tau functions of integrable hierarchies,
\item[$(iii)$] matrix integrals,
\item[$(iv)$] higher quantum Airy structures and their partition functions,
\item[$(v)$] differential constraints and partition function obtained from periods on a spectral curve,
\end{itemize}
one can ask for an equivalent description in the four other contexts.

\subsection{\texorpdfstring{$r$}{r}-spin intersection numbers}
\label{Srspin}
The Witten $r$-spin partition function is one of the only examples completely understood from the five points of view.  Its construction was sketched by Witten in \cite{Wittenr}, where he proposed several conjectures which have been resolved since then.

\subsubsection*{(i) - Enumerative geometry}

Let $r,g,n$ be nonnegative integers such that $r \geq 2$, $n \geq 1$ and $2g - 2 + n > 0$. Let $i_1,\ldots,i_n \in \mathbb{Z}$ such that $2g - 2 - \sum_{l = 1}^n (i_l - 1) \in r\mathbb{Z}$. An $r$-spin structure on a smooth curve $\mathcal{C}$ with punctures $p_1,\ldots,p_n$ is the data of a line bundle $L$ and an isomorphism
$$
L^{\otimes r} \simeq K\Big(-\sum_{l = 1}^n (i_l - 1)p_l\Big)\,,
$$
where $K$ is the cotangent line bundle. Jarvis \cite{Jarvis1} constructed the compactified moduli stack of (isomorphism classes of) $r$-spin structures $\overline{\mathcal{M}}_{g,n}(r;\mathbf{i})$. Polishchuk and Vaintrob \cite{Polischuk2,Polischuk1}, and later Chiodo \cite{Chiodo} using different means, constructed a Chow cohomology class $\Omega_{g,n}(r,\mathbf{i})$ of pure dimension in $\overline{\mathcal{M}}_{g,n}(r;\mathbf{i})$ which has the basic properties expected by \cite{Wittenr} and called it the Witten $r$-spin class. 

One can then introduce the $r$-spin partition function
\beq
\label{Zxgfgun}Z_{r{\rm spin}} = \exp\left\{\sum_{2g - 2 + n > 0} \frac{\hbar^{g - 1}}{n!}  \sum_{\substack{1 \leq i_1,\ldots,i_n \leq r \\ k_1,\ldots,k_n \geq 0}} \bigg(\int_{\overline{\mathcal{M}}_{g,n}(r;\mathbf{i})} \Omega_{g,n}(r;\mathbf{i}) \prod_{l = 1}^n \psi_l^{k_l} \bigg) \prod_{l = 1}^n t_{k_l}^{i_{l}}\right\}
\eeq
on the formal variables $t_{k}^{i}$ indexed by $k \geq 0$ and $i \in \{1,\ldots,r\}$. In fact, $\Omega_{g,n}(r;\mathbf{i})$ is zero when one of the $i_{l}$ is equal to $r$, so one could restrict to $i \in \{1,\ldots,r - 1\}$.

In particular for $r = 2$, we only have to consider the value $i = 1$ and we obtain the usual intersection numbers of $\psi$-classes on the moduli space of curves
\beq
\label{Z2spin} Z_{2{\rm spin}} = \exp\left\{\sum_{2g - 2 + n > 0} \frac{\hbar^{g - 1}}{n!} \sum_{k_1,\ldots,k_n \geq 0} \bigg(\int_{\overline{\mathcal{M}}_{g,n}} \prod_{l = 1}^n \psi_{l}^{k_l}\bigg) \prod_{l = 1}^n t_{k_l} \right\}\, .
\eeq

\subsubsection*{(ii) - Integrability}

$Z_{r{\rm spin}}$ is a tau function of the $r$-KdV (also called $r$-Gelfand-Dickey) hierarchy with respect to the times
\beq
\label{changex} x_{rk + i} = \frac{(-1)^{k}\,t_{k}^{i}}{\ii \sqrt{r} \prod_{m = 0}^{k} \big(m + \frac{i - 1}{r}\big)}\,.
\eeq
For $r = 2$, this is a famous theorem of Kontsevich \cite{Kontsevich}. For general $r$, it was established via a less direct path. Adler and van Moerbeke proved in \cite{Adler} the existence of a unique tau function $Z_{r{\rm tau}}$ of the $r$-KdV hierarchy solving the string equation (it is unique up to a constant prefactor, which can be set equal to $1$ and will not be mentioned anymore). Givental showed in \cite{Giventalr} that the total descendant potential of the $A_{r - 1}$-singularity has the same property, and therefore coincides with $Z_{r{\rm tau}}$. Later, Faber, Shadrin and Zvonkine established in \cite{Faber} that $Z_{r{\rm spin}}$ is equal to the total descendant potential of the $A_{r - 1}$-singularity, as application of their proof that the Givental group preserves the notion of cohomological field theories in all genera. Therefore $Z_{r{\rm spin}} = Z_{r{\rm tau}}$.

\subsubsection*{(iii) - Formal matrix integrals}

In \cite{Adler} it is proved that $Z_{r{\rm tau}}$ admits a matrix model representation as follows. To be more precise, let $\mathcal{H}_{N}$ be the space of hermitian matrices of size $N$, and $Y \in \mathcal{H}_{N}$. We first introduce the formal matrix integral
\beq
\label{Rzrty}Z_{r{\rm tau}}^{(N)} = \frac{\int_{\mathcal{H}_{N}^{{\rm formal}}} \dd M\,e^{\hbar^{-1/2}{\rm Tr}[-V(Y + M) + V(M) + YV'(M)]}}{\int_{\mathcal{H}_{N}}^{{\rm formal}} \dd M\,e^{-\hbar^{-1/2}\,{\rm Tr}\,V_2(Y,M)}}\,,
\eeq
where
$$
V(M) = \frac{\ii \sqrt{r}}{r + 1}\,M^{r + 1},\qquad V_2(Y,M) = \frac{\ii\sqrt{r}}{2}\,\sum_{m = 0}^{r - 1} Y^{m}MY^{r - 1 - m}M\,.
$$
It is possible to define the $N \rightarrow \infty$ limit of \eqref{Rzrty} as a formal series in the variables
$$
x_{k} = \frac{\hbar^{1/2}}{k}\,{\rm Tr}\,Y^{-k}\, ,\qquad k > 0\,,
$$
which takes the form
$$
Z_{r{\rm tau}} = \exp\bigg(\sum_{2g - 2 + n > 0} \frac{\hbar^{g - 1}}{n!} \sum_{k_1,\ldots,k_n \geq 0} F_{g,n}^{r{\rm tau}}(k_1,\ldots,k_n) \prod_{l = 1}^n x_{k_l}\bigg)\,.
$$

\subsubsection*{(iv) - \texorpdfstring{$\mathcal{W}$}{W} constraints}

Another side result of \cite{Adler} says that $Z_{r{\rm tau}}$ satisfies $\mathcal{W}(\mathfrak{gl}_{r})$ constraints determining it uniquely. As a matter of fact, these constraints coincide with the differential operators of Theorem \ref{WHAS} with $s=r+1$, and $Z_{r{\rm tau}} = Z_{(r,r + 1)}$ the partition function of this quantum $r$-Airy structure.

\subsubsection*{(v) - Periods}

The $\mathcal{W}$ constraints for the total descendant potential of the $A_{r - 1}$ singularity were expressed in terms of period computations by Bakalov and Milanov \cite{BakalovMilanov1}. Milanov \cite{Milanov} later established their equivalence with the Bouchard--Eynard topological recursion on a certain higher genus curve (not the $r$-Airy spectral curve $y=-z$, $x= \frac{z^r}{r}$, but one that locally looks like it near the branch points). By different methods, using the theory of Dubrovin superpotentials, \cite{DNOPS} later proved the equivalence with the Bouchard--Eynard topological recursion on the $r$-Airy spectral curve.  In our context, this theorem of \cite{DNOPS} is equivalent to the identification of the higher quantum Airy structure of Theorem \ref{WHAS} with $s=r+1$ with the Bouchard--Eynard topological recursion as in Theorem \ref{1cccc}, for the particular case of the $r$-Airy spectral curve $y=-z$, $x=\frac{z^r}{r}$.

\subsection{Br\'ezin--Gross--Witten theory}
\label{s:BGW}

Consider the formal matrix integral introduced in \cite{BGROSS,GWitten}
\beq
\label{detqY} Z_{r{\rm BGW}}^{(N)} = \frac{\int_{\mathcal{H}_{N}}^{{\rm formal}} \dd M\,(\det M)^{-N}\,e^{\hbar^{-1/2}\,{\rm Tr}[YM + \frac{M^{1 - r}}{r- 1}]}}{{\rm Det}^{-1/2}(-Q_Y/2\pi\hbar^{1/2})\,(\det Y)^{N/r}\,e^{\hbar^{-1/2}{\rm Tr}[\frac{r}{r - 1}\,Y^{1 - 1/r}]}}\,,
\eeq
where $Q_Y$ is the Hessian of $M \mapsto {\rm Tr}\,\frac{M^{r - 1}}{r - 1}$ at the point $Y^{-1/r}$, seen as an endomorphism in $\mathcal{H}_{N}$. It is possible to define the large $N$ limit of \eqref{detqY}  as a formal series $Z_{r{\rm BGW}}$ in the times $x_{k} =  \frac{\hbar^{1/2}}{k}\,{\rm Tr}\,Y^{-k/r}$, which takes the form
$$
Z_{r{\rm BGW}} = \exp\left(\sum_{2g - 2 + n > 0} \frac{\hbar^{g - 1}}{n!} \sum_{k_1,\ldots,k_n \geq 0} F_{g,n}^{r{\rm BGW}}(k_1,\ldots,k_n) \prod_{l = 1}^n x_{k_l}\right)\,.
$$
The work of \cite{BGtau} proves that $Z_{r{\rm BGW}}$ is a tau function of the KdV hierarchy with respect to the times $x_{k} = \frac{1}{k}\,{\rm Tr}\,Y^{-k/r}$.

If we focus on $Z_{2{\rm BGW}}$, \cite{BGtau} proves that it satisfies Virasoro constraints (see also \cite{AlexandrovBGW}). These constraints are equivalent to the statement that
$$
\omega_{g,n}^{2{\rm BGW}}(z_1,\ldots,z_n) = \sum_{k_1,\ldots,k_n \geq 0} F_{g,n}^{2{\rm BGW}}(k_1,\ldots,k_n) \prod_{l = 1}^n \frac{(2k_{l} + 1)!!\,\dd z_{l}}{z_{l}^{2k_{l} + 2}}
$$
is computed by the Chekhov--Eynard--Orantin topological recursion for the Bessel spectral curve \cite{NorburyDo}
$$
x(z) = \frac{z^2}{2},\qquad y(z) = -\frac{1}{z},\qquad \omega_{0,2}(z_1,z_2) = \frac{\dd z_1\dd z_2}{(z_1 - z_2)^2}\,.
$$
This is also equivalent \cite{ABCD} to saying that $Z_{2{\rm BGW}}$ is the partition function of a quantum Airy structure. Here it corresponds to the $\mathcal{W}(\mathfrak{sl}_{2})$ quantum Airy structure associated with the dilaton shift $x_{1} \rightarrow x_{1} - 1$, \textit{i.e.} after reduction of \eqref{QHAS2W} to $x_{2m} = 0$ for $m > 0$ (see Section \ref{sec:reduction}). In the notation of Theorem~\ref{THT1} this means $Z_{2{\rm BGW}} = Z_{(2,1)}$.

Norbury \cite{Norbclass} constructed a cohomology class $\Theta_{g,n} \in H^{4g - 4 + 2n}(\overline{\mathcal{M}}_{g,n})$, and proved  that
\beq
\label{Bgw}F_{g,n}^{2{\rm BGW}}(k_1,\ldots,k_n) = \int_{\overline{\mathcal{M}}_{g,n}} \Theta_{g,n} \prod_{l = 1}^n \psi_{l}^{k_l}\, .
\eeq
The construction of the cohomology class (which is similar to a construction of Chiodo \cite{Chiodo2}) starts on the moduli space of spin structures $\overline{\mathcal{M}}_{g,n}^{(2)} = \overline{\mathcal{M}}_{g,n}(2,(0,\ldots,0))$. One constructs a vector bundle $E_{g,n}$ over $\overline{\mathcal{M}}_{g,n}^{(2)}$ whose fiber at a smooth point is $H^1(L^{\vee})^{\vee}$ where $\vee$ indicates the dual. Concretely, one looks at the bundle of the universal spin structure $\mathcal{E}$ on the universal curve $\pi\,:\,\mathcal{C} \rightarrow \overline{\mathcal{M}}_{g,n}^{(2)}$, and defines $E_{g,n} = R^1\pi_*\mathcal{E}^\vee_{g,n}$. This $ E_{g,n} $ is a vector bundle of rank $2g - 2 + n$ and one can consider its Euler class and push it forward through the forgetful map  $p\,:\,\overline{\mathcal{M}}_{g,n}^{(2)} \longrightarrow \overline{\mathcal{M}}_{g,n}$. This is up to a normalization factor the desired class
$$
\Theta_{g,n} = (-2)^{n}\,p_*c_{2g - 2 + n}(E_{g,n})\, ,
$$
where $c_{i}$ is the $i$-th Chern class.

Therefore, we have a description of $Z_{2{\rm BGW}}$ from all five points of view. It is a natural question to  ask for such a complete understanding for the higher quantum Airy structure based on the $\mathcal{W}(\mathfrak{gl}_{r})$-module obtained by twisting by a Coxeter element and the dilaton shift $x_{s} \rightarrow x_{s} - \frac{1}{s}$ for $s \in \{1,\ldots,r - 1\}$ coprime with $r$ and such that $r = \pm 1 \mod s$. We know that its partition function is computed by the topological recursion for the $(r,s)$-spectral curve
$$
x(z) = \frac{z^r}{r},\qquad y(z) = -\frac{1}{z^{r - s}},\qquad \omega_{0,2}(z_1,z_2) = \frac{\dd z_1\dd z_2}{(z_1 - z_2)^2}\, .
$$

\begin{ques}
\label{QN2} Is there a matrix model description of the partition function for the $(r,s)$-spectral curve? Can one find a $\Theta_{g,n}^{(r,s)} \in H^{\bullet}(\overline{\mathcal{M}}_{g,n})$ whose intersection with $\psi$-classes is encoded by the $F_{g,n}$?
\end{ques}

In fact, $Z_{r{\rm BGW}}$ is a natural candidate for a matrix integral/tau function representation for the partition function of the $(r,s)$-spectral curve, for either $s=1$ or $s=r-1$. It is indeed known that $Z_{r{\rm BGW}}$ satisfies a set of $\mathcal{W}$ constraints. However they are not easy to write down explicitly, and should be compared to \eqref{QHAS2W} to complete the identification. 

\begin{ques}
\label{QN1} Is $Z_{r{\rm BGW}}$ the matrix model description of the partition function for the $(r,1)$ or $(r,r-1)$ spectral curve? 
\end{ques}

Preliminary work in this direction, in the case of $r=3$, was completed by Robert Maher in his MSc thesis; it seems to indicate that $Z_{r{\rm BGW}}$ may correspond to the case $s=r-1$, but the calculations were not conclusive  \cite{Maher}.

\subsection{Open intersection theory}
\label{Sopen}

Open intersection theory studies the enumerative geometry of bordered Riemann surfaces with marked points on the boundaries and in the interior, possibly carrying $r$-spin structures. Pandharipande, Solomon and Tessler first proposed in \cite{Tessleropen} an appropriate construction of this moduli space and the associated numerical invariants in genus $0$ for $r = 2$ (\textit{i.e.} in the absence of spin structures). The extension of this construction to all genera for $r = 2$ was announced in \cite{SolomonTessler}, and some details of this already appeared in \cite[Section 2]{Tesslercomb}. The case $r \geq 2$ for genus $0$ was settled by \cite{Tessleropen2} together with conjectures about the integrability property of the (yet not constructed) partition function for all genera. We refer to those articles for precise statements about the state of the art, and will henceforth assume that all the necessary constructions exist.

Let us first focus on $r = 2$. One considers, for $2g - 2 + m + 2n > 0$, the moduli space $\mathcal{M}_{g,n,m}$ of bordered Riemann surfaces with $m$ marked points on the boundary, $n$ marked points in the interior, such that the genus of the doubled surface is $g$. This moduli space is a real orbifold of dimension $3g - 3 + m + 2n$, which admits a compactification $\overline{\mathcal{M}}_{g,n,m}$. There exist cotangent line bundles $\mathbb{L}_{l}$ at the interior punctures $p_l$ for which relative orientations and boundary conditions can be constructed. Therefore they admit relative Euler classes and one can define a partition function for open intersection numbers as follows
\beq
\label{Zopenm} Z_{{\rm open}} = Z_{{\rm closed}}\cdot \exp\left\{\sum_{2g - 2 + m + 2n > 0} \frac{\hbar^{(g - 1)/2}}{m!n!} \sum_{k_1,\ldots,k_n \geq 0} \bigg(\int_{\overline{\mathcal{M}}_{g,n,m}} e\Big(\bigoplus_{l = 1}^n \mathbb{L}_l^{k_l}\Big)\bigg)\,\prod_{l = 1}^n t_{k_l}^{{\rm B}}\,(s_0^{{\rm B}})^{m}\right\}\, ,
\eeq
where $Z_{{\rm closed}} = Z_{2{\rm spin}}$ is \eqref{Z2spin}. It was conjectured in \cite{Tessleropen} that $Z_{{\rm open}}$ is a tau function of the open KdV hierarchy. Buryak and Tessler proved this conjecture \cite{BuryakTessler} based on a combinatorial model for the moduli space of bordered Riemann surface developed in \cite{Tesslercomb}, while \cite{Buryakopen1} showed the equivalence between the open KdV equations with suitable initial data and a set of open Virasoro constraints (previously conjectured by \cite{Tessleropen}). In fact, $Z_{{\rm open}}$ is the specialization at $s_{k}^{{\rm B}} = 0$ for $k > 0$ of a partition function  $Z_{{\rm open}}^{{\rm ext}}$ depending on all the variables $(t_{k}^{{\rm B}},s_{k}^{{\rm B}})_{k > 0}$ and which is a tau function of a larger integrable hierarchy \cite{Buryakopen2}, later identified \cite{Alexandrovopen2} with the modified KdV hierarchy. According to \cite{SolomonTessler}, the variables $s_{k}^{{\rm B}}$ for $k \geq 2$ admit an enumerative interpretation as indexing insertions of boundary descendant classes, paralleling the fact that $t_{k}^{{\rm B}}$ are coupled to insertions of $\psi^{k}$.

Alexandrov studied the formal hermitian matrix integral, called the Kontsevich-Penner matrix model
\begin{equation}\label{eq:KPMM}
Z_{{\rm open},(N)}^{{\rm ext}}(q) = \frac{(\det Y)^{q}\,\int_{\mathcal{H}_{N}}^{{\rm formal}} \dd M\,(\det M)^{-q}\,e^{\hbar^{-1/2}\,{\rm Tr}[-\frac{M^3}{6} + \frac{Y^2M}{2}]}}{\int \dd M\,e^{\hbar^{-1/2}\,{\rm Tr}[-\frac{YM^2}{2} + {\rm Tr} \frac{Y^3}{3}]}}\,,
\end{equation}

It is possible to define the $N \rightarrow \infty$ limit of \eqref{eq:KPMM} as a formal series in $t_{k} = \frac{\hbar^{1/2}}{k}\,{\rm Tr}\,Y^{-k}$ for $k > 0$ which has the form
$$
Z_{{\rm open}}^{{\rm ext}}(q) = \exp\bigg(\sum_{\substack{2h - 2 + n > 0 \\ h \in \mathbb{N}/2}} \frac{\hbar^{h - 1}}{n!} \sum_{k_1,\ldots,k_n \geq 0} F_{h,n}^{{\rm open},{\rm ext}}(q;k_1,\ldots,k_n) \prod_{l = 1}^n t_{k_l}\bigg)\,.
$$
He proves in \cite{Alexandrovopen2} that for $q = 1$ it is a tau function of the modified KdV hierarchy \cite{MKP} which coincides with the open extended partition function of Buryak with identification
$$
t_k^{{\rm B}} = (2k + 1)!!\,t_{2k + 1}\,,\qquad s_{k}^{{\rm B}} = 2^{k + 1}(k + 1)!\,t_{2k + 2}\,.
$$
For general $q$ Alexandrov derives in \cite{Alexandrovopen} a set of $\mathcal{W}(\mathfrak{sl}_3)$ constraints annihilating $Z_{{\rm open}}^{{\rm ext}}(q)$. Safnuk gave in \cite{Safnukopen} an equivalent description in terms of periods on the spectral curve $x = \frac{z^2}{2}$, $y=-z$, which is an unusual modification of the Bouchard--Eynard topological recursion \cite{BE2} that involves half-integer genera. Our work in fact reproduces Alexandrov's constraints.
\begin{prop}
$Z_{{\rm open}}^{{\rm ext}}(q)$ is the partition function of the higher Airy structure of Proposition~\ref{t:ropen} with $r = 3$ and $s = 3$ considered as a function of $t_{2k} = \frac{1}{2} x^1_{2k} - x^2_k $  and $ t_{2k+1} = x^1_{2k+1} $ for $k > 0$ with the identification $J_0^1 = \hbar^{1/2}q$.
\end{prop}
\begin{proof} To make the comparison, we need to perform a reduction from  $ \mathcal{W}(\mathfrak{gl}_3)$ to $\mathcal{W}(\mathfrak{sl}_{3})$. In order to do this, we define a set of independent (commuting) Heisenberg fields:
\begin{equation*}
\begin{split}
J^+_k &:= J^1_{2k} + J^2_k ,\\
J^-_k &:= \sqrt{\frac{2}{3}} \left( \frac{1}{2} J^1_{2k} - J^2_k\right)  \qquad \forall k \in \mathbb Z
\end{split}
\end{equation*}
The normalization factor $\frac{1}{2}$ ensures that $ J^+ $ and $ J^-  $ commute, while the factor $ \sqrt{\frac{2}{3}} $ ensures that $ [J^-_k,J^-_l] = k \delta_{l+k,0} $. Here we will also  use the notation
	\[
	J_{2k+1} := J^1_{2k+1}.
	\]
	Now, we will rewrite the operators of \eqref{eq:dshiftr1} in the modes$  \{J^+_k, J^-_k, J^1_{2k+1}\} $, and then formally set $ W_k^{1} =  J^+_k$ equal to $0$ for $k \geq 0$. Then we get the following operators:
\begin{equation*}
\begin{split}
	H^2_k &= J_{2k+1}- \sum_{\substack{a+b = k-1 }}  \left( J^{-}_{a}J^{-}_{b} + \frac{1}{2} J_{2a-1}J_{2b+1} \right) - \frac{\hbar}{8} \delta_{k,1}\,, \\
	-\sqrt{\frac{3}{2}}H^3_k &= J^-_{k+1} - \sum_{b+c=k-1} J_{2b+1}J^{-}_{c}  + \sum_{a+b+c = k-2} \left(   J_{2a-1}J_{2b+1}J^{-}_{c} - \frac{2}{3} J^-_{a}J^-_{b}J^-_{c}  \right) + \frac{\hbar}{4} J^{-}_{k-2}\,. 
\end{split}
\end{equation*}
	
With the identifications
$$
J^-_{k} = \left\{\begin{array}{lll} \hbar \sqrt{\frac{3}{2}}\partial_{ t_{2k}} & & k > 0 \\ \hbar^{1/2}\sqrt{\frac{3}{2}} q & & k = 0 \\ -\sqrt{\frac{2}{3}}  k t_{-2k} & & k < 0 \end{array}\right.\,,\qquad J_{2k+1} = \left\{\begin{array}{lll} \hbar \partial_{ t_{2k+1}} & & k \geq 0  \\ (1 -2k)t_{1 - 2k} & & k < 0 \end{array}\right. \,,
$$
and $N =q$ we recognize $ H^2_k = -2 \widehat{\mathcal{L}}^{N}_k $ and $ H^3_k = -4\, \widehat{\mathcal{M}}^{N}_k $ in the notation of \cite{Alexandrovopen}. Our uniqueness result for the partition function ~\eqref{gofdugn1111} gives the statement of the Proposition.
\end{proof}

It is natural to speculate about higher $r$.
\begin{ques}
Consider the partition function associated with the higher quantum Airy structure based on $\mathcal{W}(\mathfrak{gl}_{r})$ with automorphism $\sigma = (1\,\,\cdots\,\,r - 1)$ and dilaton shift $x_{r} \rightarrow x_{r} - \frac{1}{r}$ (see Proposition~\ref{t:ropen} with $s=r$).
\begin{itemize}
\item[$\bullet$] for $q = 1$, does it coincide with the tau function of the extended open $(r - 1)$-KdV hierarchy constructed by Bertola and Yang \cite{BertolaYang}?
\item[$\bullet$] for $q = 1$, can it be expressed in terms of the generating function of (extended) open $(r - 1)$-spin intersection theory as constructed (in genus 0) by \cite{Tessleropen2}?
\item[$\bullet$] for arbitrary $q$, does it have a formal matrix integral representation generalizing \eqref{Zopenm}?
\end{itemize}
\end{ques}
The two last questions can also be asked for the dilaton shifts $x_{s} \rightarrow x_{s} - \frac{1}{s}$ with $s \in \{1,\ldots,r-1\}$ such that $s|r$. In particular, for $s = 1$ this should give an open $r$-spin generalization of Norbury's class.

\subsection{Fan--Jarvis--Ruan theories}
\label{s:FJRW}

Let $\mathbf{W} \in \mathbb{C}[x_1,\ldots,x_{t}]$ be a quasi-homogeneous polynomial, \textit{i.e.}  there exist positive integers $d,\nu_1,\ldots,\nu_t$ for which
$$ 
\forall \lambda \in \mathbb{C}^*,\qquad \mathbf{W}(\lambda^{\nu_1}x_1,\ldots,\lambda^{\nu_t}x_t) = \lambda^{d} \mathbf{W}(x_1,\ldots,x_{t})\,.
$$
Assume that $\mathbf{W}$ is non-degenerate, \textit{i.e.} $\mathrm{W} = 0$ has an isolated singularity at $0$ and $\tilde{\nu}_i = \nu_i/d$ are uniquely determined by $\mathbf{W}$. Then one can always assume that $d,\nu_1,\ldots,\nu_t$ are minimal. This assumption implies that the group of diagonal symmetries of $\mathbf{W}$
$$
\Gamma = \big\{ \alpha \in (\mathbb{C}^*)^{t}\quad |\quad \mathbf{W}(\alpha_1x_1,\ldots,\alpha_tx_t) = \mathbf{W}(x_1,\ldots,x_t)\big\}
 $$
is finite. Let us decompose $ \mathbf{W} $ into monomials
$$
\mathbf{W}(x_1,\ldots,x_t) = \sum_{\mu} c_{\mu} \prod_{u = 1}^{t} x_{u}^{\mu_u}\,.
$$
Fan, Jarvis and Ruan \cite{FJR2} constructed a compactified moduli stack of twisted spin curves, which describe isomorphism classes of orbifold curves $\mathcal{C}$ equipped with $n$ punctures $p_1,\ldots,p_n$ and line bundles $L_1,\ldots,L_t$ together with isomorphisms
$$
\phi_{\mu}\,:\,\bigotimes_{u = 1}^{t} L_u^{\otimes \mu_u} \longrightarrow K_{\mathcal{C}}\big(\sum_{i = 1}^n p_i\big)\,.
$$
They also describe a virtual fundamental class on this moduli stack. After pushing forward to $\overline{\mathcal{M}}_{g,n}$, it yields a cohomological field theory whose Frobenius algebra is given by the Jacobi ring
$$ 
{\rm Jac}(\mathrm{W}) = \frac{\mathbb{C}[x_1,\ldots,x_t]}{\big\langle \partial_{1}\mathbf{W},\ldots,\partial_{t}\mathbf{W}\big\rangle}\,.
$$
Polishchuk and Vaintrob gave another, more algebraic construction of this cohomological field theory from the category of matrix factorizations of $\mathbf{W}$ \cite{PV1,PV2}. We denote $Z^{\mathbf{W}}$ the generating series of its intersection numbers with the $\psi$-classes.
   
The most fundamental examples of such hypersurfaces $\{\mathbf{W} = 0\}$ are given by the simple singularities, of type ADE. Their total descendant potential are tau functions of an integrable hierarchy \cite{MilanovGiv,FGM}, and satisfy $\mathcal{W}$ constraints \cite{BakalovMilanov2}. These constraints coincide with the $\mathcal{W}(\mathfrak{g})$ quantum Airy structure associated with the dilaton shift $x_{r + 1} \rightarrow x_{r + 1} - \frac{1}{r+1}$ described in Section~\ref{HASW}. The uniqueness of their solution was proved in \cite{Yangunique}. The $r$-spin partition function of Section~\ref{Srspin} corresponds to the $A_{r - 1}$ case. The total descendant potential of the ADE singularity $\{\mathbf{W}_{{\rm ADE}} = 0\}$ in fact coincides with the Fan--Jarvis--Ruan partition function $Z^{\mathbf{W}_{{\rm ADE}}}$ \cite{FJRid}.

\begin{ques}
For any non degenerate quasihomogeneous polynomial in $n$ variables $\mathbf{W}$, can one write down a higher quantum Airy structure (based on $\mathcal{W}$ algebras) associated with $\mathbf{W}$ such that its partition function encodes the correlators of the cohomological field theory constructed by Fan--Jarvis--Ruan?
\end{ques}

\begin{ques}
Do the partition functions of Theorem~\ref{t:DN} for $D_N$-type and Theorem~\ref{thEEE} for $E_N$-type in the case $s = 1$ admit an interpretation in terms of Fan--Jarvis--Ruan theories?
\end{ques}

\newpage

\appendix

\section{Sums over roots of unity}
\label{Approot}

We encountered in Definition~\ref{psirootsum} the following sums for $i \in \{1,\ldots,r\}$
\beq
\label{defpsiss}\Psi(a_1,\ldots,a_i) = \frac{1}{i!} \sum_{\substack{m_1,\ldots,m_i = 0 \\ m_{l} \neq m_{l'}}}^{r - 1} \prod_{l = 1}^{i} \theta^{-m_{l}a_{l}}\,,
\eeq
where $\theta = e^{2\ii\pi/r}$ is a primitive $r$-th root of unity. This is the $j = 0$ case of the more general sum for $j \in \{0,\ldots,\lfloor i/2 \rfloor\}$
\beq
\label{Psildef}\Psi^{(j)} (a_{2j+1},\ldots, a_i)  = \frac{1}{i!} \sum_{\substack{m_1, \ldots , m_{i}=0 \\ m_{l} \neq m_{l'}}}^{r-1} \left( \prod_{l'=1}^{\ell}\frac{\theta^{m_{2l' - 1}+m_{2l'}}}{(\theta^{m_{2l'}} - \theta^{m_{2l' - 1}})^2}   \prod_{l=2j+1}^{i}\theta^{-m_{l} a_{l}}\right)\,.
\eeq
This appendix is devoted to the proof of several properties of these functions used in Section~\ref{HASW} and an explicit computation of $\Psi^{(0)}$.

\begin{lem}\label{Ajsum} For any $j \in \{0,\ldots,\lfloor i/2 \rfloor\}$, we have
\begin{equation}
\label{eq:psisum}
	\sum_{a_1, \ldots, a_{2j} = 0}^{r - 1} \Psi(a_1,\ldots, a_i)  \prod_{l'=1}^j \frac{a_{2l'-1}a_{2l'}\,\delta_{a_{2l' - 1} + a_{2l'},r}}{2r}  = \Psi^{(j)} (a_{2\ell+1},\ldots, a_i)  \,.
\end{equation}
\end{lem}
\begin{proof}
The left hand side of \eqref{eq:psisum} is equal to
\beq
\label{gfgoun}\frac{1}{i!} \sum_{\substack{m_1,\ldots,m_{i} = 0 \\ m_{l} \neq m_{l'}}}^{r - 1} \,\,\,\sum_{a_1,a_3,\ldots,a_{2j - 1} = 0}^{r - 1} \prod_{l' = 1}^{j} \theta^{(m_{2l'}-m_{2l' - 1})a_{2l' - 1}} \prod_{l = 2j + 1}^{i} \theta^{-m_{l}a_{l}}\,.
\eeq
We compute the sum
$$
\sum_{a=0}^{r-1} \frac{a(r-a)x^a}{2r} = \frac{x((r - 1)x - (r + 1))x^{r} + (r + 1)x - (r - 1)}{2r(x - 1)^3}\,.
$$
Setting $ x = \theta^{m_1-m_2}$ for distinct $m_1,m_2 \in \{0,\ldots,r - 1\}$ gives 
$$ 
\sum_{a = 0}^{r - 1}  \frac{a(r-a)\theta^{a(m_1-m_2)}}{2r} = \frac{\theta^{m_1 - m_2}}{(\theta^{m_1 - m_2} - 1)^2} = \frac{\theta^{m_1 + m_2}}{(\theta^{m_1} - \theta^{m_2})^2}\,.
$$ 
Using this formula to perform the sum over $a_1,a_3,\ldots,a_{2j - 1} = 0$ in \eqref{gfgoun} entails the claim.
\end{proof} 

We can get rid of zero entries in $\Psi^{(j)}$ in a simple way.

\begin{lem}
\label{lem:Psizero}
$$
\Psi^{(j)}(a_{2j+ 1},\ldots,a_{i - \ell},\underbrace{0,\ldots,0}_{\ell\,\,{\rm times}}) = \frac{(i - \ell)!}{i!}\,\frac{(r - i + \ell)!}{(r - i)!}\,\Psi^{(j)}(a_{2j + 1},\ldots,a_{i - \ell})\,.
$$
\end{lem}
\begin{proof}
In the sum \eqref{Psildef} defining $\Psi^{(j)}(a_{2j + 1},\ldots,a_{i - \ell},0,\ldots,0)$, the terms only depend on
$$
\mathbf{m} = (m_{1},\ldots,m_{i - \ell})
$$
and the $a$s. Therefore, we can perform the sum over the ordered $j$-tuple $(m_{l})_{l = i - \ell + 1}^{i}$ of pairwise disjoint integers in $\{0,\ldots,r - 1\}\setminus \{m_1,\ldots,m_{i - \ell}\}$ and get a global factor of $\frac{(r - i + \ell)!}{(r - i)!}$. We then recognize $\Psi^{(j)}(a_{2j + 1},\ldots,a_{i - \ell})$ up to another global factor $\frac{(i - \ell)!}{i!}$.
\end{proof}

\begin{lem}
\label{Psicom}We have $i\,\Psi(\underbrace{r - 1,\ldots,r - 1}_{i - 1\,\,{\rm times}},i - 1) = (-1)^{i - 1}r$.
\end{lem}
\begin{proof}
Let us denote for $m \in \{0,\ldots,r - 1\}$
$$
\vec{\theta} = (1,\theta,\theta^2,\ldots,\theta^{r - 1}),\qquad \vec{\theta}[m] = \vec{\theta} \setminus \theta^{m}\,.
$$
Coming back to the definition of $\Psi = \Psi^{(0)}$, we can write
\beq
\label{psieqf} i\,\Psi\big(\underbrace{r - 1,\ldots,r - 1}_{i - 1\,\,{\rm times}},i - 1\big) =\sum_{m = 0}^{r - 1} \theta^{-m(i - 1)}\bigg( \sum_{\substack{L \subseteq \{0,\ldots,r - 1\} \setminus \{m\} \\ |L| = i - 1}} \prod_{l \in L} \theta^{-l(r - 1)}\bigg) = \frac{1}{i} \sum_{m = 0}^{r - 1} \theta^{-m(i - 1)} e_{i - 1}\big(\vec{\theta}[m]\big)\,,
\eeq
where $e_{j}$ is the $j$-th elementary symmetric polynomial. Since $-\theta$ are the simple roots of the polynomial $1 + (-1)^{r}t^{r}$ we get
$$
\sum_{k = 0}^{r - 1} e_{k}(\vec{\theta})t^{k} = \prod_{a = 0}^{r - 1} (1 + t\theta^{a}) = 1 + (-1)^{r}t^r\,,
$$
from which we deduce that $e_{k}(\vec{\theta}) = \delta_{k,0}$ for $k \in \{1,\ldots,r - 1\}$. On the other hand we have by inclusion-exclusion
$$
e_{k}(\vec{\theta}) = \theta^{m}e_{k - 1}\big(\vec{\theta}[m]\big) + e_{k}(\vec{\theta}[m])\,.
$$
So we deduce by induction that
$$
\forall k \in \{1,\ldots,r - 1\},\qquad e_{k}(\vec{\theta}[m]) = (-1)^{k}\theta^{mk}\,.
$$
Inserting this result in \eqref{psieqf} gives
$$
 i\,\Psi\big(\underbrace{r - 1,\ldots,r - 1}_{i - 1\,\,{\rm times}},i - 1\big) = \sum_{m = 0}^{r - 1} (-1)^{i - 1} \theta^{-m(i - 1)} \theta^{m(i - 1)} = (-1)^{i - 1}r\,.
$$
\end{proof}

\begin{lem}
\label{Psicom2}More generally for any $s > 0$, let us introduce $d = {\rm gcd}(r,s)$ and $r' := r/d$. We have
$$ 
i\,\Psi\big(\underbrace{-s,\ldots,-s}_{i - 1\,\,{\rm times}},(i - 1)s\big) = r\,(-1)^{i - 1 + \lfloor \frac{i - 1}{r'} \rfloor} \binom{d - 1}{\lfloor \frac{i - 1}{r'} \rfloor}
$$
\end{lem}
\begin{proof}
The strategy is similar. We have
$$
\psi_{i,s} := i\,\Psi\big(\underbrace{-s,\ldots,-s}_{i - 1\,\,{\rm times}},(i - 1)s\big) = \sum_{m = 0}^{r - 1} \theta^{-m(i - 1)s}\bigg( \sum_{\substack{L \subseteq \{0,\ldots,r - 1\}\setminus \{m\} \\ |L| = i - 1}} \prod_{l \in L} \theta^{ls}\bigg) = \sum_{m = 0}^{r - 1} \theta^{-m(i - 1)s} e_{i - 1}(\vec{\theta}^{s}[m])\,,
$$
where $\vec{\theta}^s = (1^{s},\theta^{s},\ldots,\theta^{s(r - 1)})$ and $\vec{\theta}^{s}[m]$ is the sequence $\vec{\theta}^{s}$ with $\theta^{ms}$ omitted. We write by inclusion-exclusion
$$
e_{i}(\vec{\theta}^s) = \theta^{ms}e_{i - 1}(\vec{\theta}^{s}[m]) + e_{i}(\vec{\theta}^s[m])\,,
$$
with the convention $e_{-1} = 0$. We multiply this identity by $\theta^{-ims}$ and sum over $m \in \{0,\ldots,r - 1\}$ to find
\beq
\label{recun}r\delta_{r|is}\,e_{i}(\vec{\theta}^s) = \psi_{i,s} + \psi_{i + 1,s}\,,
\eeq
where $\delta_{a|b}$ is equal to $1$ is $a|b$ and to $0$ otherwise. With the value $\psi_{1,s} = r$ we will obtain by induction on $i$ a formula for $\psi_{i,s}$, provided we can compute $e_i(\vec{\theta}^s)$. For this purpose we observe that $\vec{\theta}^s$ contains each $r' := r/d$-th root of unity, with multiplicity $d$. Thus
$$
\sum_{i = 0}^{r} e_i(\vec{\theta}^s)\,t^i = (1 - (-t)^{r'})^{d} = \sum_{j = 0}^{d} (-1)^{j(r' + 1)} \binom{d}{j}\,t^{jr'}\,,
$$
therefore
\beq
\label{rerigun}r\delta_{r|is}\,e_{i}(\vec{\theta}^s) = r\delta_{is|r}\delta_{r'|i}\,(-1)^{i(1 + 1/r')} \binom{d}{i/r'} = r \delta_{r'|i}\,(-1)^{i/r'}\,{\binom{d}{i/r'}}\,.
\eeq
Solving the recursion \eqref{recun} with \eqref{rerigun} as left-hand side yields
$$
i\,\Psi\big(\underbrace{-s,\ldots,-s}_{i - 1\,\,{\rm times}},(i - 1)s\big) = (-1)^{i - 1}r \sum_{j = 0}^{\lfloor (i - 1)/r' \rfloor} (-1)^{j} \binom{d}{j}\,.
$$ 
Recall that $d \geq 2$, and let us denote
$$
\psi(k,d) := \sum_{j = 0}^{k} (-1)^{j} \binom{d}{j},\qquad k \in \{0,\ldots,d - 1\}\,,
$$
so that
$$
i\,\Psi\big(\underbrace{-s,\ldots,-s}_{i - 1\,\,{\rm times}},(i - 1)s\big) = (-1)^{i - 1}r\,\psi_{(-1)^{r' + 1}}\big(\lfloor \tfrac{i - 1}{r'} \rfloor,d\big)\,.
$$
We observe that
\beq
\label{Obvious1}\psi(k,d) - \psi(k - 1,d) = (-1)^{k} \binom{d}{k}\,,
\eeq
which is also valid for $k = 0$ with the convention $\psi(-1,d) = 0$. But Pascal's identity
$$
\binom{d}{k} = \binom{d - 1}{k} + \binom{d - 1}{k - 1}\,,
$$
implies that
\beq
\label{Pascal2} \psi(k,d) = \psi(k,d - 1) - \psi(k - 1,d - 1)\,,
\eeq
with the convention that $\binom{d}{-1} = 0$. Combining  \eqref{Pascal2} and \eqref{Obvious1} we find
$$
\psi(k,d) =  (-1)^{k}\,\binom{d - 1}{k}\,.
$$

\end{proof}

Now we focus on the evaluation of these sums for small values of $i$.

\begin{lem}
\label{LEmg}For $r \geq 1$ we have $\Psi(a_1) =  r\delta_{r|a_1}$. For $r  \geq 2$ we have
\begin{equation*}
\begin{split}
\Psi(a_1,a_2) & =  \frac{1}{2}\big(r^2\delta_{r|a_1}\delta_{r|a_2} - r\delta_{r|a_1 + a_2}\big)\, ,  \\
\Psi^{(1)}(\emptyset) & =  -\frac{r(r^2 - 1)}{24}\,.
\end{split}
\end{equation*}
For $r \geq 3$ we have
\begin{equation*}
\begin{split}
\Psi(a_1,a_2,a_3) & = \frac{1}{6}\big(\,r^3\delta_{r|a_1}\delta_{r|a_2}\delta_{r|a_3} - r^2\delta_{r|a_1}\delta_{r|a_2 + a_3}  - r^2\delta_{r|a_2}\delta_{r|a_1 + a_3}  \\
& \quad \,\,\,\,\,\,\,\,\,\, -r^2\delta_{r|a_3}\delta_{r|a_1 + a_2} + 2r\delta_{r|a_1 + a_2 + a_3}\big) \\
\Psi^{(1)}(a_3) & = -\frac{r(r - 2)(r^2 - 1)}{72}\,\delta_{r|a_3}\,.
\end{split}
\end{equation*}
\end{lem}
\begin{proof}
We have for $i = 1$ and $j = 0$
\beq
\label{gofigfg}\Psi(a_1) = \sum_{m = 0}^{r - 1} \theta^{-ma_1} = r\delta_{r|a_1}\,.
\eeq
For $i = 2$, $j$ can take the two values $0$ or $1$. We have
\begin{equation*}
\begin{split}
\Psi(a_1,a_2) & =  \frac{1}{2} \sum_{\substack{m_1,m_2 = 0 \\ m_1 \neq m_2}}^{r - 1} \theta^{-m_1a_1 - m_2a_2} = \frac{1}{2} \bigg(\sum_{m_1,m_2 = 0}^{r - 1} - \sum_{m_1 = m_2 = 0}^{r - 1}\bigg) \theta^{-m_1a_1 - m_2a_2}  \\
& =  \frac{1}{2}\big(r^2 \delta_{r|a_1}\delta_{r|a_2} - r\delta_{r|a_1 + a_2}\big)\,.
\end{split}
\end{equation*}
Using Lemma~\ref{Ajsum} and the formula we just proved, we compute
$$
\Psi^{(1)}(\emptyset) = \sum_{a_1,a_2 = 0}^{r - 1} \Psi(a_1,a_2)\,\frac{a_1a_2}{2r}\,\delta_{a_1 + a_2,r} = -\frac{1}{4} \sum_{a_1 = 0}^{r - 1} a_1(r - a_1) = -\frac{r(r^2 - 1)}{24}\,.
$$ 
For $i = 3$, we apply the same strategy
\begin{equation}
\label{Psi222}
\begin{split}
\Psi(a_1,a_2,a_3) & = \frac{1}{6} \sum_{n = 0}^{r - 1} \theta^{-na_1} \bigg(\sum_{\substack{m_2, m_3 = 0 \\ m_{2} \neq m_3 \\ m_{\alpha} \neq n}}^{r - 1} \theta^{-m_2a_2 - m_3a_3}\bigg)  \\
& =  \frac{1}{6} \sum_{n = 0}^{r - 1}\bigg( \theta^{-na_1} 2\Psi_2(a_2,a_3) -  \theta^{-n(a_1 + a_2)} \sum_{\substack{m_3 = 0 \\ m_3 \neq n}}^{r - 1} \theta^{-m_3a_3} - \theta^{-n(a_1 + a_3)} \sum_{\substack{m_2 = 0 \\ m_2 \neq n}}^{r - 1} \theta^{-m_2a_2}\bigg)  \\
& =  \frac{1}{6}\big(r\delta_{r|a_1}\,2\Psi_2(a_2,a_3) - r^2\delta_{r|a_1 + a_2}\delta_{r|a_3} - r^2\delta_{r|a_1 + a_3}\delta_{r|a_2} + 2r\delta_{r|a_1 + a_2 + a_3}\big)  \\
 & =  \frac{1}{6}\big(r^3\delta_{r|a_1}\delta_{r|a_2}\delta_{r|a_3} - r^2\delta_{r|a_1}\delta_{r|a_2 + a_3} - r^2\delta_{r|a_2}\delta_{r|a_1 + a_3}  \\
 & \quad \,\,\,\,\,\,\,\, -r^2\delta_{r|a_3}\delta_{r|a_1 + a_2} + 2r\delta_{r|a_1 + a_2 + a_3}\big)\,
\end{split}
\end{equation}
after using the result just found for $i = 2$. Using Lemma~\ref{Ajsum} we further compute
$$
\Psi^{(1)}(a_3) = \sum_{a = 0}^{r - 1} \frac{a(r - a)}{2r}\,\Psi(a,r - a,a_3) \,.
$$
The three first terms in \eqref{Psi222} do not give any contribution as they force the prefactor $a(r - a)$ to vanish. We obtain for the two last terms
$$
\Psi^{(1)}(a_3) = \frac{1}{6} \sum_{a = 0}^{r - 1} \frac{a(r - a)}{2r}\,(2r - r^2)\,\delta_{r|a_3} = -\frac{r(r - 2)(r^2 - 1)}{72}\,\delta_{r|a_3}\,.
$$
\end{proof}

We can give a general formula for $\Psi(a_1,\ldots,a_i)$, which involve the following notations. We denote $\mathbf{L} \vdash (a_1,\ldots,a_i)$ when $\mathbf{L}$ is an unordered set of $||\mathbf{L}||$ non-empty, pairwise disjoint subsequences of $(a_1,\ldots,a_n)$ whose concatenation is equal to $(a_1,\ldots,a_n)$. The length of a subsequence $L \in \mathbf{L}$ is denoted $|L|$. These notations agree with the ones used in Section~\ref{s:recursive}.
\begin{lem}
\label{Psiprinc}For general $i \in \{1,\ldots,r\}$ and $a_1,\ldots,a_r \in \mathbb{Z}$ we have the formula
$$
i!\,\Psi(a_1,\ldots,a_i) = \sum_{\mathbf{L} \vdash (a_1,\ldots,a_i)} r^{||\mathbf{L}||}\,(-1)^{i - ||{\mathbf{L}}||} \prod_{L \in \mathbf{L}} (|L| - 1)!\,\delta_{r|\sum_{l \in L} a_{l}}\,.
$$
In particular, $\Psi(a_1,\ldots,a_i) \in r\mathbb{Z}$.
\end{lem}

\begin{proof}
We continue with the strategy of the proof of Lemma~\ref{LEmg} and find by successive inclusion-exclusion
\begin{equation*}
\begin{split}
\Psi(a_1,\ldots,a_i) & =  \sum_{n = 0}^{r - 1} \sum_{L \subseteq \{2,\ldots,i\}} \theta^{-n(a_1 + \sum_{l \in L} a_l)}\,(-1)^{|L|}\,\frac{(i - |L| - 1)!|L|!}{i!}\,\Psi\big((a_{l'})_{\alpha \notin (L \cup \{1\})}\big) \ \\
& =  r \sum_{L \subseteq \{2,\ldots,i\}} (-1)^{|L|}\,\frac{(i - |L| - 1)!|L|!}{i!}\,\Psi\big((a_{l'})_{\alpha \notin (L \cup \{1\})}\big)\,\delta_{r|a_1 + \sum_{l \in L} a_l}\,,
\end{split}
\end{equation*}
where for $L = \{2,\ldots,n\}$ there appears $\Psi(\emptyset)$ which is by convention equal to $1$. This is a recursive formula for $i!\,\Psi(a_1,\ldots,a_i)$ on $i \in \{1,\ldots,r\}$, which is solved by the claimed formula.
\end{proof}

\begin{cor}
\label{psialleq} Let $i \in \{1,\ldots,r\}$ and $a \in \mathbb{Z}$ coprime with $r$. Then, for any $b \in \{0,\ldots,i - 1\}$
$$
\Psi\big(\underbrace{0,\ldots,0}_{b\,\,{\rm times}},\underbrace{a,\ldots,a}_{i - b\,\,{\rm times}}\big) = \left\{\begin{array}{ccc} \delta_{i,r}(-1)^{r - 1} & & {\rm if}\,\,b = 0 \\ 0 & & {\rm if}\,\,b \in \{1,\ldots,i - 1\} \\ 1 && {\rm if}\,\,b = i \end{array}\right.\,.
$$
\end{cor}
\begin{proof}
The case $b = i$ is obvious from the definition~\ref{defpsiss}. For $b = 0$, since $a$ is coprime with $r$, we gave $\delta_{r|\sum_{l \in L} a_l} = \delta_{r|\,|L|a} = \delta_{r|\,|L|}$. Therefore, the only non-zero contribution in the formula of Lemma~\ref{Psiprinc} occurs when $i = r$ and for the unique term which corresponds to $\mathbf{L}$ being the partition consisting of a single set, \textit{i.e.} $||\mathbf{L}|| = 1$. With $b \in \{1,\ldots,i - 1\}$ we first get rid of the zeroes thanks to Lemma~\ref{lem:Psizero}, and use the previous result to find that the expression evaluates to $0$.
\end{proof}

\begin{cor}
\label{COmg} For any $i \in \{1,\ldots,r\}$ and $j \in \{0,\ldots,\lfloor i/2 \rfloor\}$, $\Psi^{(j)}(a_{2j + 1},\ldots,a_i) \in \mathbb{Z}$ vanishes unless there exists $\mathbf{L} \vdash (a_{2j + 1},\ldots,a_i)$ such that the partial sums $\sum_{l \in L} a_{l}$ are divisible by $r$ for any $L \in \mathbf{L}$.
\end{cor}

\begin{proof}
For $j > 0$, we insert the formula of Lemma~\ref{Psiprinc} in Lemma~\ref{Ajsum}
$$
i!\,\Psi^{(j)}(a_{2j + 1},\ldots,a_{i}) = \sum_{a_1,\ldots,a_{2j} = 0}^{r - 1} \sum_{\mathbf{L} \vdash (a_1,\ldots,a_{i})} c_{\mathbf{L}} \,\delta_{r|\sum_{l \in L} a_{l}} \prod_{l' = 1}^{j} \frac{a_{2l' - 1}a_{2l'}\delta_{a_{2l' - 1} + a_{2l'},r}}{2r}\,,
$$
where
$$
c_{\mathbf{L}} = r^{||\mathbf{L}||} (-1)^{i - ||\mathbf{L}||}\,\prod_{L \in \mathbf{L}}  (|L| - 1)!\,.
$$
We first focus on the sum over the first ordered pair $(a_1,a_2)$ such that $a_1 + a_2 = r$, and meet two types of terms. If $a_1$ and $a_2$ are in the same subsequence $L$, we will have a contribution of the form
$$
\sum_{a = 0}^{r - 1} \frac{a(r - a)}{2r}\,\delta_{r|b} = \frac{(r^2 - 1)}{12}\,\delta_{r|b}\,,
$$
while if they are in two different subsequences, we rather have a contribution of the form
$$
\sum_{a = 0}^{r - 1} \frac{a(r - a)}{2r}\,\delta_{r|b_1 + a}\delta_{r|b_2 - a} = \frac{\langle b_1 \rangle \langle b_2 \rangle}{2r}\,\delta_{r|b_1 + b_2}\,,
$$
where $\langle b_1 \rangle$ is the unique integer in $\{0,\ldots,r - 1\}$ such that $b_1 - \langle b_1 \rangle \in r\mathbb{Z}$. Considering successively the sums over the other pairs $(a_{2l' - 1},a_{2l'})$, we observe a similar phenomenon and obtain the claim.
\end{proof}

For instance, we obtain for $i = 4$ and $r \geq 4$ the following formulas
\begin{equation*}
\begin{split}
 24\Psi(a_1,a_2,a_3,a_4) & =  r\Big(r^3\delta_{r|a_1}\delta_{r|a_2}\delta_{r|a_3}\delta_{r|a_4} - r^2(\delta_{r|a_1 + a_2}\delta_{r|a_3}\delta_{r|a_4} + \,\,\cdots) + r(\delta_{r|a_1 + a_2}\delta_{r|a_3 + a_4} +\,\, \cdots)  \\
\label{aaPsi1} & + 2r(\delta_{r|a_1 + a_2 + a_3}\delta_{r|a_4} + \,\,\cdots) - 6\,\delta_{r|a_1 + a_2 + a_3 + a_4}\Big)\,,
\end{split}
\end{equation*}
where the $\cdots$ indicate other terms necessary to enforce symmetry under permutation of $a_1,a_2,a_3,a_4$. Furthermore, exploiting the method sketched in the proof of Corollary~\ref{COmg}, we find
\begin{equation}
\label{aaPsi2}
\begin{split}
24\,\Psi^{(1)}(a_3,a_4) & =  -\,\,\frac{(r + 1)r^2(r - 1)(r - 4)}{12}\,\delta_{r|a_3}\delta_{r|a_4}  \\
& \quad +\,\, \bigg(\frac{(r + 1)r(r - 1)(r - 6)}{12} + r\langle a_3\rangle \langle a_4 \rangle\bigg)\delta_{r|a_3 + a_4}\,, \\
24\,\Psi^{(2)}(\emptyset) & = \frac{(r + 1)r(r - 1)(r - 2)(r - 3)(5r + 7)}{720}\,.
\end{split}
\end{equation}
It would be interesting to find a closed formula generalizing Lemma~\ref{Psiprinc} to all $j \in \{0,\ldots,\lfloor i/2 \rfloor\}$.

\section{Characterization of admissible \texorpdfstring{$(r,s)$}{(r,s)}}
\label{a:rs}

\subsection{The values of \texorpdfstring{$s$}{s} corresponding to the intermediate subalgebras}

In Section~\ref{Sinterm}, we showed that for any partition $\lambda_1 \geq \cdots \geq \lambda_p \geq 1$ such that $\sum_{j = 1}^p \lambda_j = r$, the left ideal generated by the modes $W_{k}^i$ of the $\mathcal{W}(\mathfrak{gl}_{r})$ algebra indexed by $(i,k) \in S_{\lambda}$ is a graded Lie subalgebra. The set $S_{\lambda}$ consists of the pairs $(i,k)$ with $i \in \{1,\ldots,r\}$ and 
\begin{equation}
\label{lklk} k \geq i - \lambda(i),\qquad \lambda(i) := \min\Big\{m > 0 \,\,\,\Big|\,\,\,\sum_{j = 1}^{m} \lambda_j \geq i\Big\}\, .
\end{equation}
Besides, for any $s \in \{1,\ldots,r + 1\}$ coprime with $r$, we showed in the proof of Theorem~\ref{QHAS2W} that the family $W_k^i$ indexed by $(i,k) \in \tilde{S}_{s}$ after dilaton shift $x_{s} \rightarrow x_{s} - \frac{1}{s}$ satisfies the degree $1$ condition of Definition~\ref{def:HAS}. The set $\tilde{S}_{s}$ consists of the pairs $(i,k)$ with $i \in \{1,\ldots,r\}$ and
\begin{equation}
\label{rkrk} r(k - i + 1) + (i - 1)s \geq 0\, .
\end{equation}
This section is devoted to a characterization of the values of $s$ for which \eqref{rkrk} can be equivalently described as \eqref{lklk}. For us this implies that the $W_{k}^i$ indexed by such $(i,k) \in \tilde{\mathcal{S}}_{s}$ form a higher quantum Airy structure (Theorem~\ref{QHAS2W}).
\begin{prop}
\label{p:partition}
Let $s \in \{1,\ldots,r + 1\}$ coprime with $r \geq 2$. There exists a partition $\lambda$ such that $\tilde{S}_{s} = S_{\lambda}$ if and only if $r = \pm 1\,\,{\rm mod}\,\,s$. In this case, we can decompose $r = r's + r''$ with $r'' \in \{1,s - 1\}$ and the partition is given by
$$
\lambda_1 = \cdots = \lambda_{r''} = r' + 1,\qquad \lambda_{r'' + 1} = \cdots = \lambda_{s} = r'
$$
if $s \neq r + 1$, and by $\lambda = (1,\ldots,1)$ if $s = r + 1$.
\end{prop}
\begin{proof}
Equation~\ref{rkrk} is equivalent to $k \geq i - 1 - \lfloor \frac{(i - 1)s}{r} \rfloor$ so we are asking for the characterization of $s$ appearing as
$$
\lambda(i) = 1 + \Big\lfloor \frac{(i - 1)s}{r} \Big\rfloor\, ,
$$
where $\lambda$ is a partition. In the case $s = r + 1$, we have
$$
\forall i \in \{1,\ldots,r\},\qquad 1 + \Big\lfloor \frac{(i - 1)s}{r} \Big\rfloor = i \, ,
$$
and it is clear that it arises with $\lambda = (1,\ldots,1)$. In the case $s = 1$, we have
$$
\forall i \in \{1,\ldots,r\},\qquad 1 + \Big\lfloor \frac{(i - 1)s}{r} \Big\rfloor = 1\, ,
$$
and it is clear that it arises with $\lambda = (r)$. In the remaining of the proof we treat the cases $s \in \{2,\ldots,r - 1\}$. Let us decompose
$$
r = r's + r'',\qquad r'' \in \{1,\ldots,s - 1\},\qquad r' > 0\, .
$$
We can assume that $r'' \neq 0$ since ${\rm gcd}(r,s) = 1$ and $r' > 0$ since $s \in \{2,\ldots,r - 1\}$.

Assume that we are given a weakly increasing function $\mu\,:\,\{1,\ldots,r\} \rightarrow \mathbb{N}$ such that $\mu(1) = 1$ and $\mu(i + 1) - \mu(i) \in \{0,1\}$. Let us write down the complete list of integers for which $\mu$ jumps, namely
\begin{equation}
\label{Jlist} 1 \leq \kappa_1 < \ldots < \kappa_{p - 1} < r,\qquad \mu(\kappa_{j} + 1) = \mu(\kappa_{j}) + 1\, ,
\end{equation}
and adopt the convention $\kappa_0 = 0$ and $\kappa_{p} = r$. If we set
$$
\lambda_{j} = \kappa_{j} - \kappa_{j - 1},\qquad j \in \{1,\ldots,p\}\, ,
$$
we get a $p$-tuple of positive integers such that $\sum_{j = 1}^p \lambda_j = r$ and by construction
\begin{equation}
\label{lambddef} \mu(i) = \min\Big\{m\,\,\,\Big|\,\,\,\sum_{j = 1}^{m} \lambda_{j} \geq i\Big\}\, .
\end{equation}
We however stress that $(\lambda_j)_{j = 1}^p$ may not be weakly decreasing.
 
We apply this construction to
\begin{equation}
\label{ADGIUN}\mu(i) = 1 + \Big\lfloor \frac{(i - 1)s}{r} \Big\rfloor\, ,
\end{equation}
which clearly satisfies $\mu(1) = 1$ and $\mu(i + 1) - \mu(i) \in \{0,1\}$. We compute from the definition $\mu(r) = s$ and comparing to \eqref{lambddef} we conclude that $p = s$. To make the proof more transparent, we keep the letter $p$ to indicate the length of the sequence $(\lambda_j)_{j}$. We are going to compute the sequence $\kappa$. Since $\mu(r' + 1) = 1$ and $\mu(r' + 2) = 2$ we deduce that $\kappa_1 = \lambda_{1} = r' + 1$. For any $j \in \{1,\ldots,p - 1\}$ we can decompose
$$
\kappa_js = \beta_j r + \gamma_j,\qquad \gamma_j \in \{0,\ldots,s - 1\},\qquad \beta_j \in \mathbb{N}\, .
$$
For instance we have $\gamma_{1} = s - r''$. Notice that $(\kappa_j + r')s = (\beta_j + 1)r  + \gamma_j - r''$. If $\gamma_j < r''$ we deduce
$$
(\kappa_j + r')s < (\beta_j + 1)r \leq (\kappa_j + r' + 1)s\, ,
$$
and thus $\kappa_{j + 1} = \kappa_{j} + r' + 1$ which implies $\lambda_{j + 1} = r' + 1$ and $\gamma_{j + 1} = \gamma_j + s - r''$.  If $\gamma_j \geq r''$ we rather have
$$
(\kappa_j + r' - 1)s < (\beta_j + 1)r \leq (\kappa_j + r')s\, ,
$$
and thus $\kappa_{j + 1} = \kappa_{j} + r'$ which implies $\lambda_{j + 1} = r'$ and $\gamma_{j + 1} = \gamma_j - r''$. To summarize, we always have $\lambda_{j} \in \{r',r' + 1\}$. We start with $\lambda_{1} = r' + 1$ and $\gamma_1 = s - r''$. Let $\ell > 0$ be the minimum integer such that $\lambda_{\ell + 1} = r'$. It means that $\gamma_{\ell} \geq r''$. According to the previous rules, we have
$$
\gamma_{j} = j(s - r''),\qquad j \in \{1,\ldots,\ell\}\, ,
$$
and
\begin{equation}
\label{llfl} \ell = \Big\lceil \frac{r''}{s - r''} \Big\rceil \, .
\eeq

Assume that $(\lambda_j)_{j}$ is weakly decreasing. It is equivalent to the existence of $\ell \in \{1,\ldots,p\}$ such that
\begin{equation} 
\label{LAMBDA} \lambda_j = \left\{\begin{array}{lll} r' + 1 & & {\rm if}\,\,j \leq \ell\, , \\ r' & & {\rm if}\,\,j > \ell\, .\end{array}\right. 
\end{equation}
We can compute
$$
r = \sum_{j = 1}^p \lambda_j = (r' + 1)\ell + (p - \ell)r' = pr' + \ell\, .
$$
Remembering that $p = s$ it shows that $\ell = r''$. So \eqref{llfl} yields
\begin{equation}
\label{rprpr}r'' = \Big\lceil \frac{r''}{s - r''} \Big\rceil\, .
\end{equation}
The latter is equivalent to
$$
1 + (r'' - 1)(s - r'') \leq r'' \leq r''(s - r'')\, .
$$
The upper bound always holds, while the lower bound can be rewritten as
$$
(r'' - 1)(s - r'' - 1) \leq 0\, .
$$
So \eqref{rprpr} is equivalent to $r'' = 1$ or $r'' = s - 1$. This shows that $r'' \in \{1,s - 1\}$ is a necessary condition for $\lambda$ to be of the form \eqref{LAMBDA}.

Conversely, if we assume that $r'' \in \{1,s - 1\}$, the equivalence we just stressed shows that \eqref{rprpr} holds, so that $\ell$ defined in \eqref{llfl} is equal to $r''$. Then, for any $j \in \{\ell,\ldots,s - 1\}$ we have
$$
\gamma_{\ell} - (j - \ell)r'' = r''(s - j) \geq r''\, ,
$$
hence $\gamma_{j} = \gamma_{\ell} - jr'' \geq r''$ and we must have $\lambda_{j + 1} = r'$. This shows that $(\lambda_j)_j$ is of the form \eqref{LAMBDA}, in particular it is weakly decreasing.
\end{proof}

\subsection{Computation of \texorpdfstring{$F_{0,3}$}{F03} and characterization of symmetry}
 
We consider the modes $W_{k}^i$ of the $\mathcal{W}(\mathfrak{gl}_{r})$ algebra using the twist by the Coxeter element whose expression is given in \eqref{eq:Wmodes}. For $s \in \{1,\ldots,r + 1\}$ coprime with $r$, we are going to evaluate
$$
H_{k_1}^{i_1} = - \frac{1}{2} \sum_{q_2,q_3} C^{(0)}[q_1|-q_2,-q_3]\,J_{-q_2}J_{-q_3} - \hbar\,C^{(1)}[q_1|\emptyset] + \cdots
$$
where $\cdots$ includes monomials which are different from the ones emphasized here, and we remind that the correspondence between positive integers $q$ and indices $(i,k) \in S_{s}$ for the mode $W_{k}^i$ equal to $J_{q} + O(2)$ is
\beq
\label{Pisi} q = \Pi_s(i,k) =  r(k - i + 1) + s(i_j - 1)\, ,
\eeq
and the set $\mathcal{S}_{s}$ consists precisely of those $(i,k)$ that yields a positive $q = \Pi_s(i,k)$. The notation we use is $q_j = \Pi_s(i_j,k_j)$.

If $(H_{k_1}^{i_1})_{(i_1,k_1) \in  S_{s}}$ forms a higher quantum Airy structure, then
\begin{equation}
\label{poun} F_{0,3}[q_1,q_2,q_3] = q_2q_3\,C^{(0)}[q_1|-q_2,-q_3]
\end{equation}
must be invariant under permutation of $(q_1,q_2,q_3)$. On the one hand, the representation theoretic arguments in Section~\ref{Sinterm} allowed us in Theorem~\ref{WHAS} the conclusion that if $r = \pm 1\,\,{\rm mod}\,\,s$, $(H_k^i)_{(i,k) \in S_{s}}$ is indeed a higher quantum Airy structure, so  \eqref{poun} is \emph{a priori} fully symmetric. Here we compute explicitly $F_{0,3}$ and indeed check that it is fully symmetric. On the other hand, when $r \neq \pm 1\,\,{\rm mod}\,\,s$, our explicit computation shows that the right-hand side of \eqref{poun} is not fully symmetric. Therefore, the left ideal generated by the $(H_{k}^i)_{(i,k) \in S_{s}}$ is not a graded Lie subalgebra and the results of Section~\ref{s:subalgebra} are in this sense optimal.

\begin{prop}
\label{p:symmetry}
Let $s \in \{1,\ldots,r + 1\}$ be coprime with $r$.
\begin{itemize}
\item[$\bullet$] If $r = r's + s - 1$ for $r' \geq 0$, we have $F_{0,3}[q_1,q_2,q_3] = (r' + 1)q_1q_2q_3\,\delta_{q_1 + q_2 + q_3,s}$.
\item[$\bullet$] If $r = r's + 1$ for $r' \geq 0$, we have $F_{0,3}[q_1,q_2,q_3] = -r'q_1q_2q_3\,\delta_{q_1 + q_2 + q_3,s}$.
\item[$\bullet$] In all other cases, there exists $q_1,q_2,q_3 > 0$ such that
$$
q_2q_3\,C^{(0)}[q_1|-q_2,-q_3] \neq q_1q_3\,C^{(0)}[q_2|-q_1,-q_3]\,.
$$
\end{itemize}
\end{prop}
\begin{proof}
Starting from the expression \eqref{eq:Wmodes} for the differential operators $W_k^i$, due to the dilaton shift we must get $(i - 2)$ variables $p_j$s equal to $-s$ and the two last $p$s must be equal to $q_1$ and $q_2$. So
$$
H_{k_1}^{i_1} = \sum_{q_2,q_3 > 0} \frac{(-1)^{i_1 - 2}i_1(i_1 - 1)}{2r}\,\delta_{q_2 + q_3 + (i_1 - 2)s + r(k_1 - i_1 + 1),0}\,\Psi(q_2,q_3,\underbrace{s,\ldots,s}_{i_1 - 2\,\,{\rm times}})\,J_{-q_2}J_{-q_3} + \cdots 
$$
Remember that $H_{k_1}^{i_1} = J_{q_1} + O(2)$. So the Kronecker delta imposes
\begin{equation}
\label{condqqq} q_1 + q_2 + q_3 = s\,.
\end{equation}
Since $q_j > 0$ for $j \in \{1,2,3\}$ and $1 \leq s \leq r + 1$ this imposes $q_j < r$. Let us evaluate $\Psi$ under this condition using Lemma~\ref{Psiprinc}. Since $s$ and $r$ are coprime, $r$ cannot divide $sm$ for $m \in \{1,\ldots,i_1 - 2\}$ therefore the only partitions $\mathbf{L}$ that contribute are those for which each $L \in \mathbf{L}$ contains $q_1$ or $q_2$
\begin{equation}
\label{psi222}
\begin{split}
\Psi(q_2,q_3,\underbrace{s,\ldots,s}_{i_1 - 2\,\,{\rm times}}) & =  \frac{(-1)^{i_1 - 1}\,r}{i_1}\,\delta_{r|q_2 + q_3 + (i_1 - 2)s}  \\
& \quad + \sum_{\substack{m_2,m_3 \geq 0 \\ m_2 + m_3 = i_1 - 2}} \frac{(i_1 - 2)!}{m_2!m_3!}\,\frac{(-1)^{i_1 - 2}\,r^2\,m_2!m_3! }{i_1!}\,\delta_{r|q_2 + m_2s} \delta_{r|q_3 +m_3s}\,.
\end{split}
\end{equation}
The extra combinatorial factor $\frac{(i_1 - 2)!}{m_1!m_2!}$ is the number of ways of splitting the sequence $(s,\ldots,s)$ of length $i_1 - 2$ into two subsequences of length $m_1$ and $m_2$. If there exists two $m_2$ and $m'_2$ in $\{0,\ldots,i_1 - 2\}$ such that $r|q_2 + m_2s$ and $r|q_2 + m'_2s$, then $r|(m_2 - m'_2)$.  Since $i_1 - 2 < r$ we must have $m_2 = m'_2$. Therefore, the sum over $m_2$ contains at most one term, and if it contains one term it is equal to $1$. Under the condition \eqref{condqqq}, we have for any $m \in \mathbb{Z}$
$$
-(k_1 - i_1 + 1)r = (q_2 + ms) + (q_3 + (i_1 - 2 - m)s)\, .
$$
Therefore, we can omit the factor $\delta_{r|q_3 + m_3s}$ in \eqref{psi222}. Finally, notice that we can write $q_2 + s(r - i_2 + 1) = r(k_2 - i_2 + 1 + s)$ with $r - i_2 + 1 \in \{1,\ldots,r\}$. So the existence of $b_2 > 0$ and $m_2 \geq 0$ such that $b_2r = q_2 + sm_2$ with $m_2 \leq i_1 - 2$ is equivalent to $i_2 - 1 = r - m_2 \geq r - (i_1 - 2)$ that is $i_1 + i_2 \geq r + 3$. As a consequence
\begin{equation} 
\label{C0000} C^{(0)}[q_1|-q_2,-q_3] = \big((i_1- 1) - r\,\epsilon(q_1,q_2,q_3)\big)\,\delta_{q_1 + q_2 + q_3,s}\,,
\end{equation} 
where $\epsilon(q_1,q_2,q_3) = 1$ if $q_1 + q_2 + q_3 = s$ and $i_1 + i_2 - 3 \geq r$, and $\epsilon(q_1,q_2,q_3) = 0$ otherwise.

\noindent $\bullet$ If $s = 1$ we always have $F_{0,3} = 0$.

\noindent $\bullet$ Assume $s = r + 1$. It is not possible for $q_2 \leq s - 2 < r$ to be divisible by $r$ so $\epsilon(q_1,q_2,q_3) = 0$. Likewise $q_1 < r$ so we must have $k_1 = 0$ and $i_1 - 1 = q_1$. Therefore
$$
C^{(0)}[q_1|-q_2,-q_3] = q_1\,\delta_{q_1 + q_2 + q_3,s}\,,
$$
and $F_{0,3}[q_1,q_2,q_3] = q_1q_2q_3\,\delta_{q_1 + q_2 + q_3,s}$, which is manifestly symmetric.

We now turn to the values $s \in \{2,\ldots,r - 1\}$.

\noindent $\bullet$ Assume $r = r's + s - 1$ with $r' > 0$. We multiply by $q_1$ and get
$$
q_1 = q_1((r' + 1)s - r) = -rq_1 + (r' + 1)q_1s\,.
$$
Thus $q_1$ corresponds to $i_1 = (r' + 1)q_1 + 1$ and $k_1 =r'q_1$. In particular $i_1 - 1 = (r' + 1)q_1$. Assume there exists $b_2 > 0$  and $m_2 \in \{0,\ldots,i_1 - 2\}$ such that $rb_2 = q_2 + m_2s$. Since $q_2 = -rq_2 + (r' + 1)q_2s$, there must exist $l_2 \in \mathbb{Z}$ such that
$$
b_2 = -q_2 + l_2s,\qquad m_2 = -(r' + 1)q_2 + l_2r\,.
$$ 
Since $q_1 + q_2 + q_3 = s$ with $q_3 > 0$, we have $q_1 + q_2 \leq s - 1$, hence $i_1 - 2 \leq (r' + 1)(s - 1 - q_2)$. Together with $0 \leq m_2 \leq i_1 - 2$ it leads to the inequality
$$
0 < (r' + 1)q_1 \leq l_2r \leq (r' + 1)(s - 1 - q_2) + (r' + 1)q_2 = (r' + 1)(s - 1) = r - r' < r\,,
$$
which contradicts the existence of $l_2$. Hence there are no such $m_2$, meaning that $\epsilon(q_1,q_2,q_3) = 0$. Consequently
$$
C^{(0)}[q_1|-q_2,-q_3] = (i_1 - 1)\,\delta_{q_1 + q_2 + q_3,s} = (r' + 1)q_1\,\delta_{q_1 + q_2 + q_3,s}\,,
$$
and $F_{0,3}[q_1,q_2,q_3] = (r' + 1)q_1q_2q_3\,\delta_{q_1 + q_2 + q_3,s}$, which is manifestly symmetric.

\noindent $\bullet$ Assume $r = r's + 1$ with $r' \geq 0$. We see that
$$
s - q_1 = s + q_1(r's - r) = -rq_1 + (r'q_1+ 1)s\,,
$$
so $q_1$ corresponds to $i_1- 1 = r'(s - q_1) + 1 = r - r'q_1$ and $k_1 = (s-q_1)(r' - 1) + 1$. We deduce that $i_1 - 2 = r'(q_2 + q_3)$. Since $q_2 + r'q_2s = rq_2$ is divisible by $r$, the choice $m = r'q_2$ satisfies $0 \leq m \leq i_1 - 2$ and therefore $\epsilon(q_1,q_2,q_3) = 1$. We deduce that
$$
C^{(0)}[q_1|-q_2,-q_3] = (i_1 - 1 - r)\delta_{q_1 + q_2 + q_3,s} = -r'q_1\,\delta_{q_1 + q_2 + q_3,s}\,.
$$
This implies that $F_{0,3}[q_1,q_2,q_3] = -r'q_1q_2q_3\,\delta_{q_1 + q_2 + q_3,s}$, which is manifestly symmetric.

\noindent $\bullet$ Now assume that $r'' \in \{2,\ldots,s - 2\}$. We are going to show that assuming the symmetry
\beq
\label{relsss}q_2 C^{(0)}[q_1|-q_2,-q_3] = q_1 C^{(0)}[q_2|-q_1,-q_3]
\eeq
for any $q_1,q_2,q_3 > 0$ such that $q_1 + q_2 + q_3 = s$ leads to a contradiction. We first remark from the definition that $\epsilon(q_1,q_2,q_3) = \epsilon(q_2,q_1,q_3)$ in \eqref{C0000}. We set $q_1 = 1$, which we decompose as usual $q_1 = (k_1 - i_1 + 1)r + (i_1 - 1)s$ for some $i_1 \in \{1,\ldots,r\}$. Choosing $q_2 = s - r''$ corresponds to $i_2 = r' + 2$ and we can take $q_3 = s - (q_1 + q_2) = r'' - 1 > 0$. Denoting $\varepsilon = \epsilon(1,s - r'',r'' - 1)$, the condition \eqref{relsss} implies
$$
(s - r'')(1 - i_1 + \varepsilon r) = - 1 - r' + \varepsilon r\, .
$$
Choosing $q_2 = r''$ corresponds to $i_2 = r - r' + 1$ and we can take $q_3 = s - (q_1 + q_2) = s - r'' - 1 > 0$. Denoting $\varepsilon' = \epsilon(1,r'',s - r'' - 1)$ the condition \eqref{relsss} implies
$$
r''(1 - i_1 + \varepsilon'r) = -r + r' + \varepsilon' r\, .
$$
Summing the two equations gives
\beq
\label{plplp}s(1 - i_1 + \varepsilon r) + 1 = r\big((\varepsilon + \varepsilon' - 1) + r''(\varepsilon - \varepsilon')\big)\, .
\eeq
From the definition we have that
\beq
\varepsilon = \begin{cases}1 & \text{if $i_1 \geq r-r'+1$,} \\ 0 & \text{otherwise} \end{cases}, \qquad \varepsilon' = \begin{cases} 1 & \text{if $i_1 \geq r'+2$,} \\ 0 & \text{otherwise} \end{cases}\,.
\eeq
Since $r - r' + 1 -(r'+2) = r'(s - 2) + r'' - 1 > 0$, we see that the only possible values of $(\varepsilon,\varepsilon')$ are $(1,1)$, $(0,0)$ and $(0,1)$. In the two first cases, reducing \eqref{plplp} modulo $s$ gives
$$
r'' = \pm 1\,\,{\rm mod}\,\,s\, ,
$$
which is impossible since we assumed $2 \leq r'' \leq s - 2$ from the beginning. So we must be in the case $(\varepsilon,\varepsilon') = (0,1)$, which means that
\beq
\label{ineqdf}r' + 1 \leq i_1 - 1 \leq r - r' - 1\, .
\eeq
The equation \eqref{plplp} then implies
\beq
\label{1eq} 1 = (i_1 - 1)s - rr''\, .
\eeq
Now we express the symmetry
\beq
\label{oethe} q_3 C^{(0)}[q_1|-q_3,-q_2] = q_1 C^{(0)}[q_3|-q_1,-q_2]\, ,
\eeq
with the choice $(q_1,q_2,q_3) = (1,s - r'',r'' - 1)$. We can write
$$ 
r'' - 1 = r - r's - \big((i_1 - 1)s - rr''\big) = \big(r - r' - (i_1 - 1)\big)s + r(r'' + 1 - s)\, .
$$ 
Due to the upper bound in \eqref{ineqdf} we have $r - r' - (i_1 - 1) > 0$ therefore $i_3 - 1= r - r'  - (i_1 - 1)$. Notice that $i_1 + i_3 =  2 + r - r' < r + 3$ thus $\epsilon(q_1,q_3,q_2) = \epsilon(q_3,q_1,q_2) = 0$. Using \eqref{C0000}, the equality \eqref{oethe} becomes
$$
r' + (i_1 - 1) - r = (r'' - 1)(1 - i_1)\, ,
$$
that is $r''b = r - r'$. If we use this result when multiplying \eqref{1eq} by $r''$, we find that $r'' = rs - (r - r'') - r(r'')^2$. This implies
\begin{equation}
\label{ressqur} (r'')^2 = s - 1,
\end{equation}
which combined with \eqref{1eq} forces
\begin{equation}
\label{i1nodun}i_1 - 1 = 1 + r'r''.
\end{equation}
We now would like to specialize $(q_1,q_2,q_3) = (1,s - r'' - 1,r'')$ in the symmetry relation \eqref{relsss}. Recall that there exists $p \in \mathbb{Z}$ such that $s - r'' = (r' + 1)s + pr$. Using again \eqref{1eq} shows that $s - r'' - 1 = (r + r' + 2 - i_1)s + p'r$ for some $p' \in \mathbb{Z}$. Besides
\[
r + r' + 2 - i_1 = r - r'(r'' - 1)
\]
If we had $r' = 0$, we would have $r'' = r \leq s - 2$ which is ruled out by our assumption. So $r' > 0$ and we then know that $r + r' + 2 - i_1 \in \{0,\ldots,r - 1\}$ is equal to $(i_2 - 1)$. Besides, since have $r'' = (r - r')s + r(1 - s)$, we know that $r - r' \in \{0,\ldots,r - 1\}$ is equal to $(i_3 - 1)$. As $i_1 + i_2 - 3 = r + r' > r$, we have $\epsilon(q_1,q_2,q_3) = 1$. Hence, the symmetry relation \eqref{relsss} implies
\[
r + r' + 2 - i_1 - r = (s - r'' - 1)(i_1 - 1 - r)
\]
Inserting \eqref{ressqur} and \eqref{i1nodun}, we get $r'(1 - r'') = r''(r'' - 1)(1 + r'r'' - r)$, hence
\[
r' = r''(r - 1 - r'r'') = r''(r - 1) - r'(s - 1)
\]
Computing modulo $s$ and remembering $(r'')^2 = s - 1$ this gives $r'' = -1\,\,{\rm mod}\,\,s$ which contradicts our assumption $r'' \in \{2,\ldots,s - 2\}$. In other words, we have proved that if $r'' \in \{2,\ldots,s - 2\}$, we cannot have the symmetry \eqref{C0000} for all $q_1,q_2,q_3 > 0$.
\end{proof}

\subsection{Computation of \texorpdfstring{$F_{1,1}$}{F11}}
\label{F11comput}
Let us compute $F_{1,1}$ for the quantum Airy structure of Theorem~\ref{WHAS} --- that is, assuming $r = \pm 1\,\,{\rm mod}\,\,s$.

\begin{lem}
\label{p11:symmetry} We have $F_{1,1}[q] = \frac{r^2 - 1}{24}\,\delta_{q,s}$.
\end{lem}
\begin{proof}
We need to isolate the term
$$
H_{k}^{i} = -\hbar\,F_{1,1}[\Pi_{s}(i,k)] + \cdots
$$
where $F_{1,1}[q]$ is a scalar, $\Pi_{s}(i,k)$ is given in \eqref{Pisi} and $\cdots$ represent the other monomials which we are not interested in. We recall that $H_k^i$ is obtained from the generators $W_k^i$ given in \eqref{eq:Wmodes} after the shift $J_{-s} \rightarrow J_{-s} - 1$. The only contribution to this term comes from $j = 1$ and $p_{\ell} = -s$ for all $l \in \{3,\ldots,i\}$ such that $-s(i - 2) = r(k - i + 1)$. Since $r$ and $s$ are coprime, such a contribution can only appear for $i = 2$ and $k = 1$, and in that case $\Pi_{s}(2,1) = s$. So we have
$$
F_{1,1}[q] = -\frac{1}{r}\,\Psi^{(1)}(\emptyset)\,\delta_{q,s}\,,
$$
which is equal to $\frac{r^2 - 1}{24}\,\delta_{q,s}$ using Lemma~\ref{LEmg} to evaluate $\Psi^{(1)}$.
\end{proof}

\section{Proof of uniqueness of the solution to the higher abstract loop equations}

\label{a:proof}

In this appendix we show that if a solution to the higher abstract loop equations that respects the polarization exists, then it is uniquely constructed by the Bouchard--Eynard topological recursion of \cite{BE2,BHLMR}. The argument follows along similar lines to what was presented in \cite{BSblob,BE}. For clarity we will only present the proof here for local spectral curves  with one component, but it is straightforward to generalize it to local spectral curves with $\ell$ components.

Consider a spectral curve with one component, as defined in Definition \ref{d:lsc1}. Let
$$
\omega_{0,1}(z) = \sum_{l > 0} \tau_l\,\dd\xi_l(z) = y(z) \dd\xi_r(z)\,,
$$
where  $y(z) = \sum_{l > 0}^\infty \tau_l z^{l-r}$. Let us assume that the spectral curve is admissible (Definition \ref{d:admissible1}), that is, $1 \leq s \leq r + 1$ and $r = \pm 1\,\,{\rm mod}\,\,s$ with
$$
s := \min\big\{l > 0\quad |\quad \tau_{l} \neq 0\,\,\,{\rm and}\,\,\,r \nmid l\big\}\, .
$$
Let $\omega_{0,2}(z_1,z_2)$ be the formal bidifferential that encodes the choice of polarization
$$
\omega_{0,2}(z_1,z_2) = \frac{\dd z_1 \otimes \dd z_2 }{(z_1-z_2)^2} + \sum_{l,m > 0} \phi_{l,m}\,\dd\xi_l(z_1) \otimes \dd\xi_m(z_2)\,.
$$
Then the following result holds. 
\begin{prop}\label{theeeq}
Fix an admissible\footnote{To be precise, in the proof here we only need the requirement that $s$ is coprime with $r$, we do not need the stronger admissibility requirement that $r = \pm 1 \mod s$.} local spectral curve with one component. Assume there exists a sequence of formal symmetric differentials $\omega_{g,n} \in \bigotimes_{j=1}^n V^-_{z_j}$ satisfying the higher abstract loop equations
$$
P_{\mathbf{w} | t} \left( \mathcal{E}^{(i)}_{g,n}(\mathbf{w} | \mathbf{z}) \right)  \in t^{-r \mathfrak{d}^i }\, \mathbb{C} [\![ t^{r} ]\!] \otimes V^-_{z_2} \otimes \ldots \otimes V^-_{z_n}\,,
$$
for all $g \geq 0$, $n \geq 1$, $2g-2+n > 0$, and $i \in \{1,\ldots, r\}$, where $\mathbf{z} = (z_2, \ldots, z_n)$ and $\mathbf{w} = (w_1, \ldots, w_i)$. Here, $\mathfrak{d}^i = i-1-  \big\lfloor \frac{s(i-1)}{r} \big\rfloor$ and  $\mathcal{E}^{(i)}_{g,n}(A |B)$ was introduced in Definition \ref{d:combin}, and the projection map $P_{\mathbf{w} | t}$ in Definition \ref{d:projection1}.

Then, this sequence of formal symmetric differentials is constructed by the Bouchard--Eynard topological recursion formula
\begin{multline}
\omega_{g,n}(z_1, \mathbf{z}) = \Res_{t = 0}\left(  \int^t_0 \omega_{0,2}(\cdot, z_1) \right)\sum_{i=1}^{r-1} (-1)^{i+1}  \frac{1}{i!}  \\
\times \sum_{\substack{a_1, \ldots, a_{i}=1 \\ a_{m} \neq a_l}}^{r-1}  \left( \prod_{m=1}^{i} \frac{1}{(y(t) - y(\theta^{a_{m}} t))} \right) \frac{\mathcal{R}^{(i+1)}_{g,n}(t, \theta^{a_{1}} t, \ldots, \theta^{a_{i}} t | \mathbf{z}) }{(\dd \xi_r(t))^{k}}\,. 
\label{eq:toprec}
\end{multline}
\end{prop}

\begin{proof}
Suppose that there exists a sequence of $\omega_{g,n} \in \bigotimes_{j=1}^n V^-_{z_j}$ such that 
$$
P_{\mathbf{w} | t} \left( \mathcal{E}^{(i)}_{g,n}(\mathbf{w} | \mathbf{z}) \right)  \in \frac{1}{t^{r \mathfrak{d}^i}} \mathbb{C} [\![ t^{r} ]\!] \otimes V^-_{z_2} \otimes \ldots \otimes V^-_{z_n}\,.
$$
Let us now argue that the expression
\begin{equation}\label{eq:stop}
\frac{\dd\xi_r(t)}{\prod_{m=1}^{r-1} ( y(t) - y(\theta^m t) )} \sum_{i=1}^r (-1)^i (y(t))^{r-i} P_{\mathbf{w} | t} \left( \mathcal{E}^{(i)}_{g,n}(\mathbf{w} | \mathbf{z}) \right)
\end{equation}
lives in $ \mathbb{C} [\![ t ]\!] \dd t \otimes V^-_{z_2} \otimes \ldots \otimes V^-_{z_n}$. On the one hand, if $y(t) \in \mathbb{C}[\![ t]\!]$, then this is clear as long as $y'(0) \neq 0$, that is $\tau_{r+1} \neq 0$. This is the regular case. In the irregular case, let us suppose that $\tau_s \neq 0$ with $s \in \{1,\ldots,r-1 \}$. Then $y(t) \sim \tau_{s}\,t^{s - r}$ when $t \rightarrow 0$. Assume that $s$ is coprime with $r$.  We rewrite $\mathfrak{d}^i = i-1-  \lfloor \frac{s(i-1)}{r} \rfloor = i - \lceil \frac{1+s(i-1)}{r} \rceil$. We have when $t \rightarrow 0$
\begin{equation*}
\begin{split}
\frac{\dd \xi_r(t)}{\prod_{m=1}^{r-1} ( y(t) - y(\theta^m t) )} & \sim  \tau_s^{1 - r}\,t^{(r-1)(r-s+1)} \dd t\,,  \\
 (y(t))^{r-i} P_{\mathbf{w} | t} \left( \mathcal{E}^{(i)}_{g,n}(\mathbf{w} | \mathbf{z}) \right)  & \sim \frac{\tau_s^{1 - r}\,t^{(r-i)(s-r)}}{t^{r \big( i - \lceil\frac{1+s(i-1)}{r} \rceil \big)}} = \frac{\tau_s^{1 - r}}{t^{\delta}}\,, 
\end{split}
\end{equation*}
with
$$
\delta = r i  - r \, \Big\lceil \frac{1+s(i-1)}{r} \Big\rceil  - (r-i)(s-r) \leq (r-1)(r-s+1)\,,
$$
and the result follows.

We can rewrite this statement as a residue condition
$$
\Res_{t = 0}\left(  \int^t_0 \omega_{0,2}(\cdot,z_1) \right) \frac{\dd\xi_r(t)}{\prod_{m=1}^{r-1} ( y(t) - y(\theta^m t) )} \left(\sum_{i=1}^r (-1)^i\,y(t)^{r-i}\,P_{\mathbf{w} | t} \Big( \mathcal{E}^{(i)}_{g,n}(\mathbf{w} | \mathbf{z}) \Big)\right)  = 0\,.
$$
Now let us manipulate this expression a little bit, by writing down the projection map explicitly. By definition,
$$
\sum_{i=1}^r (-1)^i (y(t))^{r-i} P_{\mathbf{w} | t} \left( \mathcal{E}^{(i)}_{g,n}(\mathbf{w} | \mathbf{z}) \right) = \sum_{i=1}^r \frac{(-1)^i}{i!}\,y(t)^{r - i}  \sum_{\substack{a_1, \ldots, a_i=0 \\ a_{m} \neq a_l}}^{r-1} \frac{\mathcal{E}^{(i)}_{g,n}(\theta^{a_1} t, \ldots, \theta^{a_i} t | \mathbf{z})}{(\dd\xi_r(t))^i}\,.
$$
We can extract the $\omega_{0,1}$ contributions out of the $ \mathcal{E}^{(i)}_{g,n}(\mathbf{w} | \mathbf{z})$ using Lemma \ref{l:er}. We get
\begin{equation*}
\begin{split}
\sum_{i=1}^r& (-1)^i (y(t))^{r-i} P_{\mathbf{w} | t} \left( \mathcal{E}^{(i)}_{g,n}(\mathbf{w} | \mathbf{z}) \right)  \\
=& \sum_{i=1}^r (-1)^i \sum_{\substack{a_1, \ldots, a_i=0 \\ a_m \neq a_l}}^{r-1} \sum_{j=1}^{i}\frac{1}{j! (i-j)! } y(t)^{r-i} \left( \prod_{l=1}^{i-j} y(\theta^{a_{j+l}} t) \right)  \frac{\mathcal{R}^{(j)}_{g,n}(\theta^{a_{1}} t, \ldots, \theta^{a_{j}} t | \mathbf{z}) }{(\dd \xi_r(t))^{j}}  \\
=&\sum_{\substack{a_1, \ldots, a_r=0 \\ a_m \neq a_l}}^{r-1} \sum_{i=1}^r (-1)^{i} \sum_{j=1}^i    \frac{1}{j! (i-j)! (r-i)!}y(t)^{r-i} \left ( \prod_{l=1}^{i-j} y(\theta^{a_{j+l}} t)  \right) \frac{\mathcal{R}^{(j)}_{g,n}(\theta^{a_{1}} t, \ldots, \theta^{a_{j}} t | \mathbf{z}) }{(\dd \xi_r(t))^{j}}  \\
=&\sum_{\substack{a_1, \ldots, a_r=0 \\ a_m \neq a_l}}^{r-1}  \sum_{\ell=1}^r  \sum_{m=0}^{r-\ell}  \frac{(-1)^{\ell+m}}{\ell! (r-\ell-m)! m!} y(t)^{r-\ell-m} \left( \prod_{l=1}^{m} y(\theta^{a_{\ell+l}} t)  \right) \frac{\mathcal{R}^{(\ell)}_{g,n}(\theta^{a_{1}} t, \ldots, \theta^{a_{\ell}} t | \mathbf{z}) }{(\dd \xi_r(t))^{\ell}}  \\
=&\sum_{\substack{a_1, \ldots, a_r=0 \\ a_m \neq a_l}}^{r-1}  \sum_{\ell=1}^r  \frac{(-1)^\ell}{\ell! (r-\ell)!}\left( \prod_{l=1}^{r-\ell} (y(t) - y(\theta^{a_{\ell+l}} t)) \right)  \frac{\mathcal{R}^{(\ell)}_{g,n}(\theta^{a_{1}} t, \ldots, \theta^{a_{\ell}} t | \mathbf{z}) }{(\dd \xi_r(t))^{\ell}}  \\
=& \sum_{\substack{a_1, \ldots, a_{r-1}=1 \\ a_l \neq a_m}}^{r-1}  \sum_{\ell=1}^r \frac{(-1)^{\ell} }{(\ell - 1)!(r-\ell)!}\left( \prod_{l=1}^{r-\ell} (y(t) - y(\theta^{a_{\ell-1+l}} t)) \right)  \frac{\mathcal{R}^{(\ell)}_{g,n}(t, \theta^{a_{1}} t, \ldots, \theta^{a_{\ell-1}} t | \mathbf{z}) }{(\dd \xi_r(t))^{\ell}}\,. 
\end{split}
\end{equation*}
Then 
\begin{equation*}
\begin{split}
&\frac{\dd \xi_r(t)}{\prod_{m=1}^{r-1} ( y(t) - y(\theta^m t) )} \sum_{i=1}^r (-1)^i (y(t))^{r-i} P_{\mathbf{w} | t} \left( \mathcal{E}^{(i)}_{g,n}(\mathbf{w} | \mathbf{z}) \right)  \\
&=    \sum_{i=1}^r   \frac{(-1)^i}{(i-1)!} \sum_{\substack{a_1, \ldots, a_{i-1}=1 \\ a_{l} \neq a_{m}}}^{r-1}\left( \prod_{l=1}^{i-1} \frac{1}{(y(t) - y(\theta^{a_{l}} t))} \right)  \frac{\mathcal{R}^{(i)}_{g,n}(t, \theta^{a_{1}} t, \ldots, \theta^{a_{i-1}} t | \mathbf{z}) }{(\dd \xi_r(t))^{i-1}}\,.
\end{split}
\end{equation*}
Thus the higher abstract loop equations imply that
\begin{equation*}
\begin{split}
\Res_{t = 0}\left(  \int^t_0 \omega_{0,2}(\cdot, z_1) \right)\sum_{i=1}^r  \frac{(-1)^i }{(i-1)!}  \sum_{\substack{a_1, \ldots, a_{i-1}=1 \\ a_l \neq a_m}}^{r-1}  \left( \prod_{l=1}^{i-1} \frac{1}{(y(t) - y(\theta^{a_{l}} t))} \right) \\
\times  \frac{\mathcal{R}^{(i)}_{g,n}(t, \theta^{a_{1}} t, \ldots, \theta^{a_{i-1}} t | \mathbf{z}) }{(\dd \xi_r(t))^{i-1}} =0\,.
\end{split}
\end{equation*}
Now we can take out the term with $i=1$, which is equal to
$$
- \Res_{t = 0}\left(  \int^t_0 \omega_{0,2}(\cdot, z_1) \right) \omega_{g,n}(t, \mathbf{z})\,.
$$
Assuming that $\omega_{g,n} \in  \otimes_{l=1}^n V^-_{z_l}$, the residue simply replaces the terms with $\xi_{-l}(z)$ with the same terms but with $\xi_{-l}(z_1)$. Thus
$$
- \Res_{t = 0}\left(  \int^t_0 \omega_{0,2}(\cdot, z_1) \right) \omega_{g,n}(t, \mathbf{z}) = - \omega_{g,n}(z_1, \mathbf{z})\,,
$$
and we get the topological recursion
\begin{equation*}
\begin{split}
\omega_{g,n}(z_1, \mathbf{z}) = \Res_{t = 0}\left(  \int^t_0 \omega_{0,2}(\cdot, z_1) \right)\sum_{i=1}^{r-1}   \frac{(-1)^{i+1}}{i!} \\
\times \sum_{\substack{a_1, \ldots, a_{i}=1 \\ a_{l} \neq a_{m}}}^{r-1}  \left( \prod_{l=1}^{i} \frac{1}{(y(t) - y(\theta^{a_{l}} t))} \right) \frac{\mathcal{R}^{(i+1)}_{g,n}(t, \theta^{a_{1}} t, \ldots, \theta^{a_{i}} t | \mathbf{z}) }{(\dd \xi_r(t))^{i}}\,,
\end{split}
\end{equation*}
which uniquely determines the $\omega_{g,n}$.
\end{proof}

\begin{rem}
To prove existence of a (polarized) solution to the higher abstract loop equations, one could follow the proof above step-by-step in reverse. However, for this to work one would need to prove initially that the Bouchard--Eynard topological recursion produces symmetric $\omega_{g,n}$. This is known from \cite{BE2} for spectral curves that appear as a limit of a family of curves with simple ramification points, by an indirect argument. It is however not clear to us which spectral curves satisfy this condition. Since in Section \ref{sec:hasfromtr} we identify the higher abstract loop equations with higher quantum Airy structures for admissible spectral curves, it implies via Theorem~\ref{t:KSthm} that the solution to the higher loop equations exists, in the case when $1 \leq s \leq r+1$ and $r = \pm 1 \mod s$. Hence we have a new, more direct (and perhaps more general) proof that the Bouchard--Eynard topological recursion produces symmetric $\omega_{g,n}$ for all admissible spectral curves, independently of the arguments of \cite{BE2}, as stated in Theorem \ref{c:symmetric}.
\end{rem}

\newpage

\end{document}